\documentclass[aps,prl,reprint,aps,twocolumn,superscriptaddress,showpacs,showkeys]{revtex4-1}

\usepackage{amsmath,amsthm,amssymb,graphicx,xcolor,float}

\usepackage{esint}
\usepackage[unicode=true]{hyperref}
\hypersetup{
     colorlinks=true,       		
     linkcolor=blue,          	
     citecolor=red,            
     urlcolor=magenta,           	
 }

\usepackage{mathrsfs,mathtools,mathdots,dsfont}
\usepackage{listings}
\usepackage{subfigure}

\newtheorem*{theorem*}{Theorem}

\newtheorem*{corollary*}{Corollary}

\newtheorem*{lemma*}{Lemma}

\newtheorem*{proposition*}{Example*}

\newtheorem*{conjecture*}{Conjecture}
\theoremstyle{definition}

\newtheorem*{definition*}{Definition}
\theoremstyle{remark}
\newtheorem{remark}{Remark}
\newtheorem*{remark*}{Remark}

\newcommand{\e}{\mathrm{e}}
\renewcommand{\i}{\mathrm{i}}
\renewcommand{\d}{\mathrm{d}}

\newcommand{\ket}[1]{\left|#1\right\rangle}

\begin{document}



\title{Electric-Type Stern-Gerlach Effect}

\author{Jiang-Lin Zhou}
\affiliation{School of Physics, Nankai University, Tianjin 300071, People's Republic of China}

\author{Zou-Chen Fu}
\affiliation{School of Physics, Nankai University, Tianjin 300071, People's Republic of China}

\author{Choo Hiap Oh}
\email{phyohch@nus.edu.sg}
\affiliation{Centre for Quantum Technologies and Department of Physics, National University of Singapore, 117543, Singapore}

\author{Jing-Ling Chen}
\email{chenjl@nankai.edu.cn}
\affiliation{Theoretical Physics Division, Chern Institute of Mathematics, Nankai University, Tianjin 300071, People's Republic of
	 	China}

\date{\today}

\begin{abstract}
The Stern-Gerlach (SG) experiment is a fundamental experiment for revealing the existence of ``spin''. In such an experiment, beams of silver atoms were sent through inhomogeneous magnetic fields to observe their deflection. Thus, the conventional SG experiment can be actually viewed as a magnetic-type spin effect. In this work, we successfully generalize the SG effect from magnetic-type to electric-type by solving Dirac's equation with a potential barrier, thus revealing an extraordinary spin effect. Beams of Dirac's particles can be regarded as matter waves. Based on Dirac's equation, we obtain the explicit forms of the incident, reflected, and transmitted waves. The electric-type SG effect shows that the reflected and transmitted waves can have notable spatial shifts, which depend on the spin direction and the incident angle of incident wave. The electric-type SG effect has potential applications to separating Dirac's particles with different spin directions and to estimating the spin direction of Dirac's particles. Some discussions related to the interaction between spin and electric field are also made.
\end{abstract}

\maketitle

\emph{Introduction.---} The Stern-Gerlach (SG) experiment is undoubtedly one of fundamental experiments in quantum physics \cite{2001GreinerSpin}. This experiment was performed in 1922 at the University of Frankfurt, where beams of silver atoms were sent through inhomogeneous magnetic fields to
observe their deflection. The outcome of SG experiment firmly confirmed Niels Bohr's previous conjecture, namely, these atoms have quantized magnetic
moments that can take two values. Later on, Pauli introduced a ``two-valued'' degree of freedom for electrons, without suggesting a physical interpretation. In 1925, Uhlenbeck and Goudsmit boldly proposed the hypothesis of \emph{electronic spin} \cite{1925Uhlenbeck,1926UHLENBECK}, due to which the famous SG
experiment together with other experiments referring to the quantized magnetic moment can eventually achieved satisfactory interpretations.

In 1928, Dirac developed the theory of relativistic quantum mechanics by formulating a fundamental equation -- Dirac's equation \cite{1928Dirac}. Based on which, the electronic spin is no longer a hypothesis, but a natural corollary of the relativistic quantum theory for electrons. As a cornerstone equation, Dirac's equation can not only provide the correct interpretation for the existing relativistic quantum effects, such as the SG effect (or experiment) mentioned above, but also stimulate some novel predictions, such as predicting the anti-particles.

Dirac's equation for an electron with vector potential $\vec{\mathcal{A}}$ and scalar potential $\Phi$ is given by $H |\Psi \rangle= E |\Psi\rangle$, with $E$ is the energy, $|\Psi \rangle$ is the wave-function (i.e, matter wave), and Dirac's Hamiltonian is given by \cite{2000GreinRQM}
\begin{eqnarray}\label{eq:H-1a}
H &=& c\, \vec{\alpha}\cdot \vec{\Pi} +\beta m c^2 + q \Phi,
\end{eqnarray}
where $\vec{\alpha}$, $\beta$ are $4\times 4$ Dirac's matrices, $\vec{\Pi}=\vec{p}-(q/c)\vec{\mathcal{A}} $ is the canonical momentum, $c$ is the speed of light in vacuum, $\vec{p}$, $m$ and $q$ are linear momentum, mass and electric charge, respectively. Dirac's equation can naturally interpret the above SG effect. Explicitly, by choosing $\Phi=0$ and an appropriate $\vec{\mathcal{A}}$,  then an inhomogeneous magnetic field along the $z$-direction $\vec{\mathcal{B}}=\vec{\nabla}\times \vec{\mathcal{A}}\simeq B(\vec{r}) \, \vec{e}_z $ is obtained. In the non-relativistic limit, Dirac's Hamiltonian reduces to Pauli's Hamiltonian
$H_{\rm Pauli}= \vec{\Pi}^2/2m +\vec{\mu}\cdot\vec{\mathcal{B}}$,
with $\vec{\mu}=-({q\hbar}/{2mc}) \vec{\sigma}$ the magnetic moment vector, $\hbar$ Planck's constant, and $\vec{\sigma}=(\sigma_x, \sigma_y, \sigma_z)$ the vector of Pauli matrices \cite{1927Pauli}. Because (i) the force acting on the moment $\vec{\mu}$  reads $\vec{F}=\vec{\nabla}(\vec{\mu}\cdot\vec{\mathcal{B}})\simeq \mu_z ({\partial B(\vec{r})}/{\partial z}) \, \vec{e}_z$, and (ii) Pauli matrix $\sigma_z$ takes two possible values (i.e., $\pm 1$), as a result, silver atoms with positive/negative $\mu_z$ should be deflected upwards/downwards, thus successfully explaining the SG experiment.

If one re-examines the SG experiment from the perspective of Dirac's equation, he finds that the SG experiment is actually a pure \emph{magnetic-type spin effect} since the potentials $\{\Phi=0, \vec{\mathcal{A}}\neq 0\}$ have been chosen. For convenience, let us call the conventional SG experiment as the magnetic-type SG (MTSG) effect. The MTSG effect not only reveals the magnetic moments are quantized, but also indicates an interesting physical phenomenon: as an internal degree of freedom, the spin can indeed affect the external motion of atoms (i.e., through the vector potential $\vec{\mathcal{A}}$, or through the interaction $\vec{\mu}\cdot\vec{\mathcal{B}}$ between the spin and the magnetic field). However, the MTSG effect is only a half story of the whole family of SG effects. This naturally gives rise to a scientific question: What type of spin effect will one obtain if the potentials $\{\Phi\neq 0, \vec{\mathcal{A}}= 0\}$ are considered? In this work, we advance the study of the SG effect. By considering the potentials $\{\Phi\neq 0, \vec{\mathcal{A}}= 0\}$ in Dirac's equation, we present an extraordinary spin effect -- \emph{the electric-type Stern-Gerlach (ETSG) effect}. For simplicity, we choose the scalar potential as a one-dimensional potential barrier. After exactly solving Dirac's equation, we then obtain the explicit forms of incident, reflected, and transmitted waves. The ETSG effect shows that the reflected and transmitted waves can have notable spatial shifts, which depend on the spin direction and the incident angle of the incident wave. This electric-type spin effect has potential applications to separating Dirac's particles (i.e., the incident waves) with different spin directions and to estimating the spin direction of Dirac's particles.

\begin{figure}[t]
    \includegraphics[width=70mm]{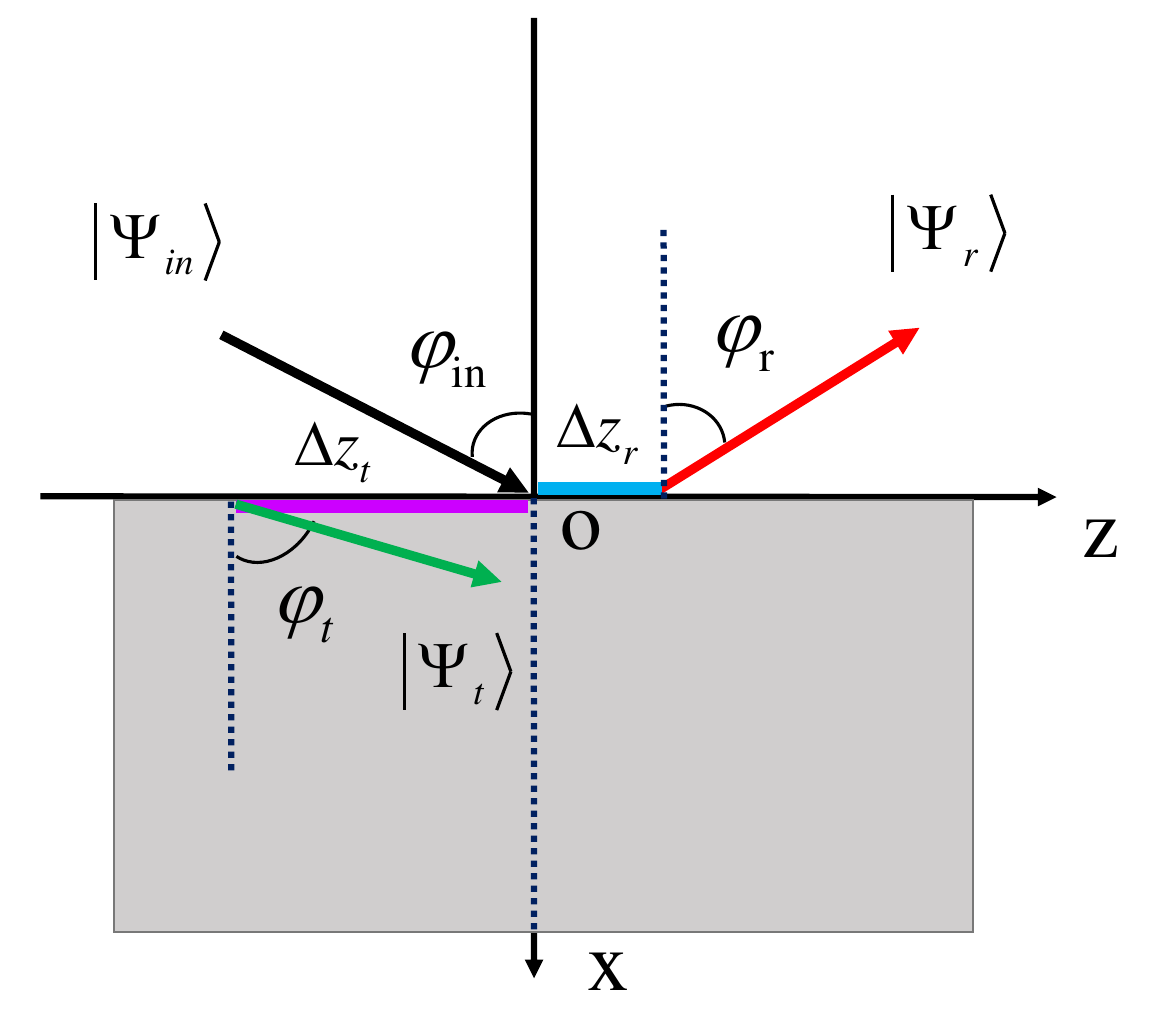}
    \caption{{\bf Illustration of the ETSG Effect}. $\ket{\Psi_{\rm in}}$ is the incident wave (the solid black arrow), $\ket{\Psi_{\rm r}}$ is the reflected wave (the solid red arrow), and $\ket{\Psi_{\rm t}}$ is the transmitted wave (the solid green arrow). $\varphi_{\rm in}$ is the incident angle, $\varphi_{\rm r}$ is the reflected angle angle, and $\varphi_{\rm t}$ is the transmitted angle. The reflected wave has a spatial shift $\Delta z_{\rm r}$ (the solid blue line), and the transmitted wave has a spatial shift $ \Delta z_{\rm t}$ (the solid purple line), which are in the order of Compton wavelength. The shifts can be positive or negative, depending on the spin direction and the incident angle of incident wave. The ETSG effect is a pure relativistic effect caused by spin, which vanishes in the non-relativistic limit.
    }\label{fig:SGH6a}
 \end{figure}

\emph{Results.---} The potential barrier is given by \cite{1999Flugge}
\begin{equation}\label{eq:poten}
   V(x)= q \Phi = \begin{cases}
        V_0 , & x > 0, \\
        0, & x \leq 0,
    \end{cases}
\end{equation}
with $V_0>0$. By solving Dirac's equation, one can determine the incident, reflected, and transmitted waves. Due to the wave-particle duality, these matter waves can be used to represent Dirac's particles themselves. In Fig. \ref{fig:SGH6a},
the reflect interface is the $yz$-plane, point $O$ is the incident point, $\ket{\Psi_{\rm in}}$, $\ket{\Psi_{\rm r}}$ and $\ket{\Psi_{\rm t}}$ are respectively the incident, reflected and transmitted waves, $\varphi_{\rm in}$, $\varphi_{\rm r}$ and $\varphi_{\rm t}$ are respectively the incident, reflected and transmitted angles. Similar to wave optics, after the incident wave arrives at the point $O$, generally the reflection and transmission phenomena will occur. However, due to the existence of the ``spin'' for Dirac's particle, there will be some spatial shifts deviating from point $O$ for the reflected wave and the transmitted wave, such as $\Delta z_{\rm r}$ and $ \Delta z_{\rm t}$.

In this work, we consider the incident finite-width wave $\ket{\Psi_{\rm in}}$. For simplicity, we consider that the momentum $ p_y=0$ (i.e., the incident plane is just the $xz$-plane as shown in Fig. \ref{fig:SGH6a}), and the effective wavefront area of the incident wave still has finite dimensions along  $z$-axis with infinite dimensions along the $y$-axis. $\ket{\Psi_{\rm in}}$ can be represented as the composition of infinite-width incident waves, namely, 
\begin{eqnarray}
   \ket{\Psi_{\rm in}}= \int \d k_z \begin{pmatrix}
      (\mathcal{E}+mc^2) \ket{\chi}\\
      c\hbar (\vec{\sigma} \cdot {\vec{k}})\ket{\chi}
  \end{pmatrix}\frac{{\rm e}^{{\rm i} \left[(k_x x+ k_z z) - \frac{1}{\hbar} \mathcal{E}t\right]}}{\sqrt{2\mathcal{E}(\mathcal{E} +mc^2)}},
\end{eqnarray}
where $\vec{k} =\vec{p}/\hbar= (k_x,0, k_z)$, 
and $\mathcal{E}=+\sqrt{\vec{p}^{\,2} c^2+m^2c^4}$ corresponds to the positive energy of Dirac's particle. This means that the energy of Dirac's particle keeps positive during its motion. The state $\ket{\chi}= \begin{pmatrix}
    \cos\frac{\theta}{2}\\
    \sin\frac{\theta}{2}\, {\rm e}^{{\rm i}\phi}
\end{pmatrix}$ can be used to indicate the spin direction $\vec{\tau}=(\tau_x, \tau_y, \tau_z)$ of the incident wave $\ket{\Psi_{\rm in}}$, with $|\chi\rangle\langle \chi|=(\openone + \vec{\sigma}\cdot \vec{\tau})/2$ and $\openone$ is the $2\times 2$ unit matrix \cite{2010Nielsen}. Here generally $\ket{\chi}$ is independent on $\vec{k}$.

We assume when $x=0$ , $\ket{\Psi_{\rm in}}$ approximately satisfies the following relation,
\begin{eqnarray}\label{pz11}
   \ket{\Psi_{\rm in}}(0,z) = \begin{cases}
       \begin{pmatrix}
      (\mathcal{E}+mc^2) \ket{\chi}\\
      c\hbar (\vec{\sigma} \cdot {\vec{k}_0})\ket{\chi}
  \end{pmatrix}\frac{{\rm e}^{{\rm i} \left[k_{z_0} z- \frac{1}{\hbar} \mathcal{E}t\right]}}{\sqrt{2\mathcal{E}(\mathcal{E} +mc^2)}}, & |z|\le a, \\
         0, & |z| > a,
     \end{cases}\nonumber
  \end{eqnarray}
where $\vec{k}_0= (k_{x_0},0, k_{z_0})$ can be regarded as the average momentum of the incident finite-width wave, and $2a$ is the beam width. Accordingly, this relation makes only the part whose $k_z$ is very near the average momentum of the incident wave  $k_{z_0}$ need to be considered.

We designate the momentums for the reflected wave and the transmitted wave as $\vec{p}_{\rm r}= \hbar \vec{k}_{\rm r}=\hbar (-k_x,0, k_z)$ and $\vec{p}_{\rm t}= \hbar \vec{k}_{\rm t}=\hbar (k'_x,0, k_z)$, respectively. They satisfy the energy-momentum relations $\mathcal{E}^2 =\vec{p}_{\rm r}^{\, 2} c^2 +m^2 c^4$ and $(\mathcal{E}-V_0)^2 =\vec{p}_{\rm t}^{\, 2} c^2 +m^2 c^4$.
Then in the $y$-basis $\{\ket{\uparrow}_y, \ket{\downarrow}_y\}$, the finite-width reflected and transmitted waves read


 \hfill
\onecolumngrid
\begin{subequations}
\begin{eqnarray}
&&  \ket{\Psi_{\rm r}}= \int \d k_z \left\{ |A'|\e^{\i \theta_a} \begin{pmatrix}
      (\mathcal{E}+mc^2) |\uparrow\rangle_y\\
      c\hbar (\vec{\sigma} \cdot {\vec{k}_{\rm r}}) |\uparrow\rangle_y
  \end{pmatrix}
  +|B'|\e^{\i \theta_b} \begin{pmatrix}
   (\mathcal{E}+mc^2) |\downarrow\rangle_y\\
   c\hbar (\vec{\sigma} \cdot {\vec{k}_{\rm r}}) |\downarrow\rangle_y
  \end{pmatrix}
  \right\}
  \dfrac{{\rm e}^{{\rm i} [(-k_x x+ k_z z) - \frac{1}{\hbar} \mathcal{E}t]}}{\sqrt{2\mathcal{E}(\mathcal{E} +mc^2)}}, \\
&& \ket{\Psi_{\rm t}}= \int \d k_z \left\{ \left|C'\right|\e^{\i \theta_c} \begin{pmatrix}
      (\mathcal{E}+mc^2) |\uparrow\rangle_y\\
      c\hbar (\vec{\sigma} \cdot {\vec{k}_{\rm t}}) |\uparrow\rangle_y
  \end{pmatrix}
  +|D'|\e^{\i \theta_d} \begin{pmatrix}
   (\mathcal{E}+mc^2) |\downarrow\rangle_y\\
   c\hbar (\vec{\sigma} \cdot {\vec{k}_{\rm t}}) |\downarrow\rangle_y
  \end{pmatrix}
  \right\}
  \dfrac{{\rm e}^{{\rm i} [(k'_x x+ k_z z) - \frac{1}{\hbar} \mathcal{E}t]}}{\sqrt{2(\mathcal{E}-V_0)(\mathcal{E} -V_0 + mc^2)}},
\end{eqnarray}
\end{subequations}
\twocolumngrid
\noindent where $A'$, $B'$, $C'$, and $D'$ are superposition coefficients. The states $\ket{\uparrow}_y$ and $\ket{\downarrow}_y$  correspond to the spin direction $\vec{\tau} = (0,1,0)$ and $(0,-1,0)$, respectively. In the $y$-basis, one has the simple relations $\theta_b=-\theta_a$, $\theta_d=-\theta_c$, and
\begin{eqnarray}
&& \tan \theta_a=-\frac{2 k_x k_z  (n-1)}{k_x^2-k_z^2(n-1)^2-k_x'^2 n^2}, \nonumber\\
&& \tan \theta_c = -\frac{k_z (1-n)}{k_x +n k_x'},
\end{eqnarray}
with $n = \frac{\mathcal{E}+mc^2 }{\mathcal{E}-V_0 +mc^2}$ representing the ``refractive index''. Similar to the case of incident wave, only the part whose $k_z$ is very near the average momentum of the reflected (and the transmitted) wave $k_{z_0}$ need to be considered. This allows us to conduct the stationary phase method \cite{1948Artmann}, which has been used for calculating various Goos-H{\"a}nchen (GH) shifts in wave optics and quantum mechanics \cite{1947Goos,maaza1997possibility,ignatovich2004neutron,beenakker2009quantum,de2010observation,cheng2012goos,zhou2015goos,
huaman2019anomalous,2023Ban,2024Zhou,xiang2024total}.

According to the stationary phase method, the spatial shift is defined by $\Delta z=-\frac{\partial \theta}{\partial k_{z}}$.  Then we can have  spatial shifts for the reflected and the transmitted waves as
\onecolumngrid
\begin{subequations}
\begin{eqnarray}
&&      \Delta {z}_{\rm r}=-\frac{1}{|A'|^2 + |B'|^2}\; \left(|A'|^2\; \frac{\partial \theta_a}{\partial k_{z}}
+|B'|^2 \;\frac{\partial \theta_b}{\partial k_{z}}\right)
=\dfrac{\hbar}{mc} \times \tau_y \, \dfrac{\sqrt{1-\mu_{\rm E} } }{\sqrt{1+\mu_{\rm E}}} \times
\dfrac{ \mu_{\rm E} \cos \varphi}{ \sin^2 \varphi + \mu_{\rm E}^2 \cos^2 \varphi }
 \left[\tan^2 \varphi -   \mu_{\rm E} \right], \label{eq:Dzr}\\
&& \Delta {z}_{\rm t}= - \frac{\hbar}{mc}\times \dfrac{\tau_y}{2 \mu_{\rm V}} \times  \dfrac{ \dfrac{\sqrt{1-\mu_{\rm E} }}{\sqrt{ 1+\mu_{\rm E} }}\dfrac{1}{ \cos\varphi}+  \dfrac{\dfrac{1}{\mu_{\rm E}}-\dfrac{1}{\mu_{\rm V}}- 1}{\sqrt{ \dfrac{1}{\mu_{\rm V}}\left(\dfrac{1}{\mu_{\rm V}}-2\dfrac{1}{\mu_{\rm E}}\right)+ \left(\dfrac{1}{\mu^2_{\rm E}}- 1\right)\cos^2\varphi }}}{\sqrt{ \dfrac{1}{\mu^2_{\rm E}}- 1} \cos\varphi \left[\sqrt{ \dfrac{1}{\mu^2_{\rm E}}- 1} \cos\varphi+ \sqrt{ \dfrac{1}{\mu_{\rm V}}(\dfrac{1}{\mu_{\rm V}}-2\dfrac{1}{\mu_{\rm E}})+ (\dfrac{1}{\mu^2_{\rm E}}- 1)\cos^2\varphi }\,\right]-\dfrac{1}{\mu_{\rm E}}\dfrac{1}{\mu_{\rm V}}}, \label{eq:Dzt}
\end{eqnarray}
\end{subequations}
\twocolumngrid
\noindent where $\mu_{\rm E} = {mc^2}/{\mathcal{E}}$, $\mu_{\rm V} = {mc^2}/{V_0}$, $\tau_y=\vec{\tau}\cdot\vec{e}_y$ is the $y$-component of spin direction of the incident wave, and $\varphi$ is the incident angle satisfying $\tan \varphi={k_z}/{k_x}$ and $\varphi \in [0, \pi/2)$.

\begin{remark}
For the incident angle, there are two critical angles, which are denoted by $\varphi_{\rm cr1}$ and $\varphi_{\rm cr2}$, respectively. $\varphi_{\rm cr1}$ is the critical angle for total reflection, which is a very common phenomenon for waves. Namely, if the incident angle $\varphi \geq \varphi_{\rm cr1}$, then it occurs the total reflection phenomenon (i.e., the transmission phenomenon vanishes). The first angle $\varphi_{\rm cr1}$ can be determined from the condition of the transmission coefficient $\mathcal{T}=0$. As a result, one has $\varphi_{\rm cr1}= \arcsin \left[\sqrt{\frac{(\mathcal{E}-V_0)^2 -  m^2 c^4}{\mathcal{E}^2 -  m^2 c^4}}\right]$. Here the energy region $\mathcal{E}-V_0 +mc^2>0$ (or $n>0$) has been considered to avoid Klein's paradox \cite{krekora_klein_2004} in relativistic quantum mechanics. The second angle $\varphi_{\rm cr2}$ is determined by $\Delta {z}_{\rm r}=0$, i.e., $\varphi_{\rm cr2} = \arctan  \sqrt{\frac{mc^2}{\mathcal{E}}}$. In the literature, the GH shifts for optic waves and matter waves occur in the region of $\varphi \geq \varphi_{\rm cr1}$. However, the ETSG effect is an extraordinary spin effect (see the explanation in Table \ref{tab:shift} below). To essentially distinguish the ETSG effect from the GH effect, we restrict our study in the region $ \varphi_{\rm cr2} \in \left[0, \varphi_{\rm cr1} \right]$.  $\blacksquare$
\end{remark}

\begin{remark}
From Eqs. (\ref{eq:Dzr}) and (\ref{eq:Dzt}), one finds that the spatial shifts $\Delta {z}_{\rm r}$ and $\Delta {z}_{\rm t}$ are both in the order of Compton wavelength $\lambda={h}/{mc}$,  with $h=2 \pi \hbar$. $\blacksquare$
\end{remark}

\begin{remark}
 Under the non-relativistic approximation, one has $|\vec{p}|\approx 0$. This yields $\mathcal{E}  - m c^2 \approx 0$, $V_0\approx \frac{\vec{p}^{\, 2}}{2m} \approx \mathcal{E} - m c^2  \approx 0$, thus $\Delta {z}_{\rm r} \; ( {\rm and }\; \Delta {z}_{\rm t}) \propto \sqrt{\mathcal{E}-mc^2} \approx 0$. This implies that, unlike the MTSG effect, the ETSG effect vanishes in the non-relativistic limit. Therefore, spatial shifts for the reflected and transmitted waves are pure relativistic effects caused by spin, which is the practical reason why the ETSG effect has not ever been observed in non-relativistic quantum mechanics before. $\blacksquare$
\end{remark}

\emph{Application.---} In Table \ref{tab:shift}, we have demonstrated the behaviors of spatial shifts for the reflected wave and the transmitted wave. One finds that the shift $\Delta {z}_{\rm r}$ can be positive, zero, or negative, depending on the choices of $\tau_y$ and the incident angle $\varphi$, while the shift $\Delta {z}_{\rm t}$ can be positive, zero, or negative, depending on merely on the choice of $\tau_y$. In  Fig. \ref{fig:SGH6a}, we have illustrated the case of $\Delta z_{\rm r}>0$ and  $\Delta z_{\rm t}<0$, which corresponds to $0<\tau_y \leq 1$ and
$\varphi \in (\varphi_{\rm cr2}, \varphi_{\rm cr1})$. Due to these findings, we can have the following possible applications:

\begin{table}[h]
\centering
\caption{The behaviors of $\Delta z_{\rm r}$ and  $\Delta z_{\rm t}$ with respect to $\varphi$ and $\tau_y$. If the spin direction is reversed, i.e., $\vec{\tau}\mapsto -\vec{\tau}$, accordingly the signs of spatial shifts are reversed, thus indicating the ETSG effect is a spin effect.}
\begin{tabular}{llccc}
\hline\hline
Shift & \;\;\;\;\; $ \varphi $   & \;$ 0<\tau_y \leq 1$  \;\;\;&  $\tau_y =0$ \;\; & $-1\leq \tau_y <0$ \\
  \hline
$ \Delta z_{\rm r}$ \;\;&  $\varphi \in [0, \varphi_{\rm cr2})$  & $-$ &   0 & $+$ \\
  \hline
 & $\varphi = \varphi_{\rm cr2}$  &   0 & 0 &  0\\
  \hline
 & $\varphi \in (\varphi_{\rm cr2}, \varphi_{\rm cr1})$  & $+$  &  0 & $-$ \\
  \hline
$ \Delta z_{\rm t}$ \;\; &   $\varphi\in [0, \varphi_{\rm cr1})$  & $-$&  0& $+$  \\
 \hline\hline
\end{tabular}\label{tab:shift}
\end{table}

\begin{figure}[t]
            \centering
            \includegraphics[width=70mm]{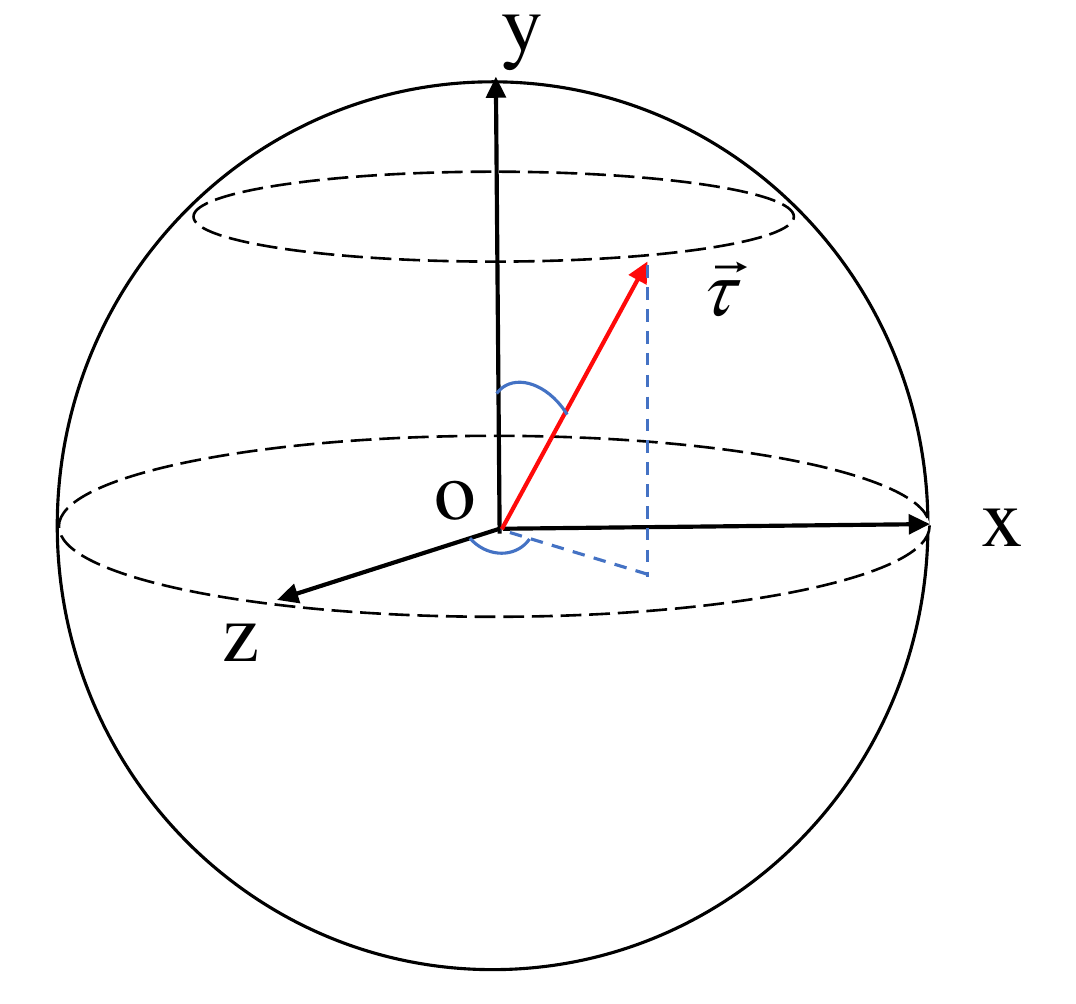}\hspace{5mm}
            \caption{\textbf{Illustration of spin direction of $\ket{\Psi_{\rm in}}$ in Bloch sphere}. $\vec{\tau}=(\tau_x, \tau_y, \tau_z)$ is  the spin direction (the red solid arrow).} \label{fig:SGH3a}
      \end{figure}

(i) Separating Dirac's particles with different spin directions. The information of spin direction of Dirac's particle is encoded in the incident wave $\ket{\Psi_{\rm in}}$ (or in the state $|\chi\rangle$). In Fig. \ref{fig:SGH3a}, we have illustrated the spin direction $\vec{\tau}$ in the Bloch sphere \cite{2010Nielsen}. For simplicity, we call the $xz$-plane as the ``equatorial plane'', the hemispherical surface above the $xz$-plane as the ``northern hemisphere'', the hemispherical surface below the $xz$-plane as the ``southern hemisphere'', the intersection point of positive $y$-axis and sphere as the north pole, and the intersection point of negative $y$-axis and sphere as the south pole. Then the result in Table \ref{tab:shift} can be used to separate Dirac's particles with different spin directions. The procedure is as follows: for the beam of input Dirac's particles, we take the incident angle as $\varphi \in [0, \varphi_{\rm cr2})$, then after reflection, particles with spin direction in ``northern hemisphere'' will have negative shifts $ \Delta z_{\rm r}$, particles with spin directions in ``southern hemisphere'' will have positive shifts $ \Delta z_{\rm r}$, while those with spin directions in ``equatorial plane'' will have no shifts, thus realizing the separation of Dirac's particles with different spin directions by the ETSG effect.

(ii) Estimating the spin direction of Dirac's particles. Inversely, one can use the result of Table  \ref{tab:shift} to estimate the spin direction of the input Dirac's particle. The procedure is as follows: we take the incident angle as $\varphi \in [0, \varphi_{\rm cr2})$, if there is a negative shift $ \Delta z_{\rm r}$, then the spin direction of Dirac's particle is in ``northern hemisphere'', conversely, it is in ``southern hemisphere''; If no shift, then it is located in the ``equatorial plane'', thus realizing the estimation of spin direction for Dirac's particles by the ETSG effect..

\emph{Some Theoretical Data.---} Here we provide some theoretical data, which may be helpful for experimental scientists to perform the possible experiments. Let us choose the parameters $\mu_{\rm E} = 1/3$, $\mu_{\rm V} = 4$, and $\tau_y\approx 0.92$,
which yield the two critical angles as $\varphi_{\rm cr1} \approx 64.92^\circ$ and $\varphi_{\rm cr2} = 30^\circ$. In Table \ref{tab:varia} (see Appendix A), we have listed the variations of the reflected angle $\varphi_{\rm r}$, the transmitted angle $\varphi_{\rm t}$, the spatial shifts $\Delta {z}_{\rm r} $ and $\Delta {z}_{\rm t} $ when the incident angle $\varphi_{\rm in}$ runs from $0^\circ$ to the critical angle $\varphi_{\rm cr1}$. With the help of current  quantum technologies on precision measurement or interferometry, scientists can perhaps observe the ETSG effect in the lab.

\emph{Conclusion and Discussion.---} In summary, we have presented an extraordinary spin effect, i.e., the electric-type Stern-Gerlach effect, by
exactly solving Dirac's equation with a vector potential $\vec{\mathcal{A}}= 0$ and a simple scalar potential $\Phi$ given in Eq. (\ref{eq:poten}). Our work is a nontrivial generalization of the conventional SG effect from magnetic-type to electric-type, thus improving and perfecting the study of the SG effect. Our work also tells another half story of the SG effect, thereby complementing the family of SG effects. Unlike the MTSG effect, the ETSG effect vanishes in the non-relativistic limit, thus the spatial shifts for the reflected and transmitted waves are pure relativistic effects caused by spin, this enhances the deep understanding for the concept of \emph{spin}.  We expect that the extraordinary ETSG effect can be observed in the near future. If the ETSG effect can be realized in the lab, then it will provide a significant technology for separating Dirac's particles with different spin directions as well as estimating the spin direction of Dirac's particles.

Eventually, let us end the work with some discussions.

(i) Seemingly, the ETSG effect is a Goos-H{\"a}nchen-like shift. However, they are essentially different, the latter occurs in the region of total reflection and the notion of ``spin'' is unnecessary, while the former can occur at the region of non-total reflection and the property of ``spin'' is required. If one investigates Dirac's equation in the region of $\varphi \in [\varphi_{\rm cr1}, \pi/2)$, then it is expected to have a mixture of the ETSG effect and the GH effect. We shall study this issue in the future work.

(ii) In the above, we have just considered a specific case and neglected the possible shift along the $y$-axis and the influence of $p_y $. For a comprehensive analysis of the ETSG effect, one can continue to study a more general case where $p_y \ne 0 $ and the effective wavefront area of
the incident wave has finite dimensions along both $z$-axis and the $y$-axis. In such a case, one expects to have the Imbert-Fedorov-like shifts \cite{1955Fedorov,1972Imbert,2007Li}, i.e., $\Delta \vec{l}_{\rm r}  = \Delta y_{\rm r} \vec{e}_y+ \Delta z_{\rm r} \vec{e}_z$ and $\Delta \vec{l}_{\rm t}  = \Delta y_{\rm t} \vec{e}_y+ \Delta z_{\rm t} \vec{e}_z$ (see Appendix B).

(iii) The potential barrier shown in Eq. (\ref{eq:poten}) is a model with a step-like feature. Though, in reality, we consider matter waves in continuous fields more often, the potential barrier model can also give us the inspiration of the spatial shifts.  In addition,
to analyze the continuous field, we can assume that it consists of an infinite number of infinitesimal intervals with infinitesimal potential barrier. Then we can calculate the spatial shifts for each small interval one by one and describe the propagation of matter waves.

For example, we consider the following type of potential barrier
\begin{equation}\label{eq:linear}
    V(x)= q \Phi(x) = \begin{cases}
         -q \mathbb{E}_0 \, x  >0, & \; x > 0, \\
         0, & \; x \leq 0,
     \end{cases}
 \end{equation}
where $q=-e$ (here $e=|e|>0$) is the electric charge of the Dirac's electron, $\Phi(x)=-\mathbb{E}_0 \, x$ is the scalar potential with $\mathbb{E}_0>0$. Accordingly, the electric field is given by
$ \vec{\mathbb{E}} = -\vec{\nabla}\Phi(x)= \mathbb{E}_0\,  \vec{e}_x$, which is uniform electric field along the positive $x$-axis.
However, it is hard to exactly solve Dirac's equation with the potential barrier shown in Eq. (\ref{eq:linear}). Alternatively, in order to analyze the spatial shifts, we try to divide the whole space into an infinite number of infinitesimal intervals along the $x$-axis, whose width is $\d x \rightarrow 0 $. It is necessary to mention that this method is valid where $p_x$ is not so small. In this case, we can assume a matter plane wave goes through many infinitesimal-width potential barriers and adopt our above approach to attain the spatial shifts when the wave goes through any barrier.

We consider one interval. The energy of the incident wave is still $\mathcal{E}$, and the  potential can be $V_0=- q \mathbb{E}_0 \d x \approx 0$, which means $(n-1) \rightarrow 0$ and $k_x' \approx k_x$. Then we find the transmission coefficient $\mathcal{T}=1$ and the reflection coefficient $\mathcal{R}=0$, which means there is no reflected wave. The spatial shift for transmitted wave can be calculated as
\begin{eqnarray}\label{eq:tt}
    \Delta \vec{l}_{\rm t} & \approx &
 \frac{q \, \d x }{2 k_x^3(\mathcal{E}+mc^2)}
     \biggr\{ [(\vec{\tau}\times\vec{k})\cdot\vec{\mathbb{E}}] \, \vec{\eta} - k_x^2 (\vec{\tau} \times \vec{\mathbb{E}})\biggr\},
\end{eqnarray}
with $\eta=(0, k_y, k_z)=\vec{k}-k_x \,(\vec{\mathbb{E}}/\mathbb{E}_0)$ and $k_x=\vec{k}\cdot(\vec{\mathbb{E}}/\mathbb{E}_0)$.
For the next interval, the energy $\mathcal{E}  \rightarrow  \mathcal{E} -V= \mathcal{E} + q{\mathbb{E}}_0 \d x $.
Then we can attain the corresponding spatial shift in any interval, and describe the propagation trajectory of waves.
Because this method just focuses on the infinitesimal internal, it can be extended to any one-dimensional continuous field.
Finally, from Eq. (\ref{eq:tt}) one observes that the ETSG effect is related to the interaction between spin and electric field. For example, the term $(\vec{\tau} \times \vec{k}) \cdot \vec{\mathbb{E}}$ represents the Rashba-like spin-orbit coupling \cite{2015Manchon}, which can cause the well-known spin Hall effect in an electric field \cite{hirsch_spin_1999,sinova_universal_2004,nagaosa_anomalous_2010}. It is interesting to study the possible relation between the ETSG effect and spin Hall effect, which we shall investigate subsequently.

\begin{acknowledgments}
   J.L.C. is supported by the National Natural Science Foundation of China (Grant No. 12275136).
\end{acknowledgments}



\vspace{8mm}
\onecolumngrid

\vspace{8mm}

\noindent\textbf{Appendix A:}

\begin{table}[h]
\centering
\caption{The variations of the reflected angle $\varphi_{\rm r}$, the transmitted angle $\varphi_{\rm t}$, the shift $\Delta {z}_{\rm r} $, and the shift $\Delta {z}_{\rm t} $ when the incident angle $\varphi$ runs from $0^\circ$ to the critical angle $\varphi_{\rm cr1}\approx 64.92^\circ$. Here $\mu_{\rm E} = 1/3$, $\mu_{\rm V} = 4$, $\tau_y\approx 0.92$, $\varphi_{\rm r}=\varphi_{\rm in}$, and $\varphi_{\rm t}$ satisfies $\tan \varphi_{\rm t}={k_z}/{k'_x}$ with $  k_x'^2= \left[(\mathcal{E}-V_0)^2 - \left(\hbar^2 k_z^2 c^2 +m^2 c^4\right)\right]/(\hbar c)^2$. We have taken the unit of $\Delta {z}_{\rm r} $ and $\Delta {z}_{\rm t} $ as $h/mc$.}
\begin{tabular}{cccccccccccccc}
\hline\hline\hline
$\varphi_{\rm in}$ & $0^\circ$  &  $2^\circ$  &  $4^\circ$& $6^\circ$& $8^\circ$& $10^\circ$ & $12^\circ$ & $14^\circ$ & $16^\circ$ & $18^\circ$
& $20^\circ$ & $22^\circ$  \\
  \hline
$\varphi_{\rm r}$  & $0^\circ$  &  $2^\circ$  & $4^\circ$& $6^\circ$& $8^\circ$ &$10^\circ$ & $12^\circ$ & $14^\circ$ & $16^\circ$ & $18^\circ$
& $20^\circ$ & $22^\circ$  \\
  \hline
$\varphi_{\rm t}$ & $0^\circ$  &  $2.21^\circ$  &  $4.42^\circ$&  $6.62^\circ$ & $8.84^\circ$ & $11.05^\circ$& $13.27^\circ$ & $15.49^\circ$ & $17.72^\circ$ & $19.95^\circ$
& $22.19^\circ$ & $24.43^\circ$ \\
  \hline
$\Delta {z}_{\rm r}$ & $-0.1035$  & $-0.1021$  &  $-0.0980$&  $-0.0916$ & $-0.0835$& $-0.0745$ & $-0.0650$ & $-0.0556$ & $-0.0466$ & $-0.0381$
& $-0.0302$ & $-0.0231$\\
  \hline
$\Delta {z}_{\rm t}$ & $-0.0302$  &  $-0.0302$  &  $-0.0304$&  $-0.0308$&  $-0.0312$ & $-0.0318$& $-0.0326$ & $-0.0335$ & $-0.0347$ & $-0.0360$
& $-0.0376$ & $-0.0394$ \\
 \hline\hline
$\varphi_{\rm in}$& $24^\circ$ & $26^\circ$  &  $28^\circ$  &  $30^\circ$& $32^\circ$& $34^\circ$& $36^\circ$ & $38^\circ$ & $40^\circ$ & $42^\circ$ & $44^\circ$ & $46^\circ$ \\
  \hline
$\varphi_{\rm r}$& $24^\circ$ & $26^\circ$  &  $28^\circ$  &  $30^\circ$& $32^\circ$& $34^\circ$& $36^\circ$ & $38^\circ$ & $40^\circ$ & $42^\circ$ & $44^\circ$ & $46^\circ$ \\
  \hline
$\varphi_{\rm t}$ & $26.68^\circ$ & $28.95^\circ$  &  $31.22^\circ$  &  $33.51^\circ$& $35.81^\circ$& $38.13^\circ$& $40.46^\circ$ & $42.82^\circ$ & $45.21^\circ$ & $47.63^\circ$ & $50.08^\circ$ & $52.58^\circ$ \\
  \hline
$\Delta {z}_{\rm r}$ & $-0.0165$  & $-0.0105$  &  $-0.0050$  &  $0$& $0.0046$& $0.0089$& $0.0130$ & $0.0168^\circ$ & $0.0205^\circ$ & $0.0241^\circ$ & $0.0275^\circ$ & $0.0310^\circ$ \\
  \hline
$\Delta {z}_{\rm t}$ & $-0.0416$ & $-0.0441^\circ$  &  $-0.0470^\circ$  &  $-0.0505^\circ$& $-0.0546^\circ$& $-0.0595^\circ$& $-0.0653^\circ$ & $-0.0723^\circ$ & $0.0809^\circ$ & $-0.0914^\circ$ & $-0.1046^\circ$ & $-0.1214^\circ$ \\
 \hline\hline
$\varphi_{\rm in}$ & $48^\circ$ & $50^\circ$  &  $52^\circ$  &  $54^\circ$& $56^\circ$& $58^\circ$& $59^\circ$ & $60^\circ$ & $61^\circ$ & $62^\circ$ & $63^\circ$ & $64^\circ$ \\
  \hline
$\varphi_{\rm r}$& $48^\circ$ & $50^\circ$  &  $52^\circ$  &  $54^\circ$& $56^\circ$& $58^\circ$& $59^\circ$ & $60^\circ$ & $61^\circ$ & $62^\circ$ & $63^\circ$ & $64^\circ$ \\
  \hline
$\varphi_{\rm t}$ & $55.14^\circ$ & $57.76^\circ$  &  $60.46^\circ$  &  $63.28^\circ$& $66.25^\circ$& $69.44^\circ$& $71.16^\circ$ & $72.98^\circ$ & $74.94^\circ$ & $77.13^\circ$ & $79.66^\circ$ & $82.91^\circ$\\
  \hline
$\Delta {z}_{\rm r}$& $0.0345$ & $0.0381$  &  $0.0418$  &  $0.0457$& $0.0498$& $0.0543$& $0.0567$ & $0.0592$ & $0.0617$ & $0.0646$ & $0.0675$ & $0.0706$ \\
  \hline
$\Delta {z}_{\rm t}$& $-0.1433$ & $-0.1725$  &  $-0.2132$  &  $-0.2722$& $-0.3639$& $-0.5207$& $-0.6475$ & $-0.8373$ & $-1.1481$ & $-1.7366$ & $-3.2110$ & $-11.6888$ \\
 \hline\hline\hline
\end{tabular}\label{tab:varia}
\end{table}

\vspace{5mm}

\noindent\textbf{Appendix B:}

\vspace{8mm}

The spatial shifts for the reflected wave and the transmitted wave are given by 
\begin{subequations}
\begin{eqnarray} 
\Delta \vec{l}_{\rm r} &=& \Delta y_{\rm r} \, \vec{e}_y +\Delta z_{\rm r}\,  \vec{e}_z = \frac{\hbar^2 c^2 }
 { k_x (\mathcal{E} + mc^2 )^2 [\mathcal{E}^2 -{c^2\hbar^2 }  k_x^2]}\times\nonumber\\
&&    \biggr\{ \mathcal{E} (\mathcal{E} + mc^2 ) [(\vec{\tau}\times \vec{k})\cdot\vec{e}_x] \vec{\eta}
 + (k_x^2) mc^2 (\mathcal{E} + mc^2 ) (\vec{\tau} \times \vec{e}_x) + {c^2\hbar^2 }  k_x^2 (\vec{k}\cdot \vec{\tau}) (\vec{k} \times \vec{e}_x) \biggr\}, \label{shifts-a}\\
    \Delta \vec{l}_{\rm t}
       & = & \Delta y_{\rm t} \, \vec{e}_y +\Delta z_{\rm t}\,  \vec{e}_z =- \kappa_3(n-1) \left\{\kappa_1 \left[(\vec{\tau} \times  \vec{k})\cdot \vec{e}_x\right] \vec{\eta} - \kappa_2 (\vec{\tau} \times \vec{e}_x) + (n-1) (\vec{\tau} \cdot \vec{e}_x) (\vec{e}_x \times \vec{k})\right\},
       \label{shifts-b}
\end{eqnarray}
\end{subequations}
with 
$$\vec{\eta}= k_y \vec{e}_y+k_z \vec{e}_z, \;\;\;\; \kappa_1=\dfrac{1}{k_x}+\dfrac{{n}}{k'_x}, \;\;\;\; \kappa_2=k_x +{n}k_x', \;\;\;\; \kappa_3=\dfrac{1}{(k_x +{n}k_x' )^2 + (n-1)^2 (k_y^2 +k_z^2 )}.$$
One may observe that when the spin direction is reversed, i.e, $\vec{\tau} \mapsto -\vec{\tau}$, the spatial shifts $\Delta \vec{l}_{\rm r}$ and $\Delta \vec{l}_{\rm t}$ are also reversed. The scalar potential (\ref{eq:poten}) corresponds to an electric field $\vec{\mathbb{E}}$ along the $x$-axis, if one denotes $\vec{e}_x=\vec{\mathbb{E}}/|\vec{\mathbb{E}}|$, he may observe that the above spatial shifts are related to the interaction between spin and the electric field. When $k_y=0$, Eqs. (\ref{shifts-a}) and (\ref{shifts-b}) reduce to Eqs. (\ref{eq:Dzr}) and (\ref{eq:Dzt}).

\twocolumngrid

\end{document}




\title{Supplemental Material for
``Electric-Type Stern-Gerlach Effect''}

\author{Jiang-Lin Zhou}
\affiliation{School of Physics, Nankai University, Tianjin 300071, People's Republic of China}

\author{Zou-Chen Fu}
\affiliation{School of Physics, Nankai University, Tianjin 300071, People's Republic of China}

\author{Choo Hiap Oh}
\email{phyohch@nus.edu.sg}
\affiliation{Centre for Quantum Technologies and Department of Physics, National University of Singapore, 117543, Singapore}

\author{Jing-Ling Chen}
\email{chenjl@nankai.edu.cn}
\affiliation{Theoretical Physics Division, Chern Institute of Mathematics, Nankai University, Tianjin 300071, People's Republic of
	 	China}

\date{\today}

\maketitle

\onecolumngrid

\tableofcontents

\newpage



\section{The Eigen-Problem of a Free Dirac's Electron}

In relativistic quantum mechanics, the Hamiltonian of a free Dirac's electron is given by
\begin{eqnarray}\label{eq:H-1a}
H &=& c\, \vec{\alpha}\cdot \vec{p} +\beta m c^2,
\end{eqnarray}
where $m$ is the mass of the electron, $c$ is the speed of light in vacuum,
\begin{eqnarray}
\vec{p}=(p_x, p_y, p_z)=-{\rm i}\hbar \vec{\nabla}=(-{\rm i}\hbar \frac{\partial }{\partial x}, -{\rm i}\hbar \frac{\partial }{\partial y}, -{\rm i}\hbar \frac{\partial }{\partial z})\end{eqnarray}
is the linear momentum, $\hbar=h/2\pi$, $h$ is Planck's constant, and $\vec{\alpha}$ and $\beta$ are the well-known Dirac's matrices, i.e.,
      \begin{equation}
            \vec{\alpha}=
            \left(\begin{array}{cc}
                    0 & \vec{\sigma} \\
                  \vec{\sigma} & 0 \\
                  \end{array}
                \right)=\sigma_x\otimes\vec{\sigma},\;\;\;\;\;
             \beta=
             \left(\begin{array}{cc}
                  \openone & 0 \\
                  0 & -\openone \\
                  \end{array}
                \right)
           =\sigma_z\otimes\openone,
      \end{equation}
which satisfy the following relations
\begin{subequations}
\begin{eqnarray}
  && {\alpha}_x^2={\alpha}_y^2={\alpha}_z^2=\beta^2 =\mathbb{I}, \\
       && \{{\alpha}_j, \beta\}={\alpha}_j \beta+\beta{\alpha}_j=0, \;\;\;\; (j=x, y, z),  \label{z1}
\end{eqnarray}
\end{subequations}
with $\vec{\sigma}=(\sigma_x, \sigma_y, \sigma_z)$ being the vector of Pauli matrices, $\openone$ being the $2\times 2$ identity matrix, and $\mathbb{I}=\openone\otimes\openone$ being the $4\times 4$ identity matrix.

\begin{remark}\textcolor{blue}{The Eigen-Problem of a Free Dirac's Electron.}
The time-dependent Dirac equation is given by
\begin{eqnarray}\label{eq:H-1c}
{\rm i}\hbar\, \dfrac{\partial |\Psi(\vec{r}, t)\rangle}{\partial t} = H |\Psi(\vec{r}, t)\rangle,
\end{eqnarray}
where $|\Psi(\vec{r}, t)\rangle$ is the time-dependent wavefunction, $t$ is time, and $\vec{r}$ is the position. After setting
\begin{eqnarray}
|\Psi(\vec{r}, t)\rangle=|\Psi(\vec{r})\rangle\, {\rm e}^{-\frac{{\rm i}}{\hbar} E t},
\end{eqnarray}
one has the stationary Dirac equation as
\begin{eqnarray}\label{eq:eigen}
H |\Psi(\vec{r})\rangle= E |\Psi(\vec{r})\rangle,
\end{eqnarray}
where $|\Psi(\vec{r})\rangle$ is the time-independent wavefunction and $E$ is the corresponding energy.
Then the eigen-problem of a free Dirac's electron is given by
\begin{eqnarray}
\biggr[c\, \vec{\alpha}\cdot \vec{p} +\beta m c^2 \biggr] \Psi(\vec{r})= E \Psi(\vec{r}).
\end{eqnarray}

To solve the eigen-problem, as shown in the textbook \cite{flugge_relativistic_1971}, one usually introduce the \emph{helicity} operator
\begin{equation}
\hat{\Lambda}=\vec{S} \cdot \hat{p}=\dfrac{\hbar}{2}\, \left(\vec{\Sigma} \cdot \hat{p}\right),
\end{equation}
where $\hat{p}=\vec{p}/p$, $p=|\vec{p}|$,
\begin{equation}
 \vec{S}=\dfrac{\hbar}{2}\, \vec{\Sigma}
 \end{equation}
is the spin-$1/2$ operator, which satisfies the definition of angular momentum operator as follows
\begin{equation}
 \vec{S} \times \vec{S}= {\rm i}\hbar\, \vec{S},
 \end{equation}
and
      \begin{equation}
            \vec{\Sigma}=
            \left(\begin{array}{cc}
                    \vec{\sigma} & 0 \\
                  0 & \vec{\sigma} \\
                  \end{array}
                \right)=\openone \otimes\vec{\sigma}.
      \end{equation}
Because
\begin{equation}
\bigl[\vec{\Sigma}\cdot\hat{p},\ \vec{\alpha}\cdot\vec{p}\,\bigr]=0, \;\;\;\;\;\bigl[\vec{\Sigma},\ \beta\bigr]=0,
\end{equation}
then it is easy to prove that  helicity operator commutes with Hamiltonian $ H$, i.e.,
\begin{equation}
[H, \, \hat{\Lambda}]=0.
\end{equation}
Moreover, one has
\begin{equation}
\bigl[H, \vec{p}\,\bigr]=0, \;\;\;\;\;\bigl[\vec{\Sigma},\ \vec{p} \,\bigr]=0,
\end{equation}
thereby, the set $\{H, \hat{\Lambda}, \vec{p}\}$ constitutes a complete set of commuting observables for the eigen-problem shown in Eq. (\ref{eq:eigen}). Let us denote $|{{\Psi}_{\epsilon,\Lambda}}(\vec{r})\rangle$ as the  simultaneous eigenstates of $\{H, \hat{\Lambda}, \vec{p}\}$, then we have
\begin{subequations}
\begin{eqnarray}
&&H\,|{{\Psi}_{\epsilon,\Lambda}}(\vec{r})\rangle=E\,|{{\Psi}_{\epsilon,\Lambda}}(\vec{r})\rangle, \\
&& \hat{\Lambda}|{{\Psi}_{\epsilon,\Lambda}}(\vec{r})\rangle
=\Lambda\,|{{\Psi}_{\epsilon,\Lambda}}(\vec{r})\rangle,\quad \Lambda=\pm\dfrac{\hbar}{2}, \\
&& \vec{p}\,|{{\Psi}_{\epsilon,\Lambda}}(\vec{r})\rangle = {\hbar}\,\vec{k}\,|{{\Psi}_{\epsilon,\Lambda}}(\vec{r})\rangle.
\end{eqnarray}
\end{subequations}
Therefore, after setting
\begin{eqnarray}
|{{\Psi}_{\epsilon,\Lambda}}(\vec{r})\rangle= |{\Psi}(\vec{k})\rangle\, {{\rm e}^{{\rm i} \left( k_x x+k_y y+k_z z\right)},}
\end{eqnarray}
and due to
\begin{eqnarray}
&&p_x |{\Psi}(\vec{r})\rangle =  -{\rm i}\hbar \frac{\partial |{\Psi}(\vec{r})\rangle }{\partial x}= {\hbar}\, k_x \, |{\Psi}(\vec{r})\rangle,\nonumber\\
&&p_y |{\Psi}(\vec{r})\rangle =  -{\rm i}\hbar \frac{\partial |{\Psi}(\vec{r})\rangle }{\partial y}= {\hbar}\, k_y \, |{\Psi}(\vec{r})\rangle,\nonumber\\
&&p_z |{\Psi}(\vec{r})\rangle =  -{\rm i}\hbar \frac{\partial |{\Psi}(\vec{r})\rangle }{\partial z}= {\hbar}\, k_z \, |{\Psi}(\vec{r})\rangle,
\end{eqnarray}
we have explicitly the four normalized eigenstates as follows
\begin{equation}\label{eq:mom}
\begin{split}
 & E=E_+=+ \mathcal{E},\;\; \begin{cases}
                                          & |{\Psi}_1\rangle
                                          =\dfrac{1}{2\sqrt{\mathcal{E}\,k(k+k_z)}} \left[
                                                u_+ \Bigl(k+k_z ,\ k_+\Bigr)
                                                ,\dfrac{c{\hbar}\,k}{u_+}\Bigl(k+k_z,\ k_+ \Bigr)\right]^{\rm T},
                                          \quad \;\;\;\;\;\;\Lambda=+\dfrac{\hbar}{2}, \\
                                          & |{\Psi}_2\rangle
                                          =\dfrac{1}{2\sqrt{\mathcal{E}\,k(k+k_z)}}\left[
                                                u_+ \Bigl({-k_-}, k+k_z\Bigr)
                                                ,{-}\dfrac{c{\hbar}\,k}{u_+}\Bigl({-k_-}, k+k_z \Bigr)\right]^{\rm T},
                                          \quad \Lambda=-\dfrac{\hbar}{2},
                                    \end{cases} \\
 & E=E_-=-\mathcal{E},\;\; \begin{cases}
                                          & |{\Psi}_3\rangle
                                          =\dfrac{1}{2\sqrt{\mathcal{E}\,k(k+k_z)}} \left[
                                                u_- \Bigl(k+k_z ,\ k_+\Bigr)
                                                ,{-}\dfrac{c{\hbar}\,k}{u_-}\Bigl(k+k_z,\ k_+ \Bigr)\right]^{\rm T},
                                          \quad \;\;\;\Lambda=+\dfrac{\hbar}{2}, \\
                                          & |{\Psi}_4\rangle
                                          =\dfrac{1}{2\sqrt{\mathcal{E}\,k(k+k_z)}}\left[
                                                u_- \Bigl(-k_-,k+k_z\Bigr)
                                                ,\dfrac{c{\hbar}\,k}{u_-}\Bigl(-k_-, k+k_z \Bigr)\right]^{\rm T},
                                          \quad \;\;\;\Lambda=-\dfrac{\hbar}{2},
                                    \end{cases}
\end{split}
\end{equation}
with
\begin{eqnarray}
\mathcal{E}=\sqrt{p^2c^2+m^2c^4}, \;\;\;\; u_{\pm} = \sqrt{\mathcal{E} \pm mc^2}, \;\;\;\; k_\pm =k_x \pm {\rm i} k_y, \;\;\;\; k=\sqrt{k_x^2+k_y^2+k_z^2}.
\end{eqnarray}
Here $E_+$ and $E_-$ are the positive and negative energies, respectively. Note that there are two eigenstates (i.e., $\{|{\Psi}_1\rangle, |{\Psi}_2\rangle \}$) correspond to $E_+$, and there are also other two eigenstates (i.e., $\{|{\Psi}_3\rangle, |{\Psi}_4\rangle \}$) correspond to $E_-$, thus $E_+$ and $E_-$ are both two-fold degeneracies. $\blacksquare$
\end{remark}
\begin{remark}\textcolor{blue}{The Diagonalization Matrix for the Hamiltonian.} Based on the four eigenstates $\{|{\Psi}_1\rangle, |{\Psi}_2\rangle, |{\Psi}_3\rangle, |{\Psi}_4\rangle \}$, one can directly have the diagonalization matrix for the Hamiltonian $H$ as
    \begin{eqnarray}
     W &=& (|{\Psi}_1\rangle, |{\Psi}_2\rangle, |{\Psi}_3\rangle, |{\Psi}_4\rangle)\nonumber\\
     &=&\dfrac{1}{2\sqrt{\mathcal{E}\,k(k+k_z)}} \begin{pmatrix}
                u_+ \bigl(k+k_z\bigr) & -u_+ k_-&  u_- \bigl(k+k_z\bigr)& -u_- k_-\\
                u_+  k_+ & u_+ \bigl(k+k_z\bigr) & u_- k_+  & u_- \bigl(k+k_z\bigr) \\
                \dfrac{c{\hbar}\,k}{u_+}\bigl(k+k_z\bigr) & \dfrac{c{\hbar}\,k}{u_+}k_- & -\dfrac{c{\hbar}\,k}{u_-}\bigl(k+k_z \bigr) & -\dfrac{c{\hbar}\,k}{u_-}k_-  \\
                \dfrac{c{\hbar}\,k}{u_+} k_+  & -\dfrac{c{\hbar}\,k}{u_+}\bigl(k+k_z \bigr) & -\dfrac{c{\hbar}\,k}{u_-}k_+ &  \dfrac{c{\hbar}\,k}{u_-}\bigl( k+k_z \bigr)
     \end{pmatrix},
     \end{eqnarray}
     i.e.,
     \begin{eqnarray}
     W &=&\dfrac{1}{2\sqrt{\mathcal{E}\,k(k+k_z)}} \begin{pmatrix}
            u_+ \left[(k+k_z) \openone +{\rm i} (k_y \sigma_x-k_x\sigma_y)\right] \;\;\;   & u_- \left[(k+k_z) \openone +{\rm i} (k_y \sigma_x-k_x\sigma_y)\right]   \\
            \dfrac{c{\hbar}\,k}{u_+} \left[(k+k_z)\sigma_z+ k_x \sigma_x+ k_y \sigma_y \right]   & -\dfrac{c{\hbar}\,k}{u_-} \left[(k+k_z)\sigma_z+ k_x \sigma_x+ k_y \sigma_y \right]
               \end{pmatrix}\nonumber\\
     &=&\dfrac{1}{2\sqrt{\mathcal{E}\,k(k+k_z)}} \begin{pmatrix}
            u_+ \left[(k+k_z) \openone +{\rm i} (k_y \sigma_x-k_x\sigma_y)\right] \;\;\;   & u_- \left[(k+k_z) \openone +{\rm i} (k_y \sigma_x-k_x\sigma_y)\right]   \\
            \dfrac{c{\hbar}\,k}{u_+} \left[k\sigma_z+ \vec{k}\cdot \vec{\sigma} \right]   & -\dfrac{c{\hbar}\,k}{u_-} \left[k\sigma_z+ \vec{k}\cdot \vec{\sigma} \right]
               \end{pmatrix}.
     \end{eqnarray}
    By replacing $\vec{k}$ by $\vec{p}$ (sometimes we view $\vec{k}$ just as $\vec{p}$ without causing confusion), one has the diagonalization matrix $W(\vec{p})$ as
     \begin{eqnarray}\label{eq:W-1}
     W(\vec{p}) &=&\dfrac{1}{2\sqrt{\mathcal{E}\,p(p+p_z)}} \begin{pmatrix}
            u_+ \left[(p+p_z) \openone +{\rm i} (p_y \sigma_x-p_x\sigma_y)\right] \;\;\;   & u_- \left[(p+p_z) \openone +{\rm i} (p_y \sigma_x-p_x\sigma_y)\right]   \\
            \dfrac{c\,p}{u_+} \left[p\sigma_z+ \vec{p}\cdot \vec{\sigma} \right]   & -\dfrac{c\,p}{u_-} \left[p\sigma_z+ \vec{p}\cdot \vec{\sigma} \right]
               \end{pmatrix}.
     \end{eqnarray}
     Based on which, one can have
     \begin{eqnarray}
     W(\vec{p})^\dag &=& \dfrac{1}{2\sqrt{\mathcal{E}\,p(p+p_z)}} \begin{pmatrix}
            u_+ \left[(p+p_z) \openone -{\rm i} (p_y \sigma_x-p_x\sigma_y)\right] \;\;\;   &   \dfrac{c\,p}{u_+} \left[p\sigma_z+ \vec{p}\cdot \vec{\sigma} \right]  \\
           u_- \left[(p+p_z) \openone -{\rm i} (p_y \sigma_x-p_x\sigma_y)\right]   & -\dfrac{c\,p}{u_-} \left[p\sigma_z+ \vec{p}\cdot \vec{\sigma} \right]
               \end{pmatrix}.
     \end{eqnarray}
    One may easily check that $W(\vec{p})$ is unitary, i.e.,
    \begin{eqnarray}
     W(\vec{p}) W(\vec{p})^\dag = W(\vec{p})^\dag W(\vec{p})=\mathbb{I}=\openone\times\openone.
     \end{eqnarray}

    The Hamiltonian $H$ can be written in the matrix form as follows
    \begin{eqnarray}
    H &=& c\, \vec{\alpha}\cdot \vec{p} +\beta m c^2=
    \begin{pmatrix}
              m c^2 \openone & c \left(\vec{\sigma}\cdot\vec{p}\right) \\
              c \left(\vec{\sigma}\cdot\vec{p}\right) & -m c^2 \openone
            \end{pmatrix}.
    \end{eqnarray}
    With the help of the diagonalization matrix $W(\vec{p})$, one has
    \begin{eqnarray}
    H &=&  W(\vec{p})\, G\,  W(\vec{p})^\dag,
    \end{eqnarray}
    or
    \begin{eqnarray}\label{eq:HD-1}
    W(\vec{p})^\dag \, H \, W(\vec{p}) &=& G,
    \end{eqnarray}
    where $G$ is diagonalized matrix of the Hamiltonian $H$, i.e.,
     \begin{eqnarray}
    G &=& \begin{pmatrix}
           + \mathcal{E}\, \openone &   0  \\
            0   & - \mathcal{E}\, \openone
               \end{pmatrix}= \mathcal{E} \begin{pmatrix}
           \openone &   0  \\
            0   & -  \openone
               \end{pmatrix}.
     \end{eqnarray}
    Simultaneously, the helicity operator is diagonalized by the same diagonalization matrix $W(\vec{p})$ as
    \begin{eqnarray}
    W(\vec{p})^\dag \, \hat{\Lambda} \, W(\vec{p}) &=& Y,
    \end{eqnarray}
    with
     \begin{eqnarray}
    Y &=& \dfrac{\hbar}{2}\begin{pmatrix}
           1 &   0  & 0  & 0  \\
           0 &   -1  & 0  & 0  \\
            0 &   0  & 1  & 0  \\
             0 &   0  & 0  & -1
               \end{pmatrix}
               = \dfrac{\hbar}{2} \begin{pmatrix}
           \sigma_z &   0  \\
            0   &  \sigma_z
               \end{pmatrix}.
     \end{eqnarray}
    $\blacksquare$
    \end{remark}

    \begin{remark}\textcolor{blue}{The More General Diagonalization Matrix.} Let us focus on Eq. (\ref{eq:HD-1}), i.e.,
    \begin{eqnarray}\label{eq:HD-2}
    W(\vec{p})^\dag \, H \, W(\vec{p}) &=& G =  \mathcal{E} \begin{pmatrix}
           \openone &   0  \\
            0   & -  \openone
               \end{pmatrix},
    \end{eqnarray}
    whose meaning is that the Hamiltonian $H$ can be diagonalized by the diagonalization matrix $W(\vec{p})$. However, the matrix $W(\vec{p})$ given in Eq. (\ref{eq:W-1}) is only a special diagonalization matrix. To obtain the more general diagonalization matrix (denoted by $\mathcal{U}$), one need to consider the two-fold degeneracies of the energy. Let $U_1$ and $U_2$ be two arbitrary $2\times 2$ unitary matrices, then one has
    \begin{eqnarray}
     U_1 U_1^\dag = U_1^\dag  U_1=\openone, \;\;\;\;\; U_2 U_2^\dag = U_2^\dag  U_2=\openone.
     \end{eqnarray}
    Furthermore, one easily has
    \begin{eqnarray}
     U_1 \,\openone \,U_1^\dag = \openone, \;\;\;\;\; U_2\, \openone\,  U_2^\dag =\openone,
     \end{eqnarray}
    which means that the $2\times 2$ identity matrix $\openone$ is unchanged under the unitary transformations $U_j$ ($j=1, 2$).

    Based on such an observation, the matrix $G$ can be rewritten as
    \begin{eqnarray}
     G &=&  \mathcal{E} \begin{pmatrix}
           \openone &   0  \\
            0   & -  \openone
               \end{pmatrix}= \begin{pmatrix}
           U_1 &   0  \\
            0   &  U_2
               \end{pmatrix}
               \left[\mathcal{E} \begin{pmatrix}
           \openone &   0  \\
            0   & -  \openone
               \end{pmatrix}\right] \begin{pmatrix}
           U_1^\dag &   0  \\
            0   &  U_2^\dag
               \end{pmatrix} = U_3 \, G\, U_3^\dag,
    \end{eqnarray}
    where $U_3$ is a $4\times 4$ unitary matrix defined by
    \begin{eqnarray}
     U_3 &=&  \begin{pmatrix}
           U_1 &   0  \\
            0   &  U_2
               \end{pmatrix}.
    \end{eqnarray}
    Then Eq. (\ref{eq:HD-2}) can be recast to
    \begin{eqnarray}
    W(\vec{p})^\dag \, H \, W(\vec{p}) &=& U_3 \, G\, U_3^\dag,
    \end{eqnarray}
    i.e.,
    \begin{eqnarray}
    \left[W(\vec{p})U_3\right]^\dag \, H \, \left[W(\vec{p})U_3\right] &=& G,
    \end{eqnarray}
    i.e.,
    \begin{eqnarray}\label{eq:HD-3}
    \mathcal{U}^\dag \, H \, \mathcal{U} &=& G,
    \end{eqnarray}
    with
    \begin{eqnarray}\label{eq:U-1a}
    \mathcal{U}= W(\vec{p})U_3.
    \end{eqnarray}
    Based on Eq. (\ref{eq:HD-3}), we obtain the more general diagonalization matrix $\mathcal{U}$ for the Hamiltonian $H$. $\blacksquare$
    \end{remark}

    \begin{remark}\textcolor{blue}{A More Simpler Set of Eigenstates $(|{\Psi}'_1\rangle, |{\Psi}'_2\rangle, |{\Psi}'_3\rangle, |{\Psi}'_4\rangle)$.}
    In the above, we have studied the eigenstates of the Hamiltonian $H$, and have obtained the standard set of four eigenstates, i.e.,  $(|{\Psi}_1\rangle, |{\Psi}_2\rangle, |{\Psi}_3\rangle, |{\Psi}_4\rangle)$. This set of eigenstates is the common eigenstates of the set $\{H, \hat{\Lambda}, \vec{p}\}$. One may observe that, for each $|{\Psi}_j\rangle$ ($j=1, 2, 3, 4$), all the four components are not zero (supposed $p_x$, $p_y$, $p_z$ are not zero). Here, we would like to obtain a more simpler set of eigenstates $(|{\Psi}'_1\rangle, |{\Psi}'_2\rangle, |{\Psi}'_3\rangle, |{\Psi}'_4\rangle)$, such that (i) each $|{\Psi}'_j\rangle$ ($j=1, 2, 3, 4$) is still eigenstate of the Hamiltonian $H$ (but not longer of the helicity operator $\hat{\Lambda}$); and (ii) one of the components of each $|{\Psi}'_j\rangle$ ($j=1, 2, 3, 4$) can be zero, thus looking more simpler.

    \emph{Case 1.---}Because $|{\Psi}_1\rangle$ and $|{\Psi}_2\rangle$ are eigenstates of $H$ with positive energy, thus $|{\Psi}'_1\rangle=f_1 \, |{\Psi}_1\rangle + f_2\, |{\Psi}_2\rangle$ is also the eigenstate of $H$ with positive energy for any coefficients $f_1$ and $f_2$. For simplicity, we denote
     \begin{eqnarray}
    f_1 =\cos\eta, \;\;\;\; f_2=\sin\eta\, e^{{\rm i}\tau}.
     \end{eqnarray}
    If
     \begin{eqnarray}
    f_1 (u_+^2k_+)+ f_2 [u_+^2 (k+k_z)]=0,
     \end{eqnarray}
    we have
     \begin{eqnarray}
    && f_1=\cos\eta= \dfrac{k+k_z}{\sqrt{|k_+|^2+(k+k_z)^2}}=\dfrac{k+k_z}{\sqrt{2k(k+k_z)}}, \nonumber\\
    && f_2=\sin\eta\, e^{{\rm i}\tau}= -\dfrac{k_+}{\sqrt{|k_+|^2+(k+k_z)^2}}=-\dfrac{k_+}{\sqrt{2k(k+k_z)}}.
     \end{eqnarray}
    Then we obtain the superposition state $|{\Psi}'_1\rangle$ as
     \begin{eqnarray}\label{eq:psi1'}
    |{\Psi}'_1\rangle &=& f_1 \, |{\Psi}_1\rangle + f_2\, |{\Psi}_2\rangle\nonumber\\
    &=& \dfrac{1}{2\sqrt{\mathcal{E}\,k(k+k_z)}}\frac{1}{u_+} \left\{ f_1 \, \begin{pmatrix}
                u_+^2(k+k_z) \\
                u_+^2k_+ \\
                c{\hbar} k(k+k_z) \\
                c{\hbar} k k_+
            \end{pmatrix}+
            f_2\, \begin{pmatrix}
                -u_+^2k_- \\
                u_+^2 (k+k_z)\\
                c{\hbar} k k_- \\
                -c{\hbar} k (k+k_z)
            \end{pmatrix}\right\} \nonumber\\
    &=& \dfrac{1}{2\sqrt{\mathcal{E}\,k(k+k_z)}}\frac{1}{u_+} \left\{ \begin{pmatrix}
                u_+^2 \left[f_1 \,(k+k_z)- f_2\,k_- \right] \\
                0 \\
                c{\hbar} k\left[f_1 \,(k+k_z)+f_2\,k_- \right] \\
                c{\hbar} k \left[f_1 \,k_+ -f_2\,(k+k_z) \right]
            \end{pmatrix}\right\} \nonumber\\
    &=& \dfrac{1}{2\sqrt{\mathcal{E}\,k(k+k_z)}}\frac{1}{u_+} \dfrac{1}{\sqrt{2k(k+k_z)}} \left\{ \begin{pmatrix}
                u_+^2 \left[(k+k_z)^2+ k_+k_- \right] \\
                0 \\
                c{\hbar} k\left[(k+k_z)^2-k_+k_- \right] \\
                c{\hbar} k \left[(k+k_z) k_+ +k_+(k+k_z) \right]
            \end{pmatrix}\right\} \nonumber\\
    &=& \dfrac{1}{2\sqrt{2\mathcal{E}}k(k+k_z)}\frac{1}{u_+}   \begin{pmatrix}
                u_+^2 \left[2k(k+k_z)\right] \\
                0 \\
                c{\hbar} k\left[2k_z(k+k_z) \right] \\
                c{\hbar} k \left[2k_+(k+k_z) \right]
            \end{pmatrix} \nonumber\\
    &=& \dfrac{1}{\sqrt{2\mathcal{E}}}\frac{1}{u_+}  \begin{pmatrix}
                u_+^2  \\
                0 \\
                c{\hbar} k_z \\
                c{\hbar} k_+
            \end{pmatrix} \nonumber\\
    &=& \dfrac{1}{\sqrt{2\mathcal{E}(\mathcal{E} +mc^2)}}  \begin{pmatrix}
                \mathcal{E} +mc^2  \\
                0 \\
                k_z c{\hbar} \\
                (k_x+{\rm i} k_y) c{\hbar}
            \end{pmatrix},
     \end{eqnarray}
    namely, the second component of $|{\Psi}'_1\rangle$ is zero.

    Similarly, we can have another superposition state $|{\Psi}'_2\rangle$ (which is orthogonal to $|{\Psi}'_1\rangle$ and corresponds to positive energy) as
     \begin{eqnarray}\label{eq:psi2'}
    |{\Psi}'_2\rangle &=& -f_2^* \, |{\Psi}_1\rangle + f_1^*\, |{\Psi}_2\rangle\nonumber\\
    &=& \dfrac{1}{2\sqrt{\mathcal{E}\,k(k+k_z)}}\frac{1}{u_+} \left\{ \dfrac{k_-}{\sqrt{2k(k+k_z)}} \, \begin{pmatrix}
                u_+^2(k+k_z) \\
                u_+^2k_+ \\
                c{\hbar} k(k+k_z) \\
                c{\hbar} k k_+
            \end{pmatrix}
            +\dfrac{k+k_z}{\sqrt{2k(k+k_z)}}\, \begin{pmatrix}
                -u_+^2k_- \\
                u_+^2 (k+k_z)\\
                c{\hbar} k k_- \\
                -c{\hbar} k (k+k_z)
            \end{pmatrix}\right\} \nonumber\\
    &=& \dfrac{1}{\sqrt{2\mathcal{E}(\mathcal{E} +mc^2)}}  \begin{pmatrix}
                0  \\
                \mathcal{E} +mc^2 \\
                (k_x-{\rm i} k_y) c{\hbar}\\
                -k_z c{\hbar}
            \end{pmatrix},
     \end{eqnarray}
     namely, the first component of $|{\Psi}'_2\rangle$ is zero.

    \emph{Case 2.---}Because $|{\Psi}_3\rangle$ and $|{\Psi}_4\rangle$ are eigenstates of $H$ with negative energy, thus $|{\Psi}'_3\rangle=f_1 \, |{\Psi}_3\rangle + f_2\, |{\Psi}_4\rangle$ is also the eigenstate of $H$ with negative energy for any coefficients $f_1$ and $f_2$.
    Then we obtain the superposition state $|{\Psi}'_3\rangle$ as
     \begin{eqnarray}
    |{\Psi}'_3\rangle &=& f_1 \, |{\Psi}_3\rangle + f_2\, |{\Psi}_4\rangle\nonumber\\
    &=& \dfrac{1}{2\sqrt{\mathcal{E}\,k(k+k_z)}}\frac{1}{u_-} \left\{ f_1 \,
    \begin{pmatrix}
                u_-^2(k+k_z) \\
                u_-^2k_+ \\
                -c{\hbar} k(k+k_z) \\
                -c{\hbar} k k_+
            \end{pmatrix}+
            f_2\, \begin{pmatrix}
                -u_-^2k_- \\
                u_-^2 (k+k_z)\\
                -c{\hbar} k k_- \\
                c{\hbar} k (k+k_z)
            \end{pmatrix}\right\} \nonumber\\
    &=& \dfrac{1}{2\sqrt{\mathcal{E}\,k(k+k_z)}}\frac{1}{u_-} \left\{ \begin{pmatrix}
                u_-^2 \left[f_1 \,(k+k_z)- f_2\,k_- \right] \\
                 u_-^2 \left[f_1 \,k_++ f_2\, (k+k_z)\right] \\
                -c{\hbar} k\left[f_1 \,(k+k_z)+f_2\,k_- \right] \\
                -c{\hbar} k\left[f_1 \,k_+- f_2\,(k+k_z) \right]
            \end{pmatrix}\right\} \nonumber\\
    &=& \dfrac{1}{2\sqrt{\mathcal{E}\,k(k+k_z)}}\frac{1}{u_-} \dfrac{1}{\sqrt{2k(k+k_z)}} \left\{ \begin{pmatrix}
                u_-^2 \left[(k+k_z)^2+ k_+k_- \right] \\
                0 \\
                -c{\hbar} k\left[(k+k_z)^2 - k_+k_- \right] \\
                -c{\hbar} k\left[2(k+k_z)k_+\right]
            \end{pmatrix}\right\} \nonumber\\
    &=& \dfrac{1}{2\sqrt{2\mathcal{E}}k(k+k_z)}\frac{1}{u_-}   \begin{pmatrix}
                u_-^2 \left[2k(k+k_z)\right] \\
                0 \\
                -c{\hbar} k\left[2k_z(k+k_z) \right] \\
                -c{\hbar} k\left[2(k+k_z)k_+\right]
            \end{pmatrix} \nonumber\\
    &=& \dfrac{1}{\sqrt{2\mathcal{E}}}\frac{1}{u_-}  \begin{pmatrix}
                u_-^2  \\
                0 \\
                -c{\hbar} k_z \\
                -c{\hbar} k_+
            \end{pmatrix} \nonumber\\
    &=& \dfrac{1}{\sqrt{2\mathcal{E}(\mathcal{E} - mc^2)}}  \begin{pmatrix}
                \mathcal{E} -mc^2  \\
                0 \\
                -k_z c{\hbar} \\
                -(k_x+{\rm i} k_y) c{\hbar}
            \end{pmatrix},
     \end{eqnarray}
    namely, the second component of $|{\Psi}'_3\rangle$ is zero.

    Similarly, we can have another superposition state $|{\Psi}'_4\rangle$ (which is orthogonal to $|{\Psi}'_3\rangle$ and corresponds to positive energy) as
     \begin{eqnarray}
    |{\Psi}'_4\rangle &=& -f_2^* \, |{\Psi}_3\rangle + f_1^*\, |{\Psi}_4\rangle\nonumber\\
    &=& \dfrac{1}{2\sqrt{\mathcal{E}\,k(k+k_z)}}\frac{1}{u_-} \left\{ \dfrac{k_-}{\sqrt{2k(k+k_z)}} \,
    \begin{pmatrix}
                u_-^2(k+k_z) \\
                u_-^2k_+ \\
                -ck{\hbar} (k+k_z) \\
                -c{\hbar} k k_+
            \end{pmatrix}
            +\dfrac{k+k_z}{\sqrt{2k(k+k_z)}}\,
            \begin{pmatrix}
                -u_-^2k_- \\
                u_-^2 (k+k_z)\\
                -c{\hbar} k k_- \\
                c{\hbar} k (k+k_z)
            \end{pmatrix}\right\} \nonumber\\
    &=& \dfrac{1}{\sqrt{2\mathcal{E}(\mathcal{E} -mc^2)}}  \begin{pmatrix}
                0  \\
                \mathcal{E} -mc^2 \\
                -(k_x-{\rm i} k_y) c{\hbar}\\
                k_z c{\hbar}
            \end{pmatrix},
     \end{eqnarray}
     namely, the first component of $|{\Psi}'_4\rangle$ is zero.

    Based on the new set of four eigenstates $\{|{\Psi}'_1\rangle, |{\Psi}'_2\rangle, |{\Psi}'_3\rangle, |{\Psi}'_4\rangle \}$, one can directly have the diagonalization matrix for the Hamiltonian $H$ as
    \begin{eqnarray}
     W' &=& (|{\Psi}'_1\rangle, |{\Psi}'_2\rangle, |{\Psi}'_3\rangle, |{\Psi}'_4\rangle)\nonumber\\
     &=&\dfrac{1}{\sqrt{2\mathcal{E}}} \begin{pmatrix}
               \sqrt{\mathcal{E} +mc^2}  & 0  &  \sqrt{\mathcal{E} -mc^2} & 0\\
                0 & \sqrt{\mathcal{E} +mc^2} &  0 & \sqrt{\mathcal{E} -mc^2} \\
     \dfrac{k_z {\hbar}}{\sqrt{\mathcal{E} +mc^2}} & \dfrac{(k_x-{\rm i} k_y) c{\hbar}}{\sqrt{\mathcal{E} +mc^2}} &  \dfrac{-k_z c{\hbar}}{\sqrt{\mathcal{E}-mc^2}} & \dfrac{-(k_x-{\rm i} k_y) c{\hbar}}{\sqrt{\mathcal{E}-mc^2}}\\
     \dfrac{(k_x+{\rm i} k_y) c{\hbar}}{\sqrt{\mathcal{E} +mc^2}} & \dfrac{-k_z c{\hbar}}{\sqrt{\mathcal{E} +mc^2}} & \dfrac{-(k_x+{\rm i} k_y) c{\hbar}}{\sqrt{\mathcal{E}-mc^2}}  & \dfrac{k_z c{\hbar}}{\sqrt{\mathcal{E}-mc^2}}
    \end{pmatrix},
    \end{eqnarray}
    i.e.,
    \begin{eqnarray}
     W' &=& (|{\Psi}'_1\rangle, |{\Psi}'_2\rangle, |{\Psi}'_3\rangle, |{\Psi}'_4\rangle)\nonumber\\
     &=&\dfrac{1}{\sqrt{2\mathcal{E}}} \begin{pmatrix}
     \sqrt{\mathcal{E} +mc^2}\, \openone \;\;\; & \sqrt{\mathcal{E} -mc^2}\, \openone \\
     \dfrac{c{\hbar} (\vec{\sigma} \cdot \vec{k})}{\sqrt{\mathcal{E} +mc^2}}  &  -\dfrac{c{\hbar} (\vec{\sigma} \cdot \vec{k})}{\sqrt{\mathcal{E}-mc^2}}
    \end{pmatrix}.
    \end{eqnarray}
    By replacing ${\hbar}\vec{k}$ by $\vec{p}$, one has the diagonalization matrix $W'(\vec{p})$ as
     \begin{eqnarray}\label{eqW'}
     W'(\vec{p}) &=& \dfrac{1}{\sqrt{2\mathcal{E}}} \begin{pmatrix}
     \sqrt{\mathcal{E} +mc^2}\, \openone \;\;\; & \sqrt{\mathcal{E} -mc^2}\, \openone \\
     \dfrac{c (\vec{\sigma} \cdot \vec{p})}{\sqrt{\mathcal{E} +mc^2}}  &  -\dfrac{c (\vec{\sigma} \cdot \vec{p})}{\sqrt{\mathcal{E}-mc^2}}
    \end{pmatrix}\nonumber\\
    &=& \dfrac{1}{\sqrt{2\mathcal{E}}} \begin{pmatrix}
     u_+\, \openone \;\;\; & u_-\, \openone \\
     \dfrac{c (\vec{\sigma} \cdot \vec{p})}{{u_+}}  &  -\dfrac{c (\vec{\sigma} \cdot \vec{p})}{u_-}
    \end{pmatrix},
     \end{eqnarray}
    and
    \begin{eqnarray}
    W'(\vec{p})^\dag \, H \, W'(\vec{p}) &=& G =  \mathcal{E} \begin{pmatrix}
           \openone &   0  \\
            0   & -  \openone
               \end{pmatrix}.
    \end{eqnarray}
    This means that the Hamiltonian can be diagonalized by either $W(\vec{p})$, or $W'(\vec{p})$. Here $W(\vec{p})$ and $W'(\vec{p})$ are two special cases of the more general diagonalization matrix $\mathcal{U}$ as shown in Eq. (\ref{eq:U-1a}). $\blacksquare$
    \end{remark}

    \begin{remark}\textcolor{blue}{The Spectrum Decomposition of the Hamiltonian.} For the set of four eigenstates $\{|{\Psi}_1\rangle, |{\Psi}_2\rangle, |{\Psi}_3\rangle, |{\Psi}_4\rangle \}$, one has
     \begin{eqnarray}
     H\, |{\Psi}_1\rangle= +\mathcal{E}\, |{\Psi}_1\rangle, \;\;\; H\, |{\Psi}_2\rangle= +\mathcal{E}\, |{\Psi}_2\rangle,\;\;\;
     H\, |{\Psi}_3\rangle= -\mathcal{E}\, |{\Psi}_3\rangle, \;\;\; H\, |{\Psi}_4\rangle= -\mathcal{E}\, |{\Psi}_4\rangle,
     \end{eqnarray}
     then the Hamiltonian $H$ can be expressed by the following spectrum decomposition
     \begin{eqnarray}
     H= \mathcal{E}\, \left(|{\Psi}_1\rangle\langle {\Psi}_1|+ |{\Psi}_2\rangle\langle {\Psi}_2|\right)-
     \mathcal{E}\, \left(|{\Psi}_3\rangle\langle {\Psi}_3|+ |{\Psi}_4\rangle\langle {\Psi}_4|\right).
     \end{eqnarray}
     Similarly, for the new set of four eigenstates $\{|{\Psi}'_1\rangle, |{\Psi}'_2\rangle, |{\Psi}'_3\rangle, |{\Psi}'_4\rangle \}$, one has
     \begin{eqnarray}
     H\, |{\Psi}'_1\rangle= +\mathcal{E}\, |{\Psi}'_1\rangle, \;\;\; H\, |{\Psi}'_2\rangle= +\mathcal{E}\, |{\Psi}'_2\rangle,\;\;\;
     H\, |{\Psi}'_3\rangle= -\mathcal{E}\, |{\Psi}'_3\rangle, \;\;\; H\, |{\Psi}'_4\rangle= -\mathcal{E}\, |{\Psi}'_4\rangle,
     \end{eqnarray}
     then the same Hamiltonian operator $H$ can be expressed by the following spectrum decomposition
     \begin{eqnarray}\label{eq:dc-1}
     H= \mathcal{E}\, \left(|{\Psi}'_1\rangle\langle {\Psi}'_1|+ |{\Psi}'_2\rangle\langle {\Psi}'_2|\right)-
     \mathcal{E}\, \left(|{\Psi}'_3\rangle\langle {\Psi}'_3|+ |{\Psi}'_4\rangle\langle {\Psi}'_4|\right).
     \end{eqnarray}

     Let us denote
     \begin{eqnarray}
     |1\rangle= \begin{pmatrix}
     1\\
     0\\
     0\\
     0
     \end{pmatrix}, \;\;\;\;
     |2\rangle= \begin{pmatrix}
     0\\
     1\\
     0\\
     0
     \end{pmatrix}, \;\;\;\;
     |3\rangle= \begin{pmatrix}
     0\\
     0\\
     1\\
     0
     \end{pmatrix}, \;\;\;\;
     |4\rangle= \begin{pmatrix}
     0\\
     0\\
     0\\
     1
     \end{pmatrix},
      \end{eqnarray}
     then it is easy to know that
     \begin{eqnarray}
     |{\Psi}_1\rangle=W(\vec{p}) |1 \rangle, \;\;\; |{\Psi}_2\rangle=W(\vec{p}) |2 \rangle, \;\;\;
     |{\Psi}_3\rangle=W(\vec{p}) |3 \rangle, \;\;\; |{\Psi}_4\rangle=W(\vec{p}) |4 \rangle,
      \end{eqnarray}
      \begin{eqnarray}
     |{\Psi}'_1\rangle=W'(\vec{p}) |1 \rangle, \;\;\; |{\Psi}'_2\rangle=W'(\vec{p}) |2 \rangle, \;\;\;
     |{\Psi}'_3\rangle=W'(\vec{p}) |3 \rangle, \;\;\; |{\Psi}'_4\rangle=W'(\vec{p}) |4 \rangle,
      \end{eqnarray}
    and
      \begin{eqnarray}
      |1 \rangle= W(\vec{p})^\dag\, |{\Psi}_1\rangle, \;\;\; |2 \rangle= W(\vec{p})^\dag\, |{\Psi}_2\rangle, \;\;\;
      |3 \rangle= W(\vec{p})^\dag\, |{\Psi}_3\rangle, \;\;\; |4 \rangle= W(\vec{p})^\dag\, |{\Psi}_4\rangle.
      \end{eqnarray}

    Let us focus on Eq. (\ref{eq:dc-1}), we can have
    \begin{eqnarray}\label{eq:dc-2}
     H &=& \mathcal{E}\, \left(|{\Psi}'_1\rangle\langle {\Psi}'_1|+ |{\Psi}'_2\rangle\langle {\Psi}'_2|\right)-
     \mathcal{E}\, \left(|{\Psi}'_3\rangle\langle {\Psi}'_3|+ |{\Psi}'_4\rangle\langle {\Psi}'_4|\right)\nonumber\\
     &=& \mathcal{E}\, \biggr[W'(\vec{p}) |1\rangle\langle 1| W'(\vec{p})^\dag+ W'(\vec{p}) |2\rangle\langle 2| W'(\vec{p})^\dag \biggr]-
     \mathcal{E}\, \biggr[W'(\vec{p}) |3\rangle\langle 3| W'(\vec{p})^\dag+ W'(\vec{p}) |4\rangle\langle 4| W'(\vec{p})^\dag \biggr]\nonumber\\
     &=& \mathcal{E}\, \biggr\{W'(\vec{p}) \,\left[W(\vec{p})^\dag\, |{\Psi}_1\rangle\langle |{\Psi}_1| W(\vec{p})\right]\,W'(\vec{p})^\dag+ W'(\vec{p}) \,\left[W(\vec{p})^\dag\, |{\Psi}_2\rangle\langle |{\Psi}_2| W(\vec{p})\right]\, W'(\vec{p})^\dag \biggr\}\nonumber\\
     && -\mathcal{E}\, \biggr\{W'(\vec{p}) \,\left[W(\vec{p})^\dag\, |{\Psi}_3\rangle\langle |{\Psi}_3| W(\vec{p})\right]\,W'(\vec{p})^\dag+ W'(\vec{p}) \,\left[W(\vec{p})^\dag\, |{\Psi}_4\rangle\langle |{\Psi}_4| W(\vec{p})\right]\, W'(\vec{p})^\dag \biggr\}\nonumber\\
     &=& \mathcal{M} \biggr\{\mathcal{E}\, \left(|{\Psi}_1\rangle\langle {\Psi}_1|+ |{\Psi}_2\rangle\langle {\Psi}_2|\right)-
     \mathcal{E}\, \left(|{\Psi}_3\rangle\langle {\Psi}_3|+ |{\Psi}_4\rangle\langle {\Psi}_4|\right)\biggr\}\mathcal{M}^\dag\nonumber\\
     &=& \mathcal{M}\, H\, \mathcal{M}^\dag,
      \end{eqnarray}
     with the unitary matrix $\mathcal{M}$ as
     \begin{eqnarray}\label{eq:dc-3}
     \mathcal{M} &=& W'(\vec{p}) \,W(\vec{p})^\dag.
     \end{eqnarray}
 Eq. (\ref{eq:dc-2}) means that the Hamiltonian $H$ is unchanged under the unitary transformation $\mathcal{M}$. Eq. (\ref{eq:dc-2}) is also equivalent to $H \mathcal{M}=\mathcal{M}H$, i.e.,
    \begin{eqnarray}
     [H, \, \mathcal{M}]=0,
     \end{eqnarray}
    which means that $H$ and $\mathcal{M}$ are commutative. In the following let us work out the explicit expression of the matrix $\mathcal{M}$. From Eq. (\ref{eq:dc-3}), one has
    \begin{eqnarray}
     \mathcal{M} &=& W'(\vec{p}) \,W(\vec{p})^\dag \nonumber\\
     &=& \dfrac{1}{\sqrt{2\mathcal{E}}} \begin{pmatrix}
     u_+\, \openone \;\;\; & u_-\, \openone \\
     \dfrac{c (\vec{\sigma} \cdot \vec{p})}{u_+}  &  -\dfrac{c (\vec{\sigma} \cdot \vec{p})}{u_-}
    \end{pmatrix}
    \dfrac{1}{2\sqrt{\mathcal{E}\,p(p+p_z)}} \begin{pmatrix}
            u_+ \left[(p+p_z) \openone -{\rm i} (p_y \sigma_x-p_x\sigma_y)\right] \;\;\;   &   \dfrac{c\,p}{u_+} \left[p\sigma_z+ \vec{p}\cdot \vec{\sigma} \right]  \\
           u_- \left[(p+p_z) \openone -{\rm i} (p_y \sigma_x-p_x\sigma_y)\right]   & -\dfrac{c\,p}{u_-} \left[p\sigma_z+ \vec{p}\cdot \vec{\sigma} \right]
               \end{pmatrix} \nonumber\\
    &=& \dfrac{1}{2 \mathcal{E}\sqrt{2p(p+p_z)}}
    \begin{pmatrix}
            (u_+^2+u_-^2 )\left[(p+p_z) \openone -{\rm i} (p_y \sigma_x-p_x\sigma_y)\right] \;\;\;   &   0 \\
              0 & \left(\dfrac{1}{u_+^2}+\dfrac{1}{u_-^2}\right)(c\,p) c (\vec{\sigma} \cdot \vec{p}) \left[p\sigma_z+ \vec{p}\cdot \vec{\sigma} \right]
               \end{pmatrix}\nonumber\\
    &=& \dfrac{1}{2 \mathcal{E}\sqrt{2p(p+p_z)}}
    \begin{pmatrix}
            (2 \mathcal{E})\left[(p+p_z) \openone -{\rm i} (p_y \sigma_x-p_x\sigma_y)\right] \;\;\;   &   0 \\
              0 & \left(\dfrac{2 \mathcal{E}}{c^2p^2}\right)(c\,p) c (\vec{\sigma} \cdot \vec{p}) \left[p\sigma_z+ \vec{p}\cdot \vec{\sigma} \right]
               \end{pmatrix}\nonumber\\
    &=& \dfrac{1}{\sqrt{2p(p+p_z)}}
    \begin{pmatrix}
            \left[(p+p_z) \openone -{\rm i} (p_y \sigma_x-p_x\sigma_y)\right] \;\;\;   &   0 \\
              0 & (\vec{\sigma} \cdot \hat{p}) \left[p\sigma_z+ \vec{p}\cdot \vec{\sigma} \right]
               \end{pmatrix}\nonumber\\
    &=& \dfrac{1}{\sqrt{2p(p+p_z)}}
    \begin{pmatrix}
            \left[(p+p_z) \openone -{\rm i} (p_y \sigma_x-p_x\sigma_y)\right] \;\;\;   &   0 \\
              0 & \left[(p+p_z) \openone +{\rm i} (p_y \sigma_x-p_x\sigma_y)\right]
               \end{pmatrix}\nonumber\\
      &=& \dfrac{1}{\sqrt{2p(p+p_z)}} \left[(p+p_z) \openone \otimes \openone -{\rm i}\sigma_z\otimes (p_y \sigma_x-p_x\sigma_y)\right]\nonumber\\
      &=& \dfrac{1}{\sqrt{2p(p+p_z)}} \left[(p+p_z) \openone \otimes \openone -{\rm i}\sigma_z\otimes (\sigma\times \vec{p})_z\right].
    \end{eqnarray}
    Alternatively, the matrix $\mathcal{M}$ can be written as
    \begin{eqnarray}\label{eq:dc-4}
     \mathcal{M} &=& \dfrac{1}{\sqrt{2p(p+p_z)}}
    \begin{pmatrix}
             \left[p\sigma_z+ \vec{p}\cdot \vec{\sigma} \right] (\vec{\sigma} \cdot \hat{p}) \;\;\;   &   0 \\
              0 & (\vec{\sigma} \cdot \hat{p}) \left[p\sigma_z+ \vec{p}\cdot \vec{\sigma} \right]
               \end{pmatrix}\nonumber\\
    \end{eqnarray}

    Based on above, one knows that
      \begin{eqnarray}
      |{\Psi}'_1\rangle= \mathcal{M} |{\Psi}_1\rangle, \;\;\; |{\Psi}'_2\rangle= \mathcal{M} |{\Psi}_2\rangle, \;\;\;
      |{\Psi}'_3\rangle= \mathcal{M} |{\Psi}_3\rangle, \;\;\; |{\Psi}'_4\rangle= \mathcal{M} |{\Psi}_4\rangle,
      \end{eqnarray}
    i.e., the unitary matrix $\mathcal{M} $ transforms $|{\Psi}_j\rangle$ to $|{\Psi}'_j\rangle$, ($j=1, 2, 3, 4$).  $\{|{\Psi}_1\rangle, |{\Psi}_2\rangle, |{\Psi}_3\rangle, |{\Psi}_4\rangle \}$ are the four eigenstates of the common set  $\{H, \hat{\Lambda}, \vec{p}\}$. Accordingly,
    $\{|{\Psi}'_1\rangle, |{\Psi}'_2\rangle, |{\Psi}'_3\rangle, |{\Psi}'_4\rangle \}$ are the four eigenstates of the common set  $\{H, \hat{\Lambda}', \vec{p}\}$, where the `rotated'' helicity operator turns to
    \begin{eqnarray}
     \hat{\Lambda}' &=& \mathcal{M}\, \hat{\Lambda}\, \mathcal{M}^\dag\nonumber\\
     &=& \dfrac{1}{\sqrt{2p(p+p_z)}}
    \begin{pmatrix}
             \left[p\sigma_z+ \vec{p}\cdot \vec{\sigma} \right] (\vec{\sigma} \cdot \hat{p}) \;\;\;   &   0 \\
              0 & (\vec{\sigma} \cdot \hat{p}) \left[p\sigma_z+ \vec{p}\cdot \vec{\sigma} \right]
               \end{pmatrix}
               \dfrac{1}{\sqrt{2p(p+p_z)}} \left\{ \dfrac{\hbar}{2}
               \begin{pmatrix}
             \vec{\sigma} \cdot \hat{p}   \;\;\;   &   0 \\
              0 & \vec{\sigma} \cdot \hat{p}
               \end{pmatrix} \right\}\times \nonumber\\
    &&\begin{pmatrix}
             (\vec{\sigma} \cdot \hat{p}) \left[p\sigma_z+ \vec{p}\cdot \vec{\sigma} \right]  \;\;\;   &   0 \\
              0 & \left[p\sigma_z+ \vec{p}\cdot \vec{\sigma} \right] (\vec{\sigma} \cdot \hat{p})
               \end{pmatrix}\nonumber\\
     &=& \dfrac{\hbar}{4p(p+p_z)}
    \begin{pmatrix}
             \left[p\sigma_z+ \vec{p}\cdot \vec{\sigma} \right] (\vec{\sigma} \cdot \hat{p})^2 \;\;\;   &   0 \\
              0 & (\vec{\sigma} \cdot \hat{p}) \left[p\sigma_z+ \vec{p}\cdot \vec{\sigma} \right] ( \vec{\sigma} \cdot \hat{p})
               \end{pmatrix}
              \begin{pmatrix}
             (\vec{\sigma} \cdot \hat{p}) \left[p\sigma_z+ \vec{p}\cdot \vec{\sigma} \right]  \;\;\;   &   0 \\
              0 & \left[p\sigma_z+ \vec{p}\cdot \vec{\sigma} \right] (\vec{\sigma} \cdot \hat{p})
              \end{pmatrix}\nonumber\\
     &=& \dfrac{\hbar}{4p(p+p_z)}
    \begin{pmatrix}
             \left[p\sigma_z+ \vec{p}\cdot \vec{\sigma} \right] (\vec{\sigma} \cdot \hat{p}) \left[p\sigma_z+ \vec{p}\cdot \vec{\sigma} \right] \;\;\;   &   0 \\
              0 & (\vec{\sigma} \cdot \hat{p}) \left[p\sigma_z+ \vec{p}\cdot \vec{\sigma} \right] ( \vec{\sigma} \cdot \hat{p})\left[p\sigma_z+ \vec{p}\cdot \vec{\sigma} \right] ( \vec{\sigma} \cdot \hat{p})
               \end{pmatrix}\nonumber\\
     &=& \dfrac{\hbar}{4p(p+p_z)}
    \begin{pmatrix}
             \left[p\sigma_z+ \vec{p}\cdot \vec{\sigma} \right] (\vec{\sigma} \cdot \hat{p}) \left[p\sigma_z+ \vec{p}\cdot \vec{\sigma} \right] \;\;\;   &   0 \\
              0 & \left[(p+p_z) \openone +{\rm i} (p_y \sigma_x-p_x\sigma_y)\right]^2 ( \vec{\sigma} \cdot \hat{p})
               \end{pmatrix}\nonumber\\
     &=& \dfrac{\hbar}{4p(p+p_z)}
    \begin{pmatrix}
             ( \vec{\sigma} \cdot \hat{p}) \left[(p+p_z) \openone +{\rm i} (p_y \sigma_x-p_x\sigma_y)\right]^2 \;\;\;   &   0 \\
              0 & \left[(p+p_z) \openone +{\rm i} (p_y \sigma_x-p_x\sigma_y)\right]^2 ( \vec{\sigma} \cdot \hat{p})
               \end{pmatrix}\nonumber\\
    &=& \dfrac{\hbar}{4p(p+p_z)}
    \begin{pmatrix}
             ( \vec{\sigma} \cdot \hat{p}) \left[2p_z(p+p_z) \openone +2 {\rm i} (p+p_z)(p_y \sigma_x-p_x\sigma_y)\right] \;\;\;   &   0 \\
              0 & \left[2p_z(p+p_z) \openone +2 {\rm i} (p+p_z)(p_y \sigma_x-p_x\sigma_y)\right] ( \vec{\sigma} \cdot \hat{p})
               \end{pmatrix}\nonumber\\
               &=& \dfrac{\hbar}{2p}
    \begin{pmatrix}
             ( \vec{\sigma} \cdot \hat{p}) \left[p_z \openone + {\rm i} (p_y \sigma_x-p_x\sigma_y)\right] \;\;\;   &   0 \\
              0 & \left[p_z \openone + {\rm i} (p_y \sigma_x-p_x\sigma_y)\right] ( \vec{\sigma} \cdot \hat{p})
               \end{pmatrix}\nonumber\\
              &=& \dfrac{\hbar}{2p}
     \begin{pmatrix}
             p \sigma_z \;\;\;   &   0 \\
              0 & (\vec{\sigma} \cdot \hat{p}) (p \sigma_z )(\vec{\sigma} \cdot \hat{p})
               \end{pmatrix}\nonumber\\
              &=& \dfrac{\hbar}{2}
     \begin{pmatrix}
             \sigma_z \;\;\;   &   0 \\
              0 & (\vec{\sigma} \cdot \hat{p}) \sigma_z (\vec{\sigma} \cdot \hat{p})
               \end{pmatrix}.
    \end{eqnarray}
    $\blacksquare$
    \end{remark}

    \begin{remark}\textcolor{blue}{Operators that Commute with the Hamiltonian.} We have the following theorem.
    \begin{theorem}For the operator $\mathcal{O}$, if it has the following form
     \begin{eqnarray}\label{eq:structure}
     \mathcal{O} &=&
     \begin{pmatrix}
             \mathbb{T} \;\;\;   &   0 \\
              0 & (\vec{\sigma} \cdot \hat{p}) \mathbb{T} (\vec{\sigma} \cdot \hat{p})
               \end{pmatrix},
    \end{eqnarray}
    then one has
    \begin{eqnarray}
     [H, \, \mathcal{O}]=0.
    \end{eqnarray}
    \end{theorem}
    \begin{proof}
    The Hamiltonian $H$ can be written in the matrix form as follows
    \begin{eqnarray}
    H &=& c\, \vec{\alpha}\cdot \vec{p} +\beta m c^2=
    \begin{pmatrix}
              m c^2 \openone & c \left(\vec{\sigma}\cdot\vec{p}\right) \\
              c \left(\vec{\sigma}\cdot\vec{p}\right) & -m c^2 \openone
            \end{pmatrix},
    \end{eqnarray}
    then one has
    \begin{eqnarray}
    [H, \, \mathcal{O}] &=& \begin{pmatrix}
              m c^2 \openone & c \left(\vec{\sigma}\cdot\vec{p}\right) \\
              c \left(\vec{\sigma}\cdot\vec{p}\right) & -m c^2 \openone
            \end{pmatrix}
            \begin{pmatrix}
             \mathbb{T} \;\;\;   &   0 \\
              0 & (\vec{\sigma} \cdot \hat{p}) \mathbb{T} (\vec{\sigma} \cdot \hat{p})
               \end{pmatrix}-
               \begin{pmatrix}
             \mathbb{T} \;\;\;   &   0 \\
              0 & (\vec{\sigma} \cdot \hat{p}) \mathbb{T} (\vec{\sigma} \cdot \hat{p})
               \end{pmatrix}
            \begin{pmatrix}
              m c^2 \openone & c \left(\vec{\sigma}\cdot\vec{p}\right) \\
              c \left(\vec{\sigma}\cdot\vec{p}\right) & -m c^2 \openone
            \end{pmatrix}\nonumber\\
           &=&  \begin{pmatrix}
              m c^2 \openone \mathbb{T}-\mathbb{T} m c^2 \openone \;\;\;& c \left(\vec{\sigma}\cdot\vec{p}\right) (\vec{\sigma} \cdot \hat{p}) \mathbb{T} (\vec{\sigma} \cdot \hat{p})-\mathbb{T}c \left(\vec{\sigma}\cdot\vec{p}\right) \\
              c \left(\vec{\sigma}\cdot\vec{p}\right) \mathbb{T} - (\vec{\sigma} \cdot \hat{p}) \mathbb{T} (\vec{\sigma} \cdot \hat{p}) c \left(\vec{\sigma}\cdot\vec{p}\right) & -m c^2 \openone (\vec{\sigma} \cdot \hat{p}) \mathbb{T} (\vec{\sigma} \cdot \hat{p})
              +(\vec{\sigma} \cdot \hat{p}) \mathbb{T} (\vec{\sigma} \cdot \hat{p}) m c^2 \openone
            \end{pmatrix}\nonumber\\
            &=& \begin{pmatrix}
               0 & 0 \\
             0 & 0
            \end{pmatrix}=0.
    \end{eqnarray}
    This ends the proof.
    \end{proof}
    One may easily verify that the operators $\mathcal{O}$ can be the following quantities:

    (i) The total angular momentum operator $\vec{J}=\vec{L}+\vec{S}$, with $\vec{L}=\vec{r}\times \vec{p}$. Due to
     \begin{eqnarray}
              \left[\vec{L}+\dfrac{\hbar}{2}\vec{\sigma},\, \left(\vec{\sigma}\cdot\hat{p}\right)\right]=0,
    \end{eqnarray}
     one may easily check that
     \begin{eqnarray}
     \vec{J} &=& \vec{L}+\vec{S}=
     \begin{pmatrix}
              \vec{L}+\dfrac{\hbar}{2}\vec{\sigma}   &   0 \\
              0 &  \vec{L}+\dfrac{\hbar}{2}\vec{\sigma}
               \end{pmatrix}=
               \begin{pmatrix}
              \vec{L}+\dfrac{\hbar}{2}\vec{\sigma}   &   0 \\
              0 &  \left(\vec{\sigma}\cdot\hat{p}\right)\left[\vec{L}+\dfrac{\hbar}{2}\vec{\sigma}\right]\left(\vec{\sigma}\cdot\hat{p}\right)
               \end{pmatrix}.
    \end{eqnarray}

    (ii) The helicity operator $\hat{\Lambda}$. One may easily check that
     \begin{eqnarray}
    \hat{\Lambda} &=& \vec{S} \cdot \hat{p}=\dfrac{\hbar}{2}\, \left(\vec{\Sigma} \cdot \hat{p}\right)=\dfrac{\hbar}{2}\,
     \begin{pmatrix}
              \vec{\sigma}\cdot\hat{p}   &   0 \\
              0 &  \vec{\sigma}\cdot\hat{p}
               \end{pmatrix}=
               \begin{pmatrix}
              \vec{\sigma}\cdot\vec{p}   &   0 \\
              0 &  \left(\vec{\sigma}\cdot\hat{p}\right)\left(\vec{\sigma}\cdot\hat{p}\right)\left(\vec{\sigma}\cdot\hat{p}\right)
               \end{pmatrix}.
    \end{eqnarray}

     (iii) The ``rotated'' helicity operator $\hat{\Lambda}'$. One may directly check that
      \begin{eqnarray}
     \hat{\Lambda}'  &=& \dfrac{\hbar}{2}
     \begin{pmatrix}
             \sigma_z \;\;\;   &   0 \\
              0 & (\vec{\sigma} \cdot \hat{p}) \sigma_z (\vec{\sigma} \cdot \hat{p})
               \end{pmatrix}
    \end{eqnarray}
     automatically satisfies the structure of Eq. (\ref{eq:structure}). If one defines
      \begin{eqnarray}
     \hat{\Lambda}_x'  &=& \dfrac{\hbar}{2}
     \begin{pmatrix}
             \sigma_x \;\;\;   &   0 \\
              0 & (\vec{\sigma} \cdot \hat{p}) \sigma_x (\vec{\sigma} \cdot \hat{p})
               \end{pmatrix}, \nonumber\\
     \hat{\Lambda}_y'  &=& \dfrac{\hbar}{2}
     \begin{pmatrix}
             \sigma_y \;\;\;   &   0 \\
              0 & (\vec{\sigma} \cdot \hat{p}) \sigma_y (\vec{\sigma} \cdot \hat{p})
               \end{pmatrix}, \nonumber\\
               \hat{\Lambda}_z'  &=& \dfrac{\hbar}{2}
     \begin{pmatrix}
             \sigma_z \;\;\;   &   0 \\
              0 & (\vec{\sigma} \cdot \hat{p}) \sigma_z (\vec{\sigma} \cdot \hat{p})
               \end{pmatrix},
     \end{eqnarray}
     and
     \begin{eqnarray}
     \vec{\Lambda}'  &=& \left(\hat{\Lambda}_x', \hat{\Lambda}_y', \hat{\Lambda}_z'\right)=\dfrac{\hbar}{2}
     \begin{pmatrix}
             \vec{\sigma} \;\;\;   &   0 \\
              0 & (\vec{\sigma} \cdot \hat{p}) \vec{\sigma} (\vec{\sigma} \cdot \hat{p})
               \end{pmatrix},
    \end{eqnarray}
    he finds that they all commute with the Hamiltonian $H$. In addition, the operator $\vec{\Lambda}'$ satisfies the definition of angular momentum, i.e.,
    \begin{eqnarray}
     \vec{\Lambda}' \times \vec{\Lambda}' ={\rm i}\hbar\, \vec{\Lambda}'.
    \end{eqnarray}
    The relation between the helicity operator and the new ``rotated'' helicity operator $\vec{\Lambda}'$ is as follows
    \begin{eqnarray}
    \hat{\Lambda} &=& \vec{\Lambda}' \cdot \hat{p} = \dfrac{\hbar}{2} \begin{pmatrix}
              \vec{\sigma}\cdot\hat{p}   &   0 \\
              0 &  \vec{\sigma}\cdot\hat{p}
               \end{pmatrix}.
    \end{eqnarray}

    (iv) The unitary matrix $\mathcal{M}$ shown in Eq. (\ref{eq:dc-4}). One may easily check that the operator $\mathcal{M}$ automatically satisfies the structure of Eq. (\ref{eq:structure}). $\blacksquare$
    \end{remark}

\newpage

\section{The Spin Direction of Dirac's Electron}

We only restrict to the case of positive energy (i.e., $ E=+ \mathcal{E}$), namely, one assumes that Dirac's electron is always in a state of positive energy during its motion. In other words, the electron does not make a transition from positive energy to negative energy. In the future work, we shall consider the more complicate case of energy transition.

It is interesting to know the spin direction of Dirac's electron when it is in a positive-energy state. In this section, we would like to discuss this issue.

\begin{remark}
   For a particle with momentum $\vec{p} =(p_x , p_y, p_z )$, its spin direction is $\vec{\tilde{s}}=\dfrac{\hbar}{2} (0,0, 1) $ in its rest frame.
    We can use ${\tilde{s}}^{\mu} = ({\tilde{s}}^{0},\vec{\tilde{s}})$ (or ${\tilde{s}}_{\mu} = ({\tilde{s}}_{0},-\vec{\tilde{s}})$ ) to represent the spin direction of the particle in the rest frame and ${\tilde{s}}^{0}=0$ \cite{greiner_projection_2000}.

    In the relativistic case, according to the textbook of quantum mechanics, we assume $s^{\mu}$ is a 4-vector and it follows the law of Lorentz transformation like the 4-coordinate vector $x^{\mu } = (ct, x, y, z)$. For example, we consider a representative case. In an inertial reference frame $x^{\mu } = ( c t , x,y,z )$, there is a particle with momentum $\vec{p}= (p,0 , 0)$ and ${\tilde{s}}^{\mu}= (0,{\tilde{s}}_{x},{\tilde{s}}_{y},0 )$, i.e., $p_y=p_z=0$ and ${\tilde{s}}_{z}=0$.
    Its velocity is {$\vec{v} = \dfrac{pc^2}{\mathcal{E}} (1, 0, 0)$}. If we designate the 4-coordinate vector in the rest frame of the particle is ${\tilde{x}}^{\mu}= (c \tilde{t}, \tilde{x},\tilde{y},\tilde{z})$, then the Lorentz transformation gives
    \begin{equation}\label{eq:Lr-1}
        \begin{cases}
            & {c t=\gamma\left(c\tilde{t}+\overline{\beta} \tilde{x}\right),}\\
              & x=\gamma(\tilde{x}+ \overline{\beta} c \,\tilde{t}), \\
              & y=\tilde{y}, \\
              & z={\tilde{z}},
        \end{cases}
    \end{equation}
 with {$\overline{\beta} =\dfrac{v}{c}= \dfrac{pc}{\mathcal{E}}$}, $\gamma = \dfrac{1}{\sqrt{ 1- \overline{\beta}^{\,2}} }  = \dfrac{\mathcal{E}}{mc^2}$. Here, we have used $\overline{\beta}$ to denote $\dfrac{v}{c}$, in order to distinguish it from the previous Dirac matrix $\beta$.

For the spin direction $s^{\mu}$, we similarly have
\begin{equation}\label{eq:Lr-2}
    \begin{cases}
        & s^0=\gamma\left({\tilde{s}}^{0}{+}\overline{\beta} {\tilde{s}}_{x}\right)=\gamma\overline{\beta} {\tilde{s}}_{x},\\
          & s_x=\gamma\left({\tilde{s}}_{x}+ {\overline{\beta} {\tilde{s}}^{0}}\right)=\gamma{\tilde{s}}_{x}, \\
          & s_y={\tilde{s}}_{y}, \\
          & s_z={\tilde{s}}_{z}= 0.
    \end{cases}
\end{equation}
In general, one has
\begin{eqnarray}
    s^0 = \gamma \overline{\beta} (\hat{p} \cdot \vec{\tilde{ s}}), \;\;\;\;  \vec{s} = (\gamma-1) (\hat{p} \cdot \vec{\tilde{ s}}) \hat{p} + \vec{\tilde{ s}},
\end{eqnarray}
and
\begin{eqnarray}
    s^{\mu} p_{\mu}= s_{\mu} p^{\mu } = s^0 p_0 - \vec{s } \cdot \vec{p} = \gamma \overline{\beta} (\hat{p} \cdot \vec{\tilde{ s}}) \frac{\mathcal{E}}{c} -  \gamma (\hat{p} \cdot \vec{\tilde{ s}}) p=\gamma  (\hat{p} \cdot \vec{\tilde{ s}}) p-  \gamma (\hat{p} \cdot \vec{\tilde{ s}}) p=0,
\end{eqnarray}
where $p^{\mu}= (p_0 , \vec{p})$ and $p_0= \dfrac{\mathcal{E}}{c}$. $\blacksquare$
\end{remark}

\begin{remark}
    $\ket{\Psi_1'}$ can represent a Dirac's particle with momentum $\vec{p} =(p_x , p_y, p_z )$, whose spin direction is $\vec{\tilde{s}}=\dfrac{\hbar}{2} (0,0, 1) $ in its rest frame. However, it is easily to prove that $\ket{\Psi_1'}$ is not the eigenstate of
\begin{eqnarray}
S_z= \dfrac{\hbar  }{2} {\Sigma}_z = \left(\begin{array}{cc}
        \sigma_z & 0 \\
      0 & \sigma_z \\
      \end{array}
    \right),
\end{eqnarray}
i.e.,
\begin{eqnarray}
S_z \, \ket{\Psi_1'} \neq  \dfrac{\hbar  }{2} \ket{\Psi_1'}.
\end{eqnarray}
This means that $\vec{S}$ is not appropriate to point the spin direction of Dirac's electron. Therefore, we need to choose another operator to reach this purpose. Fortunately, the following operator
\begin{eqnarray}
\Gamma=\gamma_5 \gamma^{\mu} s_{\mu }
\end{eqnarray}
is a good choice \cite{greiner_projection_2000}, where
    \begin{eqnarray}
 &&       \gamma_5 = \begin{pmatrix}
            0&\openone\\
            \openone&0
        \end{pmatrix}, \nonumber\\
&& \gamma^{\mu}= (\beta,\, \beta \vec{\alpha}) = \left(
    \begin{pmatrix}
        \openone &0 \\
        0 & -\openone
\end{pmatrix},
\begin{pmatrix}
    0 & \vec{\sigma}\\
    -\vec{\sigma} & 0
\end{pmatrix}
\right).
    \end{eqnarray}
With the help of
\begin{eqnarray}
    \gamma_5 \gamma^{\mu} = \left(
\begin{pmatrix}
    0 & -\openone \\
    \openone  &0
\end{pmatrix},
\begin{pmatrix}
    -\vec{\sigma}  & 0\\
    0 &  \vec{\sigma}
\end{pmatrix}
    \right).
\end{eqnarray}
one can have
\begin{eqnarray}
    \Gamma=\gamma_5 \gamma^{\mu} s_{\mu } = \left\{
\begin{pmatrix}
    0 & -\openone s^0\\
    \openone s^0 &0
\end{pmatrix}+
\begin{pmatrix}
    \vec{\sigma} \cdot \vec{s} & 0\\
    0 & - \vec{\sigma} \cdot \vec{s}
\end{pmatrix}
    \right\}
    =\begin{pmatrix}
        \vec{\sigma} \cdot \vec{s} & -\openone s^0\\
        \openone s^0& -\vec{\sigma} \cdot \vec{s}
    \end{pmatrix},
\end{eqnarray}
where $s_{\mu }=(s_0, -\vec{s})$.

If we designate
\begin{eqnarray}
|\chi_1\rangle = \begin{pmatrix}
    1\\
    0
\end{pmatrix}=|\uparrow\rangle_z, \;\;\;\;
|\chi_2\rangle = \begin{pmatrix}
    0\\
    1
\end{pmatrix}=|\downarrow\rangle_z,
\end{eqnarray}
which represent the spin up and spin down along the $z$-axis respectively, then
according to \Eq{eqW'}, we have
\begin{eqnarray}
    \ket{\Psi_1'} =\frac{1}{\sqrt{2\mathcal{E}(\mathcal{E} +mc^2)}}  \begin{pmatrix}
        (\mathcal{E}+mc^2) |\chi_1\rangle\\
        c (\vec{\sigma} \cdot \vec{p}) |\chi_1\rangle
    \end{pmatrix},
\end{eqnarray}
\begin{eqnarray}
    \ket{\Psi_2'} = \frac{1}{\sqrt{2\mathcal{E}(\mathcal{E} +mc^2)}} \begin{pmatrix}
        (\mathcal{E}+mc^2) |\chi_2\rangle\\
        c (\vec{\sigma} \cdot \vec{p}) |\chi_2\rangle
    \end{pmatrix}.
\end{eqnarray}

For $\ket{\Psi_1'}$, we have
\begin{eqnarray}\label{eq:gam-1}
    \Gamma \ket{\Psi_1'}=\gamma_5 \gamma^{\mu} s_{\mu } \ket{\Psi_1'} = \frac{1}{\sqrt{2\mathcal{E}(\mathcal{E} +mc^2)}} \begin{pmatrix}
        (\mathcal{E}+mc^2) (\vec{\sigma} \cdot \vec{s}) |\chi_1\rangle-c s^0(\vec{\sigma} \cdot \vec{p}) |\chi_1\rangle\\
        (\mathcal{E}+mc^2) s^0 |\chi_1\rangle-c (\vec{\sigma} \cdot \vec{s})(\vec{\sigma} \cdot \vec{p}) |\chi_1\rangle
    \end{pmatrix}.
\end{eqnarray}
\begin{eqnarray}
    (\mathcal{E}+mc^2) (\vec{\sigma} \cdot \vec{s}) |\chi_1\rangle -c s^0(\vec{\sigma} \cdot \vec{p}) |\chi_1\rangle
   &=&\left\{(\mathcal{E}+mc^2) (\gamma-1) (\hat{p} \cdot \vec{\tilde{ s}}) (\vec{\sigma} \cdot \hat{p}) + (\mathcal{E}+mc^2) (\vec{\sigma} \cdot \vec{\tilde{ s}})-c \gamma \overline{\beta} (\hat{p} \cdot \vec{\tilde{ s}})(\vec{\sigma} \cdot \vec{p}) \right\} |\chi_1\rangle\notag\\
   &=&\left\{\frac{\mathcal{E}^2 - m^2c^4}{mpc^2} (\hat{p} \cdot \vec{\tilde{ s}}) (\vec{\sigma} \cdot \vec{p}) + (\mathcal{E}+mc^2) (\vec{\sigma} \cdot \vec{\tilde{ s}})-c \frac{\mathcal{E} pc}{mc^2 \mathcal{E}} (\hat{p} \cdot \vec{\tilde{ s}})(\vec{\sigma} \cdot \vec{p}) \right\} |\chi_1\rangle \notag\\
   &=&\left\{\frac{p^2c^2}{mpc^2}-c \frac{\mathcal{E} pc}{mc^2 \mathcal{E}} \right\}(\hat{p} \cdot \vec{\tilde{ s}}) (\vec{\sigma} \cdot \vec{p})|\chi_1\rangle + (\mathcal{E}+mc^2) (\vec{\sigma} \cdot \vec{\tilde{ s}}) |\chi_1\rangle\notag\\
   &=&(\mathcal{E}+mc^2) (\vec{\sigma} \cdot \vec{\tilde{ s}}) |\chi_1\rangle,
\end{eqnarray}
\begin{eqnarray}
    (\mathcal{E}+mc^2) s^0\chi_1-c(\vec{\sigma} \cdot \vec{s})(\vec{\sigma} \cdot \vec{p}) |\chi_1\rangle
  &=&\left\{\gamma \overline{\beta} (\hat{p} \cdot \vec{\tilde{ s}})(\mathcal{E}+mc^2) - c(\gamma-1) (\hat{p} \cdot \vec{\tilde{ s}})(\vec{\sigma}\cdot\hat{p})(\vec{\sigma} \cdot \vec{p}) -c(\vec{\sigma}\cdot\vec{\tilde{ s}})(\vec{\sigma} \cdot \vec{p})\right\} |\chi_1\rangle \notag\\
  &=&\left\{\frac{(\mathcal{E}+mc^2)p}{mc} (\hat{p} \cdot \vec{\tilde{ s}})- \frac{(\mathcal{E}-mc^2)p}{mc}(\hat{p} \cdot \vec{\tilde{ s}}) -c(\vec{\sigma}\cdot\vec{\tilde{ s}})(\vec{\sigma} \cdot \vec{p})\right\} |\chi_1\rangle\notag\\
  &=&\left\{\frac{(2mc^2)p}{mc} (\hat{p} \cdot \vec{\tilde{ s}}) -c(\vec{\sigma}\cdot\vec{\tilde{ s}})(\vec{\sigma} \cdot \vec{p})\right\} |\chi_1\rangle\nonumber\\
  &=&\left\{2c (\vec{p} \cdot \vec{\tilde{ s}}) -c(\vec{\sigma}\cdot\vec{\tilde{ s}})(\vec{\sigma} \cdot \vec{p})\right\} |\chi_1\rangle.
\end{eqnarray}
Because
\begin{eqnarray}
    (\vec{\sigma}\cdot\vec{\tilde{ s}})(\vec{\sigma} \cdot \vec{p})= \vec{\tilde{ s}}\cdot \vec{p} -{\rm i} (\vec{\tilde{ s}} \times \vec{p}) \cdot \vec{\sigma},
\end{eqnarray}
\begin{eqnarray}
    (\vec{\sigma} \cdot \vec{p}) (\vec{\sigma}\cdot\vec{\tilde{ s}}) = \vec{\tilde{ s}}\cdot \vec{p} + {\rm i} (\vec{\tilde{ s}} \times \vec{p}) \cdot \vec{\sigma},
\end{eqnarray}
\begin{eqnarray}
    (\vec{\sigma}\cdot\vec{\tilde{ s}})(\vec{\sigma} \cdot \vec{p}) = 2\vec{\tilde{ s}}\cdot \vec{p}- (\vec{\sigma} \cdot \vec{p}) (\vec{\sigma}\cdot\vec{\tilde{ s}}),
\end{eqnarray}
then we have
\begin{eqnarray}
    (\mathcal{E}+mc^2) s^0 |\chi_1\rangle -c(\vec{\sigma} \cdot \vec{s})(\vec{\sigma} \cdot \vec{p}) |\chi_1\rangle
   &=&\left\{2c (\vec{p} \cdot \vec{\tilde{ s}}) -c(\vec{\sigma}\cdot\vec{\tilde{ s}})(\vec{\sigma} \cdot \vec{p})\right\} |\chi_1\rangle
   = c(\vec{\sigma} \cdot \vec{p}) (\vec{\sigma}\cdot\vec{\tilde{ s}}) |\chi_1\rangle.
\end{eqnarray}
Finally, Eq. (\ref{eq:gam-1}) becomes
\begin{eqnarray}\label{eq:gam-2}
    \Gamma \ket{\Psi_1'}&=&\gamma_5 \gamma^{\mu} s_{\mu } \ket{\Psi_1'} = \frac{1}{\sqrt{2\mathcal{E}(\mathcal{E} +mc^2)}} \begin{pmatrix}
        (\mathcal{E}+mc^2) (\vec{\sigma} \cdot \vec{s}) |\chi_1\rangle -c s^0(\vec{\sigma} \cdot \vec{p}) |\chi_1\rangle \\
        (\mathcal{E}+mc^2) s^0 |\chi_1\rangle-c (\vec{\sigma} \cdot \vec{s})(\vec{\sigma} \cdot \vec{p})|\chi_1\rangle
    \end{pmatrix}\nonumber\\
 &=&\frac{1}{\sqrt{2\mathcal{E}(\mathcal{E} +mc^2)}} \begin{pmatrix}
        (\mathcal{E}+mc^2) (\vec{\sigma}\cdot\vec{\tilde{ s}}) |\chi_1\rangle  \\
        c(\vec{\sigma} \cdot \vec{p}) (\vec{\sigma}\cdot\vec{\tilde{ s}}) |\chi_1\rangle
    \end{pmatrix}.
\end{eqnarray}
Obviously, when {$\vec{\tilde{s}} = \dfrac{\hbar}{2}(0,0,1)$,} one has
\begin{eqnarray}
    (\vec{\sigma}\cdot\vec{\tilde{ s}}) \chi_1= \frac{\hbar}{2}\sigma_z\chi_1= \frac{\hbar}{2} \chi_1,
\end{eqnarray}
and
\begin{eqnarray}
   \Gamma \ket{\Psi_1'}&=& \gamma_5 \gamma^{\mu} s_{\mu } \ket{\Psi_1'} =\frac{1}{\sqrt{2\mathcal{E}(\mathcal{E} +mc^2)}} \begin{pmatrix}
        (\mathcal{E}+mc^2) (\vec{\sigma}\cdot\vec{\tilde{ s}}) |\chi_1\rangle  \\
        c(\vec{\sigma} \cdot \vec{p}) (\vec{\sigma}\cdot\vec{\tilde{ s}}) |\chi_1\rangle
    \end{pmatrix}\nonumber\\
    &=&\frac{1}{\sqrt{2\mathcal{E}(\mathcal{E} +mc^2)}} \frac{\hbar}{2}\begin{pmatrix}
        (\mathcal{E}+mc^2) |\chi_1\rangle  \\
        c(\vec{\sigma} \cdot \vec{p}) |\chi_1\rangle
    \end{pmatrix}=
    \frac{\hbar}{2} \ket{\Psi_1'}.
\end{eqnarray}
Therefore $\ket{\Psi_1'}$ can represent a particle with momentum $\vec{p} =(p_x , p_y, p_z )$ with spin direction is $\vec{\tilde{s}}=\dfrac{\hbar}{2} (0,0, 1)$. Similarly, $\ket{\Psi_2'}$ can represent a particle with momentum $\vec{p} =(p_x , p_y, p_z )$ with spin direction is $\vec{\tilde{s}}=\dfrac{\hbar}{2} (0,0, -1)$.


For a general superposition state $\ket{\Psi'}$, i.e.,
\begin{eqnarray}\label{psi10}
    \ket{\Psi'} &=&  \ell_1 \ket{\Psi_1'}+ \ell_2 \ket{\Psi_2'}\notag\\
    &=& \frac{1}{\sqrt{2\mathcal{E}(\mathcal{E} +mc^2)}} \begin{pmatrix}
        (\mathcal{E}+mc^2) (\ell_1 |\chi_1\rangle+\ell_2 |\chi_2\rangle)  \\
        c(\vec{\sigma} \cdot \vec{p}) (\ell_1 |\chi_1\rangle+\ell_2 |\chi_2\rangle)
    \end{pmatrix},
\end{eqnarray}
if we designate
\begin{eqnarray}\label{chi1}
    \chi=\ell_1 |\chi_1\rangle+\ell_2 |\chi_2\rangle =\begin{pmatrix}
    \ell_1\\
    \ell_2
\end{pmatrix},
\end{eqnarray}
we can attain
\begin{eqnarray}\label{eq:psichi-1}
    \ket{\Psi'}=\frac{1}{\sqrt{2\mathcal{E}(\mathcal{E} +mc^2)}}  \begin{pmatrix}
        (\mathcal{E}+mc^2) |\chi\rangle  \\
        c(\vec{\sigma} \cdot \vec{p}) |\chi\rangle
    \end{pmatrix}.
\end{eqnarray}
Similarly, one can have
\begin{eqnarray}
  \Gamma \ket{\Psi'}&=&  \gamma_5 \gamma^{\mu} s_{\mu } \ket{\Psi'}
    = \frac{1}{\sqrt{2\mathcal{E}(\mathcal{E} +mc^2)}}  \begin{pmatrix}
        (\mathcal{E}+mc^2) (\vec{\sigma}\cdot\vec{\tilde{ s}})  |\chi\rangle  \\
        c(\vec{\sigma} \cdot \vec{p}) (\vec{\sigma}\cdot\vec{\tilde{ s}}) |\chi\rangle
    \end{pmatrix},
\end{eqnarray}
When $\chi $ is the eigenstate of $ (\vec{\sigma}\cdot\vec{\tilde{ s}})$, i.e.,
\begin{eqnarray}\label{eq:sdire}
    (\vec{\sigma}\cdot\vec{\tilde{ s}}) |\chi\rangle = \frac{\hbar}{2} |\chi\rangle,
\end{eqnarray}
we also can have
\begin{eqnarray}
  \Gamma \ket{\Psi'}&=&  \gamma_5 \gamma^{\mu} s_{\mu } \ket{\Psi'}= \frac{\hbar}{2}\ket{\Psi'}.
\end{eqnarray}
Now, one asks a question: For the given state $\ket{\Psi'}$, what is its spin direction? The following is the answer. If $\ket{\Psi'}$ is given, then the coefficients $\ell_1$ and $\ell_2$ are fixed. For simplicity, we parameterized them as
\begin{eqnarray}
 \ell_1=\cos\frac{\theta}{2}, \;\;\;  \ell_2=\sin\frac{\theta}{2}\, {\rm e}^{{\rm i}\phi},
\end{eqnarray}
then we have
\begin{eqnarray}
    |\chi\rangle =\begin{pmatrix}
    \cos\frac{\theta}{2}\\
    \sin\frac{\theta}{2}\, {\rm e}^{{\rm i}\phi}
\end{pmatrix}.
\end{eqnarray}
From Eq. (\ref{eq:sdire}) one can work out the spin direction as
\begin{eqnarray}
\vec{\tilde{ s}} = \frac{\hbar}{2}\, \vec{\tau},
\end{eqnarray}
with
\begin{eqnarray}
\vec{\tau}=(\sin\theta\cos\phi, \sin\theta\sin\phi, \cos\theta).
\end{eqnarray}
When $\theta=0$, one has $\vec{\tilde{ s}}=\frac{\hbar}{2} (0, 0, 1)$ or $\vec{\tau}=(0, 0, 1)$; and when $\theta=\pi$, one has $\vec{\tilde{ s}}=\frac{\hbar}{2} (0, 0, -1)$, or $\vec{\tau}=(0, 0, -1)$.

Let us consider the non-relativistic limit, i.e., $(\mathcal{E}+mc^2 ) \openone \gg( \sigma \cdot \vec{p} )$, then one has
\begin{eqnarray}
    \ket{\Psi'_1}=\frac{1}{\sqrt{2\mathcal{E}(\mathcal{E} +mc^2)}} \begin{pmatrix}
        (\mathcal{E}+mc^2) |\chi_1\rangle  \\
        c(\vec{\sigma} \cdot \vec{p}) |\chi_1\rangle
    \end{pmatrix}\approx \frac{1}{\sqrt{2\mathcal{E}(\mathcal{E} +mc^2)}} \begin{pmatrix}
        (\mathcal{E}+mc^2) |\chi_1\rangle  \\
        0
    \end{pmatrix},
\end{eqnarray}
\begin{eqnarray}
\ket{\Psi'_2}=\frac{1}{\sqrt{2\mathcal{E}(\mathcal{E} +mc^2)}} \begin{pmatrix}
        (\mathcal{E}+mc^2) |\chi_2\rangle  \\
        c(\vec{\sigma} \cdot \vec{p}) |\chi_2\rangle
    \end{pmatrix}\approx\frac{1}{\sqrt{2\mathcal{E}(\mathcal{E} +mc^2)}} \begin{pmatrix}
        (\mathcal{E}+mc^2) |\chi_2\rangle  \\
       0
    \end{pmatrix} \,,\,
\end{eqnarray}
\begin{eqnarray}\label{psi11}
\ket{\Psi'}=\frac{1}{\sqrt{2\mathcal{E}(\mathcal{E} +mc^2)}}  \begin{pmatrix}
        (\mathcal{E}+mc^2) |\chi \rangle  \\
        c(\vec{\sigma} \cdot \vec{p}) |\chi\rangle
    \end{pmatrix}\approx \frac{1}{\sqrt{2\mathcal{E}(\mathcal{E} +mc^2)}} \begin{pmatrix}
        (\mathcal{E}+mc^2) |\chi\rangle  \\
       0
    \end{pmatrix}.
\end{eqnarray}
\Eq{psi11} is very similar to the representation of spin in the non-relativistic quantum mechanics. When $|\chi\rangle $ is the eigenstate of $\vec{\sigma} \cdot \vec{\tilde{s}}$ (which is the spin  operator with a specific direction in non-relativistic case), i.e.,
\begin{eqnarray}
    \vec{\sigma} \cdot \vec{\tilde{s}}\, |\chi\rangle = \frac{\hbar }{2} |\chi\rangle,
\end{eqnarray}
we also can have
\begin{eqnarray}
    \left(\openone\otimes \vec{\sigma} \cdot \vec{\tilde{s}} \right)\, \ket{\Psi'} \approx \frac{\hbar }{2} \ket{\Psi'},
\end{eqnarray}
\begin{eqnarray}
    \Gamma \ket{\Psi'}=\gamma_5 \gamma^{\mu} s_{\mu } \ket{\Psi'}= \frac{\hbar}{2}\ket{\Psi'}.
\end{eqnarray}
It indicates that $\Gamma= \gamma_5 \gamma^{\mu} s_{\mu } $ can indeed point out the spin direction of Dirac's electron in relativistic case.  $\blacksquare$
\end{remark}

\begin{remark} We continue to talk about the relation between $\Gamma=\gamma_5 \gamma^{\mu} s_{\mu } $ and the Hamiltonian $H$.
We can attain that
\begin{eqnarray}
    \Gamma=\gamma_5 \gamma^{\mu}s_{\mu} = \left\{\begin{pmatrix}
        0& -\openone \\
        \openone & 0
    \end{pmatrix}s_0 +
     \begin{pmatrix}
        \vec{\sigma} \cdot \vec{s} &0 \\
        0 &-  \vec{\sigma} \cdot \vec{s}
     \end{pmatrix}\right\}
\end{eqnarray}
For $ \gamma_5 \gamma^{0}$ , we can have that
\begin{eqnarray}
  \left[ \begin{pmatrix}
        0& -\openone \\
        \openone & 0
    \end{pmatrix}, \, \vec{\alpha} \cdot \vec{p}\right]
    =
    2\begin{pmatrix}
        - \vec{\sigma} \cdot \vec{p} &0\\
       0& \vec{\sigma} \cdot \vec{p}
    \end{pmatrix},
\end{eqnarray}
\begin{eqnarray}
    \left[ \begin{pmatrix}
          0& -\openone \\
          \openone & 0
      \end{pmatrix},  \beta\right]
      =
      2\begin{pmatrix}
         0&\openone\\
         \openone& 0
      \end{pmatrix}.
  \end{eqnarray}
So we have
\begin{eqnarray}
    \left[ \begin{pmatrix}
          0& -\openone \\
          \openone & 0
      \end{pmatrix},  H\right]
      =
      2c\begin{pmatrix}
          - \vec{\sigma} \cdot \vec{p} &0\\
         0& \vec{\sigma} \cdot \vec{p}
      \end{pmatrix}
      + 2mc^2\begin{pmatrix}
        0&\openone\\
        \openone& 0
     \end{pmatrix}
 \end{eqnarray}

 For $ \gamma_5 \gamma^{i}$, we have
 \begin{eqnarray}
   \left[ \begin{pmatrix}
        \vec{\sigma} \cdot \vec{s} &0 \\
        0 &-  \vec{\sigma} \cdot \vec{s}
     \end{pmatrix}, H\right] =
     2c\begin{pmatrix}
        0&-2 s_i p^i\\
        2 s_i p^i& 0
     \end{pmatrix}.
 \end{eqnarray}
Then we have
 \begin{eqnarray}
   [\gamma_5 \gamma^{\mu} s_{\mu} , H ] =   2s_0c\begin{pmatrix}
    - \vec{\sigma} \cdot \vec{p} &0\\
   0& \vec{\sigma} \cdot \vec{p}
\end{pmatrix}
+ 2s_0mc^2\begin{pmatrix}
  0&\openone\\
  \openone& 0
\end{pmatrix} +
2c\begin{pmatrix}
    0&-2s_ip^i\\
    2s_ip^i& 0
 \end{pmatrix} .
 \end{eqnarray}
By using $s_{\mu} p^{\mu} =s_{0} p^{0}+s_{i} p^{i} =s_{0} p^{0}-\vec{s}\cdot\vec{p} =0$ and $p_0 c = \mathcal{E} $, we finally have
 \begin{eqnarray}\label{eq:gamH-1}
    [\Gamma, H]=[\gamma_5 \gamma^{\mu} s_{\mu} , H ] =
    2s_0 \begin{pmatrix}
        - c(\vec{\sigma} \cdot \vec{p})  & \openone (\mathcal{E}+mc^2)\\
        \openone (-\mathcal{E}+mc^2)& c(\vec{\sigma} \cdot \vec{p})
    \end{pmatrix}.
 \end{eqnarray}
Alternatively, one may calculate the commutator $[\Gamma, H]$ directly. One can have
\begin{eqnarray}\label{eq:gamH-2}
    [\Gamma, H]&=& \left[ \begin{pmatrix}
        \vec{\sigma} \cdot \vec{s} & -\openone s^0\\
        \openone s^0& -\vec{\sigma} \cdot \vec{s}
    \end{pmatrix}, \, \begin{pmatrix}
              m c^2 \openone & c \left(\vec{\sigma}\cdot\vec{p}\right) \\
              c \left(\vec{\sigma}\cdot\vec{p}\right) & -m c^2 \openone
            \end{pmatrix}\right]\nonumber\\
    &=& \begin{pmatrix}
        \vec{\sigma} \cdot \vec{s} & -\openone s^0\\
        \openone s^0& -\vec{\sigma} \cdot \vec{s}
    \end{pmatrix}\begin{pmatrix}
              m c^2 \openone & c \left(\vec{\sigma}\cdot\vec{p}\right) \\
              c \left(\vec{\sigma}\cdot\vec{p}\right) & -m c^2 \openone
            \end{pmatrix}
            -
    \begin{pmatrix}
              m c^2 \openone & c \left(\vec{\sigma}\cdot\vec{p}\right) \\
              c \left(\vec{\sigma}\cdot\vec{p}\right) & -m c^2 \openone
            \end{pmatrix}\begin{pmatrix}
        \vec{\sigma} \cdot \vec{s} & -\openone s^0\\
        \openone s^0& -\vec{\sigma} \cdot \vec{s}
    \end{pmatrix}\nonumber\\
&=& \begin{pmatrix}
        (\vec{\sigma} \cdot \vec{s}) m c^2 \openone- \openone s^0c \left(\vec{\sigma}\cdot\vec{p}\right) &
        (\vec{\sigma} \cdot \vec{s}) c \left(\vec{\sigma}\cdot\vec{p}\right)+ \openone s^0  m c^2 \openone \\
        -(\vec{\sigma} \cdot \vec{s}) c \left(\vec{\sigma}\cdot\vec{p}\right)+ \openone s^0  m c^2 \openone &
        (\vec{\sigma} \cdot \vec{s}) m c^2 \openone + \openone s^0c \left(\vec{\sigma}\cdot\vec{p}\right)
    \end{pmatrix}-\nonumber\\
&& \begin{pmatrix}
        (\vec{\sigma} \cdot \vec{s}) m c^2 \openone+ \openone s^0c \left(\vec{\sigma}\cdot\vec{p}\right) &
        - c \left(\vec{\sigma}\cdot\vec{p}\right)(\vec{\sigma} \cdot \vec{s})- \openone s^0  m c^2 \openone \\
         c \left(\vec{\sigma}\cdot\vec{p}\right)(\vec{\sigma} \cdot \vec{s})- \openone s^0  m c^2 \openone &
        (\vec{\sigma} \cdot \vec{s}) m c^2 \openone - \openone s^0c \left(\vec{\sigma}\cdot\vec{p}\right)
    \end{pmatrix}\nonumber\\
&=& \begin{pmatrix}
        -2 s^0c \left(\vec{\sigma}\cdot\vec{p}\right) &
         (\vec{\sigma} \cdot \vec{s}) c \left(\vec{\sigma}\cdot\vec{p}\right)+ c \left(\vec{\sigma}\cdot\vec{p}\right)(\vec{\sigma} \cdot \vec{s})+  2 s^0  m c^2 \openone \\
        -(\vec{\sigma} \cdot \vec{s}) c \left(\vec{\sigma}\cdot\vec{p}\right)-c \left(\vec{\sigma}\cdot\vec{p}\right)(\vec{\sigma} \cdot \vec{s})+  2 s^0  m c^2 \openone &
       2  s^0c \left(\vec{\sigma}\cdot\vec{p}\right)
    \end{pmatrix}\nonumber\\
&=& \begin{pmatrix}
        -2 s^0c \left(\vec{\sigma}\cdot\vec{p}\right) &
         2c (\vec{s}\cdot \vec{p})\openone+  2 s^0  m c^2 \openone \\
        -2c (\vec{s}\cdot \vec{p})\openone+  2 s^0  m c^2 \openone &
       2  s^0c \left(\vec{\sigma}\cdot\vec{p}\right)
    \end{pmatrix}\nonumber\\
    &=& \begin{pmatrix}
        -2 s^0c \left(\vec{\sigma}\cdot\vec{p}\right) &
         2c (s^0p_0)\openone+  2 s^0  m c^2 \openone \\
        -2c (s^0p_0)\openone+  2 s^0  m c^2 \openone &
       2  s^0c \left(\vec{\sigma}\cdot\vec{p}\right)
    \end{pmatrix}\nonumber\\
        &=& \begin{pmatrix}
        -2 s^0c \left(\vec{\sigma}\cdot\vec{p}\right) &
         2 (s^0 \mathcal{E})\openone+  2 s^0  m c^2 \openone \\
        -2 (s^0 \mathcal{E})\openone+  2 s^0  m c^2 \openone &
       2  s^0c \left(\vec{\sigma}\cdot\vec{p}\right)
    \end{pmatrix}\nonumber\\
            &=& 2s^0\begin{pmatrix}
        -c \left(\vec{\sigma}\cdot\vec{p}\right) &
          (\mathcal{E}+   m c^2) \openone \\
        (-\mathcal{E}+   m c^2) \openone &
       c \left(\vec{\sigma}\cdot\vec{p}\right)
    \end{pmatrix}\nonumber\\
                &=& 2s_0\begin{pmatrix}
        -c \left(\vec{\sigma}\cdot\vec{p}\right) &
          \openone(\mathcal{E}+   m c^2)  \\
        \openone(-\mathcal{E}+   m c^2)  &
       c \left(\vec{\sigma}\cdot\vec{p}\right)
    \end{pmatrix},
 \end{eqnarray}
 which coincides with Eq. (\ref{eq:gamH-1}).

 If the wavefunction is arbitrary,   $[\Gamma, H]=[\gamma_5 \gamma^{\mu} s_{\mu} , H ] \ne 0 $ is generally valid. However, if we just consider the eighfuncion of $H$, e.g.,
 \begin{eqnarray}
    H \ket{\Psi} = E \ket{\Psi}= \begin{pmatrix}
        E \openone & 0\\
        0 &  E \openone
    \end{pmatrix} \ket{\Psi}.
 \end{eqnarray}
 For this eigenfunction $\ket{\Psi}$, we can have
 \begin{eqnarray}
    \left\{H - \begin{pmatrix}
        E \openone & 0\\
        0 &  E \openone
    \end{pmatrix} \right\}\ket{\Psi} =
\begin{pmatrix}
    \openone (mc^2 -E) & c(\vec{\sigma} \cdot \vec{p})\\
    c(\vec{\sigma} \cdot \vec{p}) &  \openone (-mc^2 -E)
\end{pmatrix} \ket{\Psi} = 0.
 \end{eqnarray}
Through multiplying $\gamma_5 \beta  $ on the left, it becomes
\begin{eqnarray}\label{eq:gamH-3}
    \gamma_5 \beta  \begin{pmatrix}
        \openone (mc^2 -E) & c(\vec{\sigma} \cdot \vec{p})\\
        c(\vec{\sigma} \cdot \vec{p}) & \openone (-mc^2 -E)
    \end{pmatrix} \ket{\Psi} &=& \gamma_5 \begin{pmatrix}
        \openone (mc^2 -E) & c(\vec{\sigma} \cdot \vec{p})\\
        -c(\vec{\sigma} \cdot \vec{p})&   \openone (mc^2 +E)
    \end{pmatrix} \ket{\Psi}
    \notag\\
    &=&\begin{pmatrix}
        - c(\vec{\sigma} \cdot \vec{p})  & \openone (E+mc^2)\\
        \openone (-E+mc^2)& c(\vec{\sigma} \cdot \vec{p})
    \end{pmatrix}\ket{\Psi} =0.
\end{eqnarray}
Let us compare Eq. (\ref{eq:gamH-2}) and Eq. (\ref{eq:gamH-3}), if the energy $E$ is chosen as the positive energy (i.e., $E=+\mathcal{E}$, and $\ket{\Psi}$ is the eigenstate corresponding to positive energy), then Eq. (\ref{eq:gamH-3}) becomes
\begin{eqnarray}\label{eq:gamH-4}
    [\Gamma, H] \ket{\Psi} =0,
\end{eqnarray}
or
\begin{eqnarray}\label{eq:gamH-5}
    \langle \Psi |[\Gamma, H] \ket{\Psi} =0.
\end{eqnarray}
The physical meaning of Eq. (\ref{eq:gamH-5}) is that, in the sub-space of positive energy (or in the Hilbert space only spanned by $|\Psi_1'\rangle$ and $|\Psi_2'\rangle$), the commutator $[\Gamma, H]$ is equivalent to 0. Thus the operator $\Gamma=\gamma_5 \gamma^{\mu} s_{\mu}$  can function as the spin direction operator for $\ket{\Psi}$. $\blacksquare$
\end{remark}

\begin{remark} \textcolor{blue}{Bloch's Representation of the Spin Direction.} Let us focus on Eq. (\ref{eq:psichi-1}), i.e.,
\begin{eqnarray}\label{eq:psichi-2}
    \ket{\Psi'}=\frac{1}{\sqrt{2\mathcal{E}(\mathcal{E} +mc^2)}}  \begin{pmatrix}
        (\mathcal{E}+mc^2) |\chi\rangle  \\
        c(\vec{\sigma} \cdot \vec{p}) |\chi\rangle
    \end{pmatrix},
\end{eqnarray}
which is the superposition of $\ket{\Psi_1'} $ and $\ket{\Psi_2'} $, i.e.,
\begin{eqnarray}
    \ket{\Psi_1'} =\frac{1}{\sqrt{2\mathcal{E}(\mathcal{E} +mc^2)}}  \begin{pmatrix}
        (\mathcal{E}+mc^2) |\chi_1\rangle\\
        c (\vec{\sigma} \cdot \vec{p}) |\chi_1\rangle
    \end{pmatrix},
\end{eqnarray}
\begin{eqnarray}
    \ket{\Psi_2'} = \frac{1}{\sqrt{2\mathcal{E}(\mathcal{E} +mc^2)}} \begin{pmatrix}
        (\mathcal{E}+mc^2) |\chi_2\rangle\\
        c (\vec{\sigma} \cdot \vec{p}) |\chi_2\rangle
    \end{pmatrix},
\end{eqnarray}
with
\begin{eqnarray}
|\chi_1\rangle = \begin{pmatrix}
    1\\
    0
\end{pmatrix} \equiv |\uparrow\rangle_z, \;\;\;\;
|\chi_2\rangle = \begin{pmatrix}
    0\\
    1
\end{pmatrix}\equiv|\downarrow\rangle_z.
\end{eqnarray}
Here $\ket{\Psi_1'} $ and $\ket{\Psi_2'} $ are two linear-independent eigenstates of the Hamiltonian $H$ with positive energy. $|\uparrow\rangle_z$
denotes the ``spin up'' state along the $z$ direction, and $|\downarrow\rangle_z$ denotes the ``spin down'' state along the $z$ direction, which satisfy
\begin{eqnarray}
  \sigma_z |\uparrow\rangle_z =+1 |\uparrow\rangle_z, \;\;\;\;   \sigma_z |\downarrow\rangle_z = -1 |\downarrow\rangle_z.
\end{eqnarray}

One may easily observe that there is a one-to-one correspondence between the state $\ket{\Psi'}$ (i.e., a $1\times 4$ column) and the state
$|\chi\rangle$ (i.e., a $1\times 2$ column), i.e.,
\begin{eqnarray}
    \ket{\Psi'}=\frac{1}{\sqrt{2\mathcal{E}(\mathcal{E} +mc^2)}}  \begin{pmatrix}
        (\mathcal{E}+mc^2) |\chi\rangle  \\
        c(\vec{\sigma} \cdot \vec{p}) |\chi\rangle
    \end{pmatrix} \;\;\; \Leftrightarrow \;\;\;   |\chi\rangle =\begin{pmatrix}
    \cos\frac{\theta}{2}\\
    \sin\frac{\theta}{2}\, {\rm e}^{{\rm i}\phi}
\end{pmatrix}.
\end{eqnarray}
In quantum information theory, the quantum state $|\chi\rangle$ is very well-known, which is nothing but the a two-qubit pure state. The state $|\chi\rangle$ can have a Bloch's representation by using Bloch's sphere, which can provide a clear picture of spin direction of the qubit.
To see clearly the geometrical picture, one may use the density matrix to express the state $|\chi\rangle$. Then he has
\begin{eqnarray}
\rho &=& |\chi\rangle \langle \chi|=\begin{pmatrix}
    \cos\frac{\theta}{2}\\
    \sin\frac{\theta}{2}\, {\rm e}^{{\rm i}\phi}
\end{pmatrix} \begin{pmatrix}
    \cos\frac{\theta}{2} &
    \sin\frac{\theta}{2}\, {\rm e}^{-{\rm i}\phi}
\end{pmatrix}=
\begin{pmatrix}
    \left(\cos\frac{\theta}{2}\right)^2 & \cos\frac{\theta}{2} \sin\frac{\theta}{2}\, {\rm e}^{-{\rm i}\phi} \\
    \cos\frac{\theta}{2} \sin\frac{\theta}{2}\, {\rm e}^{{\rm i}\phi} & \left(\sin\frac{\theta}{2}\right)^2
\end{pmatrix} \nonumber\\
&=& \dfrac{1}{2}\begin{pmatrix}
    1+\cos\theta & \sin \theta\, {\rm e}^{-{\rm i}\phi} \\
    \sin\theta\, {\rm e}^{{\rm i}\phi} & 1-\cos\theta
\end{pmatrix}= \dfrac{1}{2} \left[ \begin{pmatrix}
    1 & 0 \\
    0 & 1
\end{pmatrix}+ \begin{pmatrix}
    \cos\theta & \sin \theta\, {\rm e}^{-{\rm i}\phi} \\
    \sin\theta\, {\rm e}^{{\rm i}\phi} & -\cos\theta
\end{pmatrix}\right],
\end{eqnarray}
i.e.,
\begin{eqnarray}
\rho &=& \dfrac{1}{2}\left(\openone + \vec{\tau}\cdot \vec{\sigma} \right),
\end{eqnarray}
with
\begin{eqnarray}\label{eq:spindire}
\vec{\tau} = (\sin\theta\cos\phi, \sin\theta\sin\phi, \cos\theta).
\end{eqnarray}
Here $\vec{\tau}$ is a unit vector, i.e., $\left|\vec{\tau}\right|=1$. When the angle $\theta$ runs from $0$ to $\pi$, and the angle $\phi$ runs from $0$ to $2\pi$, the endpoint of the vector $\vec{\tau}$ will draw a spherical surface with a unit of 1. The spherical surface is called Bloch's sphere. Consequently, for a given state $\ket{\Psi'}$, its spin direction is encoded in its one-to-one correspondence state $|\chi\rangle$; furthermore, based on Bloch's representation of $|\chi\rangle$, the spin direction is shown in Eq. (\ref{eq:spindire}). $\blacksquare$
\end{remark}

\newpage

\section{The Incident, Reflected and Transmitted Wave for Dirac's Electron}

Let us consider a Dirac's electron incidents on a one-dimensional potential barrier without a vector potential (i.e., the vector potential $\vec{\mathcal{A}}=0$). In general, we consider the incident particle with momentum $\vec{p}=(p_x,p_y ,p_z)$. The potential is given by
\begin{equation}\label{eq:poten}
   V(x)= q \Phi(x) = \begin{cases}
        V_0 , & x > 0, \\
        0, & x \leq 0,
    \end{cases}
\end{equation}
where $q=-e$ is the electric charge of the Dirac's electron, $\Phi(x)$ can be viewed as a scalar potential, $V_0$ is a constant, and $V_0 >0$. Accordingly, the Hamiltonian of the Dirac electron is given by
\begin{eqnarray}\label{eq:H-1b}
H &=& c\, \vec{\alpha}\cdot \vec{p} +\beta m c^2 + V(x).
\end{eqnarray}
In this work, we consider only the case of positive energy, i.e., the energy of Dirac's electron is always positive during its motion.

For the region of $x\leq 0$, since $V(x)=0$, thus the Hamiltonian represents a free Dirac's electron.  Let $\ket{\Psi_{\rm in}}$ denote the state of the incident wave. Because $|{\Psi}'_1\rangle$ and $|{\Psi}'_2\rangle$ are eigenstates of $H$ with positive energy, thus $\ket{\Psi_{\rm in}}$ is also the eigenstate of $H$ with positive energy for any superposition coefficients $\ell_1$ and $\ell_2$, which satisfy the following the normalized condition
\begin{eqnarray}
|\ell_1|^2+|\ell_2|^2=1.
\end{eqnarray}
The incident wave is give by
\begin{align}\label{eq:in-1}
    \ket{\Psi_{\rm in}} = \biggr[ \ell_1 |{\Psi}'_1\rangle+\ell_2|{\Psi}'_2\rangle\biggr]\, \e^{\frac{\i}{\hbar} [{\hbar}(k_x x +k_y y +  k_z z) - Et]},
\end{align}
which is a superposition state of $|{\Psi}'_1\rangle$ and $|{\Psi}'_2\rangle$. Here the energy $E$ takes the positive energy, i.e., $E\equiv\mathcal{E}=+\sqrt{\vec{p}^{\,2} c^2+m^2c^4}>0$. Explicitly, based on Eq. (\ref{eq:psi1'}) and Eq. (\ref{eq:psi2'}) we have the incident wave $\ket{\Psi_{\rm in}}$ as
\begin{eqnarray}\label{eq:inwave-1}
 &&   \ket{\Psi_{\rm in}}\equiv \dfrac{1}{\sqrt{2\mathcal{E}(\mathcal{E} +mc^2)}} \begin{pmatrix}
            \ell_1(\mathcal{E} + mc^2) \\
            \ell_2(\mathcal{E} + mc^2) \\
            [\ell_1 k_z + \ell_2\,(k_x- \i k_y)]c{\hbar} \\
            [\ell_1(k_x+\i k_y) - \ell_2\,k_z]c{\hbar}
        \end{pmatrix}\e^{\frac{\i}{\hbar} [{\hbar}(k_x x +k_y y +  k_z z) - Et]}.
\end{eqnarray}

Let the interface is the $yz$-plane. The Dirac electron will reflect and transmit when it arrives at the interface. Then, due to the following transformation
\begin{eqnarray}
k_x \rightarrow -k_x, \;\;\; k_y \rightarrow k_y, \;\;\; k_z \rightarrow k_z,
\end{eqnarray}
from $\ket{\Psi_{\rm in}}$ we accordingly have the form of the reflected wave as
\begin{eqnarray}
    &&    \ket{\Psi_{\rm r}}\equiv \dfrac{1}{\sqrt{2\mathcal{E}(\mathcal{E} +mc^2)}}\left\{A \begin{pmatrix}
                \mathcal{E} + mc^2 \\
                0 \\
                k_z c {\hbar}\\
                (-k_x+ {\rm i} k_y) c {\hbar}
            \end{pmatrix}
            + B \begin{pmatrix}
                0 \\
                \mathcal{E} + mc^2 \\
                (-k_x- \i k_y)c {\hbar} \\
                -k_z c {\hbar}
            \end{pmatrix}\right\}{\rm e}^{\frac{\i}{\hbar} [{\hbar}(-k_x x +k_y y+ k_z z) - Et]}.
\end{eqnarray}
Here $A$ and $B$ need not satisfy the normalized condition (i.e., $|A|^2+|B|^2=1$). One will find that the magnitude of $|A|^2+|B|^2$ is related to the reflection coefficient.

Similarly, due to the following replacement
\begin{eqnarray}
\mathcal{E} \rightarrow \mathcal{E}-V_0, \;\;\;\;\; \vec{p}=\hbar \vec{k}=\hbar(k_x, k_y, k_z) \rightarrow \vec{p}{\,'}=\hbar \vec{k}'
=\hbar(k_x', k_y, k_z),
\end{eqnarray}
in the region of $x>0$, we have the transmitted wave $\ket{\Psi_{\rm t}}$ as
    \begin{equation}\label{eq:tran-1}
    \ket{\Psi_{\rm t}} \equiv \dfrac{1}{\sqrt{2(\mathcal{E}-V_0)(\mathcal{E} -V_0 + mc^2)}}\left\{C \begin{pmatrix}
            \mathcal{E}-V_0 + mc^2 \\
            0 \\
            k_z c {\hbar} \\
            (k_x'+ {\rm i} k_y) c {\hbar}
        \end{pmatrix}
        + D \begin{pmatrix}
            0 \\
            \mathcal{E}-V_0 + mc^2 \\
            (k_x'- {\rm i} k_y) c {\hbar} \\
            -k_z c {\hbar}
        \end{pmatrix}\right\} {\rm e}^{\frac{\i}{\hbar} [{\hbar}(k_x' x +k_y y+ k_z z) - Et]},
    \end{equation}
Here $C$ and $D$ need not satisfy the normalized condition (i.e., $|C|^2+|D|^2=1$), one will find that the magnitude of $|C|^2+|D|^2$ is related to the transmission coefficient.

Notice that in the region of $x\leq 0$, the energy-momentum relation reads
\begin{equation}\label{eq:E1}
    \mathcal{E}^2 =\vec{p}^{\, 2} c^2 +m^2 c^4 =(p_x^2 +p_y^2 + p_z^2)c^2 +m^2 c^4,
\end{equation}
or
\begin{equation}\label{eq:E1'}
    \mathcal{E}^2 =\hbar^2 \vec{k}^2 c^2 +m^2 c^4 =\hbar^2(k_x^2 +k_y^2 + k_z^2)c^2 +m^2 c^4,
\end{equation}
and in the region of $x>0$, the energy-momentum relation becomes
\begin{equation}\label{eq:EV}
 (\mathcal{E}-V_0)^2 =\vec{p}{\,'}^2 c^2 +m^2 c^4 =(p_x'^2 +p_y^2 +p_z^2)c^2 +m^2 c^4,
\end{equation}
or
\begin{equation}\label{eq:EV'}
 (\mathcal{E}-V_0)^2 =\hbar^2 \vec{k}{\,'}^2 c^2 +m^2 c^4 =\hbar^2 (k_x'^2 +k_y^2 +k_z^2)c^2 +m^2 c^4.
\end{equation}

According to the continuity condition at $x=0$, one has
\begin{eqnarray}\label{eq:rt}
    {\ket{\Psi_{\rm in}} +  \ket{\Psi_{\rm r}} =  \ket{\Psi_{\rm t}},}
\end{eqnarray}
which leads to
\begin{equation}\label{eq:ContinConds}
    \begin{cases}
    (\mathcal{E} + mc^2)({\ell_1} +  A) = \dfrac{\sqrt{\mathcal{E}(\mathcal{E} +mc^2)}}{\sqrt{(\mathcal{E}-V_0)(\mathcal{E} -V_0+mc^2)}}C({\mathcal{E}-V_0}+ mc^2), \\
    (\mathcal{E} + mc^2)({\ell_2}+  B) =\dfrac{\sqrt{\mathcal{E}(\mathcal{E} +mc^2)}}{\sqrt{(\mathcal{E}-V_0)(\mathcal{E} -V_0+mc^2)}} D({\mathcal{E}-V_0}+ mc^2), \\
    [\ell_1 k_z + \ell_2\,(k_x- \i k_y)]c {\hbar} + [  A k_z c {\hbar}+ B (-k_x- \i k_y)c {\hbar} ]=\dfrac{\sqrt{\mathcal{E}(\mathcal{E} +mc^2)}}{\sqrt{(\mathcal{E}-V_0)(\mathcal{E} -V_0+mc^2)}}[C k_z {\hbar} c+D (k'_x- \i k_y)c {\hbar} ], \\
    [\ell_1(k_x+\i k_y) - \ell_2\,k_z]c {\hbar} + [ A (-k_x+ {\rm i} k_y) c {\hbar}+ B ( -k_z c {\hbar})] =\dfrac{\sqrt{\mathcal{E}(\mathcal{E} +mc^2)} }{\sqrt{(\mathcal{E}-V_0)(\mathcal{E} -V_0+mc^2)}} [C (k'_x+ {\rm i} k_y) c {\hbar}+ D ( -k_z c {\hbar})],
    \end{cases}
\end{equation}
i.e.,
\begin{equation}
    \begin{cases}
    (\mathcal{E} + mc^2)({\ell_1} +  A) = \dfrac{\sqrt{\mathcal{E}(\mathcal{E} +mc^2)}}{\sqrt{(\mathcal{E}-V_0)(\mathcal{E} -V_0+mc^2)}}C({\mathcal{E}-V_0}+ mc^2), \\
    (\mathcal{E} + mc^2)({\ell_2}+  B) =\dfrac{\sqrt{\mathcal{E}(\mathcal{E} +mc^2)}}{\sqrt{(\mathcal{E}-V_0)(\mathcal{E} -V_0+mc^2)}} D({\mathcal{E}-V_0}+ mc^2), \\
    [\ell_1 k_z + \ell_2\,(k_x- \i k_y)]  + [  A k_z + B (-k_x- \i k_y) ]=\dfrac{\sqrt{\mathcal{E}(\mathcal{E} +mc^2)}}{\sqrt{(\mathcal{E}-V_0)(\mathcal{E} -V_0+mc^2)}}[C k_z +D (k'_x- \i k_y) ], \\
    [\ell_1(k_x+\i k_y) - \ell_2\,k_z] + [ A (-k_x+ {\rm i} k_y)+ B ( -k_z )] =\dfrac{\sqrt{\mathcal{E}(\mathcal{E} +mc^2)}}{\sqrt{(\mathcal{E}-V_0)(\mathcal{E} -V_0+mc^2)}} [C (k'_x+ {\rm i} k_y) + D ( -k_z )],
    \end{cases}
\end{equation}
One can have
\begin{equation}\label{eq:MCEqD}
M\,\vec{c}=\vec{d},
\end{equation}
where
\begin{eqnarray}
&& M=\begin{pmatrix}
    (\mathcal{E} + mc^2) & 0 & -({\mathcal{E}-V_0}+ mc^2) & 0\\
    0 & (\mathcal{E} + mc^2) & 0 & -({\mathcal{E}-V_0}+ mc^2)\\
    k_z & (-k_x- \i k_y) & -k_z & -(k'_x- \i k_y) \\
    (-k_x+ {\rm i} k_y) & {-k_z} & -(k'_x+ {\rm i} k_y) & {k_z}
    \end{pmatrix},\nonumber\\
&&
\vec{c}=\begin{pmatrix}
     A\\
     B\\
    \dfrac{\sqrt{\mathcal{E}(\mathcal{E} +mc^2)}}{\sqrt{(\mathcal{E}-V_0)(\mathcal{E} -V_0+mc^2)}} C\\
    \dfrac{\sqrt{\mathcal{E}(\mathcal{E} +mc^2)}}{\sqrt{(\mathcal{E}-V_0)(\mathcal{E} -V_0+mc^2)}} D\\
\end{pmatrix},\qquad
\vec{d}=\begin{pmatrix}
    -(\mathcal{E} + mc^2){\ell_1}\\
    -(\mathcal{E} + mc^2){\ell_2} \\
    -[\ell_1 k_z + \ell_2\,(k_x- \i k_y)] \\
    -[\ell_1(k_x+\i k_y) - \ell_2\,k_z]
\end{pmatrix}.
\end{eqnarray}
By denoting the ``refractive index'' as
\begin{eqnarray}
n = \frac{\mathcal{E}+mc^2 }{\mathcal{E}-V_0 +mc^2},
\end{eqnarray}
one has
\begin{eqnarray}
&& M=\begin{pmatrix}
    n & 0 & -1 & 0 \\
    0 & n & 0 & -1 \\
    k_z & (-k_x- \i k_y) & -k_z & -(k'_x- \i k_y) \\
    (-k_x+ {\rm i} k_y) & {-k_z} & -(k'_x+ {\rm i} k_y) & {k_z}
\end{pmatrix},\nonumber\\
&&
\vec{c}=\begin{pmatrix}
    A\\
    B\\
    \dfrac{\sqrt{\mathcal{E}(\mathcal{E} +mc^2)} }{\sqrt{(\mathcal{E}-V_0))(\mathcal{E} -V_0+mc^2)}} C\\
    \dfrac{\sqrt{\mathcal{E}(\mathcal{E} +mc^2)} }{\sqrt{(\mathcal{E}-V_0)(\mathcal{E} -V_0+mc^2)}} D\\
\end{pmatrix},\qquad
\vec{d}=\begin{pmatrix}
    -{\ell_1} n\\
    -{\ell_2} n \\
    -[\ell_1 k_z + \ell_2\,(k_x- \i k_y)] \\
    -[\ell_1(k_x+\i k_y) - \ell_2\,k_z]
\end{pmatrix}.
\end{eqnarray}
Based on which, we can attain the coefficients $A$, $B$, $C$ and $D$ as follows
\begin{subequations}
   \begin{align}
&A=\frac{k_x^2 \ell_1+2 k_x (-i k_y \ell_1+k_z \ell_2) (-1+n)-\ell_1 \left[k_y^2 (-1+n)^2+k_z^2 (-1+n)^2+k_x'^2 n^2\right]}{k_x^2+k_z^2+k_y^2 (-1+n)^2+2 k_x k_x' n-2 k_z^2 n+k_x'^2 n^2+k_z^2 n^2}, \label{eq:2-A}\\
&B=\frac{k_x^2 \ell_2-2 k_x (k_z \ell_1-i k_y \ell_2) (-1+n)-\ell_2 \left[k_y^2 (-1+n)^2+k_z^2 (-1+n)^2+k_x'^2 n^2\right]}{k_x^2+k_z^2+k_y^2 (-1+n)^2+2 k_x k_x' n-2 k_z^2 n+k_x'^2 n^2+k_z^2 n^2}, \label{eq:2-B} \\
&C=\dfrac{\sqrt{(\mathcal{E}-V_0)(\mathcal{E} -V_0+mc^2)}}{\sqrt{\mathcal{E}(\mathcal{E} +mc^2)}}\frac{2 k_x n \left[k_x \ell_1-k_z \ell_2-i k_y \ell_1 (-1+n)+k_x' \ell_1 n+k_z \ell_2 n\right]}{k_x^2+k_z^2+k_y^2 (-1+n)^2+2 k_x k_x' n-2 k_z^2 n+k_x'^2 n^2+k_z^2 n^2}, \label{eq:2-C}\\
& D=\dfrac{\sqrt{(\mathcal{E}-V_0)(\mathcal{E} -V_0+mc^2)}}{\sqrt{\mathcal{E}(\mathcal{E} +mc^2)}}\frac{2 k_x n \left[\ell_2 (k_x+i k_y (-1+n)+k_x' n)+k_z (\ell_1-\ell_1 n)\right]}{k_x^2+k_z^2+k_y^2 (-1+n)^2+2 k_x k_x' n-2 k_z^2 n+k_x'^2 n^2+k_z^2 n^2}, \label{eq:2-D}
    \end{align}
\end{subequations}
i.e.,
\begin{subequations}
   \begin{align}
&A=\frac{k_x^2 \ell_1+2 k_x (-i k_y \ell_1+k_z \ell_2) (n-1)-\ell_1 \left[(k_y^2+k_z^2)(n-1)^2+k_x'^2 n^2\right]}
{k_x^2+(k_y^2+k_z^2)(n-1)^2+2 k_x k_x' n +k_x'^2 n^2}, \\
&B= \frac{k_x^2 \ell_2-2 k_x (k_z \ell_1-i k_y \ell_2) (n-1)-\ell_2 \left[(k_y^2+k_z^2)(n-1)^2+k_x'^2 n^2\right]}
{k_x^2+(k_y^2+k_z^2)(n-1)^2+2 k_x k_x' n +k_x'^2 n^2}, \\
&C=\dfrac{\sqrt{(\mathcal{E}-V_0)(\mathcal{E} -V_0+mc^2)}}{\sqrt{\mathcal{E}(\mathcal{E} +mc^2)}}\frac{2 k_x n \left[k_x \ell_1+k_z \ell_2(n-1)-i k_y \ell_1 (n-1)+k_x' \ell_1 n\right]}
{k_x^2+(k_y^2+k_z^2)(n-1)^2+2 k_x k_x' n +k_x'^2 n^2},\\
& D=\dfrac{\sqrt{(\mathcal{E}-V_0)(\mathcal{E} -V_0+mc^2)}}{\sqrt{\mathcal{E}(\mathcal{E} +mc^2)}}\frac{2 k_x n \left[k_x \ell_2+  i k_y \ell_2 (n-1)- k_z \ell_1(n-1)+ k_x' \ell_2 n\right]}
{k_x^2+(k_y^2+k_z^2)(n-1)^2+2 k_x k_x' n +k_x'^2 n^2}.
    \end{align}
\end{subequations}


For simplicity, in this work, we let the linear momentum of the incident particle be $\vec{p}=(p_x, 0, p_z)$, i.e., the $y$-component of momentum is zero ($p_y=\hbar k_y=0$). In this situation, the four coefficients $A$, $B$, $C$ and $D$ can be simplified to
\begin{subequations}
   \begin{align}\label{eqa1}
&A=\frac{k_x^2 \ell_1+2 k_x k_z \ell_2 (n-1)-\ell_1 \left[k_z^2(n-1)^2+k_x'^2 n^2\right]}
{k_x^2+k_z^2(n-1)^2+2 k_x k_x' n +k_x'^2 n^2}, \\
&B=\frac{k_x^2 \ell_2-2 k_x k_z \ell_1 (n-1)-\ell_2 \left[k_z^2(n-1)^2+k_x'^2 n^2\right]}
{k_x^2+k_z^2(n-1)^2+2 k_x k_x' n +k_x'^2 n^2}, \\
&C=\dfrac{\sqrt{(\mathcal{E}-V_0)(\mathcal{E} -V_0+mc^2)}}{\sqrt{\mathcal{E}(\mathcal{E} +mc^2)}} \frac{2 k_x n \left[k_x \ell_1+k_z \ell_2(n-1)+k_x' \ell_1 n\right]}
{k_x^2+k_z^2(n-1)^2+2 k_x k_x' n +k_x'^2 n^2},\\
& D=\dfrac{\sqrt{(\mathcal{E}-V_0)(\mathcal{E} -V_0+mc^2)}}{\sqrt{\mathcal{E}(\mathcal{E} +mc^2)} } \frac{2 k_x n \left[k_x \ell_2 - k_z \ell_1(n-1)+ k_x' \ell_2 n\right]}
{k_x^2+k_z^2(n-1)^2+2 k_x k_x' n +k_x'^2 n^2},
    \end{align}
\end{subequations}
i.e.,
\begin{subequations}
   \begin{align}
&A=\frac{\ell_1 \left[k_x^2-k_z^2(n-1)^2-k_x'^2 n^2\right]+ \ell_2 \, 2 k_x k_z  (n-1)}
{k_x^2+k_z^2(n-1)^2+2 k_x k_x' n +k_x'^2 n^2}, \label{eq:1-A}\\
&B=\frac{-\ell_1\, 2 k_x k_z  (n-1)+\ell_2\, \left[k_x^2 -k_z^2(n-1)^2-k_x'^2 n^2\right]}
{k_x^2+k_z^2(n-1)^2+2 k_x k_x' n +k_x'^2 n^2}, \label{eq:1-B}\\
&C=\dfrac{\sqrt{(\mathcal{E}-V_0)(\mathcal{E} -V_0+mc^2)}}{\sqrt{\mathcal{E}(\mathcal{E} +mc^2)}} \frac{2 k_x n \left\{\ell_1\, \left[k_x +k_x' n\right]+\ell_2\, k_z (n-1)\right\}}
{k_x^2+k_z^2(n-1)^2+2 k_x k_x' n +k_x'^2 n^2},\label{eq:1-C}\\
& D=\dfrac{\sqrt{(\mathcal{E}-V_0)(\mathcal{E} -V_0+mc^2)}}{\sqrt{\mathcal{E}(\mathcal{E} +mc^2)}}
\frac{2 k_x n \left\{- \ell_1\, k_z (n-1)+  \ell_2\, \left[k_x +k_x' n \right]\right\}}
{k_x^2+k_z^2(n-1)^2+2 k_x k_x' n +k_x'^2 n^2}. \label{eq:1-D}
    \end{align}
\end{subequations}

\begin{remark} \textcolor{blue}{The Reflection Coefficient and the Transmission Coefficient.} The $x$-component of particle current density is defined by
\begin{equation}
j_x = \bra{\Psi}v_x \ket{\Psi}=c\bra{\Psi}\alpha_x \ket{\Psi},
\end{equation}
here we have used the ``velocity operator'' of Dirac's electron, i.e.,
\begin{eqnarray}
\vec{v}=\dfrac{{\rm d} \vec{r}}{{\rm d} t}=\dfrac{1}{{\rm i}\hbar}\left[\vec{r}, H\right]=c\vec{\alpha}.
\end{eqnarray}
Based on which, we have the $x$-component of particle current density for the incident wave as
\begin{eqnarray}
j_{x}^{\rm in} &=&c\bra{\Psi_{\rm in}}\alpha_x \ket{\Psi_{\rm in}} = \dfrac{1}{{2\mathcal{E}(\mathcal{E} +mc^2)}}\left[2c( \mathcal{E} + mc^2) c \hbar k_x(\recv{|\ell_1|^2 + |\ell_2|^2})\right]=\dfrac{\hbar k_x c^2}{\mathcal{E}},
\end{eqnarray}
and the $x$-component of particle current density for the reflected wave as
\begin{eqnarray}
j_{x}^{\rm r} &=&c\bra{\Psi_{\rm r}}\alpha_x \ket{\Psi_{\rm r}} = \dfrac{1}{{2\mathcal{E}(\mathcal{E} +mc^2)}}\left[2c( \mathcal{E} + mc^2)
(-c \hbar k_x)(|A|^2 + |B|^2)\right]=-\dfrac{\hbar k_x c^2}{\mathcal{E}}(|A|^2 + |B|^2).
\end{eqnarray}
Similarly, we have the $x$-component of particle current density for the transmission wave as
\begin{eqnarray}
j_{x}^{\rm t} &=&c\bra{\Psi_{\rm t}}\alpha_x \ket{\Psi_{\rm t}} = \dfrac{1}{{2(\mathcal{E}-V_0)(\mathcal{E}-V_0 +mc^2)}}\left[2c( \mathcal{E}-V_0 + mc^2)
(c\hbar k'_x)(|C|^2 + |D|^2)\right]\nonumber\\
&=& \dfrac{\hbar k'_x c^2}{\mathcal{E}-V_0}(|C|^2 + |D|^2).
\end{eqnarray}
The reflection coefficient is defined as
\begin{eqnarray}
\mathcal{R}= \left|\dfrac{j_{x}^{\rm r}}{j_{x}^{\rm in}}\right| &=& |A|^2 + |B|^2,
\end{eqnarray}
and the transmission coefficient is defined as
\begin{eqnarray}
\mathcal{T}= \left|\dfrac{j_{x}^{\rm t}}{j_{x}^{\rm in}}\right| &=& \dfrac{|k'_x {\mathcal{E}}|}{|k_x(\mathcal{E}-V_0)|}\left(|C|^2 + |D|^2\right).
\end{eqnarray}

\emph{Analysis 1.---} The case of
\begin{equation}\label{eq:EV'-1}
 (\mathcal{E}-V_0)^2 - \left[{\hbar^2 (k_y^2 +k_z^2)c^2} +m^2 c^4\right]=\hbar^2 k_x'^2 \geq 0.
\end{equation}
In this case, we easily know that $k_x'$ is a real number. We can have the reflection coefficient as
\begin{eqnarray}\label{eq:R1-a}
\mathcal{R} &=&  |A|^2 + |B|^2\nonumber\\
&=& \frac{\left|\ell_1 \left[k_x^2-k_z^2(n-1)^2-k_x'^2 n^2\right]+ \ell_2 \, 2 k_x k_z  (n-1)\right|^2}
{[k_x^2+k_z^2(n-1)^2+2 k_x k_x' n +k_x'^2 n^2]^2}+
\frac{\left|-\ell_1\, 2 k_x k_z  (n-1)+\ell_2\, \left[k_x^2 -k_z^2(n-1)^2-k_x'^2 n^2\right]\right|^2}
{[k_x^2+k_z^2(n-1)^2+2 k_x k_x' n +k_x'^2 n^2]^2}\nonumber\\
&=&\frac{1}{[k_x^2+k_z^2(n-1)^2+2 k_x k_x' n +k_x'^2 n^2]^2} \times \nonumber\\
&& \biggr\{\left[  \left[k_x^2-k_z^2(n-1)^2-k_x'^2 n^2\right]^2+ \left[2 k_x k_z  (n-1)\right]^2 \right] \left(|\ell_1|^2+|\ell_2|^2\right) \nonumber\\
&&+ \left[  \left[k_x^2-k_z^2(n-1)^2-k_x'^2 n^2\right] \left[2 k_x k_z  (n-1)\right] \right] \left(\ell_1 \ell_2^*+\ell_1^* \ell_2\right)\nonumber\\
&&-\left[  \left[k_x^2-k_z^2(n-1)^2-k_x'^2 n^2\right] \left[2 k_x k_z  (n-1)\right] \right] \left(\ell_1 \ell_2^*+\ell_1^* \ell_2\right) \biggr\}\nonumber\\
&=& \frac{\left[  \left[k_x^2-k_z^2(n-1)^2-k_x'^2 n^2\right]^2+ \left[2 k_x k_z  (n-1)\right]^2 \right]
}{[k_x^2+k_z^2(n-1)^2+2 k_x k_x' n +k_x'^2 n^2]^2} \nonumber\\
&=& \frac{(k_x^2-k_x'^2 n^2)^2 -2k_z^2(n-1)^2(k_x^2-k_x'^2 n^2)+k_z^4(n-1)^4+4k_z^2k_x^2(n-1)^2}{[(k_x+nk_x')^2+k_z^2(n-1)^2]^2}\notag\\
&=& \frac{[(k_x+nk_x')^2+k_z^2(n-1)^2][(k_x-nk_x')^2+k_z^2(n-1)^2]}{[(k_x+nk_x')^2+k_z^2(n-1)^2]^2}\notag\\
&=&\frac{(k_x-nk_x')^2+k_z^2(n-1)^2}{[(k_x+nk_x')^2+k_z^2(n-1)^2]}.
\end{eqnarray}
Similarly, we can have the transmission coefficient as
\begin{eqnarray}
\mathcal{T} &=&  \dfrac{|k'_x {\mathcal{E}}|}{|k_x(\mathcal{E}-V_0)|}\left(|C|^2 + |D|^2\right) \nonumber\\
 &=& \dfrac{4k_x^2 n^2 {(\mathcal{E}-V_0)(\mathcal{E} -V_0+mc^2)}}{{\mathcal{E}(\mathcal{E} +mc^2)}} \dfrac{|k'_x {\mathcal{E}}|}{|k_x(\mathcal{E}-V_0)|}\frac{1}{[k_x^2+k_z^2(n-1)^2+2 k_x k_x' n +k_x'^2 n^2]^2} \times \notag\\
 &&\biggr\{[(k_x +k_x' n)^2 +  k_z^2 (n-1)^2](|\ell_1|^2+|\ell_2|^2) + \left[k_z (n-1)(k_x +k_x' n)\right](\ell_1 \ell_2^*+\ell_1^* \ell_2)\nonumber\\
 &&-\left[k_z (n-1)(k_x +k_x' n)\right](\ell_1 \ell_2^*+\ell_1^* \ell_2)\biggr\}\notag\\
&=& \dfrac{{(\mathcal{E} -V_0+mc^2)}}{(\mathcal{E} +mc^2)}\frac{4n^2k_x k'_x}{[k_x^2+k_z^2(n-1)^2+2 k_x k_x' n +k_x'^2 n^2]^2} [(k_x +k_x' n)^2 +  k_z^2 (n-1)^2](|\ell_1|^2+|\ell_2|^2) \notag\\
&=& \frac{4nk_x k'_x}{[k_x^2+k_z^2(n-1)^2+2 k_x k_x' n +k_x'^2 n^2]^2} [(k_x +k_x' n)^2 +  k_z^2 (n-1)^2]\notag\\
&=& \frac{4nk_x k'_x}{[(k_x+nk_x')^2+k_z^2(n-1)^2]}.
\end{eqnarray}
Then
\begin{eqnarray}\label{eq:RT-1}
    \mathcal{R}+\mathcal{T}&=& \frac{(k_x-nk_x')^2+k_z^2(n-1)^2}{[(k_x+nk_x')^2+k_z^2(n-1)^2]}+\frac{4nk_x k'_x}{[(k_x+nk_x')^2+k_z^2(n-1)^2]} =1.
\end{eqnarray}
The underlying physical reason for Eq. (\ref{eq:RT-1}) is the conservation of current density, i.e.,
\begin{eqnarray}\label{eq:RT-2}
    j_{x}^{\rm in }+   j_{x}^{\rm r }=   j_{x}^{\rm t },
\end{eqnarray}
Actually, based on Eq. (\ref{eq:RT-2}), one easily finds that Eq. (\ref{eq:RT-1}) is automatically satisfied.

In addition, let us discuss the ``refractive index''. If
\begin{eqnarray}
n = \frac{\mathcal{E}+mc^2 }{\mathcal{E}-V_0 +mc^2} <0,
\end{eqnarray}
i.e., the negative ``refractive index'', because $k_x>0$, $k_x'>0$ (i.e., both of them propagate along the positive $x$-axis), then from Eq. (\ref{eq:R1-a}) we know that
\begin{eqnarray}
\mathcal{R}&=&\frac{(k_x-nk_x')^2+k_z^2(n-1)^2}{[(k_x+nk_x')^2+k_z^2(n-1)^2]}>1,
\end{eqnarray}
and $\mathcal{T}=1-\mathcal{R}<0$. This is known as Klein's paradox.

In this work, we would like to avoid Klein's paradox, which means we require $n>0$. Then from $n>0$, we have
\begin{eqnarray}\label{eq:con-1a}
\mathcal{E}-V_0 +mc^2>0.
\end{eqnarray}
From Eq. (\ref{eq:tran-1}) one has known that the transmitted wave $\ket{\Psi_{\rm t}}$ is a superposition of two quantum states, i.e.,
    \begin{equation}\label{eq:tran-1a}
    \ket{\Psi_{\rm t}} \equiv C \ket{\Psi''_1} + D \ket{\Psi''_2},
    \end{equation}
with
    \begin{eqnarray}
  &&  \ket{\Psi''_1}= \dfrac{1}{\sqrt{2(\mathcal{E}-V_0)(\mathcal{E} -V_0 + mc^2)}}\begin{pmatrix}
            \mathcal{E}-V_0 + mc^2 \\
            0 \\
            k_z c {\hbar} \\
            (k_x'+ {\rm i} k_y) c {\hbar}
        \end{pmatrix} {\rm e}^{\frac{\i}{\hbar} [{\hbar}(k_x' x +k_y y+ k_z z) - Et]}, \nonumber\\
  &&  \ket{\Psi''_2}=
         \dfrac{1}{\sqrt{2(\mathcal{E}-V_0)(\mathcal{E} -V_0 + mc^2)}}\begin{pmatrix}
            0 \\
            \mathcal{E}-V_0 + mc^2 \\
            (k_x'- {\rm i} k_y) c {\hbar} \\
            -k_z c {\hbar}
        \end{pmatrix} {\rm e}^{\frac{\i}{\hbar} [{\hbar}(k_x' x +k_y y+ k_z z) - Et]}.
    \end{eqnarray}
    Based on the normalization factor
\begin{eqnarray}
  &&  (\mathcal{E}-V_0)(\mathcal{E} -V_0 + mc^2)>0,
      \end{eqnarray}
and due to $(\mathcal{E} -V_0 + mc^2)>0$, one can have
\begin{eqnarray}\label{eq:con-1b}
  &&  \mathcal{E}-V_0>0.
      \end{eqnarray}
Obviously, Eq. (\ref{eq:con-1b}) automatically guarantees Eq. (\ref{eq:con-1a}). Thus, in the work we shall discuss the scientific problem in the energy region of
\begin{eqnarray}
  &&  {\mathcal{E}-V_0>0}.
\end{eqnarray}

\emph{Analysis 2.---} The case of
\begin{equation}
 (\mathcal{E}-V_0)^2 - \left[{\hbar^2 (k_y^2 +k_z^2)c^2} +m^2 c^4\right]=\hbar^2 k_x'^2 < 0.
\end{equation}
In this case, we easily know that $k_x'$ is a pure imaginary number. For convenience, we let
\begin{equation}
k_x'={\rm i} q.
\end{equation}
Accordingly, the wave part of the transmitted wavefunction becomes
    \begin{equation}
 \ket{\Psi_{\rm t}} \equiv C \ket{\Psi''_1} + D \ket{\Psi''_2},
     \end{equation}
with
    \begin{eqnarray}
        &&  \ket{\Psi''_1}= \dfrac{1}{\sqrt{2(\mathcal{E}-V_0)(\mathcal{E} -V_0 + mc^2)}}\begin{pmatrix}
                  \mathcal{E}-V_0 + mc^2 \\
                  0 \\
                  k_z c{\hbar} \\
                  ({\rm i} q+ {\rm i} k_y) c {\hbar}
              \end{pmatrix}  {\rm e}^ {{-} q x} {\rm e}^{\frac{\i}{\hbar} [{\hbar}(+k_y y+ k_z z) - Et]}, \nonumber\\
        &&  \ket{\Psi''_2}=
               \dfrac{1}{\sqrt{2(\mathcal{E}-V_0)(\mathcal{E} -V_0 + mc^2)}}\begin{pmatrix}
                  0 \\
                  \mathcal{E}-V_0 + mc^2 \\
                  ({\rm i} q- {\rm i} k_y) c {\hbar} \\
                  -k_z c {\hbar}
              \end{pmatrix}  {\rm e}^ {{-} q x} {\rm e}^{\frac{\i}{\hbar} [{\hbar}(+k_y y+ k_z z) - Et]}.
          \end{eqnarray}
and
          \begin{eqnarray}
            &&  \bra{\Psi''_1}= \dfrac{1}{\sqrt{2(\mathcal{E}-V_0)(\mathcal{E} -V_0 + mc^2)}}\begin{pmatrix}
                      \mathcal{E}-V_0 + mc^2 ,
                     & 0,
                      &k_z c {\hbar} ,
                     & -({\rm i} q+ {\rm i} k_y) c {\hbar}
                  \end{pmatrix}  {\rm e}^ {{-} q x} {\rm e}^{\frac{-\i}{\hbar} [{\hbar}(+k_y y+ k_z z) - Et]}, \nonumber\\
            &&  \bra{\Psi''_2}=
                   \dfrac{1}{\sqrt{2(\mathcal{E}-V_0)(\mathcal{E} -V_0 + mc^2)}}\begin{pmatrix}
                      0,
                    &  \mathcal{E}-V_0 + mc^2 ,
                     & -({\rm i} q- {\rm i} k_y) c {\hbar} ,
                     & -k_z c {\hbar},
                  \end{pmatrix}  {\rm e}^ {{-} q x} {\rm e}^{-\frac{\i}{\hbar} [{\hbar}(+k_y y+ k_z z) - Et]}.
              \end{eqnarray}
One can calculate that
              \begin{eqnarray}
                &&  \alpha_x \ket{\Psi''_1}= \dfrac{1}{\sqrt{2(\mathcal{E}-V_0)(\mathcal{E} -V_0 + mc^2)}}\begin{pmatrix}
                      ({\rm i} q+ {\rm i} k_y) c {\hbar}    \\
                        k_z c {\hbar}    \\
                       0   \\
                          \mathcal{E}-V_0 + mc^2
                      \end{pmatrix}  {\rm e}^ {{-} q x} {\rm e}^{\frac{\i}{\hbar} [{\hbar}(+k_y y+ k_z z) - Et]}, \nonumber\\
                && \alpha_x  \ket{\Psi''_2}=
                       \dfrac{1}{\sqrt{2(\mathcal{E}-V_0)(\mathcal{E} -V_0 + mc^2)}}\begin{pmatrix}
                         -k_z c {\hbar}  \\
                          ({\rm i} q- {\rm i} k_y) c {\hbar}  \\
                          \mathcal{E}-V_0 + mc^2\\
                         0
                      \end{pmatrix}  {\rm e}^ {{-} q x} {\rm e}^{\frac{\i}{\hbar} [{\hbar}(+k_y y+ k_z z) - Et]}.
                  \end{eqnarray}
Then he has
            \begin{eqnarray}
              &&  \bra{\Psi''_2} \alpha_x \ket{\Psi''_1}=  \dfrac{1}{{2(\mathcal{E}-V_0)(\mathcal{E} -V_0 + mc^2)}}{\rm e}^ {{-} 2q x} \left\{( \mathcal{E}-V_0 + mc^2)k_z c {\hbar}- (\mathcal{E}-V_0 + mc^2)k_z c {\hbar}\right\}=0,\notag\\
               && \bra{\Psi''_1} \alpha_x \ket{\Psi''_2}=  \dfrac{1}{{2(\mathcal{E}-V_0)(\mathcal{E} -V_0 + mc^2)}}{\rm e}^ {{-} 2q x} \left\{-( \mathcal{E}-V_0 + mc^2)k_z c {\hbar}+ (\mathcal{E}-V_0 + mc^2)k_z c {\hbar}\right\}=0,
            \end{eqnarray}
and
\begin{eqnarray}
   && \bra{\Psi''_1} \alpha_x \ket{\Psi''_1}=  \dfrac{{\rm e}^ {{-} 2q x}}{{2(\mathcal{E}-V_0)(\mathcal{E} -V_0 + mc^2)}} \left\{( \mathcal{E}-V_0 + mc^2) ({\rm i} q+ {\rm i} k_y) c {\hbar}  - (\mathcal{E}-V_0 + mc^2) ({\rm i} q+ {\rm i} k_y) c {\hbar}  \right\}=0, \notag\\
   && \bra{\Psi''_2} \alpha_x \ket{\Psi''_2}=  \dfrac{{\rm e}^ {{-} 2q x}}{{2(\mathcal{E}-V_0)(\mathcal{E} -V_0 + mc^2)}} \left\{( \mathcal{E}-V_0 + mc^2)({\rm i} q- {\rm i} k_y) c {\hbar} - (\mathcal{E}-V_0 + mc^2)({\rm i} q- {\rm i} k_y) c {\hbar} \right\}=0.
\end{eqnarray}
This fact indicates
\begin{eqnarray}
    j_x^{\rm t}=  c\bra{\Psi_{\rm t}} \alpha_x \ket{\Psi_{\rm t}}=0,
\end{eqnarray}
and thus the transmission coefficient becomes
\begin{eqnarray}
    \mathcal{T}=0.
\end{eqnarray}

For the reflection coefficient, one has,
\begin{eqnarray}\label{eq:R1-a'}
    \mathcal{R} &=&  |A|^2 + |B|^2\nonumber\\
    &=& \frac{\left|\ell_1 \left[k_x^2-k_z^2(n-1)^2-k_x'^2 n^2\right]+ \ell_2 \, 2 k_x k_z  (n-1)\right|^2+
    \left|-\ell_1\, 2 k_x k_z  (n-1)+\ell_2\, \left[k_x^2 -k_z^2(n-1)^2-k_x'^2 n^2\right]\right|^2}
    {[k_x^2+k_z^2(n-1)^2+2 k_x k_x' n +k_x'^2 n^2][k_x^2+k_z^2(n-1)^2-2 k_x k_x' n +k_x'^2 n^2]}\nonumber\\
    &=&\frac{1}{[(k_x+nk_x')^2+k_z^2(n-1)^2][(k_x-nk_x')^2+k_z^2(n-1)^2]} \times \nonumber\\
    && \biggr\{\left[  \left[k_x^2-k_z^2(n-1)^2-k_x'^2 n^2\right]^2+ \left[2 k_x k_z  (n-1)\right]^2 \right] \left(|\ell_1|^2+|\ell_2|^2\right) \nonumber\\
    &&+ \left[  \left[k_x^2-k_z^2(n-1)^2-k_x'^2 n^2\right] \left[2 k_x k_z  (n-1)\right] \right] \left(\ell_1 \ell_2^*+\ell_1^* \ell_2\right)\nonumber\\
    &&-\left[  \left[k_x^2-k_z^2(n-1)^2-k_x'^2 n^2\right] \left[2 k_x k_z  (n-1)\right] \right] \left(\ell_1 \ell_2^*+\ell_1^* \ell_2\right) \biggr\}\nonumber\\
    &=& \frac{\left[  \left[k_x^2-k_z^2(n-1)^2-k_x'^2 n^2\right]^2+ \left[2 k_x k_z  (n-1)\right]^2 \right]
    }{[(k_x+nk_x')^2+k_z^2(n-1)^2][(k_x-nk_x')^2+k_z^2(n-1)^2]} \nonumber\\
    &=& \frac{(k_x^2-k_x'^2 n^2)^2 -2k_z^2(n-1)^2(k_x^2-k_x'^2 n^2)+k_z^4(n-1)^4+4k_z^2k_x^2(n-1)^2}{[(k_x+nk_x')^2+k_z^2(n-1)^2][(k_x-nk_x')^2+k_z^2(n-1)^2]}\notag\\
    &=& \frac{[(k_x+nk_x')^2+k_z^2(n-1)^2][(k_x-nk_x')^2+k_z^2(n-1)^2]}{[(k_x+nk_x')^2+k_z^2(n-1)^2][(k_x-nk_x')^2+k_z^2(n-1)^2]}\notag\\
    &=&1.
    \end{eqnarray}
    So the condition of the conservation of current density can be satisfied,
    i.e. \begin{eqnarray}
        \mathcal{R}+\mathcal{T}=1.
    \end{eqnarray}


In summary, one has
\begin{equation}\label{R11}
\mathcal{R}=\begin{cases}
    |A|^2 + |B|^2=\dfrac{(k_x-nk_x')^2+k_z^2(n-1)^2}{[(k_x+nk_x')^2+k_z^2(n-1)^2]}, &\;\;\; (\mathcal{E}-V_0)^2 - \left[{\hbar^2 (k_y^2 +k_z^2)c^2} +m^2 c^4\right]=\hbar^2 k_x'^2 \geq 0, \\
    1, & \;\;\; (\mathcal{E}-V_0)^2 - \left[{\hbar^2 (k_y^2 +k_z^2)c^2} +m^2 c^4\right]=\hbar^2 k_x'^2 < 0,
\end{cases}
\end{equation}
and
\begin{equation}\label{R22}
    \mathcal{T} = \begin{cases}
        \dfrac{|k'_x {\mathcal{E}}|\left(|C|^2 + |D|^2\right)}{|k_x(\mathcal{E}-V_0)|}=\dfrac{4nk_x k'_x}{[(k_x+nk_x')^2+k_z^2(n-1)^2]}, & \;\;\; (\mathcal{E}-V_0)^2 - \left[{\hbar^2 (k_y^2 +k_z^2)c^2} +m^2 c^4\right]=\hbar^2 k_x'^2 \geq 0, \\
        0, & \;\;\; (\mathcal{E}-V_0)^2 - \left[{\hbar^2 (k_y^2 +k_z^2)c^2} +m^2 c^4\right]=\hbar^2 k_x'^2 < 0.
    \end{cases}
\end{equation}

$\blacksquare$
\end{remark}

\newpage

\section{The Electric-Type Stern-Gerlach Effect for Dirac's Electron }\label{s11}

In this work, we would like to study the electric-type Stern-Gerlach (MTSG) effect for Dirac's electron. The ETSG effect is a Goos-H{\"a}nchen-like shift. However, they are essentially different, the latter occurs in the region of total reflection and the notion of ``spin'' is unnecessary, while the former can occur at the region of non-total reflection and the property of ``spin'' is required. To essentially distinguish the ETSG effect from the GH effect, we restrict our study in the region of non-total reflection.


From the previous section, the non-total reflection corresponds to the following energy region
\begin{equation}
 (\mathcal{E}-V_0)^2 - \left[\hbar^2 (k_y^2 +k_z^2)c^2 +m^2 c^4\right] \geq 0.
\end{equation}
And to avoid Klein's paradox, we have considered the condition in Eq. (\ref{eq:con-1b}). Thus, we have the constraint for $E$ and $V_0$ as
\begin{equation}
    \begin{cases}
(\mathcal{E}-V_0)^2 - \left[\hbar^2 (k_y^2 +k_z^2)c^2 +m^2 c^4\right] \geq 0, \\
\mathcal{E}-V_0>0,
    \end{cases}
\end{equation}
which is equivalent to
\begin{equation}
 \mathcal{E}-V_0 > \sqrt{\hbar^2 (k_y^2 +k_z^2)c^2 +m^2 c^4}.
\end{equation}
Note that in this work, the energy $\mathcal{E}>0$ and the potential $V_0>0$.

To explain the MTSG effect clearly, we introduce a finite-width incident wave and then consider a specific case $\dfrac{\ell_2}{\ell_1} = \i $ and  $p_y=0$, which means the spin direction $\vec{\tau}= (0,1,0)$. Sequently, we analyze the case with arbitrary spin direction.

\hfill
\subsection{The Finite-Width Incident Wave with Finite Dimensions just along the $z$ Direction}

First, for the finite-width incident wave, it can be regarded as the composition of infinite-width waves, which is the solution of Dirac's equation. For simplicity, we just consider that it has finite dimensions just along the $z$ direction, i.e.,
\begin{equation}\label{12}
    \ket{\Psi_{\rm in}(0, z)}= \begin{cases}
        |\Psi_0\rangle {\rm e}^{{\rm i} \left[k_{z_0} z- \frac{1}{\hbar} \mathcal{E}t\right]}, & |z| \leq a, \\
        0, & {\rm other\; case },
    \end{cases}
\end{equation}
with
\begin{equation}\label{eq:Psi0}
    |\Psi_0\rangle = \begin{pmatrix}
        \ell_1(\mathcal{E} + mc^2) \\
        \ell_2(\mathcal{E} + mc^2) \\
        [\ell_1 k_{z_0} + \ell_2\,k_{x_0}]c {\hbar}\\
        [\ell_1 k_{x_0} - \ell_2\, k_{z_0}] c {\hbar}
    \end{pmatrix}.
\end{equation}
Here we define the average momentum of the beam as $ \vec{p}_0 = p_{x_0} \,\vec{e}_x
+p_{z_0} \,\vec{e}_z= \hbar(k_{x_0} \,\vec{e}_x
+k_{z_0} \,\vec{e}_z) $, and the subscripts $x_0$ and $z_0$ correspond the $x$-component and $z$-component of the average momentum, respectively. The probability of Dirac's electron appearing outside the beam is negligible.

We consider a finite-width matter wave, i.e.,
    \begin{align}\label{incident wave fouriered}
         \ket{\Psi_{\rm in}(x,z)} &= \frac{1}{\sqrt{2 \pi}} \int_{-\infty}^{\infty}\mathrm{\Theta}(k^2 - k_z^2) \ket{\psi_k(k_z,x,z)}  \d k_z \notag \\
        &= \frac{1}{\sqrt{2 \pi}} \int_{-\infty}^{\infty}
        \mathrm{\Theta}(k^2  -k_z^2) |\psi_k( k_z)\rangle
         \exp\bigg[{\rm i} \big( k_z z + x \sqrt{k^2 - k_z^2}\big)\bigg]  \d k_z,
        \end{align}
       where
         \begin{eqnarray}
    \ket{\psi_{{k}}(k_z,x,z)}=|\psi_{{k}}(k_z)\rangle \exp\bigg[{\rm i}\big(k_z z + x \sqrt{k^2 - k_z^2}\big)\bigg],
      \end{eqnarray}
   and
           \begin{eqnarray}
                \mathrm{\Theta}(k^2 -k_z^2)=\begin{cases}
                    1,\quad k^2 -k_z^2\ge 0, \\
                    0,\quad k^2 -k_z^2 < 0.
                \end{cases}.
            \end{eqnarray}
   Here $\ket{\psi_{{k}}(k_z,x,z)}$ is a solution of Dirac's equation, and $\ket{\psi_k(k_z)}$ is the inverse Fourier transform of
    $ \ket{\Psi_{\rm in}(0,z)}$. {The subscript $k$ not only distinguishes the infinite-width wave and the finite-width wave $\ket{\Psi_{\rm in}(x,z)}$, but also represents that $\ket{\psi_{{k}}(k_z,x,z)}$ refers to the infinite-width wave with a certain momentum $\vec{k}$.}

   At $x = 0$, one has
       \begin{equation}
        |\psi_k(k_z)\rangle = \frac{1}{ { \sqrt{2 \pi}}} \int_{-\infty}^{\infty}
            \ket{\Psi_{\rm in}(0,z)}\exp\left[-{\rm i} (k_z\,z)\right]\d z.
        \end{equation}
It follows that
     \begin{align}\label{eq:PsiKPz}
        \ket{\psi_k(k_z)}=& \frac{1}{{  \sqrt{2 \pi}}} \int_{-\infty}^{\infty}
        \ket{\Psi_{\rm in}(0,z)}\exp\left[-{\rm i}(k_z\,z)
            \right]  {\rm d}z \notag \\
        =& \frac{1}{ {\sqrt{2 \pi}}}
        \ket{\Psi_0}\int_{-a}^{a} \exp\left[
            {\rm i} (k_{z0} -k_z)z\right]{\rm d}z \notag \\
        =& \ket{\Psi_0}\sqrt{\frac{2}{\pi}}\,\dfrac{\sin\left[(k_{z0} - k_z)a\right]}{(k_{z_0} - k_z)} \notag\\
        =&F(k_z)\ket{\Psi_0},
        \end{align}
where
\begin{eqnarray}
    F(k_z)=\sqrt{\frac{2}{\pi}}\,\dfrac{\sin\left[(k_{z0} - k_z)a\right]}{(k_{z_0} - k_z)}.
\end{eqnarray}

Obviously, $\ket{\psi_k(k_z )}$ cannot be the solution of Dirac's equation. However, the function $F(k_z)$ indicates that only the $ \ket{\psi_k(k_z )}$ with $k_z $ very near $k_{z_0}$ can have influence. When $k_{z_0} a$ is large, we only need to consider $k_z$ near $k_{z_0}$. Let $k_z=k_{z0}+\delta k_z$, by considering $\delta k_z \rightarrow 0$, one then has
\begin{eqnarray}
    \ket{\Psi_0} \approx \ket{\Psi},
\end{eqnarray}
with
\begin{equation}\label{eq:Psi0-a}
    |\Psi\rangle = \begin{pmatrix}
        \ell_1\mathcal{E} + mc^2 \\
        \ell_2(\mathcal{E} + mc^2) \\
        [\ell_1 k_z + \ell_2\,k_x]c {\hbar} \\
        [\ell_1 k_x - \ell_2\,k_z]c {\hbar}
    \end{pmatrix}.
\end{equation}
Then one obtains
\begin{eqnarray}
    \ket{\psi_k(k_z)}=F(k_z)\ket{\Psi_0} \approx F(k_z)\ket{\Psi}.
\end{eqnarray}
Eventually the incident wave is equal to
\begin{eqnarray}
    \ket{\Psi_{\rm in}(x,  z)} &=& \frac{1}{\sqrt{2 \pi}} \int_{k_{z_0}-\delta k_z}^{k_{z_0}+\delta k_z} \ket{\psi_k( k_z, {x, z})} \d  k_z\nonumber\\
    &=& \frac{1}{\sqrt{2 \pi}} \int_{k_{z_0}-\delta k_z}^{k_{z_0}+\delta k_z} \ket{\psi_k( k_z, {x, z})} \d k_z.
\end{eqnarray}
Here $ \ket{\Psi_{\rm in}(x,  z)}$ can be regarded as the composition of the solution of Dirac's equation approximately.

\subsubsection{The Case of $\;{\ell_2}/{\ell_1}= {\rm i}$}

Let us consider the simple case of $\dfrac{\ell_2}{\ell_1}= \i$, which means the spin direction is along the positive $y$-axis, i.e.,
\begin{eqnarray}
\vec{\tau}=(0,1,0).
\end{eqnarray}
Now, for a infinite-width wave, from Eq. (\ref{eq:1-A}) one can find

\begin{eqnarray}
    A &=&\ell_1\dfrac{[k_x^2-n^2{k_x'}^2 - k_z^2(1 - n)^2 ]- 2{\rm i} k_x k_z(1 - n)}{(nk_x' + k_x)^2 + (1 - n)^2k_z^2}, \label{eq:A'}
\end{eqnarray}
and
\begin{eqnarray}
    B= {\rm i } A, \;\;\;\;\;    D= {\rm i }C.
\end{eqnarray}
This fact means that the reflected wave and the transmitted wave have the same spin direction $\vec{\tau}=(0,1,0)$ in its rest frame as the incident wave. It also indicates
\begin{eqnarray}
    A = \ell_1 \dfrac{\sqrt{[k_x^2-n^2{k_x'}^2 - k_z^2(1 - n)^2 ]^2+ [2k_x k_z(1 - n)]^2 }}{(nk_x' + k_x)^2 + (1 - n)^2k_z^2} \, e^{-{\rm i } \theta_{\rm r}(k_z) } =\ell_1 \mathcal{M}_1(k_z ) e^{-{\rm i } \theta_{\rm r}},
\end{eqnarray}
where
\begin{eqnarray}
  \label{theta1}  \tan \theta_{\rm r} = \frac{2k_x k_z(1 - n)}{[k_x^2-n^2{k_x'}^2 - k_z^2(1 - n)^2 ]},
\end{eqnarray}
\begin{eqnarray}
    \mathcal{M}_1(k_z )=\dfrac{\sqrt{[k_x^2-n^2{k_x'}^2 - k_z^2(1 - n)^2 ]^2+ [2k_x k_z(1 - n)]^2 }}{(nk_x' + k_x)^2 + (1 - n)^2k_z^2}.
\end{eqnarray}

According to the stationary phase method \cite{1948Artmann}, this additional phase $\theta_{\rm r}$ can cause a spatial shift for a finite-width wave. Though the existence of matrix structure in the wavefunction of Dirac's electron makes the conclusion not valid strictly, some reasonable approximations can lead to the same conclusion for Dirac's electron. For the reflected wave, we may introduce a reflection operator ${\cal R} (k_z)$, i.e.,
\begin{equation}\label{reflection wave'}
    \ket{\Psi_{\rm r} (x, z)}
    = \dfrac{1}{\sqrt{2 \pi}} \int_{k_{z_0}-\delta k_z}^{k_{z_0}+\delta k_z}
    {\cal R}(k_{z}) |\psi_k(k_z)\rangle   \exp\left\{{\rm i} \big[k_z z - x \sqrt{k^2 - k_z^2}
                - \theta_{\rm r}(k_z)\big]\right\} {\rm d}k_z,
    \end{equation}
where
\begin{align}
    {\cal R} (k_z) |\psi_k(k_z)\rangle =& {\cal R} (k_z) \ell_1\begin{pmatrix}
        \mathcal{E} + mc^2 \\
        {\rm i } [\mathcal{E} + mc^2] \\
        k_z c {\hbar} + \i k_x c {\hbar}\\
        k_x c {\hbar}- \i k_z c {\hbar}
    \end{pmatrix}
    =  \ell_1 \mathcal{M}_1(k_z )\begin{pmatrix}
        \mathcal{E} + mc^2 \\
        {\rm i } [\mathcal{E} + mc^2] \\
        k_z c {\hbar}-\i k_x c {\hbar}\\
        -k_x c {\hbar}- \i k_z c {\hbar}
    \end{pmatrix}\notag\\
    \approx &
    \ell_1 \mathcal{M}_1(k_{z_0} )\begin{pmatrix}
        \mathcal{E} + mc^2 \\
        {\rm i } [\mathcal{E} + mc^2] \\
        k_{z_0} c {\hbar} -\i k_{x_0} c {\hbar}\\
        -k_{x_0} c {\hbar}- \i k_{z_0} c {\hbar}
    \end{pmatrix}
    ={\cal R} (k_{z_0}) |\psi_k(k_{z_0})\rangle.
\end{align}
These methods show that the matrix structure can be approximately constant. It still means only the change of phase term should be considered. This case is the same as the one where stationary phase method can be conducted.

Because the numerator of $A$ and $B$ provide a new phase $-\theta_{\rm r}$ , due to the stationary phase method, we can similarly attain a new shift $\Delta z_{\rm r}$, i.e.,
\begin{align}
    \ket{\Psi_{\rm r} (x, z)} \approx & \dfrac{{\cal R}(k_{z_0})}{\sqrt{2 \pi}}
        \int_{k_{z_0}-\delta k_z}^{k_{z_0}+\delta k_z} |\psi_k(k_z)\rangle
        \exp\left\{{\rm i} \big[
            k_z z - x \sqrt{k^2 - k_z^2} - \theta_{\rm r}(k_z)\big]\right\}{\rm d}k_z
        \notag \\
    \approx & \dfrac{{\cal R}(k_{z_0})}{\sqrt{2 \pi}}
    \int_{k_{z_0}-\delta k_z}^{k_{z_0}+\delta k_z} |\Psi_0\rangle \,F(k_z)\,
    \exp\Bigg({\rm i}
            \bigg\{k_z z - x \sqrt{k^2 - k_z^2} - \Big[\theta_{\rm r\, 0}
                +\frac{\partial\,\theta_{\rm r}}{\partial k_z} \Big|_{k_z = k_{z_0}} (k_z - k_{z_0})
                    \Big]\bigg\}\Bigg){\rm d}k _z \notag \\
    =& \dfrac{1}{\sqrt{2 \pi}} {\cal R}(k_{z_0})\,|\Psi_0\rangle \int_{k_{z_0}-\delta k_z}^{k_{z_0}+\delta k_z} F(k_z)
        \,
        \exp\bigg({\rm i} \Big\{(\bar{k}_z+k_{z_0})z - x \sqrt{k^2 - k_z^2} -\left[
            \theta_{\rm r\,0} +\frac{\partial\,\theta_{\rm r}}{\partial k_z} \Big|_{k_z = k_{z_0}} \bar{k}_z\right]\Big\}\bigg){\rm d}k_z \notag \\
    =& \frac{1}{\sqrt{2 \pi}} \e^{\i(-\theta_{\rm r\,0} +k_{z_0} z)}
        {\cal R}(k_{z_0}) |\Psi_0\rangle \int_{k_{z_0}-\delta k_z}^{k_{z_0}+\delta k_z} F(k_z)\,\exp\left\{
            {\rm i} \left[\bar{k}_z\left(z -\frac{\partial\,\theta_{\rm r}}{\partial k_z} \Big|_{k_z = k_{z0}}\right) - x\sqrt{k^2 - k_z^2}
                \right]\right\}{\rm d}k_z,
    \end{align}
    where \begin{eqnarray}
        \bar{k}_z = k_z- k_{z_0}.
    \end{eqnarray}
So we can attain the shift along the $z$-axis as
\begin{eqnarray}\label{eq:shift-1a}
    \Delta z_{\rm r} = - \frac{\partial (-\theta_{\rm r}) }{\partial k_z} =\frac{\partial \theta_{\rm r} }{\partial k_z}= \frac{\partial  }{\partial k_z} \left[\arctan \left(\frac{2k_x k_z(1 - n)}{[k_x^2-n^2{k_x'}^2 - k_z^2(1 - n)^2 ]}\right) \right].
    \end{eqnarray}
Now we come to treat Eq. (\ref{eq:shift-1a}). We can have
\begin{eqnarray}\label{zr}
\Delta z_{\rm r}& =&  \frac{\partial  }{\partial k_z} \left[\arctan \left(\frac{2k_x k_z(1 - n)}{[k_x^2-n^2{k_x'}^2 - k_z^2(1 - n)^2 ]}\right) \right]\nonumber\\
&=& \dfrac{1}{1+ \left(\dfrac{2k_x k_z(1 - n)}{[k_x^2-n^2{k_x'}^2 - k_z^2(1 - n)^2 ]}\right)^2} \times
\frac{\partial  }{\partial k_z} \left[\frac{2k_x k_z(1 - n)}{[k_x^2-n^2{k_x'}^2 - k_z^2(1 - n)^2 ]} \right]\nonumber\\
&=& \dfrac{1}{1+ \left(\dfrac{2k_x k_z(1 - n)}{[k_x^2-n^2{k_x'}^2 - k_z^2(1 - n)^2 ]}\right)^2} \times \nonumber\\
&&\frac{\dfrac{\partial [2k_x k_z(1 - n)] }{\partial k_z}[k_x^2-n^2{k_x'}^2 - k_z^2(1 - n)^2 ]-[2k_x k_z(1 - n)]
\dfrac{\partial [k_x^2-n^2{k_x'}^2 - k_z^2(1 - n)^2 ] }{\partial k_z}    }{[k_x^2-n^2{k_x'}^2 - k_z^2(1 - n)^2 ]^2}\nonumber\\
&=& \frac{\dfrac{\partial [2k_x k_z(1 - n)] }{\partial k_z}[k_x^2-n^2{k_x'}^2 - k_z^2(1 - n)^2 ]-[2k_x k_z(1 - n)]
\dfrac{\partial [k_x^2-n^2{k_x'}^2 - k_z^2(1 - n)^2 ] }{\partial k_z}    }{[k_x^2-n^2{k_x'}^2 - k_z^2(1 - n)^2 ]^2+[2k_x k_z(1 - n)]^2}.
\end{eqnarray}
In our consideration, the energy $\mathcal{E}$ of the incidence Dirac's particle is fixed, and the potential $V_0$ is fixed, and because $mc^2$ is a constant, then we have
\begin{eqnarray}
\frac{\partial \mathcal{E} }{\partial k_z}=0, \;\;\;\;\; \frac{\partial V_0 }{\partial k_z}=0, \;\;\;\;\; \frac{\partial \, (mc^2) }{\partial k_z}=0,
\;\;\;\;\; \frac{\partial n }{\partial k_z}=0.
\end{eqnarray}
Due to the energy-momentum relation
\begin{eqnarray}
 &&   \mathcal{E}^2 =\hbar^2(k_x^2  + k_z^2)c^2 +m^2 c^4,\nonumber\\
 &&(\mathcal{E}-V_0)^2 =\hbar^2 (k_x'^2  +k_z^2)c^2 +m^2 c^4,
\end{eqnarray}
we have
\begin{eqnarray}
&& k_x= \frac{\sqrt{{\mathcal{E}}^2-m^2c^4 -c^2 \hbar^2\, k_z^2}}{c\hbar }, \nonumber\\
&& k'_x= \frac{\sqrt{\left(\mathcal{E}-V_0\right)^2-m^2c^4 -c^2 \hbar^2\, k_z^2}}{c\hbar }.
\end{eqnarray}
Based on which, one has
\begin{eqnarray}
&& k_x^2-n^2{k_x'}^2 - k_z^2(1 - n)^2  = k_x^2-n^2{k_x'}^2 - k_z^2(1 - 2n+n^2)=k_x^2+ (2n-1)  k_z^2-n^2 ({k_x'}^2+k_z^2)\nonumber\\
&=&\frac{{{\mathcal{E}}^2-m^2c^4 -c^2 \hbar^2\, k_z^2}}{c^2\hbar^2 }+ (2n-1)  k_z^2
-n^2 \frac{\left(\mathcal{E}-V_0\right)^2-m^2c^4}{c^2\hbar^2 }\nonumber\\
&=&\frac{{\mathcal{E}}^2-m^2c^4 }{c^2\hbar^2 }+ 2(n-1)  k_z^2
-n^2 \frac{\left(\mathcal{E}-V_0\right)^2-m^2c^4}{c^2\hbar^2 }\nonumber\\
&=&\frac{{\mathcal{E}}^2-m^2c^4 }{c^2\hbar^2 }+ 2(n-1)  k_z^2
-\frac{(\mathcal{E}+mc^2)^2 }{(\mathcal{E}-V_0 +mc^2)^2} \frac{\left(\mathcal{E}-V_0\right)^2-m^2c^4}{c^2\hbar^2 }\nonumber\\
&=&\frac{{\mathcal{E}}^2-m^2c^4 }{c^2\hbar^2 }+ 2(n-1)  k_z^2
-\frac{(\mathcal{E}+mc^2)^2 }{(\mathcal{E}-V_0 +mc^2)} \frac{\left(\mathcal{E}-V_0\right)-mc^2}{c^2\hbar^2 }\nonumber\\
&=&2(n-1)  k_z^2+
\frac{({\mathcal{E}}^2-m^2c^4)(\mathcal{E}-V_0 +mc^2)-(\mathcal{E}+mc^2)^2\left(\mathcal{E}-V_0-mc^2\right) }{(\mathcal{E}-V_0 +mc^2)c^2\hbar^2 } \nonumber\\
&=&2(n-1)  k_z^2+(\mathcal{E}+mc^2) \times
\frac{({\mathcal{E}}-mc^2)(\mathcal{E}-V_0 +mc^2)-(\mathcal{E}+mc^2)\left(\mathcal{E}-V_0-mc^2\right) }{(\mathcal{E}-V_0 +mc^2)c^2\hbar^2 } \nonumber\\
&=&2(n-1)  k_z^2+(\mathcal{E}+mc^2) \times
\frac{{\mathcal{E}} \times 2mc^2 -mc^2 \times  2(\mathcal{E}-V_0)  }{(\mathcal{E}-V_0 +mc^2)c^2\hbar^2 } \nonumber\\
&=&2\left(\frac{\mathcal{E}+mc^2 }{\mathcal{E}-V_0 +mc^2}-1\right)  k_z^2+(\mathcal{E}+mc^2) \times
\frac{2 V_0 mc^2 }{(\mathcal{E}-V_0 +mc^2)c^2\hbar^2 } \nonumber\\
&=&\frac{2 V_0 }{\mathcal{E}-V_0 +mc^2}  k_z^2+(\mathcal{E}+mc^2) \times
\frac{2 V_0 mc^2 }{(\mathcal{E}-V_0 +mc^2)c^2\hbar^2 } \nonumber\\
&=&\frac{2 V_0 }{\mathcal{E}-V_0 +mc^2} \left[ k_z^2+
\frac{(\mathcal{E}+mc^2)  mc^2 }{c^2\hbar^2 }\right],
\end{eqnarray}

\begin{eqnarray}
[2k_x k_z(1 - n)]^2 &=& 4 k^2_x k^2_z(1 - n)^2= 4 \frac{{\mathcal{E}}^2-m^2c^4 -c^2 \hbar^2\, k_z^2}{c^2\hbar^2 } k^2_z \left(\frac{\mathcal{E}+mc^2 }{\mathcal{E}-V_0 +mc^2}-1\right)^2\nonumber\\
&=& 4 \frac{{\mathcal{E}}^2-m^2c^4 -c^2 \hbar^2\, k_z^2}{c^2\hbar^2 } k^2_z \left(\frac{V_0 }{\mathcal{E}-V_0 +mc^2}\right)^2\nonumber\\
&=& \left(\frac{2 V_0 }{\mathcal{E}-V_0 +mc^2}\right)^2 \frac{{\mathcal{E}}^2-m^2c^4 -c^2 \hbar^2\, k_z^2}{c^2\hbar^2 } k^2_z,
\end{eqnarray}

\begin{eqnarray}
&&[k_x^2-n^2{k_x'}^2 - k_z^2(1 - n)^2]^2+[2k_x k_z(1 - n)]^2 \nonumber\\
&=&
\left[\frac{2 V_0 }{\mathcal{E}-V_0 +mc^2} \left[ k_z^2+
\frac{(\mathcal{E}+mc^2)  mc^2 }{c^2\hbar^2 }\right]\right]^2+\left(\frac{2 V_0 }{\mathcal{E}-V_0 +mc^2}\right)^2 \frac{{\mathcal{E}}^2-m^2c^4 -c^2 \hbar^2\, k_z^2}{c^2\hbar^2 } k^2_z \nonumber\\
&=&
\left(\frac{2 V_0 }{\mathcal{E}-V_0 +mc^2}\right)^2 \biggr\{ \left[ k_z^2+
\frac{(\mathcal{E}+mc^2)  mc^2 }{c^2\hbar^2 }\right]^2+ \frac{{\mathcal{E}}^2-m^2c^4 -c^2 \hbar^2\, k_z^2}{c^2\hbar^2 } k^2_z\biggr\} \nonumber\\
&=&
\frac{4 V_0^2 }{(\mathcal{E}-V_0 +mc^2)^2} \frac{1}{c^4 \hbar^4} \biggr\{ \left[ c^2\hbar^2 k_z^2+(\mathcal{E}+mc^2)  mc^2
\right]^2+ ({{\mathcal{E}}^2-m^2c^4 -c^2 \hbar^2\, k_z^2}) {c^2\hbar^2 } k^2_z\biggr\} \nonumber\\
&=&
\frac{4 V_0^2 }{(\mathcal{E}-V_0 +mc^2)^2} \frac{1}{c^4 \hbar^4} \biggr\{ \left[ 2 c^2\hbar^2 k_z^2(\mathcal{E}+mc^2)  mc^2+ (\mathcal{E}+mc^2)^2  m^2c^4
\right]+ ({{\mathcal{E}}^2-m^2c^4 }) {c^2\hbar^2 } k^2_z\biggr\} \nonumber\\
&=&
\frac{4 V_0^2 }{(\mathcal{E}-V_0 +mc^2)^2} \frac{(\mathcal{E}+mc^2) }{c^4 \hbar^4} \biggr\{ \left[ 2 c^2\hbar^2 k_z^2 mc^2+ (\mathcal{E}+mc^2)  m^2c^4
\right]+ ({{\mathcal{E}}-mc^2 }) {c^2\hbar^2 } k^2_z\biggr\} \nonumber\\
&=&
\frac{4 V_0^2 }{(\mathcal{E}-V_0 +mc^2)^2} \frac{(\mathcal{E}+mc^2) }{c^4 \hbar^4} \biggr\{  (\mathcal{E}+mc^2)  m^2c^4
+ ({{\mathcal{E}}+mc^2 }) {c^2\hbar^2 } k^2_z\biggr\} \nonumber\\
&=&
\frac{4 V_0^2 }{(\mathcal{E}-V_0 +mc^2)^2} \frac{(\mathcal{E}+mc^2)^2 }{c^4 \hbar^4} \biggr\{    m^2c^4
+  {c^2\hbar^2 } k^2_z\biggr\}, \nonumber\\
\end{eqnarray}

\begin{eqnarray}
[2k_x k_z(1 - n)]&=& \left(\frac{\hilight{-}2 V_0 }{\mathcal{E}-V_0 +mc^2}\right) \frac{\sqrt{{\mathcal{E}}^2-m^2c^4 -c^2 \hbar^2\, k_z^2}}{c\hbar } k_z= \left(\frac{\hilight{-} 2 V_0 }{\mathcal{E}-V_0 +mc^2}\right) k_x  k_z,
\end{eqnarray}

\begin{eqnarray}
 \frac{\partial [k_x^2-n^2{k_x'}^2 - k_z^2(1 - n)^2]}{\partial k_z}
&=& \frac{4 V_0 }{\mathcal{E}-V_0 +mc^2} k_z,
\end{eqnarray}

\begin{eqnarray}
 \frac{\partial [2k_x k_z(1 - n)]}{\partial k_z}
&=& 2 k_z(1 - n) \frac{\partial k_x}{\partial k_z}-2k_x (n-1) \nonumber\\
&=& 2 k_z(1 - n) \frac{\partial }{\partial k_z} \left[\frac{\sqrt{{\mathcal{E}}^2-m^2c^4 -c^2 \hbar^2\, k_z^2}}{c\hbar }\right]
-2k_x \left(\frac{\mathcal{E}+mc^2 }{\mathcal{E}-V_0 +mc^2}-1\right) \nonumber\\
&=& 2 k_z(1 - n) \dfrac{1}{c\hbar} \left[\frac{-2 c^2 \hbar^2\, k_z}{2}\frac{1}{\sqrt{{\mathcal{E}}^2-m^2c^4 -c^2 \hbar^2\, k_z^2}}\right]
-2k_x \left(\frac{\mathcal{E}+mc^2 }{\mathcal{E}-V_0 +mc^2}-1\right) \nonumber\\
&=& 2 k^2_z( n-1) c\hbar  \left[\frac{1}{\sqrt{{\mathcal{E}}^2-m^2c^4 -c^2 \hbar^2\, k_z^2}}\right]
-2k_x \frac{V_0 }{\mathcal{E}-V_0 +mc^2} \nonumber\\
&=& \frac{2 V_0 }{\mathcal{E}-V_0 +mc^2} \left[ \frac{ k^2_z c\hbar }{\sqrt{{\mathcal{E}}^2-m^2c^4 -c^2 \hbar^2\, k_z^2}} -k_x\right]  \nonumber\\
&=& \frac{2 V_0 }{\mathcal{E}-V_0 +mc^2} \left[ \frac{ k^2_z c\hbar }{c\hbar k_x} -k_x\right]  \nonumber\\
&=& \frac{2 V_0 }{\mathcal{E}-V_0 +mc^2}  \frac{ k^2_z -k^2_x }{ k_x}.
\end{eqnarray}
Then we have
\begin{eqnarray}
\Delta z_{\rm r}&=& \frac{\dfrac{\partial [2k_x k_z(1 - n)] }{\partial k_z}[k_x^2-n^2{k_x'}^2 - k_z^2(1 - n)^2 ]-[2k_x k_z(1 - n)]
\dfrac{\partial [k_x^2-n^2{k_x'}^2 - k_z^2(1 - n)^2 ] }{\partial k_z}    }{[k_x^2-n^2{k_x'}^2 - k_z^2(1 - n)^2 ]^2+[2k_x k_z(1 - n)]^2}\nonumber\\
&=& \dfrac{\left[ \dfrac{2 V_0 }{\mathcal{E}-V_0 +mc^2}  \dfrac{ k^2_z -k^2_x }{ k_x}\right]
\left[\dfrac{2 V_0 }{\mathcal{E}-V_0 +mc^2} \left[ k_z^2+
\dfrac{(\mathcal{E}+mc^2)  mc^2 }{c^2\hbar^2 }\right]\right]
- \left[\left(\dfrac{{-}2 V_0 }{\mathcal{E}-V_0 +mc^2}\right) k_x  k_z\right]\left[\dfrac{4 V_0 }{\mathcal{E}-V_0 +mc^2} k_z\right]}{\dfrac{4 V_0^2 }{(\mathcal{E}-V_0 +mc^2)^2} \dfrac{(\mathcal{E}+mc^2)^2 }{c^4 \hbar^4} \biggr\{    m^2c^4
+  {c^2\hbar^2 } k^2_z\biggr\}}\nonumber\\
&=& \dfrac{\left[   \dfrac{ k^2_z -k^2_x }{ k_x}\right]
\left[  k_z^2+
\dfrac{(\mathcal{E}+mc^2)  mc^2 }{c^2\hbar^2 }\right]
{+} \left[ k_x  k_z\right]\left[2 k_z\right]}{ \dfrac{(\mathcal{E}+mc^2)^2 }{c^4 \hbar^4} \biggr\{    m^2c^4
+  {c^2\hbar^2 } k^2_z\biggr\}}\nonumber\\
&=& \dfrac{c^2\hbar^2\left[ k^2_z -k^2_x \right]
\left[ c^2\hbar^2 k_z^2+(\mathcal{E}+mc^2)  mc^2
\right]{+} 2 c^4 \hbar^4 k_x^2 k_z^2}
{ k_x (\mathcal{E}+mc^2)^2  \biggr\{    m^2c^4
+  {c^2\hbar^2 } k^2_z\biggr\}}\nonumber\\
&=& \dfrac{c^2\hbar^2\left[ k^2_z +k^2_x \right] c^2\hbar^2 k_z^2
+ c^2\hbar^2\left[ k^2_z -k^2_x \right]
\left[ (\mathcal{E}+mc^2)  mc^2\right]}
{ k_x (\mathcal{E}+mc^2)^2  \left(    m^2c^4
+  {c^2\hbar^2 } k^2_z \right)}.
\end{eqnarray}
With the help of
\begin{eqnarray}
k_x= \frac{1}{c\hbar }\sqrt{{\mathcal{E}}^2-m^2c^4} \, \cos \varphi, \;\;\;\;\;
 k_z= \frac{1}{c\hbar}\sqrt{{\mathcal{E}}^2-m^2c^4}\,  \sin \varphi,
\end{eqnarray}
where $\varphi$ represents the incident angle, then one has
\begin{eqnarray}
\Delta z_{\rm r}&=& \dfrac{c^2\hbar^2\left[ k^2_z +k^2_x \right] c^2\hbar^2 k_z^2
+ c^2\hbar^2\left[ k^2_z -k^2_x \right]
\left[ (\mathcal{E}+mc^2)  mc^2\right]}
{ k_x (\mathcal{E}+mc^2)^2  \left(    m^2c^4
+  {c^2\hbar^2 } k^2_z \right)}\nonumber\\
&=& \dfrac{ ({\mathcal{E}}^2-m^2c^4) ({\mathcal{E}}^2-m^2c^4)\sin^2 \varphi
\hilight{-} ({\mathcal{E}}^2-m^2c^4) \cos(2\varphi)
\left[ (\mathcal{E}+mc^2)  mc^2\right]}
{ k_x (\mathcal{E}+mc^2)^2  \left(    m^2c^4
+  {c^2\hbar^2 } k^2_z \right)}\nonumber\\
&=& \dfrac{ ({\mathcal{E}}-mc^2)^2 \sin^2 \varphi
\hilight{-} ({\mathcal{E}}-mc^2)  mc^2 \cos(2\varphi)}
{ k_x   \left(    m^2c^4
+  {c^2\hbar^2 } k^2_z \right)}\nonumber\\
&=& \dfrac{ ({\mathcal{E}}-mc^2)^2 \sin^2 \varphi
\hilight{-} ({\mathcal{E}}-mc^2)  mc^2 \cos(2\varphi)}
{  \dfrac{1}{c\hbar }\sqrt{{\mathcal{E}}^2-m^2c^4} \, \cos \varphi  \left[    m^2c^4
+ ({\mathcal{E}}^2-m^2c^4)\sin^2 \varphi  \right]}\nonumber\\
&=& \dfrac{c\hbar ({\mathcal{E}}-mc^2) }{\sqrt{{\mathcal{E}}^2-m^2c^4} \, \cos \varphi}
\dfrac{ ({\mathcal{E}}-mc^2) \sin^2 \varphi
\hilight{-}  mc^2 \cos(2\varphi)}
{   \left[    m^2c^4 + ({\mathcal{E}}^2-m^2c^4)\sin^2 \varphi  \right]}\nonumber\\
&=& \dfrac{c\hbar \sqrt{\mathcal{E}-mc^2} }{\sqrt{{\mathcal{E}}+mc^2} \, \cos \varphi}
\dfrac{ ({\mathcal{E}}-mc^2) \sin^2 \varphi
\hilight{-}   mc^2 \cos(2\varphi)}
{   \left[ {\mathcal{E}}^2 \sin^2 \varphi + m^2c^4 \cos^2 \varphi \right]}\nonumber\\
&=& \dfrac{\sqrt{\mathcal{E}-mc^2} }{\sqrt{{\mathcal{E}}+mc^2}}
\dfrac{c\hbar }{\cos \varphi}
\dfrac{ {\mathcal{E}}\sin^2 \varphi \hilight{-}   mc^2 (\cos2\varphi+\sin^2 \varphi)}
{   \left[ {\mathcal{E}}^2 \sin^2 \varphi + m^2c^4 \cos^2 \varphi \right]}\nonumber\\
&=& \dfrac{\sqrt{\mathcal{E}-mc^2} }{\sqrt{{\mathcal{E}}+mc^2}}
\dfrac{c\hbar }{\cos \varphi}
\dfrac{ {\mathcal{E}}\sin^2 \varphi -   mc^2 \cos^2 \varphi}
{   \left[ {\mathcal{E}}^2 \sin^2 \varphi + m^2c^4 \cos^2 \varphi \right]}.
\end{eqnarray}
Alternatively, one has
\begin{eqnarray}
\Delta z_{\rm r} &=& \dfrac{\sqrt{\mathcal{E}-mc^2} }{\sqrt{{\mathcal{E}}+mc^2}}
\dfrac{c\hbar }{\cos \varphi}
\dfrac{ {\mathcal{E}}\sin^2 \varphi -   mc^2 \cos^2 \varphi}
{   \left[ {\mathcal{E}}^2 \sin^2 \varphi + m^2c^4 \cos^2 \varphi \right]}\nonumber\\
&=& \dfrac{\sqrt{\mathcal{E}-mc^2} }{\sqrt{{\mathcal{E}}+mc^2}}
\dfrac{c\hbar {\mathcal{E}} \cos \varphi}{\left[ {\mathcal{E}}^2 \sin^2 \varphi + m^2c^4 \cos^2 \varphi \right]}
 \left[\tan^2 \varphi -   \dfrac{mc^2}{\mathcal{E}} \right].
\end{eqnarray}
One can set the incident angle
\begin{eqnarray}
\varphi\in \left[0, \dfrac{\pi}{2}\right),
\end{eqnarray}
because
\begin{eqnarray}
\dfrac{\sqrt{\mathcal{E}-mc^2} }{\sqrt{{\mathcal{E}}+mc^2}} >0, \;\;\;
c\hbar {\mathcal{E}}>0, \;\;\;  \cos \varphi >0, \;\;\;
 {\mathcal{E}}^2 \sin^2 \varphi + m^2c^4 \cos^2 \varphi >0,
\end{eqnarray}
thus the sign of $\Delta z_{\rm r}$ depends on the sign of $\left[\tan^2 \varphi -   \dfrac{mc^2}{\mathcal{E}} \right]$, i.e.,
\begin{equation}
    \Delta z_{\rm r}\begin{cases}
        >0, \;\;\; & \tan \varphi > \sqrt{\dfrac{mc^2}{\mathcal{E}}}, \\
        =0, \;\;\; & \tan \varphi = \sqrt{\dfrac{mc^2}{\mathcal{E}}},\\
        <0, \;\;\; & \tan \varphi < \sqrt{\dfrac{mc^2}{\mathcal{E}}}.
    \end{cases}
\end{equation}
In the non-relativistic limit, i.e., $\mathcal{E}-mc^2 \rightarrow 0$, one will have $\Delta z_{\rm r} \rightarrow 0$, which means the spatial shift vanishes. This is the reason why the ``spin'' is hard to cause the spatial shift in Schr{\" o}dinger equation.


For the transmitted wave, from
   \begin{eqnarray}
&&C=\dfrac{\sqrt{(\mathcal{E}-V_0)(\mathcal{E} -V_0+mc^2)}}{\sqrt{\mathcal{E}(\mathcal{E} +mc^2)}} \frac{2 k_x n \left\{\ell_1\, \left[k_x +k_x' n\right]+\ell_2\, k_z (n-1)\right\}}
{k_x^2+k_z^2(n-1)^2+2 k_x k_x' n +k_x'^2 n^2},
    \end{eqnarray}
one has
\begin{eqnarray}
    C=\ell_1 \dfrac{\sqrt{(\mathcal{E}-V_0)(\mathcal{E} -V_0+mc^2)}}{\sqrt{\mathcal{E}(\mathcal{E} +mc^2)}} 2n k_x \dfrac{\sqrt{(k_x + nk_x')^2 +[k_z(1 - n)]^2}}{(nk_x' + k_x)^2 + (1 - n)^2k_z^2}\, e^{-{\rm i} \theta_{\rm t}(k_z)}=\ell_1 \mathcal{M}_2(k_z ) e^{-{\rm i } \theta_{\rm t}},
\end{eqnarray}
where
\begin{eqnarray}
    {\rm tan} \theta_{\rm t} = \frac{k_z (1-n)}{k_x +n k_x'},
\end{eqnarray}
\begin{eqnarray}
    \mathcal{M}_2(k_z )= \dfrac{\sqrt{(\mathcal{E}-V_0)(\mathcal{E} -V_0+mc^2)}}{\sqrt{\mathcal{E}(\mathcal{E} +mc^2)}}
    2nk_x \dfrac{\sqrt{(k_x + nk_x')^2 +[k_z(1 - n)]^2}}{(nk_x' + k_x)^2 + (1 - n)^2k_z^2}.
\end{eqnarray}
Similar to the reflected wave, we can introduce a transmitted operator ${\cal{T}} (k_z)$, then the transmitted wave can be written as
\begin{equation}\label{tra wave'}
    \ket{\Psi_{\rm t} (x, z)}
    = \dfrac{1}{\sqrt{2 \pi}} \int_{k_{z_0}-\delta k_z}^{k_{z_0}+\delta k_z}
    {\cal T}(k_{z}) |\psi_k(k_z)\rangle  \exp\left\{{\rm i}\big[k_z z + x \sqrt{(\mathcal{E}-V_0)^2 -m^2c^4 - k_z^2c^2 \hbar^2}
                - \theta_{\rm t}(k_z)\big]\right\} {\rm d}k_z,
    \end{equation}
    where
    \begin{align}
        {\cal T} (k_z) |\psi_k(k_z)\rangle=& {\cal T} (k_z) \ell_1\begin{pmatrix}
            \mathcal{E} + mc^2 \\
            {\rm i } [\mathcal{E} + mc^2] \\
            k_z c {\hbar}+ \i k_x c {\hbar}\\
            k_x c {\hbar}- \i k_z c {\hbar}
        \end{pmatrix}
        =  \ell_1 \mathcal{M}_2(k_z )\begin{pmatrix}
            \mathcal{E} + mc^2 \\
            {\rm i } [\mathcal{E} + mc^2] \\
            k_z c {\hbar}+\i k_x' c {\hbar}\\
            k_x' c {\hbar}- \i k_z c {\hbar}
        \end{pmatrix}\notag\\
        \approx \;\; &
        \ell_1 \mathcal{M}_2(k_{z_0} )\begin{pmatrix}
            \mathcal{E} + mc^2 \\
            {\rm i } [\mathcal{E} + mc^2] \\
            k_{z_0} c +\i k'_{x_0} c\\
            k'_{x_0} c- \i k_{z_0} c
        \end{pmatrix}
        ={\cal T} (k_{z_0})\ket{\psi_k(k_{z_0})},
    \end{align}
    where
     \begin{eqnarray}
        |\psi_k(k_z)\rangle= \ell_1\begin{pmatrix}
            \mathcal{E} + mc^2 \\
            {\rm i } [\mathcal{E} + mc^2] \\
            k_z c {\hbar}+ \i k_x c {\hbar}\\
            k_x c {\hbar}- \i k_z c {\hbar}
        \end{pmatrix}
    \end{eqnarray}
    So we can still attain
    \begin{align}
        & \ket{\Psi_{\rm t} (x, z)} \nonumber\\
        \approx& \dfrac{{\cal T}(k_{z_0})}{\sqrt{2 \pi}}
            \int_{k_{z_0}-\delta k_z}^{k_{z_0}+\delta k_z} |\psi_k(k_z)\rangle \exp\left\{ {\rm i} \big[k_z z + x \sqrt{(\mathcal{E}-V_0)^2 -m^2c^4
            - k_z^2c^2\hbar^2} - \theta_{\rm t}(k_z)\big]\right\}{\rm d} k_z
            \notag \\
        \approx& \dfrac{{\cal T}(k_{z0})}{\sqrt{2 \pi}}
        \int_{k_{z_0}-\delta k_z}^{k_{z_0}+\delta k_z} |\Psi_0\rangle \,F(k_z)\,\exp\Bigg({\rm i}
                \bigg\{k_z z +x \sqrt{(\mathcal{E}-V_0)^2 -m^2c^4 - k_z^2c^2 \hbar^2} - \Big[\theta_{\rm t\,0}
                    +\frac{\partial\,\theta_{\rm t}}{\partial k_z} \Big|_{k_z = k_{z_0}} (k_z - k_{z_0})
                        \Big]\bigg\}\Bigg){\rm d}k_z \notag \\
        =& \dfrac{{\cal T}(k_{z0})\,|\Psi_0\rangle}{\sqrt{2 \pi}}  \int_{k_{z_0}-\delta k_z}^{k_{z_0}+\delta k_z} F(k_z)
            \,\exp\bigg({\rm i} \Big\{(\bar{k}_z+k_{z_0})z +x \sqrt{(\mathcal{E}-V_0)^2 -m^2c^4 - k_z^2c^2\hbar^2}
            -\left[\theta_{\rm t\,0} +\frac{\partial\,\theta_{\rm t}}{\partial k_z} \Big|_{k_z = k_{z_0}} \bar{k}_z\right]\Big\}\bigg){\rm d}k_z \notag \\
        =& \e^{\i(-\theta_{\rm t\,0} +{k}_{z_0} z)} \frac{{\cal T}(k_{z_0}) |\Psi_0\rangle}{\sqrt{2 \pi}}
             \int_{k_{z_0}-\delta k_z}^{k_{z_0}+\delta k_z} F(k_z)\,\exp\left\{
                {\rm i} \left[\bar{k}_z \left(z -\frac{\partial\,\theta_{\rm t}}{\partial k_z} \Big|_{k_z = k_{z_0}}\right) +x \sqrt{(\mathcal{E}-V_0)^2 -m^2c^4 - k_z^2c^2\hbar^2}
                    \right]\right\}{\rm d}k_z,
        \end{align}
        where \begin{eqnarray}
            \bar{k}_z  = k_z- k_{z_0}.
        \end{eqnarray}
So we can attain the new shift
\begin{eqnarray}\label{eq:shift-1b}
      \Delta z_{\rm t} = - \frac{\partial (-\theta_{\rm t}) }{\partial k_z}=\frac{\partial \theta_{\rm t} }{\partial k_z}=  \frac{\partial  }{\partial k_z} \left[\arctan \left(\frac{k_z (1-n)}{k_x +n k_x'}\right) \right].
  \end{eqnarray}
Now we come to treat Eq. (\ref{eq:shift-1b}). We can have
\begin{eqnarray}
    \label{z2'}  \Delta z_{\rm t} &=&  \frac{\partial  }{\partial k_z} \left[\arctan \left(\frac{k_z (1-n)}{k_x +n k_x'}\right) \right]\nonumber\\
    &=& \dfrac{1}{1+\left(\dfrac{k_z (1-n)}{k_x +n k_x'}\right)^2}\times  \dfrac{\left[\dfrac{\partial [k_z (1-n)]}{\partial k_z}\right][k_x +n k_x']-[k_z (1-n)]\left[\dfrac{\partial [k_x +n k_x']}{\partial k_z}\right]}{\left(k_x +n k_x'\right)^2}\nonumber\\
    &=&  \dfrac{\left[\dfrac{\partial [k_z (1-n)]}{\partial k_z}\right][k_x +n k_x']-[k_z (1-n)]\left[\dfrac{\partial [k_x +n k_x']}{\partial k_z}\right]}{\left(k_x +n k_x'\right)^2+\left(k_z (1-n)\right)^2}\nonumber\\
    &=&  \dfrac{ (1-n)[k_x +n k_x']-[k_z (1-n)]\left[\dfrac{\partial k_x}{\partial k_z}+n\dfrac{\partial k_x'}{\partial k_z}\right]}{\left(k_x +n k_x'\right)^2+\left(k_z (1-n)\right)^2}\nonumber\\
    &=&  \dfrac{ (1-n)[k_x +n k_x']-[k_z (1-n)]\left[\dfrac{\partial \dfrac{\sqrt{{\mathcal{E}}^2-m^2c^4 -c^2 \hbar^2\, k_z^2}}{c\hbar }}{\partial k_z}+n\dfrac{\partial \dfrac{\sqrt{\left(\mathcal{E}-V_0\right)^2-m^2c^4 -c^2 \hbar^2\, k_z^2}}{c\hbar }}{\partial k_z}\right]}{\left(k_x +n k_x'\right)^2+\left(k_z (1-n)\right)^2}\nonumber\\
    &=&  \dfrac{ (1-n)[k_x +n k_x']-\dfrac{[k_z (1-n)]}{c\hbar}\left[\dfrac{\partial {\sqrt{{\mathcal{E}}^2-m^2c^4 -c^2 \hbar^2\, k_z^2}}}{\partial k_z}+n\dfrac{\partial {\sqrt{\left(\mathcal{E}-V_0\right)^2-m^2c^4 -c^2 \hbar^2\, k_z^2}}}{\partial k_z}\right]}{\left(k_x +n k_x'\right)^2+\left(k_z (1-n)\right)^2}\nonumber\\
    &=&  \dfrac{ (1-n)[k_x +n k_x']-\dfrac{[k_z (1-n)]}{c\hbar}\left[ \dfrac{1}{2}\dfrac{-2c^2 \hbar^2\, k_z}{\sqrt{{\mathcal{E}}^2-m^2c^4 -c^2 \hbar^2\, k_z^2}}+n\dfrac{1}{2}\dfrac{-2c^2 \hbar^2\, k_z}{\sqrt{{\mathcal{E}-V_0}^2-m^2c^4 -c^2 \hbar^2\, k_z^2}}\right]}{\left(k_x +n k_x'\right)^2+\left(k_z (1-n)\right)^2}\nonumber\\
    &=&  \dfrac{ (1-n)[k_x +n k_x']-\dfrac{[k_z (1-n)]}{c\hbar}\left[ \dfrac{1}{2}\dfrac{-2c^2 \hbar^2\, k_z}{c \hbar k_x}+n\dfrac{1}{2}\dfrac{-2c^2 \hbar^2\, k_z}{c \hbar k_x'}\right]}{\left(k_x +n k_x'\right)^2+\left(k_z (1-n)\right)^2}\nonumber\\
    &=&  \dfrac{ (1-n)[k_x +n k_x']+[k_z (1-n)]\left[ \dfrac{ k_z}{ k_x}+n\dfrac{k_z}{k_x'}\right]}{\left(k_x +n k_x'\right)^2+\left(k_z (1-n)\right)^2}\nonumber\\
    &=& (1-n) \dfrac{ (k_x +n k_x')+k^2_z \left( \dfrac{1}{ k_x}+n\dfrac{1}{k_x'}\right)}{\left(k_x +n k_x'\right)^2+\left[k_z (1-n)\right]^2}\nonumber\\
    &=& -(n-1) \dfrac{ (k_x +n k_x')+\dfrac{k^2_z}{k_x k_x'} \left( k_x'+n k_x\right)}{\left(k_x +n k_x'\right)^2+\left[k_z (1-n)\right]^2}.
  \end{eqnarray}
  Because

\begin{eqnarray}
k_x>0, \;\;\; k_x'>0, \;\;\; n>1,
\end{eqnarray}
thus one has
\begin{eqnarray}
\Delta z_{\rm t}<0.
\end{eqnarray}
This means that the spatial shift for the transmitted wave is negative.

\subsubsection{The Case of $\; {\ell_2}/{\ell_1}= -\i$}

Let us consider the simple case of $\dfrac{\ell_2}{\ell_1}= -\i$, which means the spin direction is along the negative $y$-axis, i.e.,
\begin{eqnarray}
\vec{\tau}= (0,-1,0).
\end{eqnarray}
One can find that
\begin{eqnarray}
    B= -{\rm i } A,  \;\;\;\;  D= -{\rm i }C,
\end{eqnarray}
and
\begin{eqnarray}
    A = \ell_1 \dfrac{\sqrt{[k_x^2-n^2{k_x'}^2 - k_z^2(1 - n)^2 ]^2+ [2k_x k_z(1 - n)]^2 }}{(nk_x' + k_x)^2 + (1 - n)^2k_z^2} \, e^{{\rm i } \theta_{\rm r} },
\end{eqnarray}
\begin{eqnarray}
    C=\ell_1 \dfrac{\sqrt{(\mathcal{E}-V_0)(\mathcal{E} -V_0+mc^2)}}{\sqrt{\mathcal{E}(\mathcal{E} +mc^2)}} 2n k_x \dfrac{\sqrt{(k_x + nk_x')^2 +[k_z(1 - n)]^2}}{(nk_x' + k_x)^2 + (1 - n)^2k_z^2}\, e^{{\rm i} \theta_{\rm t}}.
\end{eqnarray}
As a result, we similarly have the spatial shifts for the reflected wave and transmitted wave as
\begin{eqnarray}
    \Delta z_{\rm r}'= -\Delta z_{\rm r} = - \frac{\partial \theta_{\rm r} }{\partial k_z},
\end{eqnarray}
\begin{eqnarray}
    \Delta z_{\rm t}'= -\Delta z_{\rm t} =-  \frac{\partial \theta_{\rm t} }{\partial k_z}.
\end{eqnarray}
This indicates when the spin flip, the spatial shift becomes the opposite value. This phenomena shows that the ETSG effect has a strong relation with the spin direction, thus is a spin effect.

\subsubsection{The Case of $\;{\ell_2}/{\ell_1} = G_1$ ($G_1 \in$ Real Numbers)}\label{D1D2}

Here $G_1$ is a real number. One may parameterize $\ell_1$ and $\ell_2$ as
\begin{eqnarray}\label{eq:l1l2}
\ell_1=\cos\frac{\theta}{2}, \;\;\;\; \ell_2=\sin\frac{\theta}{2},
\end{eqnarray}
then
\begin{eqnarray}
G_1=\frac{\ell_2}{\ell_1}=\tan\frac{\theta}{2},
\end{eqnarray}
and the normalization condition gives
\begin{eqnarray}
{|\ell_1|}^2 \left(1+G_1^2\right)=1.
\end{eqnarray}
In this case, from Eqs. (\ref{eq:1-A})-(\ref{eq:1-D}) we easily know that $\{A, B, C, D\}/\ell_1$ are all the real numbers, i.e.,
\begin{subequations}
   \begin{align}
&A=\ell_1\,  \frac{ \left[k_x^2-k_z^2(n-1)^2-k_x'^2 n^2\right]+ G_1 \, 2 k_x k_z  (n-1)}
{k_x^2+k_z^2(n-1)^2+2 k_x k_x' n +k_x'^2 n^2}\, {\rm e}^{-{\rm i}\theta_{\rm ra}}, \;\;\;\;\; \theta_{\rm ra}=0,  \label{eq:1-A-1a}\\
&B=\ell_1\, \frac{- 2 k_x k_z  (n-1)+ G_1 \left[k_x^2 -k_z^2(n-1)^2-k_x'^2 n^2\right]}
{k_x^2+k_z^2(n-1)^2+2 k_x k_x' n +k_x'^2 n^2}\, {\rm e}^{-{\rm i}\theta_{\rm rb}}, \;\;\;\;\; \theta_{\rm rb}=0, \label{eq:1-B-1a}\\
&C=\ell_1\, \dfrac{\sqrt{(\mathcal{E}-V_0)(\mathcal{E} -V_0+mc^2)}}{\sqrt{\mathcal{E}(\mathcal{E} +mc^2)}} \frac{2 k_x n \left\{\left[k_x +k_x' n\right]+G_1 k_z (n-1)\right\}}
{k_x^2+k_z^2(n-1)^2+2 k_x k_x' n +k_x'^2 n^2}\, {\rm e}^{-{\rm i}\theta_{\rm rc}}, \;\;\;\;\; \theta_{\rm rc}=0,\label{eq:1-C-1a}\\
& D=\ell_1\, \dfrac{\sqrt{(\mathcal{E}-V_0)(\mathcal{E} -V_0+mc^2)}}{\sqrt{\mathcal{E}(\mathcal{E} +mc^2)}}
\frac{2 k_x n \left\{-  k_z (n-1)+  G_1\, \left[k_x +k_x' n \right]\right\}}
{k_x^2+k_z^2(n-1)^2+2 k_x k_x' n +k_x'^2 n^2}\, {\rm e}^{-{\rm i}\theta_{\rm rd}}, \;\;\;\;\; \theta_{\rm rd}=0. \label{eq:1-D-1a}
    \end{align}
\end{subequations}
 hence, $\{A, B, C, D\}$ cannot provide the nontrivial phases (i.e., $\theta_{\rm ra}$, $\theta_{\rm rb}$, $\theta_{\rm rc}$, and $\theta_{\rm ra}$) that depend on $k_z$. In this situation, the reflected wave becomes ($p_y=\hbar k_y=0$)
 \begin{eqnarray}\label{eq:z-1a}
    &&    \ket{\Psi_{\rm r}}\equiv \dfrac{1}{\sqrt{2\mathcal{E}(\mathcal{E} +mc^2)}}\left\{A \begin{pmatrix}
                \mathcal{E} + mc^2 \\
                0 \\
                k_z c {\hbar}\\
                -k_x c {\hbar}
            \end{pmatrix}
            + B \begin{pmatrix}
                0 \\
                \mathcal{E} + mc^2 \\
                -k_x c {\hbar} \\
                -k_z c {\hbar}
            \end{pmatrix}\right\}{\rm e}^{\frac{\i}{\hbar} [{\hbar}(-k_x x+ k_z z) - Et]},
\end{eqnarray}
i.e.,
 \begin{eqnarray}
    &&    \ket{\Psi_{\rm r}}\equiv \dfrac{1}{\sqrt{2\mathcal{E}(\mathcal{E} +mc^2)}}\left\{\dfrac{\ell_1}{|\ell_1|} |A| \, {\rm e}^{-{\rm i}\theta_{\rm ra}} \begin{pmatrix}
                \mathcal{E} + mc^2 \\
                0 \\
                k_z c {\hbar}\\
                -k_x c {\hbar}
            \end{pmatrix}
            + \dfrac{\ell_1}{|\ell_1|} |B| \, {\rm e}^{-{\rm i}\theta_{\rm rb}} \begin{pmatrix}
                0 \\
                \mathcal{E} + mc^2 \\
                -k_x c {\hbar} \\
                -k_z c{\hbar}
            \end{pmatrix}\right\}{\rm e}^{{\rm i} [(-k_x x+ k_z z) - \frac{1}{\hbar}Et]},
\end{eqnarray}
i.e.,
\begin{eqnarray}
    &&    \ket{\Psi_{\rm r}}\equiv \dfrac{\dfrac{\ell_1}{|\ell_1|}}{\sqrt{2\mathcal{E}(\mathcal{E} +mc^2)}}\left\{ |A| \, {\rm e}^{-{\rm i}\theta_{\rm ra}} \begin{pmatrix}
                \mathcal{E} + mc^2 \\
                0 \\
                k_z c {\hbar}\\
                -k_x c {\hbar}
            \end{pmatrix}
            +  |B| \, {\rm e}^{-{\rm i}\theta_{\rm rb}} \begin{pmatrix}
                0 \\
                \mathcal{E} + mc^2 \\
                -k_x c {\hbar} \\
                -k_z c{\hbar}
            \end{pmatrix}\right\}{\rm e}^{{\rm i} [(-k_x x+ k_z z) - \frac{1}{\hbar}Et]},
\end{eqnarray}
The first term in the wavefunction $\ket{\Psi_{\rm r}}$ contributes a spatial shift as
\begin{eqnarray}
    \Delta z_{\rm ra}= \frac{\partial \theta_{\rm ra} }{\partial k_z}=0,
\end{eqnarray}
and the second term in the wavefunction $\ket{\Psi_{\rm r}}$ contributes a spatial shift as
\begin{eqnarray}
    \Delta z_{\rm rb}= \frac{\partial \theta_{\rm rb} }{\partial k_z}=0.
\end{eqnarray}


In the $z$-basis, for reflected wave, the total spatial shift is denoted as $\Delta z_r$, i.e.
\begin{eqnarray}\label{eq:Zab-1}
    \Delta {z}_{\rm r} =\frac{{|A| }^2}{ {|A| }^2+ {|B| }^2} \Delta z_{\rm ra}+\frac{{|B| }^2}{ {|A| }^2+ {|B| }^2}\Delta z_{\rm rb}.
\end{eqnarray}
For transmitted wave, the total spatial shift is denoted as  $\Delta z_t$, i.e.
\begin{eqnarray}
    \Delta {z}_{\rm t} =\frac{{|C| }^2}{ {|C| }^2+ {|D| }^2} \Delta z_{\rm ta}+\frac{{|D| }^2}{ {|C| }^2+ {|D| }^2}\Delta z_{\rm tb}.
\end{eqnarray}
Then at this case, the total spatial shifts for the reflected wave is given by
\begin{eqnarray}\label{eq:Zab-1'}
    \Delta {z}_{\rm r} =\frac{{|A| }^2}{ {|A| }^2+ {|B| }^2} \Delta z_{\rm ra}+\frac{{|B| }^2}{ {|A| }^2+ {|B| }^2}\Delta z_{\rm rb}=0.
\end{eqnarray}
Similarly, the total spatial shifts for the transmitted wave is also zero, i.e.,
\begin{eqnarray}
    \Delta {z}_{\rm t} =\frac{{|C| }^2}{ {|C| }^2+ {|D| }^2} \Delta z_{\rm ta}+\frac{{|D| }^2}{ {|C| }^2+ {|D| }^2}\Delta z_{\rm tb}=0.
\end{eqnarray}

\begin{remark}
\textcolor{blue}{Calculating the Zero Shifts in the $z$-Basis}. Actually, in the above the zero shifts have been calculated in the bases of $z$-direction (i.e., the $z$-basis). Let us explain this issue in the following. The incident wave is given in Eq. (\ref{eq:inwave-1}), i.e.,
\begin{eqnarray}
 &&   \ket{\Psi_{\rm in}}\equiv \dfrac{1}{\sqrt{2\mathcal{E}(\mathcal{E} +mc^2)}} \begin{pmatrix}
            \ell_1(\mathcal{E} + mc^2) \\
            \ell_2(\mathcal{E} + mc^2) \\
            [\ell_1 k_z + \ell_2\,(k_x- \i k_y)]c{\hbar} \\
            [\ell_1(k_x+\i k_y) - \ell_2\,k_z]c{\hbar}
        \end{pmatrix}\e^{\frac{\i}{\hbar} [{\hbar}(k_x x +k_y y +  k_z z) - Et]}.
\end{eqnarray}
In this section we have considered the simple case of $p_y=\hbar k_y=0$, thus the incident wave becomes
\begin{eqnarray}
 &&   \ket{\Psi_{\rm in}}\equiv \dfrac{1}{\sqrt{2\mathcal{E}(\mathcal{E} +mc^2)}} \begin{pmatrix}
            \ell_1(\mathcal{E} + mc^2) \\
            \ell_2(\mathcal{E} + mc^2) \\
            [\ell_1 k_z + \ell_2\,k_x]c{\hbar} \\
            [\ell_1 k_x - \ell_2\,k_z]c{\hbar}
        \end{pmatrix}\e^{ {\rm i} [(k_x x +  k_z z) -\frac{1}{\hbar} Et]},
\end{eqnarray}
i.e.,
\begin{eqnarray}
\ket{\Psi_{\rm in}} = \left\{\ell_1 \, \ket{{\tilde{s}}_z = +\frac{\hbar}{2}} + \ell_2 \ket{{\tilde{s}}_z = -\frac{\hbar}{2}}\right\}{\rm e}^{\frac{\i}{\hbar} [{\hbar}(k_x x+ k_z z) - Et]},
\end{eqnarray}
or
\begin{eqnarray} \label{eq:in-1a}  \ket{\Psi_{\rm in}} = \ell_1 \left\{ \ket{{\tilde{s}}_z = +\frac{\hbar}{2}} + G_1 \ket{{\tilde{s}}_z = -\frac{\hbar}{2}}\right\} {\rm e}^{\frac{\i}{\hbar} [{\hbar}(k_x x+ k_z z) - Et]}.
\end{eqnarray}
 In other words, the incident wave $\ket{\Psi_{\rm in}}$ has been written in the $z$-basis, where
\begin{eqnarray}
&&   \ket{{\tilde{s}}_z = +\frac{\hbar}{2}} \equiv \ket{\Psi_1'} =\frac{1}{\sqrt{2\mathcal{E}(\mathcal{E} +mc^2)}}  \begin{pmatrix}
        (\mathcal{E}+mc^2) |\uparrow\rangle_z\\
        c\hbar (\vec{\sigma} \cdot \vec{k}) |\uparrow\rangle_z
    \end{pmatrix},\nonumber\\
&&   \ket{{\tilde{s}}_z = -\frac{\hbar}{2}} \equiv   \ket{\Psi_2'} = \frac{1}{\sqrt{2\mathcal{E}(\mathcal{E} +mc^2)}} \begin{pmatrix}
        (\mathcal{E}+mc^2) |\downarrow\rangle_z\\
        c\hbar (\vec{\sigma} \cdot \vec{k}) |\downarrow\rangle_z
    \end{pmatrix},
\end{eqnarray}
represent the ``spin up'' and ``spin down'' along the $z$-axis respectively, and they are mutually orthogonal. Here
\begin{eqnarray}
|\uparrow\rangle_z = \begin{pmatrix}
    1\\
    0
\end{pmatrix}, \;\;\;\;
|\downarrow\rangle_z = \begin{pmatrix}
    0\\
    1
\end{pmatrix},
\end{eqnarray}
are eigenstates of $\sigma_z$, i.e.,
\begin{eqnarray}
 && \sigma_z |\uparrow\rangle_z =+1 |\uparrow\rangle_z, \;\;\;\;   \sigma_z |\downarrow\rangle_z =-1 |\downarrow\rangle_z.
\end{eqnarray}
Physically $|\uparrow\rangle_z$ and $|\downarrow\rangle_z$ (or $\ket{{\tilde{s}}_z = +\frac{\hbar}{2}}$ and $\ket{{\tilde{s}}_z = -\frac{\hbar}{2}}$) represent the ``spin up'' state and the ``spin down'' state along the $z$ direction, respectively.

Accordingly, due to the transformation
\begin{eqnarray}
&& k_x \rightarrow -k_x, \;\;\; k_y \rightarrow k_y, \;\;\; k_z \rightarrow k_z,\nonumber\\
&& \vec{k}=(k_x, k_y, k_z) \rightarrow  \vec{k}_{\rm r}=(-k_x, k_y, k_z),
\end{eqnarray}
from the incident wave one can have the form of the reflected wave $\ket{\Psi_{\rm r}}$ as shown in Eq. (\ref{eq:z-1a}). In the $z$-basis, the reflected wave $\ket{\Psi_{\rm r}}$ is given by
\begin{eqnarray}\label{eq:r-1a}
    &&    \ket{\Psi_{\rm r}} =\left\{A\, \ket{{\tilde{s}}_z = +\frac{\hbar}{2}}' + B\,  \ket{{\tilde{s}}_z = -\frac{\hbar}{2}}'\right\} {\rm e}^{\frac{\i}{\hbar} [{\hbar}(-k_x x+ k_z z) - Et]},
\end{eqnarray}
where
\begin{eqnarray}
    &&    \ket{{\tilde{s}}_z = +\frac{\hbar}{2}}'= \frac{1}{\sqrt{2\mathcal{E}(\mathcal{E} +mc^2)}}  \begin{pmatrix}
        (\mathcal{E}+mc^2) |\uparrow\rangle_z\\
        c\hbar (\vec{\sigma} \cdot \vec{k}_{\rm r}) |\uparrow\rangle_z
    \end{pmatrix}, \nonumber\\
    &&  \ket{{\tilde{s}}_z = -\frac{\hbar}{2}}'= \frac{1}{\sqrt{2\mathcal{E}(\mathcal{E} +mc^2)}}  \begin{pmatrix}
        (\mathcal{E}+mc^2) |\downarrow\rangle_z\\
        c\hbar (\vec{\sigma} \cdot \vec{k}_{\rm r}) |\downarrow\rangle_z
    \end{pmatrix}.
\end{eqnarray}

For convenience, one can choose $\ell_1$ as a real number, thus one can rewrite the reflected wave as
 \begin{eqnarray}
    &&    \ket{\Psi_{\rm r}} =\left\{|A|\, {\rm e}^{-{\rm i}\theta_{\rm ra}} \ket{{\tilde{s}}_z = +\frac{\hbar}{2}} + |B|\,  {\rm e}^{-{\rm i}\theta_{\rm rb}}\ket{{\tilde{s}}_z = -\frac{\hbar}{2}} \right\} {\rm e}^{\frac{\i}{\hbar} [{\hbar}(-k_x x+ k_z z) - Et]},
\end{eqnarray}
When $\ell_1/\ell_2$ equals to a real number, then one easily has $\theta_{\rm ra}=\theta_{\rm rb}=0$, thus $\Delta z_{\rm ra}=\Delta z_{\rm rb}=0$. Due to Eq. (\ref{eq:Zab-1}) one easily obtains the zero shift for the reflected wave. Furthermore, if we denote
 \begin{eqnarray}
    && P_1= \frac{{|A| }^2}{ {|A| }^2+ {|B| }^2}, \;\;\;\;\;   P_2= \frac{{|B| }^2}{ {|A| }^2+ {|B| }^2},
\end{eqnarray}
where $P_1$ ($P_2$) represents the probability of $\ket{{\tilde{s}}_z = +\frac{\hbar}{2}}$  ($\ket{{\tilde{s}}_z = -\frac{\hbar}{2}}$) appearing in the reflected wave $\ket{\Psi_{\rm r}}$, then the spatial shift of the reflected wave can be simply written as
\begin{eqnarray}
    \Delta {z}_{\rm r} =P_1 \Delta z_{\rm ra}+ P_2 \Delta z_{\rm rb}=0.
\end{eqnarray}
Similarly, the spatial shift of the transmitted wave can be written as
\begin{eqnarray}
    \Delta {z}_{\rm t} =P_1 \Delta z_{\rm ta}+ P_2 \Delta z_{\rm tb}=0.
\end{eqnarray}
$\blacksquare$
\end{remark}

\begin{remark}
\textcolor{blue}{Calculating the Zero Shifts in the $y$-Basis}. For the incident wave $\ket{\Psi_{\rm in}}$ and the reflected wave $\ket{\Psi_{\rm r}}$, it can be expended not only in the $z$-basis, and but also in other bases, such as the $y$-basis. Physically, the spatial shifts will be the same, namely, the shifts do not depend on the bases that one has chosen. However, for the simplified case of $k_y=0$, the $y$-basis is particularly interesting. In the following we would like to calculate the zero shifts in the $y$-basis.

Let us focus on the reflected wave in Eq. (\ref{eq:r-1a}), i.e.,
 \begin{eqnarray}
    &&    \ket{\Psi_{\rm r}} =\left\{A\, \ket{{\tilde{s}}_z = +\frac{\hbar}{2}}' + B\,  \ket{{\tilde{s}}_z = -\frac{\hbar}{2}}'\right\} {\rm e}^{\frac{\i}{\hbar} [{\hbar}(-k_x x+ k_z z) - Et]},
\end{eqnarray}
Alternatively, one can expend the reflected wave in the $y$-basis, i.e.,
 \begin{eqnarray}\label{eq:r-1c}
    &&    \ket{\Psi_{\rm r}} =\left\{A'\, \ket{{\tilde{s}}_y = +\frac{\hbar}{2}}' + B'\,  \ket{{\tilde{s}}_y = -\frac{\hbar}{2}}' \right\} {\rm e}^{\frac{\i}{\hbar} [{\hbar}(-k_x x+ k_z z) - Et]}.
\end{eqnarray}
Here
\begin{eqnarray}
&&   \ket{{\tilde{s}}_y = +\frac{\hbar}{2}}'  =\frac{1}{\sqrt{2\mathcal{E}(\mathcal{E} +mc^2)}}  \begin{pmatrix}
        (\mathcal{E}+mc^2) |\uparrow\rangle_y\\
        c \hbar (\vec{\sigma} \cdot \vec{k}_{\rm r}) |\uparrow\rangle_y
    \end{pmatrix},\nonumber\\
&&   \ket{{\tilde{s}}_y = -\frac{\hbar}{2}}' = \frac{1}{\sqrt{2\mathcal{E}(\mathcal{E} +mc^2)}} \begin{pmatrix}
        (\mathcal{E}+mc^2) |\downarrow\rangle_y\\
        c \hbar(\vec{\sigma} \cdot \vec{k}_{\rm r}) |\downarrow\rangle_y
    \end{pmatrix},
\end{eqnarray}
are the ``spin up'' and ``spin down'' states along the $y$-axis, and
\begin{eqnarray}
|\uparrow\rangle_y = \frac{1}{\sqrt{2}}\begin{pmatrix}
    1\\
    {\rm i}
\end{pmatrix}, \;\;\;\;
|\downarrow\rangle_y = \frac{1}{\sqrt{2}} \begin{pmatrix}
    1\\
    -{\rm i}
\end{pmatrix},
\end{eqnarray}
are eigenstates of $\sigma_y$, i.e.,
\begin{eqnarray}
  \sigma_y |\uparrow\rangle_y =+1 |\uparrow\rangle_y, \;\;\;\;   \sigma_y |\downarrow\rangle_y = - 1 |\downarrow\rangle_y.
\end{eqnarray}
By solving the following equation
  \begin{eqnarray}
    &&  A'\, \ket{{\tilde{s}}_y = +\frac{\hbar}{2}}' + B'\,  \ket{{\tilde{s}}_y = -\frac{\hbar}{2}}'=A\, \ket{{\tilde{s}}_z = +\frac{\hbar}{2}}' + B\,  \ket{{\tilde{s}}_z = -\frac{\hbar}{2}}',
\end{eqnarray}
one has
\begin{eqnarray}
    A' |\uparrow\rangle_y + B' |\downarrow\rangle_y =  A |\uparrow\rangle_z + B |\downarrow\rangle_z.
\end{eqnarray}
Based on which, one has
\begin{eqnarray}
    A' \frac{1}{\sqrt{2}}\begin{pmatrix}
    1\\
    {\rm i}
\end{pmatrix} + B' \frac{1}{\sqrt{2}}\begin{pmatrix}
    1\\
    -{\rm i}
\end{pmatrix} = A  \begin{pmatrix}
    1\\
    0
\end{pmatrix} + B \begin{pmatrix}
    0\\
    1
\end{pmatrix},
\end{eqnarray}
i.e.,
\begin{eqnarray}
&&    \frac{1}{\sqrt{2}} (A'+B') = A, \nonumber\\
&&    \frac{{\rm i}}{\sqrt{2}} (A'-B') =B,
\end{eqnarray}
which yields
\begin{eqnarray}
    A' = \frac{1}{\sqrt{2}}(A-{\rm i}B),\,\,\,\, B' = \frac{1}{\sqrt{2}}(A+{\rm i } B).
\end{eqnarray}

Due to Eq. (\ref{eq:1-A}) and Eq. (\ref{eq:1-B}), i.e.,
\begin{eqnarray}
&&A=\frac{\ell_1 \left[k_x^2-k_z^2(n-1)^2-k_x'^2 n^2\right]+ \ell_2 \, 2 k_x k_z  (n-1)}
{k_x^2+k_z^2(n-1)^2+2 k_x k_x' n +k_x'^2 n^2}, \nonumber\\
&&B=\frac{-\ell_1\, 2 k_x k_z  (n-1)+\ell_2\, \left[k_x^2 -k_z^2(n-1)^2-k_x'^2 n^2\right]}
{k_x^2+k_z^2(n-1)^2+2 k_x k_x' n +k_x'^2 n^2},
\end{eqnarray}
one has
\begin{eqnarray}
    A' &=& \frac{1}{\sqrt{2}}(A-{\rm i}B)= \frac{1}{\sqrt{2}} \frac{ {(\ell_1-{\rm i} \ell_2)} \left[k_x^2-k_z^2(n-1)^2-k_x'^2 n^2\right]+ {(\ell_2+{\rm i}\ell_1)} \, 2 k_x k_z  (n-1)}
{k_x^2+k_z^2(n-1)^2+2 k_x k_x' n +k_x'^2 n^2}\nonumber\\
&=&  \frac{{(\ell_1-{\rm i} \ell_2)}}{\sqrt{2}}  \frac{  \left[k_x^2-k_z^2(n-1)^2-k_x'^2 n^2\right]-{\rm i}\, 2 k_x k_z  (n-1)}
{k_x^2+k_z^2(n-1)^2+2 k_x k_x' n +k_x'^2 n^2}\nonumber\\
&=&  \frac{{(\ell_1-{\rm i} \ell_2)}}{\sqrt{2}}  \biggr|\frac{  \left[k_x^2-k_z^2(n-1)^2-k_x'^2 n^2\right]-{\rm i}\, 2 k_x k_z  (n-1)}
{k_x^2+k_z^2(n-1)^2+2 k_x k_x' n +k_x'^2 n^2}\biggr|\, {\rm e}^{-{\rm i}\theta_{\rm ra}},
\end{eqnarray}
with
\begin{eqnarray}
\tan \theta_{\rm ra}=\frac{2 k_x k_z  (n-1)}{k_x^2-k_z^2(n-1)^2-k_x'^2 n^2}.
\end{eqnarray}
Similarly, one has
\begin{eqnarray}
    B' &=&  \frac{{(\ell_1+{\rm i} \ell_2)}}{\sqrt{2}}  \biggr|\frac{  \left[k_x^2-k_z^2(n-1)^2-k_x'^2 n^2\right]+{\rm i}\, 2 k_x k_z  (n-1)}
{k_x^2+k_z^2(n-1)^2+2 k_x k_x' n +k_x'^2 n^2}\biggr|\, {\rm e}^{-{\rm i}\theta_{\rm rb}},
\end{eqnarray}
with
\begin{eqnarray}\label{eq:theta-1}
 \theta_{\rm rb}=-  \theta_{\rm ra}.
\end{eqnarray}
Based on Eq. (\ref{eq:theta-1}), one easily knows that
\begin{eqnarray}
&& \Delta z_{\rm r}(+\hat{y})\equiv \Delta z_{\rm ra}= \frac{\partial \theta_{\rm ra} }{\partial k_z} =\dfrac{\sqrt{\mathcal{E}-mc^2} }{\sqrt{{\mathcal{E}}+mc^2}}
\dfrac{c\hbar {\mathcal{E}} \cos \varphi}{\left[ {\mathcal{E}}^2 \sin^2 \varphi + m^2c^4 \cos^2 \varphi \right]}
 \left[\tan^2 \varphi -   \dfrac{mc^2}{\mathcal{E}} \right], \nonumber\\
&& \Delta z_{\rm r}(-\hat{y})\equiv \Delta z_{\rm rb}= \frac{\partial \theta_{\rm rb} }{\partial k_z}=-\frac{\partial \theta_{\rm ra} }{\partial k_z} =-\Delta z_{\rm r}(+\hat{y}).
\end{eqnarray}
Here $+\hat{y}$ denotes $\vec{e}_y=(0, 1, 0)$, $-\hat{y}$ denotes $-\vec{e}_y=(0, -1, 0)$, $\Delta z_{\rm r}(+\hat{y})$ denotes the spatial shift caused by the state $\ket{{\tilde{s}}_y = +\frac{\hbar}{2}}'$, and $\Delta z_{\rm r}(-\hat{y})$ denotes the spatial shift caused by the state $\ket{{\tilde{s}}_y = -\frac{\hbar}{2}}'$. One can verify that
\begin{eqnarray}
&& |A'|^2=|B'|^2,
\end{eqnarray}
thus the spatial shift for the reflected wave is given by
\begin{eqnarray}
    \Delta {z}_{\rm r} &=& \frac{{|A'| }^2}{ {|A'| }^2+ {|B'| }^2} \Delta z_{\rm ra}+\frac{{|B'| }^2}{ {|A'| }^2+ {|B'| }^2}\Delta z_{\rm rb}\nonumber\\
    &=& \frac{{|A'| }^2}{ {|A'| }^2+ {|B'| }^2} \Delta z_{\rm r}(+\hat{y})+\frac{{|B'| }^2}{ {|A'| }^2+ {|B'| }^2}\Delta z_{\rm r}(-\hat{y})\nonumber\\
    &=& \frac{{|A'| }^2- {|B'| }^2}{ {|A'| }^2+ {|B'| }^2} \Delta z_{\rm r}(+\hat{y})=0.
\end{eqnarray}
Thus one gets the same zero shift for the reflected wave in the $y$-basis. Similarly, one can also calculate the zero shift for the transmitted wave in the $y$-basis. $\blacksquare$
\end{remark}

\subsubsection{The Case of $\;{\ell_2}/{\ell_1} = G_1+ {\rm i} G_2$ ($G_1, G_2 \in$ Real Numbers)}\label{D2D2}

Here $G_1$ and $G_2$ are real numbers. One may parameterize $\ell_1$ and $\ell_2$ as
\begin{eqnarray}\label{eq:l1l2-1}
\ell_1=\cos\frac{\theta}{2}, \;\;\;\; \ell_2=\sin\frac{\theta}{2} {\rm e}^{{\rm i}\phi},
\end{eqnarray}
where $\theta \in [0, \pi]$ and $\phi\in [0,2\pi]$, then
\begin{eqnarray}
G_1+ {\rm i} G_2=\frac{\ell_2}{\ell_1}=\tan\frac{\theta}{2} {\rm e}^{{\rm i}\phi},
\end{eqnarray}
and the normalization condition gives
\begin{eqnarray}
{|\ell_1|}^2 \left(1+G_1^2+G_2^2\right)=1.
\end{eqnarray}
In this case, the spin direction is in a general direction, i.e.,
\begin{eqnarray}
\vec{\tau}(\theta, \phi) = (\sin\theta\cos\phi, \sin\theta\sin\phi, \cos\theta).
\end{eqnarray}
Note that if we parameterize $\ell_1$ and $\ell_2$ as
\begin{eqnarray}
&&\ell_1=\cos\frac{\theta}{2} {\rm e}^{{\rm i}\phi_1}, \;\;\;\; \ell_2=\sin\frac{\theta}{2} {\rm e}^{{\rm i}\phi_2}, \nonumber\\
&& \phi=\phi_2-\phi_1,
\end{eqnarray}
one finds that the spin direction is the same. So the phase factor $\phi_1$ is trivial, and the relative phase factor $\phi=\phi_2-\phi_1$ is non-trivial. Without loss of generality, one can always choose $\ell_1=\cos\frac{\theta}{2}$ as a real number.

The form of the incident wave is still
\begin{eqnarray}
 &&   \ket{\Psi_{\rm in}}\equiv \dfrac{1}{\sqrt{2\mathcal{E}(\mathcal{E} +mc^2)}} \begin{pmatrix}
            \ell_1(\mathcal{E} + mc^2) \\
            \ell_2(\mathcal{E} + mc^2) \\
            [\ell_1 k_z + \ell_2\,k_x]c{\hbar} \\
            [\ell_1 k_x - \ell_2\,k_z]c{\hbar}
        \end{pmatrix}\e^{ {\rm i} [(k_x x +  k_z z) -\frac{1}{\hbar} Et]},
\end{eqnarray}
or
\begin{eqnarray}\label{eq:in-3a}
    \ket{\Psi_{\rm in}} =\frac{1}{\sqrt{2\mathcal{E}(\mathcal{E} +mc^2)}}  \begin{pmatrix}
        (\mathcal{E}+mc^2) |\chi\rangle  \\
        c {\hbar}(\vec{\sigma} \cdot \vec{k}) |\chi\rangle
    \end{pmatrix},
\end{eqnarray}
with the state
\begin{eqnarray}
 |\chi\rangle= \begin{pmatrix}
    \ell_1\\
    \ell_2
\end{pmatrix}= \begin{pmatrix}
   \cos\frac{\theta}{2} \\
    \sin\frac{\theta}{2} {\rm e}^{{\rm i}\phi}
\end{pmatrix}.
\end{eqnarray}
The form of the reflected wave is still
\begin{eqnarray}
 &&   \ket{\Psi_{\rm r}}\equiv \dfrac{1}{\sqrt{2\mathcal{E}(\mathcal{E} +mc^2)}} \begin{pmatrix}
            A(\mathcal{E} + mc^2) \\
            B(\mathcal{E} + mc^2) \\
            [A k_z - B k_x]c{\hbar} \\
            [-A k_x - B k_z]c{\hbar}
        \end{pmatrix}\e^{ {\rm i} [(-k_x x +  k_z z) -\frac{1}{\hbar} Et]},
\end{eqnarray}
or
\begin{eqnarray}
    \ket{\Psi_{\rm r}} =\frac{1}{\sqrt{2\mathcal{E}(\mathcal{E} +mc^2)}}  \begin{pmatrix}
        (\mathcal{E}+mc^2) |\chi_{\rm r}\rangle  \\
        c {\hbar}(\vec{\sigma} \cdot \vec{k}_{\rm r}) |\chi_{\rm r}\rangle
    \end{pmatrix},
\end{eqnarray}
with
\begin{eqnarray}
 |\chi_{\rm r}\rangle= \begin{pmatrix}
    A\\
    B
\end{pmatrix}.
\end{eqnarray}
Here $|\chi_{\rm r}\rangle$ is unnormalized. The coefficients $A$ and $B$ are given in Eq. (\ref{eq:1-A}) and Eq. (\ref{eq:1-B}), i.e.,
\begin{eqnarray}
&&A=\frac{\ell_1 \left[k_x^2-k_z^2(n-1)^2-k_x'^2 n^2\right]+ \ell_2 \, 2 k_x k_z  (n-1)}
{k_x^2+k_z^2(n-1)^2+2 k_x k_x' n +k_x'^2 n^2}, \nonumber\\
&&B=\frac{-\ell_1\, 2 k_x k_z  (n-1)+\ell_2\, \left[k_x^2 -k_z^2(n-1)^2-k_x'^2 n^2\right]}
{k_x^2+k_z^2(n-1)^2+2 k_x k_x' n +k_x'^2 n^2}.
\end{eqnarray}

One can expand $ |\chi_{\rm r}\rangle$ in the $y$-basis. He has
\begin{eqnarray}
 |\chi_{\rm r}\rangle= A'  |\uparrow\rangle_y + B' |\downarrow\rangle_y,
\end{eqnarray}
i.e.,
\begin{eqnarray}
 \begin{pmatrix}
    A\\
    B
\end{pmatrix} = A'  \frac{1}{\sqrt{2}}\begin{pmatrix}
   1 \\
   {\rm i}
\end{pmatrix} + B' \frac{1}{\sqrt{2}} \begin{pmatrix}
   1 \\
   -{\rm i}
\end{pmatrix},
\end{eqnarray}
which leads to
\begin{eqnarray}
    A' = \frac{1}{\sqrt{2}}(A-{\rm i}B),\,\,\,\, B' = \frac{1}{\sqrt{2}}(A+{\rm i } B).
\end{eqnarray}
Then one obtains
\begin{eqnarray}
 &&   A' =  \frac{{(\ell_1-{\rm i} \ell_2)}}{\sqrt{2}}  \biggr|\frac{  \left[k_x^2-k_z^2(n-1)^2-k_x'^2 n^2\right]-{\rm i}\, 2 k_x k_z  (n-1)}
{k_x^2+k_z^2(n-1)^2+2 k_x k_x' n +k_x'^2 n^2}\biggr|\, {\rm e}^{-{\rm i}\theta_{\rm ra}}, \nonumber\\
&&   B' =  \frac{(\ell_1+{\rm i} \ell_2)}{\sqrt{2}}  \biggr|\frac{  \left[k_x^2-k_z^2(n-1)^2-k_x'^2 n^2\right]+{\rm i}\, 2 k_x k_z  (n-1)}
{k_x^2+k_z^2(n-1)^2+2 k_x k_x' n +k_x'^2 n^2}\biggr|\, {\rm e}^{-{\rm i}\theta_{\rm rb}},
\end{eqnarray}
with
\begin{eqnarray}
&&\tan \theta_{\rm ra}=\frac{2 k_x k_z  (n-1)}{k_x^2-k_z^2(n-1)^2-k_x'^2 n^2}, \nonumber\\
&& \theta_{\rm rb}=-\theta_{\rm ra}.
\end{eqnarray}
In the $y$-basis, the reflected wave reads
\begin{eqnarray}
\ket{\Psi_{\rm r}} &=& \left\{A'\, \ket{{\tilde{s}}_y = +\frac{\hbar}{2}}' + B'\,  \ket{{\tilde{s}}_y = -\frac{\hbar}{2}}' \right\} {\rm e}^{\frac{\i}{\hbar} [{\hbar}(-k_x x+ k_z z) - Et]}\nonumber\\
& \propto & \left\{ |A'|\, { {\rm e}^{{\rm i}{\rm arg}[\ell_1-{\rm i} \ell_2]}} {\rm e}^{-{\rm i}\theta_{\rm ra}} \ket{{\tilde{s}}_y = +\frac{\hbar}{2}}' + |B'|\, {{\rm e}^{{\rm i}{\rm arg}[\ell_1+{\rm i} \ell_2]}}{\rm e}^{{\rm i}\theta_{\rm ra}} \ket{{\tilde{s}}_y = -\frac{\hbar}{2}}' \right\} {\rm e}^{\frac{\i}{\hbar} [{\hbar}(-k_x x+ k_z z) - Et]}.
\end{eqnarray}
Similarly, based on Eq. (\ref{eq:theta-1}), one can have
\begin{eqnarray}
&& \Delta z_{\rm r}(+\hat{y})\equiv \Delta z_{\rm ra}= \frac{\partial \theta_{\rm ra} }{\partial k_z} =\dfrac{\sqrt{\mathcal{E}-mc^2} }{\sqrt{{\mathcal{E}}+mc^2}}
\dfrac{c\hbar {\mathcal{E}} \cos \varphi}{\left[ {\mathcal{E}}^2 \sin^2 \varphi + m^2c^4 \cos^2 \varphi \right]}
 \left[\tan^2 \varphi -   \dfrac{mc^2}{\mathcal{E}} \right], \nonumber\\
&& \Delta z_{\rm r}(-\hat{y})\equiv \Delta z_{\rm rb}= \frac{\partial \theta_{\rm rb} }{\partial k_z}=-\frac{\partial \theta_{\rm ra} }{\partial k_z} =-\Delta z_{\rm r}(+\hat{y}),
\end{eqnarray}
where $\Delta z_{\rm r}(+\hat{y})$ and $\Delta z_{\rm r}(-\hat{y})$ denote the shifts caused by the states $\ket{{\tilde{s}}_y = +\frac{\hbar}{2}}'$, and $\ket{{\tilde{s}}_y = -\frac{\hbar}{2}}'$, respectively.

Let us define the probabilities
\begin{eqnarray}
&& P_1= \frac{{|A'| }^2}{ {|A'| }^2+ {|B'| }^2},\nonumber\\
&& P_2= \frac{{|B'| }^2}{ {|A'| }^2+ {|B'| }^2},
\end{eqnarray}
then the spatial shift for the reflected wave is given by
\begin{eqnarray}
    \Delta {z}_{\rm r} =P_1 \Delta z_{\rm ra}+ P_2 \Delta z_{\rm rb},
\end{eqnarray}
i.e.,
\begin{eqnarray}\label{eq:z-1b}
    \Delta {z}_{\rm r} &=& P_1 \Delta z_{\rm r}(+\hat{y})+ P_2 \Delta z_{\rm r}(-\hat{y})=(P_1-P_2)\Delta z_{\rm r}(+\hat{y})\nonumber\\
    &=& \frac{{|A'| }^2-{|B'| }^2}{ {|A'| }^2+ {|B'| }^2}\Delta z_{\rm r}(+\hat{y}).
\end{eqnarray}
Because
\begin{eqnarray}
&& \Delta z_{\rm r}(+\hat{y})=\dfrac{\sqrt{\mathcal{E}-mc^2} }{\sqrt{{\mathcal{E}}+mc^2}}
\dfrac{c\hbar {\mathcal{E}} \cos \varphi}{\left[ {\mathcal{E}}^2 \sin^2 \varphi + m^2c^4 \cos^2 \varphi \right]}
 \left[\tan^2 \varphi -   \dfrac{mc^2}{\mathcal{E}} \right],
\end{eqnarray}
and
\begin{eqnarray}\label{eq:P1P2}
 \frac{{|A'| }^2-{|B'| }^2}{ {|A'| }^2+ {|B'| }^2}&=& \dfrac{1-\dfrac{|B'| ^2}{|A'| ^2}}{1+\dfrac{|B'| ^2}{|A'| ^2}}=
  \dfrac{1-\dfrac{|\ell_1+{\rm i} \ell_2| ^2}{|\ell_1-{\rm i} \ell_2| ^2}}{1+\dfrac{|\ell_1+{\rm i} \ell_2| ^2}{|\ell_1-{\rm i} \ell_2| ^2}}=
  \dfrac{|\ell_1-{\rm i} \ell_2| ^2-|\ell_1+{\rm i} \ell_2| ^2}{|\ell_1-{\rm i} \ell_2| ^2+|\ell_1+{\rm i} \ell_2| ^2}\nonumber\\
  &=& \dfrac{\biggr|1-{\rm i} \dfrac{\ell_2}{\ell_1}\biggr| ^2-\biggr|1+{\rm i} \dfrac{\ell_2}{\ell_1}\biggr| ^2}{\biggr|1-{\rm i} \dfrac{\ell_2}{\ell_1}\biggr| ^2+\biggr|1+{\rm i} \dfrac{\ell_2}{\ell_1}\biggr| ^2}=
  \dfrac{\biggr|1-{\rm i} (G_1+{\rm i} G_2)\biggr| ^2-\biggr|1+{\rm i} (G_1+{\rm i} G_2)\biggr| ^2}{\biggr|1-{\rm i} (G_1+{\rm i} G_2)\biggr| ^2+\biggr|1+{\rm i} (G_1+{\rm i} G_2)\biggr| ^2}\nonumber\\
&=& \dfrac{\biggr|(1+G_2)-{\rm i} G_1\biggr| ^2-\biggr|(1-G_2)+{\rm i} G_1\biggr| ^2}
{\biggr|(1+G_2)-{\rm i} G_1\biggr| ^2+\biggr|(1-G_2)+{\rm i} G_1\biggr| ^2}= \dfrac{\biggr[(1+G_2)^2+ G_1^2\biggr]-\biggr[(1-G_2)^2+ G_1^2\biggr]}
{\biggr[(1+G_2)^2+ G_1^2\biggr]+\biggr[(1-G_2)^2+ G_1^2\biggr]}\nonumber\\
&=& \dfrac{ 4 G_2}
{\biggr[(1+G_2)^2+ G_1^2\biggr]+\biggr[(1-G_2)^2+ G_1^2\biggr]}= \dfrac{ 4 G_2}
{2 \biggr[1+G_1^2+ G_2^2\biggr]}=\dfrac{ 2 |\ell_1|^2 G_2}
{|\ell_1|^2 \biggr[1+G_1^2+ G_2^2\biggr]}\nonumber\\
&=&  2 |\ell_1|^2 G_2=2 \biggr|\sin \frac{\theta}{2}\biggr|^2  \tan\frac{\theta}{2} \sin\phi = 2 \biggr(\sin \frac{\theta}{2}\biggr)^2  \tan\frac{\theta}{2} \sin\phi=2 \sin \frac{\theta}{2}  \cos\frac{\theta}{2} \sin\phi \nonumber\\
&=& \sin\theta\sin\phi.
\end{eqnarray}
By considering
\begin{eqnarray}
\vec{\tau}(\theta, \phi) = (\sin\theta\cos\phi, \sin\theta\sin\phi, \cos\theta),
\;\;\;\;\; \vec{e}_y=(0, 1, 0),
\end{eqnarray}
one has
\begin{eqnarray}
\tau_y=\vec{\tau}\cdot\vec{e}_y = \sin\theta\sin\phi,
\end{eqnarray}
thus
\begin{eqnarray}
 \frac{{|A'| }^2-{|B'| }^2}{ {|A'| }^2+ {|B'| }^2}=\tau_y.
\end{eqnarray}
Then from Eq. (\ref{eq:z-1b}) we have
\begin{eqnarray}\label{eq:z-1c}
    \Delta {z}_{\rm r} &=& \frac{{|A'| }^2-{|B'| }^2}{ {|A'| }^2+ {|B'| }^2}\Delta z_{\rm r}(+\hat{y})=
    \tau_y \Delta z_{\rm r}(+\hat{y}),
\end{eqnarray}
or
\begin{eqnarray}\label{eq:z-1d}
    \Delta {z}_{\rm r} &=& (\sin\theta\sin\phi) \, \dfrac{\sqrt{\mathcal{E}-mc^2} }{\sqrt{{\mathcal{E}}+mc^2}}
\dfrac{c\hbar {\mathcal{E}} \cos \varphi}{\left[ {\mathcal{E}}^2 \sin^2 \varphi + m^2c^4 \cos^2 \varphi \right]}
 \left[\tan^2 \varphi -   \dfrac{mc^2}{\mathcal{E}} \right].
\end{eqnarray}
Similarly, for the transmitted wave one has the spin GH shift as
\begin{eqnarray}\label{eq:z-1e}
    \Delta {z}_{\rm t} &=& \tau_y\Delta z_{\rm t}(+\hat{y}) \nonumber\\
    &=& (\sin\theta\sin\phi) \, (1-n) \dfrac{ (k_x +n k_x')+k^2_z \left( \dfrac{1}{ k_x}+n\dfrac{1}{k_x'}\right)}{\left(k_x +n k_x'\right)^2+\left[k_z (1-n)\right]^2}.
\end{eqnarray}


\begin{remark}\textcolor{blue}{Calculating the Spatial Shifts in the $z$-Basis}. If the spatial shifts are calculated in the other more general bases, one can proved that the same results are still valid. Here we provide the calculation for the case of $z$-basis. In the following, let us consider the spatial shift for the reflected wave, and the spatial shift for the transmitted wave can be obtained similarly.

In this situation, the reflected wave becomes (in the case of $k_y=0$)
 \begin{eqnarray}
    &&    \ket{\Psi_{\rm r}}\equiv \dfrac{1}{\sqrt{2\mathcal{E}(\mathcal{E} +mc^2)}}\left\{A \begin{pmatrix}
                \mathcal{E} + mc^2 \\
                0 \\
                k_z c {\hbar}\\
                -k_x c {\hbar}
            \end{pmatrix}
            + B \begin{pmatrix}
                0 \\
                \mathcal{E} + mc^2 \\
                -k_x c{\hbar} \\
                -k_z c{\hbar}
            \end{pmatrix}\right\}{\rm e}^{\frac{\i}{\hbar} [{\hbar}(-k_x x+ k_z z) - Et]},
\end{eqnarray}
i.e.,
\begin{eqnarray}
    &&    \ket{\Psi_{\rm r}}\equiv \dfrac{\dfrac{\ell_1}{|\ell_1|}}{\sqrt{2\mathcal{E}(\mathcal{E} +mc^2)}}\left\{ |A| \, {\rm e}^{-{\rm i}\theta_{\rm ra}} \begin{pmatrix}
                \mathcal{E} + mc^2 \\
                0 \\
                k_z c {\hbar}\\
                -k_x c {\hbar}
            \end{pmatrix}
            +  |B| \, {\rm e}^{-{\rm i}\theta_{\rm rb}} \begin{pmatrix}
                0 \\
                \mathcal{E} + mc^2 \\
                -k_x c{\hbar} \\
                -k_z c{\hbar}
            \end{pmatrix}\right\}{\rm e}^{{\rm i} [(-k_x x+ k_z z) - \frac{1}{\hbar}Et]},
\end{eqnarray}
where the coefficients
    \begin{eqnarray}
 &&A=\frac{\ell_1 \left[k_x^2-k_z^2(n-1)^2-k_x'^2 n^2\right]+ \ell_2 \, 2 k_x k_z  (n-1)}
 {k_x^2+k_z^2(n-1)^2+2 k_x k_x' n +k_x'^2 n^2}, \nonumber\\
 &&B=\frac{-\ell_1\, 2 k_x k_z  (n-1)+\ell_2\, \left[k_x^2 -k_z^2(n-1)^2-k_x'^2 n^2\right]}
 {k_x^2+k_z^2(n-1)^2+2 k_x k_x' n +k_x'^2 n^2}.
\end{eqnarray}

 Let us denote
    \begin{eqnarray}
    &&    \Omega_1 = k_x^2-k_z^2(n-1)^2-k_x'^2 n^2,\nonumber\\
    &&    \Omega_2 = 2 k_x k_z  (n-1),
    \end{eqnarray}
then we have
    \begin{eqnarray}
     &&   A = \ell_1 \dfrac{\Omega_1+ (G_1 +\i G_2)\Omega_2}{(n k_x' + k_x)^2 + (1 - n)^2k_z^2}, \nonumber\\
     &&   B = \ell_1\dfrac{-\Omega_2 + (G_1+\i G_2) \Omega_1}{(nk_x' + k_x)^2 + (1 - n)^2k_z^2},
    \end{eqnarray}
Therefore
    \begin{eqnarray}
      &&  |A|^2 = |\ell_1|^2  \frac{(\Omega_1+ G_1 \Omega_2 )^2 +G_2^2 \Omega_2^2}{[(n k_x' + k_x)^2 + (1 - n)^2k_z^2]^2}, \nonumber\\
      &&  |B|^2 = |\ell_1|^2  \frac{(G_1 \Omega_1 - \Omega_2 )^2 +G_2^2 \Omega_1^2}{[(n k_x' + k_x)^2 + (1 - n)^2 k_z^2]^2},
    \end{eqnarray}
and
    \begin{eqnarray}
     &&  \tan \theta_{\rm ra} =- \frac{G_2 \Omega_2}{\Omega_1+G_1 \Omega_2}, \;\;\;\;\;\;
     \tan \theta_{\rm rb} = -\frac{G_2 \Omega_1}{G_1 \Omega_1 -  \Omega_2}.
    \end{eqnarray}
As a result, the first term in the wavefunction $\ket{\Psi_{\rm r}}$ contributes a shift as
\begin{eqnarray}
    \Delta z_{\rm ra}= \frac{\partial \theta_{\rm ra} }{\partial k_z}
    = -\frac{{G_2 \left(\frac{\partial \Omega_2}{\partial k_z }\right)}(\Omega_1+G_1 \Omega_2)
    -(G_2 \Omega_2)\left(\frac{\partial \Omega_1}{\partial k_z }+G_1 \frac{\partial \Omega_2}{\partial k_z }\right)}{(G_2 \Omega_2)^2 +(\Omega_1+G_1 \Omega_2)^2 },
\end{eqnarray}
and the second term in the wavefunction $\ket{\Psi_{\rm r}}$ contributes a shift as
\begin{eqnarray}
    \Delta z_{\rm rb}= \frac{\partial \theta_{\rm rb} }{\partial k_z}=
    -\frac{{G_2 \left(\frac{\partial \Omega_1}{\partial k_z }\right)}(G_1 \Omega_1- \Omega_2)-(G_2 \Omega_1)\left(G_1\frac{\partial \Omega_1}{\partial k_z }- \frac{\partial \Omega_2}{\partial k_z }\right)}{(G_2 \Omega_1)^2 +(G_1 \Omega_1- \Omega_2)^2 }.
\end{eqnarray}

Due to
\begin{eqnarray}
&&    \frac{{|A| }^2}{ {|A| }^2+ {|B| }^2} = \frac{(\Omega_1+ G_1 \Omega_2 )^2 +G_2^2 \Omega_2^2 }{(\Omega_1+ G_1 \Omega_2 )^2 +G_2^2 \Omega_2^2 +(G_1 \Omega_1 - \Omega_2 )^2 +G_2^2 \Omega_1^2}=\frac{(\Omega_1+ G_1 \Omega_2 )^2 +G_2^2 \Omega_2^2 }{(\Omega_1^2 +\Omega_2^2)(1+G_1^2 +G_2^2 )  },\nonumber\\
&&    \frac{{|B| }^2}{ {|A| }^2+ {|B| }^2} = \frac{(G_1 \Omega_1 - \Omega_2 )^2 +G_2^2 \Omega_1^2 }{(\Omega_1+ G_1 \Omega_2 )^2 +G_2^2 \Omega_2^2 +(G_1 \Omega_1 - \Omega_2 )^2 +G_2^2 \Omega_1^2}=\frac{(G_1 \Omega_1 - \Omega_2 )^2 +G_2^2 \Omega_1^2 }{(\Omega_1^2 +\Omega_2^2)(1+G_1^2 +G_2^2 )},
\end{eqnarray}
eventually the total spatial shifts for the reflected wave is given by
\begin{eqnarray}
    \Delta {z}_{\rm r} &=& \frac{{|A| }^2}{ {|A| }^2+ {|B| }^2} \Delta z_{\rm ra}+\frac{{|B| }^2}{ {|A| }^2+ {|B| }^2}\Delta z_{\rm rb}\notag\\
    &=&-\frac{(\Omega_1+ G_1 \Omega_2 )^2 +G_2^2 \Omega_2^2 }{(\Omega_1^2 +\Omega_2^2)(1+G_1^2 +G_2^2 )  }
    \frac{{G_2 \left(\frac{\partial \Omega_2}{\partial k_z }\right)}(\Omega_1+G_1 \Omega_2)
    -(G_2 \Omega_2)\left(\frac{\partial \Omega_1}{\partial k_z }+G_1 \frac{\partial \Omega_2}{\partial k_z }\right)}{(G_2 \Omega_2)^2 +(\Omega_1+G_1 \Omega_2)^2 } \notag\\
    &&- \frac{(G_1 \Omega_1 - \Omega_2 )^2 +G_2^2 \Omega_1^2 }{(\Omega_1^2 +\Omega_2^2)(1+G_1^2 +G_2^2 )}
     \frac{{G_2 \left(\frac{\partial \Omega_1}{\partial k_z }\right)}(G_1 \Omega_1- \Omega_2)-(G_2 \Omega_1)\left(G_1\frac{\partial \Omega_1}{\partial k_z }- \frac{\partial \Omega_2}{\partial k_z }\right)}{(G_2 \Omega_1)^2 +(G_1 \Omega_1- \Omega_2)^2 }\notag\\
    &=&- \frac{{G_2 \frac{\partial \Omega_1}{\partial k_z }}(G_1 \Omega_1- \Omega_2)-(G_2 \Omega_1)(G_1\frac{\partial \Omega_1}{\partial k_z }- \frac{\partial \Omega_2}{\partial k_z }) +{G_2 \frac{\partial \Omega_2}{\partial k_z }}(\Omega_1+G_1 \Omega_2)-(G_2 \Omega_2)(\frac{\partial \Omega_1}{\partial k_z }+G_1 \frac{\partial \Omega_2}{\partial k_z }) }
    {(\Omega_1^2 +\Omega_2^2)(1+G_1^2 +G_2^2 )} \notag\\
    &=& \frac{2G_2 (\frac{\partial \Omega_1}{\partial k_z }\Omega_2-\frac{\partial \Omega_2}{\partial k_z }\Omega_1) }{(\Omega_1^2 +\Omega_2^2)(1+G_1^2 +G_2^2 )}.
\end{eqnarray}
From \Eq{zr} one has
\begin{eqnarray}
    \Delta {z}_{r}( +\hat{y}) = \frac{\frac{\partial \Omega_1}{\partial k_z }\Omega_2-\frac{\partial \Omega_2}{\partial k_z }\Omega_1 }{\Omega_1^2 +\Omega_2^2},
\end{eqnarray}
and due to
\begin{eqnarray}
   \frac{2G_2}{1+G_1^2 +G_2^2}=\frac{2 \tan \frac{\theta}{2 } \sin \phi}{ 1+\tan^2 \frac{\theta}{2 }}  =2 \tan \frac{\theta}{2 } \cos^2 \frac{\theta}{2 }\sin \phi= \sin \theta \sin \phi,
\end{eqnarray}
thus finally one has
\begin{eqnarray}
    \Delta {z}_{\rm r} = \tau_y \Delta z_{\rm r}(+\hat{y}).
\end{eqnarray}
This ends the proof. Similarly, one can obtain the same spatial shift for the transmitted wave in the $z$-basis. $\blacksquare$
\end{remark}

\begin{remark}\label{r10}\textcolor{blue}{Calculating the Spatial Shifts in the General Basis}. As one can expect, the spatial shifts will be the same in the other more general bases. In the following we would like to provide the detailed calculation for the reflected wave.

We still start from Eqs. (\ref{eq:1-A})-(\ref{eq:1-B}), which are
\begin{subequations}
   \begin{align}
&A=\frac{\ell_1 \left[k_x^2-k_z^2(n-1)^2-k_x'^2 n^2\right]+ \ell_2 \, 2 k_x k_z  (n-1)}
{k_x^2+k_z^2(n-1)^2+2 k_x k_x' n +k_x'^2 n^2}, \label{eq:1-A1}\\
&B=\frac{-\ell_1\, 2 k_x k_z  (n-1)+\ell_2\, \left[k_x^2 -k_z^2(n-1)^2-k_x'^2 n^2\right]}
{k_x^2+k_z^2(n-1)^2+2 k_x k_x' n +k_x'^2 n^2}, \label{eq:1-B1}
    \end{align}
\end{subequations}
For the reflected wave, we can have its representation as ($p_y=\hbar k_y=0$)
\begin{eqnarray}
    &&    \ket{\Psi_{\rm r}}=\dfrac{1}{\sqrt{2\mathcal{E}(\mathcal{E} +mc^2)}}
    \begin{pmatrix}
        {(\mathcal{E} + mc^2 )}\, |\chi_{\rm r}\rangle \\
        {c\hbar (-\sigma_x k_x +\sigma_z k_z )} \, |\chi_{\rm r}\rangle
    \end{pmatrix}
            {\rm e}^{\frac{\i}{\hbar} [{\hbar}(-k_x x +k_y y+ k_z z) - Et]},
\end{eqnarray}
where
\begin{eqnarray}
    |\chi_{\rm r}\rangle = \begin{pmatrix}
        A\\
        B
    \end{pmatrix},
\end{eqnarray}
Note that $|\chi_{\rm r}\rangle$ is unnormalized.
%

Now we consider a general spin direction
\begin{eqnarray}
\vec{\tau} = (\sin \theta \cos \phi , \sin \theta \sin \phi , \cos \theta),
\end{eqnarray}
and its eigenstate is
\begin{eqnarray}
    \left|\chi\left(\vec{\tau} \right)\right\rangle = \begin{pmatrix}
         \cos \frac{\theta}{2}\\
         \sin \frac{\theta}{2}  \e^{\i \phi}
    \end{pmatrix}.
\end{eqnarray}
The reversed spin direction is
\begin{eqnarray}
-\vec{\tau} &=& (\sin ( \pi -\theta) \cos (\pi+\phi) , \sin (\pi-\theta) \sin (\pi +\phi ), \cos (\pi- \theta ))\nonumber\\
&=& (-\sin \theta \cos \phi, -\sin \theta \sin \phi , -\cos \theta),
\end{eqnarray}
and its eigenstate is
\begin{eqnarray}
   \left|\chi\left(-\vec{\tau} \right)\right\rangle = \begin{pmatrix}
         \sin \frac{\theta}{2}\\
         -\cos  \frac{\theta}{2}  \e^{\i \phi}
    \end{pmatrix}.
\end{eqnarray}
For simplicity, let us denote
\begin{eqnarray}
    \ket{\uparrow}_{\vec{\tau}}=\left|\chi\left(\vec{\tau} \right)\right\rangle, \,\,\,\,
    \ket{\downarrow}_{\vec{\tau}}=\left|\chi\left(-\vec{\tau} \right)\right\rangle.
\end{eqnarray}
Obviously, $\{\ket{\uparrow}_{\vec{\tau}}, \ket{\downarrow}_{\vec{\tau}}\}$ forms an orthogonal basis.

One can expend the state $|\chi_{\rm r}\rangle$ in the general basis $\{\ket{\uparrow}_{\vec{\tau}}, \ket{\downarrow}_{\vec{\tau}}\}$ as follows
\begin{eqnarray}
    |\chi_{\rm r}\rangle = A' \ket{\uparrow}_{\vec{\tau}}+ B' \ket{\downarrow}_{\vec{\tau}},
\end{eqnarray}
i.e.,
\begin{eqnarray}
    \begin{pmatrix}
        A\\
        B
    \end{pmatrix} = A' \begin{pmatrix}
         \cos \frac{\theta}{2}\\
         \sin \frac{\theta}{2}  \e^{\i \phi}
    \end{pmatrix}+ B' \begin{pmatrix}
         \sin \frac{\theta}{2}\\
         -\cos  \frac{\theta}{2}  \e^{\i \phi}
    \end{pmatrix},
\end{eqnarray}
which means
\begin{eqnarray}
    &&A= A' \cos \frac{\theta}{2} + B' \sin \frac{\theta}{2}, \nonumber\\
    &&B= A' \sin \frac{\theta}{2} \e^{\i \phi} - B' \cos \frac{\theta}{2} \e^{\i \phi}.
\end{eqnarray}
Based on above, we can obtain
\begin{eqnarray}
    A' = \frac{A \cos \frac{\theta}{2} \e^{\i \phi}+ B   \sin \frac{\theta}{2} }{ \cos^2 \frac{\theta}{2} \e^{\i \phi}+ \sin^2 \frac{\theta}{2} \e^{\i \phi}}= \left[A \cos \frac{\theta}{2} \e^{\i \phi}+ B   \sin \frac{\theta}{2}  \right]\e^{-\i \phi},
\end{eqnarray}
\begin{eqnarray}
    B' = \frac{A \sin \frac{\theta}{2} \e^{\i \phi}- B   \cos \frac{\theta}{2} }{ \cos^2 \frac{\theta}{2} \e^{\i \phi}+ \sin^2 \frac{\theta}{2} \e^{\i \phi}}= \left[A \sin \frac{\theta}{2} \e^{\i \phi}- B   \cos \frac{\theta}{2}  \right]\e^{-\i \phi}.
\end{eqnarray}
We designate
\begin{subequations}
    \begin{eqnarray}
        A = \ell_1 \dfrac{\Omega_1+ (G_1 +\i G_2)\Omega_2}{(n k_x' + k_x)^2 + (1 - n)^2k_z^2},
    \end{eqnarray}
    \begin{eqnarray}
        B = \ell_1\dfrac{-\Omega_2 + (G_1+\i G_2) \Omega_1}{(n k_x' + k_x)^2 + (1 - n)^2 k_z^2}.
    \end{eqnarray}
\end{subequations}
\begin{subequations}
    \begin{eqnarray}
        \Omega_1 = k_x^2-n^2{k_x'}^2 - k_z^2(n-1) ,
    \end{eqnarray}
    \begin{eqnarray}
        \Omega_2 = 2 k_x k_z(n-1).
    \end{eqnarray}
\end{subequations}
    Then we have
        \begin{eqnarray}
        A'& =&  \frac{\ell_1 {\rm e}^{-{\rm i} \phi}}{(n k_x' + k_x)^2 + (1 - n)^2 k_z^2} \times \biggr\{ \left[(\Omega_1+G_1 \Omega_2) \cos \frac{\theta}{2} \cos \phi - G_2 \Omega_2 \cos \frac{\theta}{2} \sin \phi + (G_1 \Omega_1 -\Omega_2 ) \sin \frac{\theta}{2}\right]  \notag\\
        &&+ {\rm i} \left[G_2 \Omega_2 \cos \frac{\theta}{2} \cos \phi + (\Omega_1+ G_1 \Omega_2 )\cos \frac{\theta}{2} \sin \phi + G_2 \Omega_1 \sin \frac{\theta}{2}\right] \biggr\},
       \end{eqnarray}
  \begin{eqnarray}
        B'& =&  \frac{\ell_1 {\rm e}^{-{\rm i} \phi}}{(n k_x' + k_x)^2 + (1 - n)^2 k_z^2} \times \biggr\{ \left[(\Omega_1+G_1 \Omega_2) \sin \frac{\theta}{2} \cos \phi - G_2 \Omega_2 \sin \frac{\theta}{2} \sin \phi- (G_1 \Omega_1 -\Omega_2 ) \cos \frac{\theta}{2}\right]  \notag\\
        &&+ {\rm i} \left[G_2 \Omega_2 \sin \frac{\theta}{2} \cos \phi + (\Omega_1+ G_1 \Omega_2 )\sin \frac{\theta}{2} \sin \phi -G_2 \Omega_1 \cos \frac{\theta}{2}\right]\biggr\}.
    \end{eqnarray}

    We designate
    \begin{eqnarray}
        M_{A'}^2 &=& \left[(\Omega_1+G_1 \Omega_2) \cos \frac{\theta}{2} \cos \phi - G_2 \Omega_2 \cos \frac{\theta}{2} \sin \phi + (G_1 \Omega_1 -\Omega_2 ) \sin \frac{\theta}{2}\right]^2 \notag\\
        && + \left[G_2 \Omega_2 \cos \frac{\theta}{2} \cos \phi + (\Omega_1+ G_1 \Omega_2 )\cos \frac{\theta}{2} \sin \phi + G_2 \Omega_1 \sin \frac{\theta}{2}\right]^2,
    \end{eqnarray}
    \begin{eqnarray}
        M_{B'}^2 &=& \left[(\Omega_1+G_1 \Omega_2) \sin \frac{\theta}{2} \cos \phi - G_2 \Omega_2 \sin \frac{\theta}{2} \sin \phi- (G_1 \Omega_1 -\Omega_2 ) \cos \frac{\theta}{2}\right]^2 \notag\\
        && + \left[G_2 \Omega_2 \sin \frac{\theta}{2} \cos \phi + (\Omega_1+ G_1 \Omega_2 )\sin \frac{\theta}{2} \sin \phi -G_2 \Omega_1 \cos \frac{\theta}{2}\right]^2,
    \end{eqnarray}
    then we can rewrite $A'$ and $B'$ as
    \begin{eqnarray}\label{A'}
        A' =\frac{\ell_1 M_{A'}}{(n k_x' + k_x)^2 + (1 - n)^2 k_z^2} \e^{\i \theta_{A'}-\i \phi}
    \end{eqnarray}
    \begin{eqnarray}\label{B'}
        B' =\frac{\ell_1 M_{B'}}{(n k_x' + k_x)^2 + (1 - n)^2 k_z^2} \e^{\i \theta_{B'}-\i \phi}
    \end{eqnarray}
    where
    \begin{eqnarray}
        \tan \theta_{A'} = \frac{G_2 \Omega_2 \cos \frac{\theta}{2} \cos \phi + (\Omega_1+ G_1 \Omega_2 )\cos \frac{\theta}{2} \sin \phi
        + G_2 \Omega_1 \sin \frac{\theta}{2}}{(\Omega_1+G_1 \Omega_2) \cos \frac{\theta}{2} \cos \phi - G_2 \Omega_2 \cos \frac{\theta}{2} \sin \phi + (G_1 \Omega_1 -\Omega_2 ) \sin \frac{\theta}{2}},
    \end{eqnarray}
    \begin{eqnarray}
        \tan \theta_{B'} =\frac{ G_2 \Omega_2 \sin \frac{\theta}{2} \cos \phi + (\Omega_1+ G_1 \Omega_2 )\sin \frac{\theta}{2} \sin \phi -G_2 \Omega_1 \cos \frac{\theta}{2}}{ (\Omega_1+G_1 \Omega_2) \sin \frac{\theta}{2} \cos \phi - G_2 \Omega_2 \sin \frac{\theta}{2} \sin \phi- (G_1 \Omega_1 -\Omega_2 ) \cos \frac{\theta}{2}}.
    \end{eqnarray}

    Note that $\phi$ does not depend on the momentum $\hbar \vec{k}$, thus it has no contribution to the spatial shift. The spatial shift is calculated through
  \begin{eqnarray}
    \Delta {z}_{\rm r} = P_1 \Delta z_{\rm ra}+ P_2 \Delta z_{\rm rb},
\end{eqnarray}
with
    \begin{eqnarray}
&& P_1= \frac{{|A'| }^2}{ {|A'| }^2+ {|B'| }^2}, \;\;\;\;\; P_2= \frac{{|B'| }^2}{ {|A'| }^2+ {|B'| }^2},
\end{eqnarray}
 \begin{eqnarray}
        \Delta z_{\rm ra} = - \frac{ \partial \theta_{A'}}{\partial k_z },  \;\;\;\;\;    \Delta z_{\rm rb} = -\frac{ \partial \theta_{B'}}{\partial k_z }.
    \end{eqnarray}
We then have
    \begin{eqnarray}
      \Delta {z}_{\rm r} = \frac{M^2_{A'} \Delta z_{\rm ra} + M^2_{B'}\Delta z_{\rm rb}}{M^2_{A'}+M^2_{B'}}.
    \end{eqnarray}

(i) For $M^2_{A'} \Delta z_{\rm ra}$, one has
\begin{eqnarray}
    M^2_{A'} \Delta z_{\rm ra} &=& - M^2_{A'} \frac{1}{M^2_{A'}}\Biggl\{
    \left[\frac{\partial }{\partial k_z} \left(G_2 \Omega_2 \cos \frac{\theta}{2} \cos \phi + (\Omega_1+ G_1 \Omega_2 )\cos \frac{\theta}{2} \sin \phi + G_2 \Omega_1 \sin \frac{\theta}{2}\right)\right]\times  \notag\\
    &&  \left[(\Omega_1+G_1 \Omega_2) \cos \frac{\theta}{2} \cos \phi - G_2 \Omega_2 \cos \frac{\theta}{2} \sin \phi + (G_1 \Omega_1 -\Omega_2 ) \sin \frac{\theta}{2} \right] \nonumber\\
   && -\left[G_2 \Omega_2 \cos \frac{\theta}{2} \cos \phi + (\Omega_1+ G_1 \Omega_2 )\cos \frac{\theta}{2} \sin \phi + G_2 \Omega_1 \sin \frac{\theta}{2} \right]\times   \notag\\
   && \left[\frac{\partial }{\partial k_z} \left((\Omega_1+G_1 \Omega_2) \cos \frac{\theta}{2} \cos \phi - G_2 \Omega_2 \cos \frac{\theta}{2} \sin \phi + (G_1 \Omega_1 -\Omega_2 ) \sin \frac{\theta}{2}\right) \right] \Biggr\}\notag\\
&=& - \Biggl\{
    \left[\frac{\partial }{\partial k_z} \left(G_2 \Omega_2 \cos \frac{\theta}{2} \cos \phi + (\Omega_1+ G_1 \Omega_2 )\cos \frac{\theta}{2} \sin \phi + G_2 \Omega_1 \sin \frac{\theta}{2}\right)\right]\times  \notag\\
    &&  \left[(\Omega_1+G_1 \Omega_2) \cos \frac{\theta}{2} \cos \phi - G_2 \Omega_2 \cos \frac{\theta}{2} \sin \phi + (G_1 \Omega_1 -\Omega_2 ) \sin \frac{\theta}{2} \right] \nonumber\\
   && -\left[G_2 \Omega_2 \cos \frac{\theta}{2} \cos \phi + (\Omega_1+ G_1 \Omega_2 )\cos \frac{\theta}{2} \sin \phi + G_2 \Omega_1 \sin \frac{\theta}{2} \right]\times   \notag\\
   && \left[\frac{\partial }{\partial k_z} \left((\Omega_1+G_1 \Omega_2) \cos \frac{\theta}{2} \cos \phi - G_2 \Omega_2 \cos \frac{\theta}{2} \sin \phi + (G_1 \Omega_1 -\Omega_2 ) \sin \frac{\theta}{2} \right) \right] \Biggr\}.
\end{eqnarray}
For simplicity, we designate
\begin{eqnarray}
    \Omega'_{1} =  \frac{\partial  \Omega_1}{ \partial k_z}, \;\;\;\;\;\;\;  \Omega'_{2} = \frac{\partial  \Omega_2}{ \partial  k_z},
\end{eqnarray}
then
\begin{eqnarray}
 M^2_{A'} \Delta z_{\rm ra} &=& - \Biggl\{
    \left[G_2 \Omega'_2 \cos \frac{\theta}{2} \cos \phi + (\Omega'_1+ G_1 \Omega'_2 )\cos \frac{\theta}{2} \sin \phi + G_2 \Omega'_1 \sin \frac{\theta}{2}\right]\times  \notag\\
    &&  \left[(\Omega_1+G_1 \Omega_2) \cos \frac{\theta}{2} \cos \phi - G_2 \Omega_2 \cos \frac{\theta}{2} \sin \phi + (G_1 \Omega_1 -\Omega_2 ) \sin \frac{\theta}{2} \right] \nonumber\\
   && - \left[G_2 \Omega_2 \cos \frac{\theta}{2} \cos \phi + (\Omega_1+ G_1 \Omega_2 )\cos \frac{\theta}{2} \sin \phi + G_2 \Omega_1 \sin \frac{\theta}{2} \right]\times   \notag\\
   && \left[(\Omega'_1+G_1 \Omega'_2) \cos \frac{\theta}{2} \cos \phi - G_2 \Omega'_2 \cos \frac{\theta}{2} \sin \phi + (G_1 \Omega'_1 -\Omega'_2 ) \sin \frac{\theta}{2} \right] \Biggr\},
\end{eqnarray}
i.e.,
\begin{eqnarray}
 M^2_{A'} \Delta z_{\rm ra} &=&
    -\left[G_2 \Omega'_2 \cos \frac{\theta'}{2} \cos \phi + (\Omega'_1+ G_1 \Omega'_2 )\cos \frac{\theta}{2} \sin \phi + G_2 \Omega'_1 \sin \frac{\theta}{2}\right]\times  \notag\\
    &&  \left[(\Omega_1+G_1 \Omega_2) \cos \frac{\theta}{2} \cos \phi - G_2 \Omega_2 \cos \frac{\theta}{2} \sin \phi + (G_1 \Omega_1 -\Omega_2 ) \sin \frac{\theta}{2} \right] \nonumber\\
   && + \left[G_2 \Omega_2 \cos \frac{\theta}{2} \cos \phi + (\Omega_1+ G_1 \Omega_2 )\cos \frac{\theta}{2} \sin \phi + G_2 \Omega_1 \sin \frac{\theta}{2} \right]\times   \notag\\
   && \left[(\Omega'_1+G_1 \Omega'_2) \cos \frac{\theta}{2} \cos \phi - G_2 \Omega'_2 \cos \frac{\theta}{2} \sin \phi + (G_1 \Omega'_1 -\Omega'_2 ) \sin \frac{\theta}{2} \right].
\end{eqnarray}
Because
\begin{eqnarray}
    && -G_2 \Omega'_2(\Omega_1+G_1 \Omega_2) \cos \frac{\theta}{2} \cos \phi \cos \frac{\theta}{2} \cos \phi + G_2 \Omega_2 \cos \frac{\theta}{2} \cos \phi(\Omega'_1+G_1 \Omega'_2) \cos \frac{\theta}{2} \cos \phi \notag\\
    &=& G_2 (\Omega'_1 \Omega_2 - \Omega_1 \Omega'_2)\cos^2 \frac{\theta}{2} \cos^2 \phi,
\end{eqnarray}
\begin{eqnarray}
   && (\Omega'_1+ G_1 \Omega'_2 )\cos \frac{\theta}{2} \sin \phi  G_2 \Omega_2 \cos \frac{\theta}{2} \sin \phi - (\Omega_1+ G_1 \Omega_2 )\cos \frac{\theta}{2} \sin \phi  G_2 \Omega'_2 \cos \frac{\theta}{2} \sin \phi \notag\\
   &=& G_2 (\Omega'_1 \Omega_2 - \Omega_1 \Omega'_2 ) \cos^2 \frac{\theta}{2} \sin^2  \phi,
\end{eqnarray}
\begin{eqnarray}
    && - G_2 \Omega'_1 \sin \frac{\theta}{2}(G_1 \Omega_1 -\Omega_2 ) \sin \frac{\theta}{2}
     +G_2 \Omega_1 \sin \frac{\theta}{2}(G_1 \Omega'_1 -\Omega'_2 ) \sin \frac{\theta}{2} = (\Omega'_1 \Omega_2 - \Omega_1 \Omega'_2 ) G_2 \sin^2 \frac{\theta}{2},
\end{eqnarray}
\begin{eqnarray}
   && G_2 \Omega'_2 \cos \frac{\theta}{2} \cos \phi G_2 \Omega_2 \cos \frac{\theta}{2} \sin \phi -(\Omega'_1+ G_1 \Omega'_2 )\cos \frac{\theta}{2} \sin \phi (\Omega_1+G_1 \Omega_2) \cos \frac{\theta}{2} \cos \phi \notag\\
    &&-   G_2 \Omega_2 \cos \frac{\theta}{2} \cos \phi G_2 \Omega_2' \cos \frac{\theta}{2} \sin \phi +(\Omega_1+ G_1 \Omega_2 )\cos \frac{\theta}{2} \sin \phi (\Omega'_1+G_1 \Omega'_2) \cos \frac{\theta}{2} \cos \phi=0,
\end{eqnarray}
\begin{eqnarray}
    &&G_2 \Omega'_2 \cos \frac{\theta}{2} \cos \phi (G_1 \Omega_1 -\Omega_2 ) \sin \frac{\theta}{2} +G_2 \Omega'_1 \sin \frac{\theta}{2}(\Omega_1+G_1 \Omega_2) \cos \frac{\theta}{2} \cos \phi\notag\\
&&-G_2 \Omega_2 \cos \frac{\theta}{2} \cos \phi (G_1 \Omega'_1 -\Omega'_2 ) \sin \frac{\theta}{2} -G_2 \Omega_1 \sin \frac{\theta}{2}(\Omega'_1+G_1 \Omega'_2) \cos \frac{\theta}{2} \cos \phi=0,
\end{eqnarray}
\begin{eqnarray}
    &&-(\Omega'_1+ G_1 \Omega'_2 )\cos \frac{\theta}{2} \sin \phi(G_1 \Omega_1 -\Omega_2 ) \sin \frac{\theta}{2}+G_2 \Omega'_1 \sin \frac{\theta}{2}G_2 \Omega_2 \cos \frac{\theta}{2} \sin \phi\notag\\
    &&+(\Omega_1+ G_1 \Omega_2 )\cos \frac{\theta}{2} \sin \phi(G_1 \Omega'_1 -\Omega'_2 ) \sin \frac{\theta}{2}-G_2 \Omega_1 \sin \frac{\theta}{2}G_2 \Omega'_2 \cos \frac{\theta}{2} \sin \phi\notag\\
    &=&(1+G^2_1 +G^2_2)(\Omega'_1 \Omega_2 -\Omega_1 \Omega'_2)\cos \frac{\theta}{2} \sin \frac{\theta}{2} \sin \phi,
\end{eqnarray}
one can attain
\begin{eqnarray}\label{m2za}
    M^2_{A'} \Delta z_{\rm ra} & = & G_2 \left( \Omega'_1 \Omega_2 - \Omega_1 \Omega'_2 \right) \left(\cos^2 \frac{\theta}{2} \cos^2 \phi
    + \cos^2 \frac{\theta}{2} \sin^2 \phi+ \sin^2 \frac{\theta}{2} \right) \nonumber\\
    && + (1+G^2_1 +G^2_2)(\Omega'_1 \Omega_2 -\Omega_1 \Omega'_2)\cos \frac{\theta}{2} \sin \frac{\theta}{2} \sin \phi \notag\\
&=& G_2 (\Omega'_1 \Omega_2 -\Omega_1 \Omega'_2)  + (1+G^2_1 +G^2_2)(\Omega'_1 \Omega_2 - \Omega_1 \Omega'_2)\cos \frac{\theta}{2} \sin \frac{\theta}{2} \sin \phi.
\end{eqnarray}

(ii) For $M^2_{B'} \Delta z_{\rm rb}$, one has
\begin{eqnarray}
     M^2_{B'} \Delta z_{\rm rb} &=&  - M^2_{B'} \frac{1}{M^2_{B'}}\Biggl\{ \left[\frac{\partial }{\partial k_z} \left( G_2 \Omega_2 \sin \frac{\theta}{2} \cos \phi + (\Omega_1+ G_1 \Omega_2 )\sin \frac{\theta}{2} \sin \phi -G_2 \Omega_1 \cos \frac{\theta}{2}\right)\right] \times \notag\\
    &&  \left[(\Omega_1+G_1 \Omega_2) \sin \frac{\theta}{2} \cos \phi - G_2 \Omega_2 \sin \frac{\theta}{2} \sin \phi- (G_1 \Omega_1 -\Omega_2 ) \cos \frac{\theta}{2} \right] \nonumber\\
    &&- \left[ G_2 \Omega_2 \sin \frac{\theta}{2} \cos \phi + (\Omega_1+ G_1 \Omega_2 )\sin \frac{\theta}{2} \sin \phi -G_2 \Omega_1 \cos \frac{\theta}{2}\right]\times   \notag\\
   && \left[\frac{\partial }{\partial k_z} \left((t_1+G_1 t_2) \sin \frac{\theta}{2} \cos \phi - G_2 t_2 \sin \frac{\theta}{2} \sin \phi- (G_1 t_1 -t_2 ) \cos \frac{\theta}{2} \right)\right] \Biggr\}\notag\\
   &=& - \Biggl\{ \left[\frac{\partial }{\partial k_z} \left( G_2 \Omega_2 \sin \frac{\theta}{2} \cos \phi + (\Omega_1+ G_1 \Omega_2 )\sin \frac{\theta}{2} \sin \phi -G_2 \Omega_1 \cos \frac{\theta}{2}\right)\right] \times \notag\\
    &&  \left[(\Omega_1+G_1 \Omega_2) \sin \frac{\theta}{2} \cos \phi - G_2 \Omega_2 \sin \frac{\theta}{2} \sin \phi- (G_1 \Omega_1 -\Omega_2 ) \cos \frac{\theta}{2} \right] \nonumber\\
    &&- \left[ G_2 \Omega_2 \sin \frac{\theta}{2} \cos \phi + (\Omega_1+ G_1 \Omega_2 )\sin \frac{\theta}{2} \sin \phi -G_2 \Omega_1 \cos \frac{\theta}{2}\right]\times   \notag\\
   && \left[\frac{\partial }{\partial k_z} \left((\Omega_1+G_1 \Omega_2) \sin \frac{\theta}{2} \cos \phi - G_2 \Omega_2 \sin \frac{\theta}{2} \sin \phi- (G_1 \Omega_1 -\Omega_2 ) \cos \frac{\theta}{2} \right)\right],
\end{eqnarray}
i.e.,
\begin{eqnarray}
   M^2_{B'} \Delta z_{\rm rb} &= & - \left[ G_2 \Omega'_2 \sin \frac{\theta}{2} \cos \phi + (\Omega'_1+ G_1 \Omega'_2 )\sin \frac{\theta}{2} \sin \phi -G_2 \Omega'_1 \cos \frac{\theta}{2}\right] \times \nonumber\\
   && \left[(\Omega_1+G_1 \Omega_2) \sin \frac{\theta}{2} \cos \phi - G_2 \Omega_2 \sin \frac{\theta}{2} \sin \phi- (G_1 \Omega_1 -\Omega_2 ) \cos \frac{\theta}{2}\right]  \notag\\
  && + \left[ G_2 \Omega_2 \sin \frac{\theta}{2} \cos \phi + (\Omega_1+ G_1 \Omega_2 )\sin \frac{\theta}{2} \sin \phi -G_2 \Omega_1 \cos \frac{\theta}{2}\right]\times \nonumber\\
  && \left[(\Omega'_1+G_1 \Omega'_2) \sin \frac{\theta}{2} \cos \phi - G_2 \Omega'_2 \sin \frac{\theta}{2} \sin \phi- (G_1 \Omega'_1 -\Omega'_2 ) \cos \frac{\theta}{2} \right].
\end{eqnarray}

Because
\begin{eqnarray}
    && -G_2 \Omega'_2 \sin \frac{\theta}{2} \cos \phi(\Omega_1+G_1 \Omega_2) \sin \frac{\theta}{2} \cos \phi +G_2 \Omega_2 \sin \frac{\theta}{2} \cos \phi(\Omega'_1+G_1 \Omega'_2) \sin \frac{\theta}{2} \cos \phi  \notag\\
    &=& G_2 (\Omega'_1 \Omega_2 -\Omega_1 \Omega'_2) \sin^2 \frac{\theta}{2} \cos^2 \phi,
\end{eqnarray}
\begin{eqnarray}
   && (\Omega'_1+ G_1 \Omega'_2 )\sin \frac{\theta}{2} \sin \phi   G_2 \Omega_2 \sin \frac{\theta}{2} \sin \phi-   (\Omega_1+ G_1 \Omega_2 )\sin \frac{\theta}{2} \sin \phi   G_2 \Omega'_2 \sin \frac{\theta}{2} \sin \phi \notag\\
   &=&G_2(\Omega'_1 \Omega_2 - \Omega_1 \Omega'_2 )   \sin^2 \frac{\theta}{2} \sin^2  \phi,
\end{eqnarray}
\begin{eqnarray}
    &&-G_2 \Omega'_1 \cos \frac{\theta}{2}(G_1 \Omega_1 -\Omega_2 ) \cos \frac{\theta}{2} +  G_2 \Omega_1 \cos \frac{\theta}{2}(G_1 \Omega'_1 -\Omega'_2 ) \cos \frac{\theta}{2} = G_2 (\Omega'_1 \Omega_2 - \Omega_1 \Omega'_2 )  \cos^2 \frac{\theta}{2} ,
\end{eqnarray}
\begin{eqnarray}
   &&G_2 \Omega'_2 \sin \frac{\theta}{2} \cos \phi G_2 \Omega_2 \sin \frac{\theta}{2} \sin \phi -(\Omega'_1+ G_1 \Omega'_2 )\sin \frac{\theta}{2} \sin \phi (\Omega_1+G_1 \Omega_2) \sin \frac{\theta}{2} \cos \phi\notag\\
    &&-  G_2 \Omega_2 \sin \frac{\theta}{2} \cos \phi G_2 \Omega'_2 \sin \frac{\theta}{2} \sin \phi +(\Omega_1+ G_1 \Omega_2 )\sin \frac{\theta}{2} \sin \phi (\Omega'_1+G_1 \Omega'_2) \sin \frac{\theta}{2} \cos \phi=0 ,
\end{eqnarray}
\begin{eqnarray}
    &&-G_2 \Omega'_2 \sin \frac{\theta}{2} \cos \phi (G_1 \Omega_1 -\Omega_2 ) \cos \frac{\theta}{2} -G_2 \Omega'_1 \cos \frac{\theta}{2}(\Omega_1+G_1 \Omega_2) \sin \frac{\theta}{2} \cos \phi\notag\\
&&+G_2 \Omega_2 \sin \frac{\theta}{2} \cos \phi (G_1 \Omega'_1 -\Omega'_2 ) \cos \frac{\theta}{2} +G_2 \Omega_1 \cos \frac{\theta}{2}(\Omega'_1+G_1 \Omega'_2) \sin \frac{\theta}{2} \cos \phi=0,
\end{eqnarray}
\begin{eqnarray}
    &&+(\Omega'_1+ G_1 \Omega'_2 ) \sin \frac{\theta}{2} \sin \phi(G_1 \Omega_1 -\Omega_2 ) \cos \frac{\theta}{2}
    -G_2 \Omega'_1 \cos \frac{\theta}{2}G_2 \Omega_2 \sin \frac{\theta}{2} \sin \phi\notag\\
    &&-(\Omega_1+ G_1 \Omega_2 )\sin \frac{\theta}{2} \sin \phi(G_1 \Omega'_1 -\Omega'_2 ) \cos \frac{\theta}{2}
    +G_2 \Omega_1 \cos \frac{\theta}{2}G_2 \Omega'_2 \sin \frac{\theta}{2} \sin \phi\notag\\
    &=&-(1+G^2_1 +G^2_2)(\Omega'_1 \Omega_2 -\Omega_1 \Omega'_2)\cos \frac{\theta}{2} \sin \frac{\theta}{2} \sin \phi,
\end{eqnarray}
one can attain
\begin{eqnarray}\label{m2zb}
    M^2_{B'} \Delta z_{\rm rb} &=& G_2 (\Omega'_1 \Omega_2 -\Omega_1 \Omega'_2) (\cos^2 \frac{\theta}{2} \cos^2 \phi + \cos^2 \frac{\theta}{2} \sin^2 \phi+ \sin^2 \frac{\theta}{2}) \nonumber\\
    && -(1+G^2_1 +G^2_2)(\Omega'_1 \Omega_2 -\Omega_1 \Omega'_2)\cos \frac{\theta}{2} \sin \frac{\theta}{2} \sin \phi \notag\\
&=& G_2 (\Omega'_1 \Omega_2 -\Omega_1 \Omega'_2)  - (1+G^2_1 -G^2_2)(\Omega'_1 \Omega_2 -\Omega_1 \Omega'_2)\cos \frac{\theta}{2} \sin \frac{\theta}{2} \sin \phi.
\end{eqnarray}
Then from \Eq{m2za} and \Eq{m2zb}, one can attain
\begin{eqnarray}
    M^2_{A'} \Delta z_{\rm ra} +  M^2_{B'} \Delta z_{\rm rb}= 2 G_2 (\Omega'_1 \Omega_2 -\Omega_1 \Omega'_2).
\end{eqnarray}

According to \Eq{A'} and \Eq{B'}, one also can have
\begin{eqnarray}
&&|\ell_1|^2{M^2_{A'}} = [(n k_x' + k_x)^2 + (1 - n)^2 k_z^2]^2 |A'|^2 \notag\\
&&= [(n k_x' + k_x)^2 + (1 - n)^2 k_z^2]^2 \Biggl\{|A|^2 \cos^2 \frac{\theta}{2}  + |B|^2 \sin^2 \frac{\theta}{2} +A B^{*}\cos \frac{\theta}{2}\sin \frac{\theta}{2} \e^{\i \phi}+A^{*} B\cos \frac{\theta}{2}\sin \frac{\theta}{2} \e^{-\i \phi}  \Biggr\},
\end{eqnarray}
\begin{eqnarray}
    &&|\ell_1|^2{M^2_{B'}} = [(n k_x' + k_x)^2 + (1 - n)^2 k_z^2]^2 |B'|^2 \notag\\
    &&= [(n k_x' + k_x)^2 + (1 - n)^2 k_z^2]^2 \Biggl\{|A|^2 \sin^2 \frac{\theta}{2}  + |B|^2 \cos^2 \frac{\theta}{2} -A B^{*}\cos \frac{\theta}{2}\sin \frac{\theta}{2} \e^{\i \phi}-A^{*} B\cos \frac{\theta}{2}\sin \frac{\theta}{2} \e^{-\i \phi}  \Biggr\},
     \end{eqnarray}
which leads to
     \begin{eqnarray}
        |\ell_1|^2 [{M^2_{A'}}+{M^2_{B'}}]= [(n k_x' + k_x)^2 + (1 - n)^2 k_z^2]^2  \left\{|A|^2 +|B|^2\right\}.
     \end{eqnarray}
So finally the spatial shift for reflected wave is given by
\begin{eqnarray}\label{eq:zr-1}
    \Delta {z}_{\rm r} = \frac{M^2_{A'} \Delta z_{\rm ra} + M^2_{B'}\Delta z_{\rm rb}}{M^2_{A'}+M^2_{B'}}
    = \frac{|\ell_1|^2\, 2 G_2 (\Omega'_1 \Omega_2 -\Omega_1 \Omega'_2)}{[(n k_x' + k_x)^2 + (1 - n)^2 k_z^2]^2  \left\{|A|^2 +|B|^2\right\} }.
\end{eqnarray}

Because of \Eq{R11}, one also can  attain
\begin{eqnarray}
   [(n k_x' + k_x)^2 + (1 - n)^2 k_z^2]^2  \left\{|A|^2 +|B|^2\right\} &=& [(k_x+n k_x')^2+k_z^2(n-1)^2] [(k_x-n k_x')^2+ k_z^2(n-1)^2]\nonumber\\
   &=&  \left[k_x^2-k_z^2(n-1)^2-k_x'^2 n^2\right]^2+ \left[2 k_x k_z  (n-1)\right]^2 \nonumber\\
   &=& \Omega_1^2 + \Omega_2^2,
\end{eqnarray}
then Eq. (\ref{eq:zr-1}) becomes
\begin{eqnarray}
   \Delta {z}_{\rm r} = \frac{2|\ell_1|^2 G_2 (\Omega'_1 \Omega_2 - \Omega_1 \Omega'_2)}{ \Omega_1^2 + \Omega_2^2 }=2|\ell_1|^2G_2 \times \frac{ (\Omega'_1 \Omega_2 - \Omega_1 \Omega'_2)}{ \Omega_1^2 + \Omega_2^2 }.
\end{eqnarray}
It is obvious that $ \Delta {z}_{\rm r}$ has no relation with $\theta$ and $\phi$. This indicates that $\Delta {z}_{\rm r}$ is unique, no matter what spin direction the reflected wave is.

Due to \Eq{zr}, one has
\begin{eqnarray}
    \Delta {z}_{\rm r}( +\hat{y}) = \frac{ (\Omega'_1 \Omega_2 - \Omega_1 \Omega'_2)}{ \Omega_1^2 + \Omega_2^2 }.
\end{eqnarray}
Due to Eq. (\ref{eq:P1P2}), one has
\begin{eqnarray}
   \tau_y= \sin\theta\sin\phi=2|\ell_1|^2G_2.
\end{eqnarray}
Then we finally has
\begin{eqnarray}
   \Delta {z}_{\rm r} = \tau_y \,\Delta {z}_{\rm r}( +\hat{y}),
\end{eqnarray}
which coincides with Eq. (\ref{eq:z-1c}). Similarly, one can have the sptial shift for the transmitted wave as
\begin{eqnarray}
   \Delta {z}_{\rm t} = \tau_y \,\Delta {z}_{\rm t}( +\hat{y}).
\end{eqnarray}
This ends the proof.
$\blacksquare$
\end{remark}

\newpage

\subsection{Summary: Demonstrating the ETSG Effect Step by Step with Figures}

In this section, we would like to demonstrate the ETSG effect step by step with figures. This not only enables theoretical physicists to understand easily what is the ETSG effect, but also facilitates experimental physicists to perform the possible experiments.
      \begin{figure}[ht]
            \centering
            \includegraphics[width=75mm]{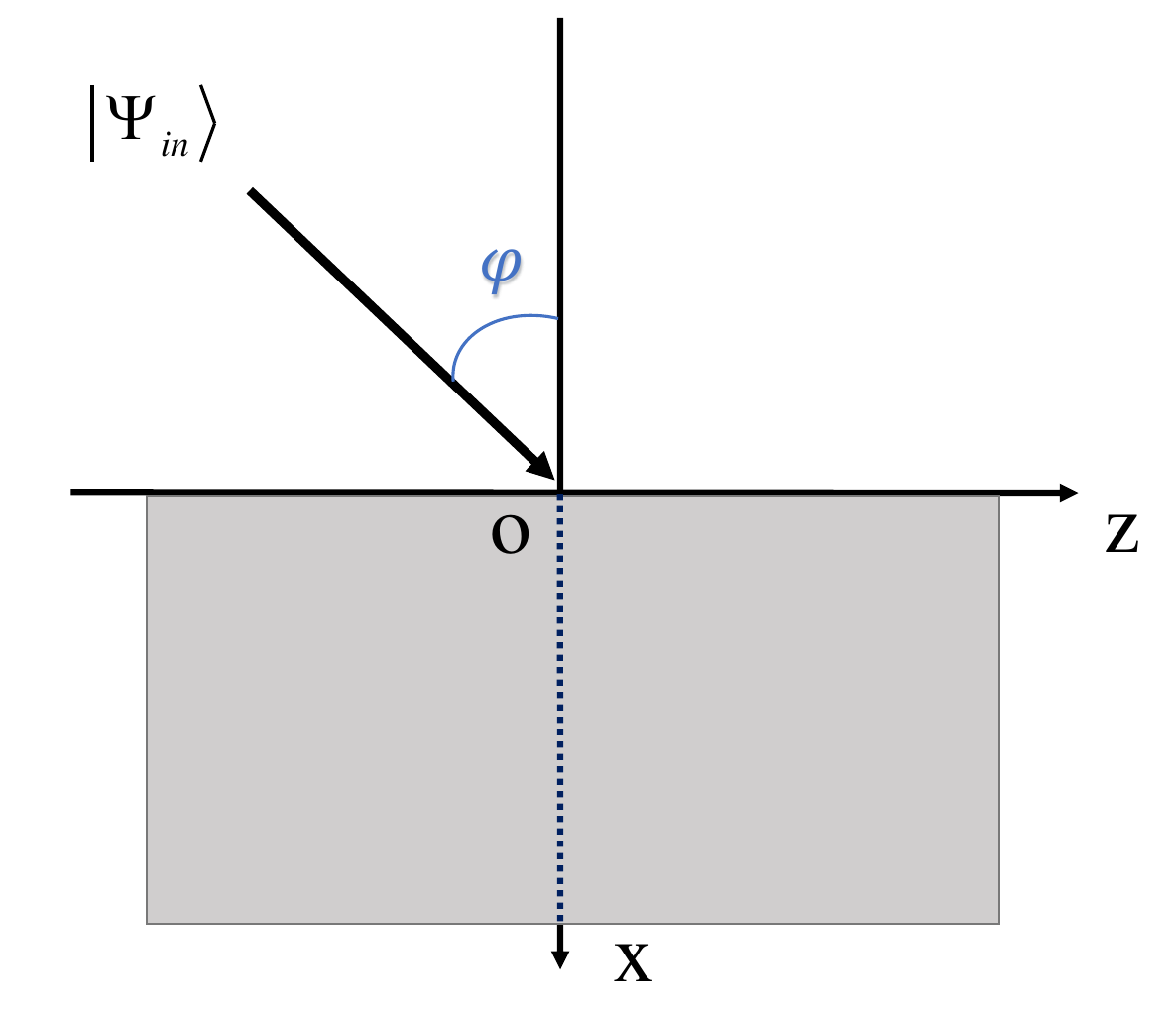}\hspace{5mm}
            \includegraphics[width=75mm]{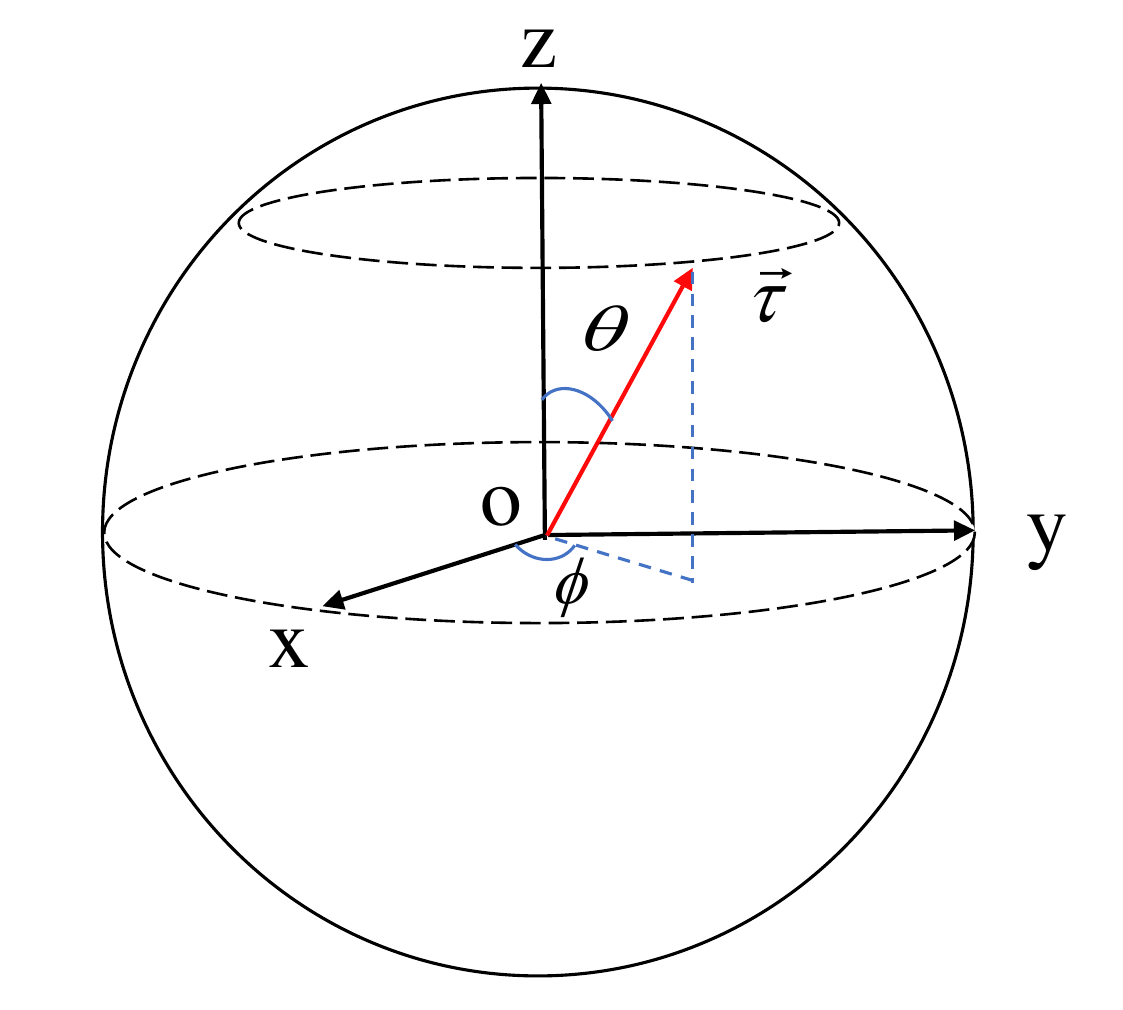}
            \centerline{(a)\;\;\;\;\;\;\;\;\;\;\;\;\;\;\;\;\;\;\;\;\;\;\;\;\;\;\;\;\;\;\;\;\;\;\;\;\;\;\;\;\;\;\;\;\;\;\;\;\;\;\;\;\;\;\;\;\;\;
            \;\;\;\;\;\;\;\;\;\;\;\;\;\;\;\;\;\;\;\;\;\;\;\;\;\;\;\;\;\; (b)}
            \caption{Illustration of the incident wave and its spin direction.
            (a) $\ket{\Psi_{\rm in}}$ is the incident wave (the black solid arrow). The incident plane is the $xz$-plane, the incident point is the origin $O$, and the incident angle is $\varphi$, and the reflect interface is the $yz$-plane. (b) The Bloch-sphere representation of $\ket{\Psi_{\rm in}}$ (or the state $|\chi\rangle$), the spin direction of $\ket{\Psi_{\rm in}}$ (or the state $|\chi\rangle$) is $\vec{\tau}$ (the red solid arrow), with $\vec{\tau}=(\tau_x, \tau_y, \tau_z)=(\sin\theta\cos\phi, \sin\theta\sin\phi, \cos\theta)$.
            }\label{fig:SGH1}
      \end{figure}

\begin{enumerate}
  \item \textcolor{blue}{The Incident Wave}. In Fig. \ref{fig:SGH1}(a), the incident wave of Dirac's particle is given by
\begin{align}
    \ket{\Psi_{\rm in}} = \biggr[ \ell_1 |{\Psi}'_1\rangle+\ell_2|{\Psi}'_2\rangle\biggr]\, \e^{{\rm i} [(k_x x +  k_z z) - \mathcal{E} t]}.
\end{align}
Here
\begin{eqnarray}
\mathcal{E}=E_+=\sqrt{p^2c^2+m^2c^4}
\end{eqnarray}
corresponds to the positive energy of Dirac's particle. This means that the energy of Dirac's particle keeps positive during its motion.
The linear momentum of Dirac's particle is given by
\begin{eqnarray}\label{eq:momin}
\vec{p}=\hbar \vec{k}= \hbar(k_x, 0, k_z),
\end{eqnarray}
i.e., the $y$-component $k_y=0$. This means that the incident plane is just the $xz$-plane. Such a choice will simplify the problem and enable people to easily grasp the central idea of the ETSG effect.

 $\;\;$ The incident $ \ket{\Psi_{\rm in}}$ is a superposition state of $|{\Psi}'_1\rangle$ and $|{\Psi}'_2\rangle$, which are two eigenstates of Dirac's particle with positive energy. In the $z$-basis, they can be written as
\begin{eqnarray}
&&  \ket{\Psi_1'} \equiv \ket{{\tilde{s}}_z = +\frac{\hbar}{2}}   = \frac{1}{\sqrt{2\mathcal{E}(\mathcal{E} +mc^2)}}  \begin{pmatrix}
        (\mathcal{E}+mc^2) |\uparrow\rangle_z\\
        c\hbar (\vec{\sigma} \cdot \vec{k}) |\uparrow\rangle_z
    \end{pmatrix},\nonumber\\
&&   \ket{\Psi_2'} \equiv  \ket{{\tilde{s}}_z = -\frac{\hbar}{2}}     = \frac{1}{\sqrt{2\mathcal{E}(\mathcal{E} +mc^2)}} \begin{pmatrix}
        (\mathcal{E}+mc^2) |\downarrow\rangle_z\\
        c\hbar (\vec{\sigma} \cdot \vec{k}) |\downarrow\rangle_z
    \end{pmatrix},
\end{eqnarray}
with
\begin{eqnarray}
|\uparrow\rangle_z = \begin{pmatrix}
    1\\
    0
\end{pmatrix}, \;\;\;\;
|\downarrow\rangle_z = \begin{pmatrix}
    0\\
    1
\end{pmatrix}.
\end{eqnarray}
 $\;\;$ One can introduce an angle $\varphi$ to express $k_x$ and $k_z$, namely,
\begin{eqnarray}
k_x =k \cos\varphi, \;\;\;\;\; k_z=k\sin\varphi, \;\;\;\;\; k=|\vec{k}|=\sqrt{k_x^2+k_z^2}.
\end{eqnarray}
The physical meaning of $\varphi$ is just the incident angle, which satisfies
\begin{eqnarray}
\tan \varphi=\frac{k_z}{k_x}.
\end{eqnarray}
Because for the incident wave,
\begin{eqnarray}
k_x>0, \;\;\;\; k_z>0,
\end{eqnarray}
then the range of $\varphi$ is
\begin{eqnarray}
 \varphi \in \left[0, \frac{\pi}{2}\right).
\end{eqnarray}

 $\;\;$ The incident wave contains the information of spin direction of Dirac's particle. The spin direction can be read out from the two superposition coefficients $\ell_1$ and $\ell_2$. The spin direction of the state $|\uparrow\rangle_z$ is the positive $z$-axis, and the spin direction of the state $|\downarrow\rangle_z$ is the negative $z$-axis. For a general superposition state
\begin{eqnarray}
&& |\chi\rangle =  \ell_1\, |\uparrow\rangle_z+\ell_2\, |\downarrow\rangle_z =
 \begin{pmatrix}
    \ell_1\\
    \ell_2
\end{pmatrix}= \begin{pmatrix}
    \cos\frac{\theta}{2}\\
    \sin\frac{\theta}{2}\, {\rm e}^{{\rm i}\phi}
\end{pmatrix}, \nonumber\\
&& \ell_1= \cos\frac{\theta}{2}, \;\;\;\;\; \ell_2= \sin\frac{\theta}{2}\, {\rm e}^{{\rm i}\phi},
\end{eqnarray}
due to
\begin{eqnarray}
&& |\chi\rangle \langle \chi| =\dfrac{1}{2} \left(\openone + \vec{\tau}\cdot \vec{\sigma} \right).
\end{eqnarray}
one has its spin direction as
\begin{eqnarray}
&& \vec{\tau}= (\tau_x, \tau_y, \tau_z)=(\sin\theta\cos\phi, \sin\theta\sin\phi, \cos\theta).
\end{eqnarray}
In terms of the state $|\chi\rangle$, the general incident wave can be written as
\begin{align}
    \ket{\Psi_{\rm in}} = \frac{1}{\sqrt{2\mathcal{E}(\mathcal{E} +mc^2)}}  \begin{pmatrix}
        (\mathcal{E}+mc^2) |\chi\rangle \\
        c\hbar (\vec{\sigma} \cdot \vec{k}) |\chi\rangle
    \end{pmatrix}.
    \end{align}
Because there is a one-to-one correspondence between the state $\ket{\Psi_{\rm in}}$ and the state
$|\chi\rangle$, thus one can adopt the unit vector $\vec{\tau}$ to represent the spin direction of the incident wave $\ket{\Psi_{\rm in}}$. In Fig. \ref{fig:SGH1}(b), we have used the Bloch sphere to describe the spin direction $\vec{\tau}$.

\item \textcolor{blue}{The Reflected Wave and the Transmitted Wave}. From Fig. \ref{fig:SGH1}(a), the reflect interface is the $yz$-plane. After the incident wave arrives at the point $O$, generally the reflection and transmission phenomena will occur. For the incident angle, there are two critical angles, which are denoted by $\varphi_{\rm cr1}$ and $\varphi_{\rm cr2}$, respectively. If the incident angle
\begin{eqnarray}
&& \varphi \geq \varphi_{\rm cr1},
\end{eqnarray}
then it occurs only the total reflection phenomenon, i.e., the transmission phenomenon vanishes. The critical value $\varphi_{\rm cr1}$ can be determined in the following way.

$\;\;$  The momentum of the incident wave is given in Eq. (\ref{eq:momin}), which is denoted as
\begin{eqnarray}\label{eq:momin-1}
\vec{p}_{\rm in}=\hbar \vec{k}= \hbar(k_x, 0, k_z).
\end{eqnarray}
Since the reflect interface is the $yz$-plane, thus the momentum of the reflected wave is given by
\begin{eqnarray}\label{eq:momrf}
\vec{p}_{\rm r}=\hbar(-k_x, 0, k_z).
\end{eqnarray}
And the momentum of the transmitted wave takes the following form
  \begin{eqnarray}\label{eq:momtr}
\vec{p}_{\rm t}=\hbar \vec{k}'= \hbar(k'_x, 0, k_z).
\end{eqnarray}
These momentums satisfy the following relativistic energy-momentum relations
\begin{eqnarray}
&&    \mathcal{E}^2 =\vec{p}_{\rm in}^{\, 2} c^2 +m^2 c^4, \nonumber\\
&&    \mathcal{E}^2 =\vec{p}_{\rm r}^{\, 2} c^2 +m^2 c^4, \nonumber\\
&& (\mathcal{E}-V_0)^2 =\vec{p}_{\rm t}^{\, 2} c^2 +m^2 c^4,
\end{eqnarray}
i.e.,
\begin{eqnarray}\label{eq:kt}
 &&   \mathcal{E}^2 =\hbar^2(k_x^2 + k_z^2)c^2 +m^2 c^4, \nonumber\\
 &&   (\mathcal{E}-V_0)^2 =\hbar^2 (k_x'^2 +k_z^2)c^2 +m^2 c^4,
\end{eqnarray}
where $V_0$ is the potential barrier.

$\;\;$  From Eq. (\ref{eq:kt}), we have
\begin{eqnarray}
  k_x'^2= \dfrac{1}{c^2 \hbar^2} \left[(\mathcal{E}-V_0)^2 - \left(\hbar^2 k_z^2 c^2 +m^2 c^4\right)\right].
\end{eqnarray}
When
\begin{eqnarray}\label{eq:kt-1}
  k_x'^2 \leq 0,
\end{eqnarray}
one can have the transmission coefficient $\mathcal{T}=0$, which means no transmission phenomenon. Thus Eq. (\ref{eq:kt-1}) is the condition for the total reflection, i.e.,
\begin{eqnarray}
  (\mathcal{E}-V_0)^2 - \left(\hbar^2 k_z^2 c^2 +m^2 c^4\right)\leq 0,
\end{eqnarray}
i.e.,
\begin{eqnarray}
  (\mathcal{E}-V_0)^2 - \left(\hbar^2 k^2 c^2 \sin^2 \varphi + m^2 c^4\right)\leq 0,
\end{eqnarray}
i.e.,
\begin{eqnarray}
  \hbar^2 k^2 c^2 \sin^2 \varphi \geq (\mathcal{E}-V_0)^2 -  m^2 c^4,
\end{eqnarray}
i.e.,
\begin{eqnarray}
   \sin^2 \varphi \geq \dfrac{(\mathcal{E}-V_0)^2 -  m^2 c^4}{\hbar^2 k^2 c^2}
   =\dfrac{(\mathcal{E}-V_0)^2 -  m^2 c^4}{\mathcal{E}^2 -  m^2 c^4}.
\end{eqnarray}
From Eq. (\ref{eq:con-1a}), to avoid Klein's paradox in our work, we have known that
\begin{eqnarray}
(\mathcal{E}-V_0)^2> m^2c^4.
\end{eqnarray}
And Eq. (\ref{eq:con-1b}) has shown that
\begin{eqnarray}
  &&  \mathcal{E}-V_0>0.
      \end{eqnarray}
By considering
\begin{eqnarray}\label{eq:con-2}
&& (\mathcal{E}-V_0)^2- m^2c^4>0, \;\;\;\; \mathcal{E}^2 -  m^2 c^4>0, \;\;\;\; \mathcal{E}>0, \;\;\;\;
V_0>0, \;\;\;\; \mathcal{E}-V_0>0, \nonumber\\
&& 0\leq \dfrac{(\mathcal{E}-V_0)^2 -  m^2 c^4}{\mathcal{E}^2 -  m^2 c^4} <1, \;\;\;\; 0\leq \varphi < \frac{\pi}{2},
\end{eqnarray}
we have the critical angle as
\begin{eqnarray}
    \varphi_{\rm cr1}= \arcsin \left[\sqrt{\dfrac{(\mathcal{E}-V_0)^2 -  m^2 c^4}{\mathcal{E}^2 -  m^2 c^4}}\right].
\end{eqnarray}

     \begin{figure}[ht]
            \centering
            \includegraphics[width=75mm]{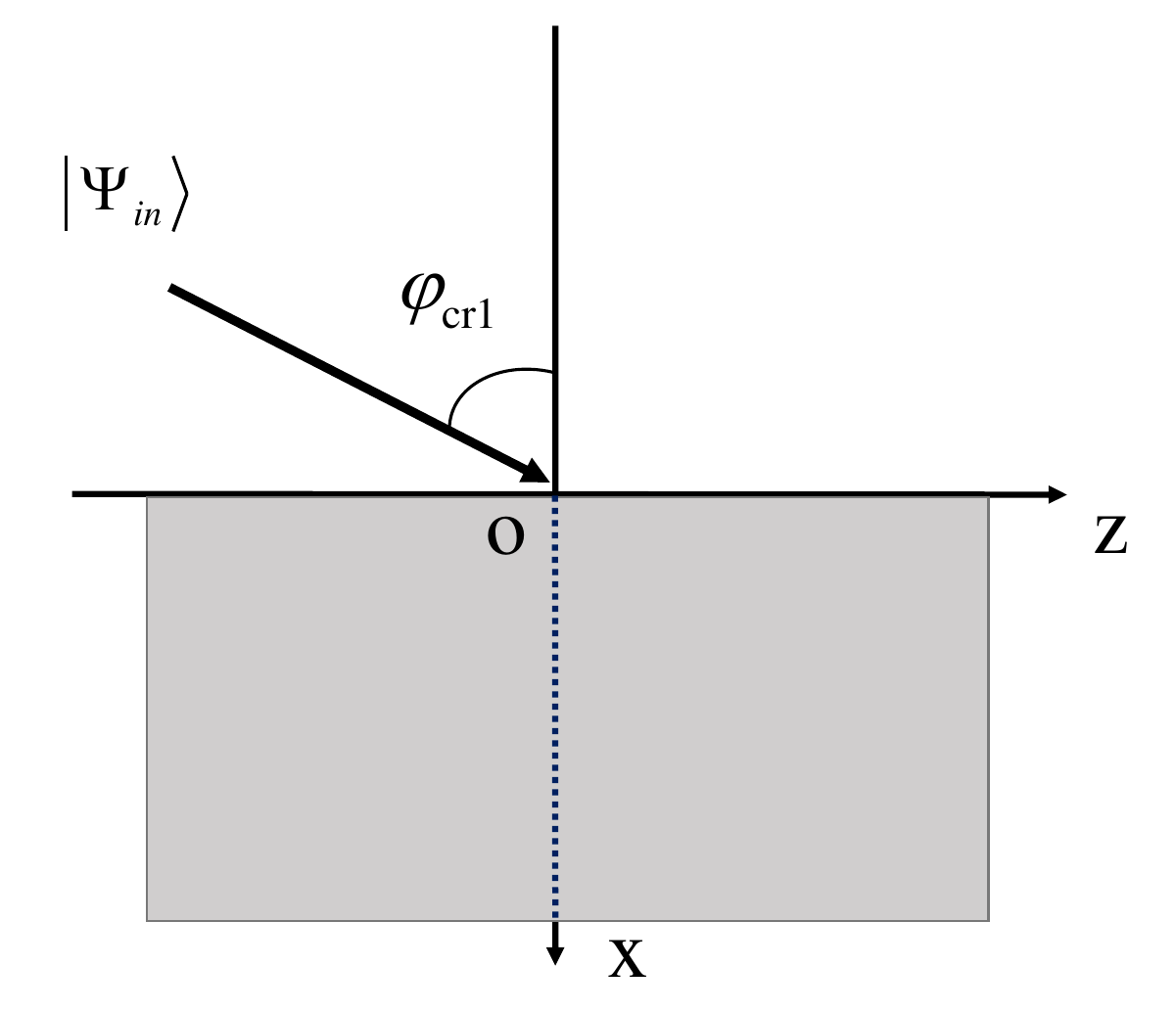}\hspace{5mm}
            \includegraphics[width=75mm]{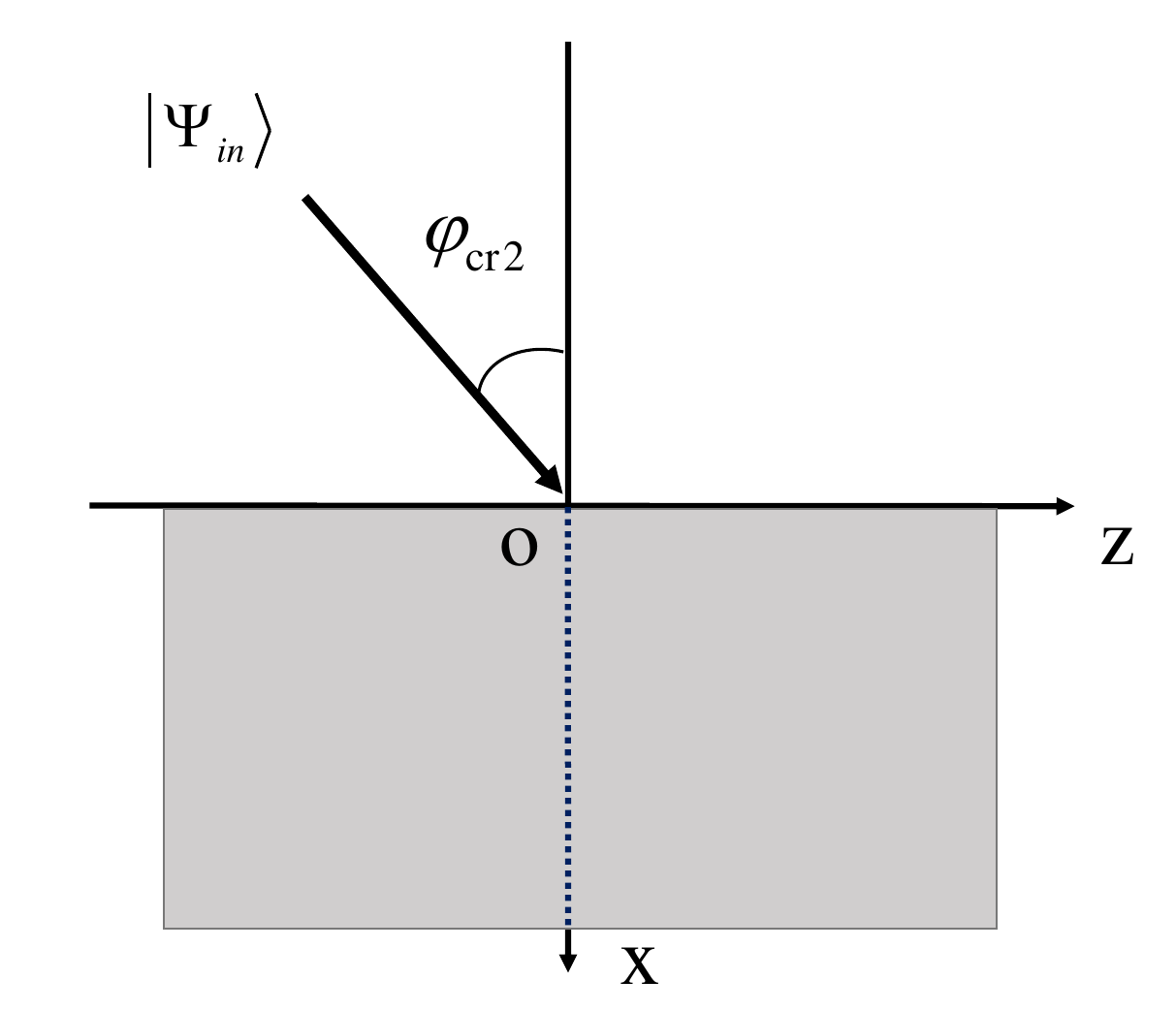}
            \centerline{(a)\;\;\;\;\;\;\;\;\;\;\;\;\;\;\;\;\;\;\;\;\;\;\;\;\;\;\;\;\;\;\;\;\;\;\;\;\;\;\;\;\;\;\;\;\;\;\;\;\;\;\;\;\;\;\;\;\;\;
            \;\;\;\;\;\;\;\;\;\;\;\;\;\;\;\;\;\;\;\;\;\;\;\;\;\;\;\;\;\; (b)}
            \caption{Illustration of two critical incident angles. (a) The angle $\varphi_{\rm cr1}$, which is the critical angle between total reflection and non-total reflection. (b) The angle $\varphi_{\rm cr2}$, which is the critical angle for $\Delta {z}_{\rm r}=0$ when $\tau_y\neq 0$.
            }\label{fig:SGH2}
      \end{figure}
In Fig. \ref{fig:SGH2}(a), we have plotted the critical angle $\varphi_{\rm cr1}$.  In this work, we shall restrict to the region of incident angle $\varphi$ as
\begin{eqnarray}
 \varphi \in \left[0, \varphi_{\rm cr1} \right),
\end{eqnarray}
where the non-total reflection phenomenon occurs.

$\;\;$ However, there is another critical angle $\varphi_{\rm cr2}$ when one considers the spatial shift of the reflected wave. The critical angle $\varphi_{\rm cr2}$ is between $0$ and $\varphi_{\rm cr1}$, i.e.,
\begin{eqnarray}
 \varphi_{\rm cr2} \in \left[0, \varphi_{\rm cr1} \right].
\end{eqnarray}
Based on Eq. (\ref{eq:z-1d}), the spatial shift of the reflected wave is given by
\begin{eqnarray}\label{eq:z-1d-1b}
    \Delta {z}_{\rm r} &=& (\sin\theta\sin\phi) \, \dfrac{\sqrt{\mathcal{E}-mc^2} }{\sqrt{{\mathcal{E}}+mc^2}}
\dfrac{c\hbar {\mathcal{E}} \cos \varphi}{\left[ {\mathcal{E}}^2 \sin^2 \varphi + m^2c^4 \cos^2 \varphi \right]}
 \left[\tan^2 \varphi -   \dfrac{mc^2}{\mathcal{E}} \right], \nonumber\\
&=& \tau_y \, \dfrac{\sqrt{\mathcal{E}-mc^2} }{\sqrt{{\mathcal{E}}+mc^2}}
\dfrac{c\hbar {\mathcal{E}} \cos \varphi}{\left[ {\mathcal{E}}^2 \sin^2 \varphi + m^2c^4 \cos^2 \varphi \right]}
 \left[\tan^2 \varphi -   \dfrac{mc^2}{\mathcal{E}} \right],
\end{eqnarray}
The critical angle $\varphi_{\rm cr2}$ satisfies the following condition
\begin{eqnarray}
\tan^2 \varphi -   \dfrac{mc^2}{\mathcal{E}}=0,
\end{eqnarray}
i.e.,
\begin{eqnarray}
\varphi_{\rm cr2} = \arctan  \sqrt{\dfrac{mc^2}{\mathcal{E}}}.
\end{eqnarray}
Note that $mc^2 \leq \mathcal{E}$, one easily knows that $mc^2 /\mathcal{E}\leq 1$ and $0 \leq \varphi_{\rm cr2}\leq \pi/4$. Thus the range of $\varphi_{\rm cr2}$ is given by
\begin{eqnarray}
 \varphi_{\rm cr2} \in \left[0, \varphi_{\rm cr1} \right] \cap \left[0, \dfrac{\pi}{4}\right].
\end{eqnarray}
Since
\begin{eqnarray}
\dfrac{\sqrt{\mathcal{E}-mc^2} }{\sqrt{{\mathcal{E}}+mc^2}}
\dfrac{c\hbar {\mathcal{E}} \cos \varphi}{\left[ {\mathcal{E}}^2 \sin^2 \varphi + m^2c^4 \cos^2 \varphi \right]}>0,
\end{eqnarray}
from Eq. (\ref{eq:z-1d-1b}) one finds that the sign of the spatial shift $\Delta {z}_{\rm r}$ is jointly determined by  the factor $\tau_y = \sin\theta\sin\phi$ and the critical angle $\varphi_{\rm cr2}$. In other words, the shift $\Delta {z}_{\rm r}$ can be positive, zero, or negative, depending on the choices of $\tau_y$ and $\varphi_{\rm cr2}$. From Eq. (\ref{eq:z-1d-1b}) one knows that the spatial shift of the reflected wave does not depend on the parameter $V_0$.

$\;\;$From Eq. (\ref{eq:z-1e}) we have known the spatial shift for the transmitted wave as
  \begin{eqnarray}
    \Delta z_{\rm t}
    &=& - \tau_y (n-1) \dfrac{ (k_x +n k_x')+\dfrac{k^2_z}{k_x k_x'} \left( k_x'+n k_x\right)}{\left(k_x +n k_x'\right)^2+\left[k_z (1-n)\right]^2}.
  \end{eqnarray}
For the non-total reflection phenomenon, one has
\begin{eqnarray}
k_x>0, \;\;\; k_x'>0, \;\;\; n>1,
\end{eqnarray}
thus
\begin{eqnarray}
\Delta z_{\rm t} \propto -\tau_y.
\end{eqnarray}
This means that the spatial shift for the transmitted wave (i.e., $\Delta z_{\rm t}$) depends on the sign of $\tau_y$. Since $k_x'$ and $n$ depend on $V_0$, hence generally $\Delta z_{\rm t}$ depends on the parameter $V_0$.

\item \textcolor{blue}{The Behaviors of the Spin Shifts}. In Table \ref{tab:shift}, we have listed the positive, zero, or negative spatial shifts for the reflected wave and the transmitted wave.
\begin{table}[h]
\centering
\caption{The behaviors of spatial shifts for the reflected wave and the transmitted wave. The shift $\Delta {z}_{\rm r}$ can be positive, zero, or negative, depending on the choices of $\tau_y$ and $\varphi_{\rm cr2}$, while the shift $\Delta {z}_{\rm t}$ can be positive, zero, or negative, depending on merely on the choice of $\tau_y$. If the spin direction is reversed, i.e., $\vec{\tau}\mapsto -\vec{\tau}$, accordingly the signs of spatial shifts are reversed, thus indicating the ETSG effect is a spin effect.}
\begin{tabular}{llccc}
\hline\hline
Shift & \;\;\;\; Incident Angle $ \varphi $\;\;\;\;   & \;\;\;$ 0<\tau_y \leq 1$  \;\;\;&  \;\;$\tau_y =0$ \;\; & \;\;$-1\leq \tau_y <0$\;\; \\
  \hline
$ \Delta z_{\rm r}$ & \;\;\;\; $\varphi \in [0, \varphi_{\rm cr2})$  & $-$ &   0 & $+$ \\
  \hline
 &\;\;\;\; $\varphi = \varphi_{\rm cr2}$  &   0 & 0 &  0\\
  \hline
 & \;\;\;\; $\varphi \in (\varphi_{\rm cr2}, \varphi_{\rm cr1})$  & $+$  &  0 & $-$ \\
  \hline
$ \Delta z_{\rm t}$ & \;\;\;\; $\varphi\in [0, \varphi_{\rm cr1})$  & $-$&  0& $+$  \\
 \hline\hline
\end{tabular}\label{tab:shift}
\end{table}

For convenience, one may re-plot the Bloch sphere (see Fig. \ref{fig:SGH1}) as shown in Fig. \ref{fig:SGH3a}. For simplicity, we call the $xz$-plane as the ``equatorial plane'', the hemispherical surface above the $xz$-plane as the ``northern hemisphere'', the hemispherical surface below the $xz$-plane as the ``southern hemisphere'', the intersection point of positive $y$-axis and sphere as the ``north pole'', and the intersection point of negative $y$-axis and sphere as the ``south pole''. Then from Table \ref{tab:shift}, the behaviors of spatial shifts for the reflected wave and the transmitted wave are clear. One may observe that

(i) If the spin direction $\vec{\tau}$ lies on the $zx$-plane, i.e., $\tau_y=0$, then the spatial shifts $ \Delta z_{\rm r}=0$ and $ \Delta z_{\rm t}=0$.

(ii) If the incident angle $\varphi = \varphi_{\rm cr2}$, then whatever the spin direction $\vec{\tau}$ is,  one always has the shifts $ \Delta z_{\rm r}=0$.

(iii) If the spin direction $\vec{\tau}$ locates on the ``northern hemisphere'', then the spatial shifts for the transmitted wave $ \Delta z_{\rm t}<0$; while the spin direction $\vec{\tau}$ locates on the ``southern hemisphere'', then the spatial shifts for the transmitted wave $ \Delta z_{\rm t}>0$.

(iv) If the spin direction $\vec{\tau}$ locates on the ``northern hemisphere'', for  $\varphi \in [0, \varphi_{\rm cr2})$, one has  $ \Delta z_{\rm r}<0$, and for $\varphi \in (\varphi_{\rm cr2}, \varphi_{\rm cr1})$, one has $ \Delta z_{\rm r}> 0$.
If the spin direction $\vec{\tau}$ locates on the ``southern hemisphere'', for  $\varphi \in [0, \varphi_{\rm cr2})$, one has  $ \Delta z_{\rm r}>0$, and for $\varphi \in (\varphi_{\rm cr2}, \varphi_{\rm cr1})$, one has $ \Delta z_{\rm r}< 0$.

(v) Given two incident waves, i.e., the incident wave $\ket{\Psi_{\rm in}}$ with spin direction $\vec{\tau}=(\tau_x, \tau_y, \tau_z)$ and the incident wave $\ket{\Psi_{\rm in}}'$ with opposite spin direction $\vec{\tau}'=-\vec{\tau}=(-\tau_x, -\tau_y, -\tau_z)$, if their incident angles are the same, then it is easy to know that $ \Delta z_{\rm r}=-\Delta' z_{\rm r}$ and $ \Delta z_{\rm t}=-\Delta' z_{\rm t}$. This interesting phenomenon can be used to separate the incident waves with opposite spin directions.

(vi) For a fixed incident angle $\varphi$ (the energy $\mathcal{E}$ and $V_0$ are also fixed), the maximal spatial shifts occur at $\vec{\tau}$ pointing to the north pole or the south pole (i.e., $\tau_y=1$ or $-1$).

     \begin{figure}[t]
            \centering
            \includegraphics[width=75mm]{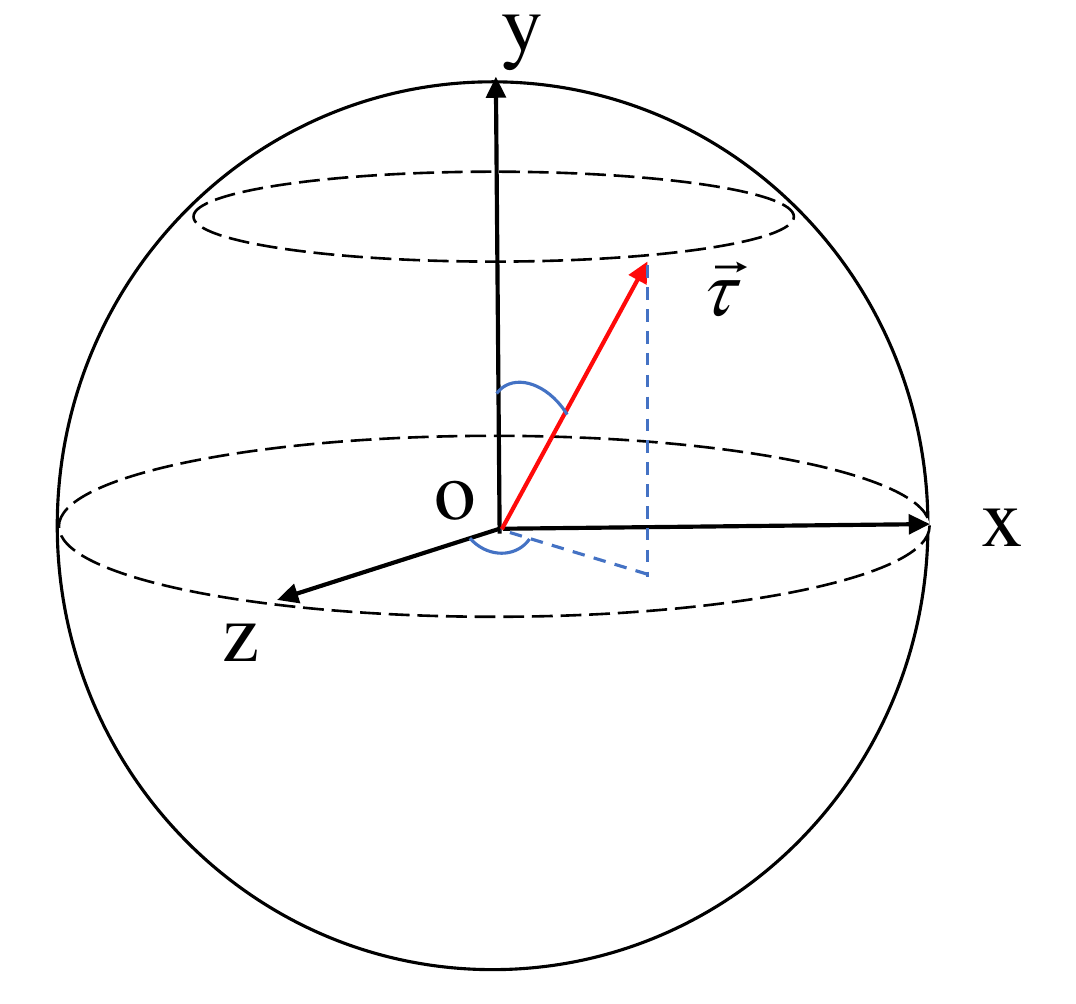}\hspace{5mm}
            \caption{The Bloch-sphere representation of $\ket{\Psi_{\rm in}}$, the spin direction of $\ket{\Psi_{\rm in}}$ is $\vec{\tau}$
            (the red solid arrow), with $\vec{\tau}=(\tau_x, \tau_y, \tau_z)$. Here the intersection point of positive $y$-axis and sphere is viewed as the ``north pole'', and the intersection point of negative $y$-axis and sphere is viewed as the ``south pole''
            }\label{fig:SGH3a}
      \end{figure}

\item \textcolor{blue}{The Non-Relativistic Approximation}. For the incident wave and the reflected wave, the energy-momentum relation is given by
\begin{eqnarray}
\mathcal{E}^2=\vec{p}^{\,2} c^2+m^2c^4.
\end{eqnarray}
The non-relativistic approximation implies that the magnitude of momentum $\vec{p}$ (or the velocity $\vec{v}$) of Dirac's particle is very small, such that
\begin{eqnarray}
\mathcal{E} \approx m c^2,
\end{eqnarray}
or
\begin{eqnarray}
\mathcal{E} - m c^2 \approx 0.
\end{eqnarray}
Based on Eq. (\ref{eq:z-1d-1b}), we find that under the non-relativistic approximation, the spatial shift for the reflected wave tends to zero, i.e.,
\begin{eqnarray}\label{eq:z-1d-1c}
    \Delta {z}_{\rm r} &=& \tau_y \, \dfrac{\sqrt{\mathcal{E}-mc^2} }{\sqrt{{\mathcal{E}}+mc^2}}
\dfrac{c\hbar {\mathcal{E}} \cos \varphi}{\left[ {\mathcal{E}}^2 \sin^2 \varphi + m^2c^4 \cos^2 \varphi \right]}
 \left[\tan^2 \varphi -   \dfrac{mc^2}{\mathcal{E}} \right] \propto \sqrt{\mathcal{E}-mc^2} \approx 0.
\end{eqnarray}
Thus, the spatial shift for the reflected wave is a pure relativistic effect caused by ``spin''.

$\;\;$ For the transmitted wave, the energy-momentum relation is given by
\begin{eqnarray}
&& (\mathcal{E}-V_0)^2 =\vec{p}_{\rm t}^{\, 2} c^2 +m^2 c^4,
\end{eqnarray}
with
\begin{eqnarray}
\vec{p}_{\rm t}=\hbar \vec{k}'= \hbar(k'_x, 0, k_z),
\end{eqnarray}
hence
\begin{eqnarray}
k'_x=\dfrac{1}{\hbar c}\sqrt{ (\mathcal{E}-V_0)^2- m^2 c^4 -\hbar^2 c^2 k_z^2}.
\end{eqnarray}
Based on Eq. (\ref{eq:z-1e}), we have known the spatial shift for the transmitted wave as
  \begin{eqnarray}
    \Delta z_{\rm t}
    &=& - \tau_y (n-1) \dfrac{ (k_x +n k_x')+\dfrac{k^2_z}{k_x k_x'} \left( k_x'+n k_x\right)}{\left(k_x +n k_x'\right)^2+\left[k_z (1-n)\right]^2}.
  \end{eqnarray}
Because $\mathcal{E} \approx m c^2$, then
\begin{eqnarray}
\vec{p}^{\,2} =\dfrac{1}{c^2} \left(\mathcal{E}^2-m^2c^4\right) \approx 0,
\end{eqnarray}
thus
\begin{eqnarray}
 k_z = \dfrac{1}{\hbar} |\vec{p}| \sin\varphi \approx 0.
\end{eqnarray}
Under the non-relativistic approximation, Dirac's equation reduces to Schr{\" o}dinger's equation, and the quantity
\begin{eqnarray}
\mathcal{E} - m c^2 &=& \frac{m c^2 }{\sqrt{1-\dfrac{v^2}{c^2}}}-m c^2 \approx m c^2 \left(1+\frac{1}{2} \dfrac{v^2}{c^2}\right)-m c^2 \nonumber\\
&=& \frac{m}{2}v^2=\dfrac{(mv)^2}{2m}=\dfrac{\vec{p}^{\, 2}}{2m} \approx 0.
\end{eqnarray}
In the Schr{\" o}dinger Hamiltonian
\begin{eqnarray}
H_{\rm Schr}=\dfrac{\vec{p}^{\, 2}}{2m} +V_0,
\end{eqnarray}
the scalar potential $V_0$ is in the order of the kinetic energy, i.e.,
\begin{eqnarray}
V_0\approx \dfrac{\vec{p}^{\, 2}}{2m} \approx \mathcal{E} - m c^2  \approx 0.
\end{eqnarray}
Thus
  \begin{eqnarray}
  n= \frac{\mathcal{E}+mc^2 }{\mathcal{E}-V_0 +mc^2}  \approx 1,
  \end{eqnarray}
and
  \begin{eqnarray}
&& k'_x= \dfrac{1}{\hbar c} \sqrt{ (\mathcal{E}-V_0)^2- m^2 c^4 -\hbar^2 c^2 k_z^2} \approx k_x.
  \end{eqnarray}

By considering
  \begin{eqnarray}
&& k'_x \approx k_x, \;\;\;\; k_z \approx 0,
  \end{eqnarray}
we have the spatial shift for the transmitted wave as
  \begin{eqnarray}
   \Delta z_{\rm t}
    &=& - \tau_y (n-1) \dfrac{ (k_x +n k_x')+\dfrac{k^2_z}{k_x k_x'} \left( k_x'+n k_x\right)}{\left(k_x +n k_x'\right)^2+\left[k_z (1-n)\right]^2}\nonumber\\
    &\approx& - \tau_y (n-1) \dfrac{ (n +1) k_x +\dfrac{k^2_z}{k_x^2} (n +1) k_x }{(n +1)^2 k_x^2+\left[k_z (1-n)\right]^2}\nonumber\\
    &\approx& - \tau_y (n-1) \dfrac{ (n +1) k_x}{(n +1)^2 k_x^2}\nonumber\\
    &=& - \tau_y (n-1) \dfrac{1}{(n +1) k_x}= - \dfrac{\tau_y}{n+1} \dfrac{n-1}{k_x} \nonumber\\
    &=&  - \dfrac{\tau_y}{n+1} \dfrac{\frac{V_0}{\mathcal{E}-V_0 +mc^2}}{\dfrac{1}{\hbar c} p_x} =
    - \dfrac{\tau_y \hbar c}{n+1} \dfrac{1}{\mathcal{E}-V_0 +mc^2}\dfrac{V_0}{p_x} \nonumber\\
    &=&      - \dfrac{\tau_y \hbar c}{2\mathcal{E}-V_0 +2mc^2} \dfrac{V_0}{|\vec{p}|\cos\varphi}\approx
       - \dfrac{\tau_y \hbar c}{2\mathcal{E}-V_0 +2mc^2} \dfrac{\mathcal{E} - m c^2}{\sqrt{\mathcal{E} - m c^2}\cos\varphi} \nonumber\\
    &=&  \sqrt{\mathcal{E} - m c^2} \times  \left[-  \dfrac{\tau_y \hbar c}{(2\mathcal{E}-V_0 +2mc^2)\cos\varphi}\right]\approx 0.
  \end{eqnarray}
Thus, the spatial shift for the transmitted wave is also a pure relativistic effect caused by spin.

 \item \textcolor{blue}{Expressing $\Delta {z}_{\rm r}$ and $\Delta {z}_{\rm t}$ in Terms of $\mathcal{E}$, $V_0$, $mc^2$, and $\varphi$}.
Based on Eq. (\ref{eq:z-1d}), the spatial shift for the reflected wave can be rewritten as
\begin{eqnarray}
    \Delta {z}_{\rm r}
&=& \tau_y \, \dfrac{\sqrt{\mathcal{E}-mc^2} }{\sqrt{{\mathcal{E}}+mc^2}}
\dfrac{c\hbar {\mathcal{E}} \cos \varphi}{\left[ {\mathcal{E}}^2 \sin^2 \varphi + m^2c^4 \cos^2 \varphi \right]}
 \left[\tan^2 \varphi -   \dfrac{mc^2}{\mathcal{E}} \right]\nonumber\\
&=& \tau_y \, \dfrac{\sqrt{1-\dfrac{mc^2}{\mathcal{E}}} }{\sqrt{1+\dfrac{mc^2}{\mathcal{E}}}} \times \dfrac{\hbar}{mc}
\dfrac{\dfrac{mc^2}{\mathcal{E}} \cos \varphi}{ \sin^2 \varphi + \left(\dfrac{mc^2}{\mathcal{E}}\right)^2 \cos^2 \varphi }
 \left[\tan^2 \varphi -   \dfrac{mc^2}{\mathcal{E}} \right],
\end{eqnarray}
thus, if we introduce the notation
\begin{eqnarray}
\mu_{\rm E} = \dfrac{mc^2}{\mathcal{E}},
\end{eqnarray}
then
\begin{eqnarray}
    \Delta {z}_{\rm r}
&=& \dfrac{\hbar}{mc} \times \tau_y \, \dfrac{\sqrt{1-\mu_{\rm E} } }{\sqrt{1+\mu_{\rm E}}} \times
\dfrac{ \mu_{\rm E} \cos \varphi}{ \sin^2 \varphi + \mu_{\rm E}^2 \cos^2 \varphi }
 \left[\tan^2 \varphi -   \mu_{\rm E} \right].
\end{eqnarray}
Note that the spatial shift $ \Delta {z}_{\rm r}$ is in the order of Compton wavelength, i.e.,
\begin{eqnarray}
    \Delta {z}_{\rm r} \sim \dfrac{h}{mc} \approx 2.426 \times 10^{-12} m.
\end{eqnarray}
Here Planck's constant $h =2\pi \hbar$.

$\;\;$ Similarly, we would like to express the spatial shift for the transmitted wave $ \Delta z_{\rm t}$ in terms of $\mathcal{E}$, $V_0$, $mc^2$, and $\varphi$.
Because
\begin{eqnarray}
&& n= \frac{\mathcal{E}+mc^2 }{\mathcal{E}-V_0 +mc^2}, \;\;\; k_x = \dfrac{1}{\hbar c} \sqrt{ \mathcal{E}^2- m^2 c^4} \cos\varphi, \;\;\; k_z = \dfrac{1}{\hbar c} \sqrt{ \mathcal{E}^2- m^2 c^4} \sin\varphi,
\end{eqnarray}
and
\begin{eqnarray}
k'_x &=&\dfrac{1}{\hbar c} \sqrt{ (\mathcal{E}-V_0)^2- m^2 c^4 -\hbar^2 c^2 k_z^2}\nonumber\\
&=&\dfrac{1}{\hbar c} \sqrt{ (\mathcal{E}-V_0)^2- m^2 c^4 -(\mathcal{E}^2- m^2 c^4 -\hbar^2 c^2 k_x^2)}\nonumber\\
&=&\dfrac{1}{\hbar c} \sqrt{ V_0(V_0-2\mathcal{E})+ \hbar^2 c^2 k_x^2}\nonumber\\
&=&\dfrac{1}{\hbar c} \sqrt{ V_0(V_0-2\mathcal{E})+ (\mathcal{E}^2- m^2 c^4)\cos^2\varphi },
\end{eqnarray}
we then have
  \begin{eqnarray}
  && (k_x +n k_x')+\dfrac{k^2_z}{k_x k_x'} \left( k_x'+n k_x\right)=(k_x +n k_x')+ \dfrac{k^2_z}{k_x}+ n \dfrac{k^2_z}{k_x'}
  = \dfrac{k_x^2+k^2_z}{k_x}+ n \dfrac{(k_x')^2+k^2_z}{k_x'}\nonumber\\
  &=& \dfrac{1}{\hbar^2 c^2} \dfrac{\mathcal{E}^2- m^2 c^4}{k_x}+ \dfrac{1}{\hbar^2 c^2} \, n \dfrac{(\mathcal{E}-V_0)^2- m^2 c^4}{k_x'}\nonumber\\
  &=& \dfrac{1}{\hbar^2 c^2}\, \dfrac{\mathcal{E}^2- m^2 c^4}{k_x}+ \dfrac{1}{\hbar^2 c^2}\, \frac{\mathcal{E}+mc^2 }{\mathcal{E}-V_0 +mc^2} \dfrac{(\mathcal{E}-V_0)^2- m^2 c^4}{k_x'}\nonumber\\
  &=&\dfrac{1}{\hbar^2 c^2}\, (\mathcal{E}+mc^2 )\times \left[ \dfrac{\mathcal{E}- m c^2}{k_x}+  \dfrac{(\mathcal{E}-V_0)- m c^2}{k_x'}\right],
  \end{eqnarray}
and
  \begin{eqnarray}
  && \left(k_x +n k_x'\right)^2+\left[k_z (1-n)\right]^2= k_x^2 + 2n k_x k_x'+n^2 (k_x')^2+ k_z^2 (1-2n+n^2)\nonumber\\
  &=& (k_x^2+k_z^2) + n^2 [(k_x')^2+k^2_z]+2n (k_x k_x' - k_z^2)\nonumber\\
  &=& \dfrac{1}{\hbar^2 c^2}\,(\mathcal{E}^2- m^2 c^4)+\dfrac{1}{\hbar^2 c^2}\,n^2 [(\mathcal{E}-V_0)^2- m^2 c^4]+2n (k_x k_x' - k_z^2)\nonumber\\
  &=& \dfrac{1}{\hbar^2 c^2}\,(\mathcal{E}^2- m^2 c^4)+ \dfrac{1}{\hbar^2 c^2}\, \left(\frac{\mathcal{E}+mc^2 }{\mathcal{E}-V_0 +mc^2}\right)^2 [(\mathcal{E}-V_0)^2- m^2 c^4]+2n (k_x k_x' - k_z^2)\nonumber\\
  &=& \dfrac{1}{\hbar^2 c^2}\,(\mathcal{E}^2- m^2 c^4)+ \dfrac{1}{\hbar^2 c^2}\,(\mathcal{E}+mc^2 )^2
  \dfrac{(\mathcal{E}-V_0)- m c^2}{(\mathcal{E}-V_0) +mc^2}+2n (k_x k_x' - k_z^2)\nonumber\\
  &=& \dfrac{1}{\hbar^2 c^2}\, (\mathcal{E}+ m c^2) \times \left[(\mathcal{E}- m c^2)+ (\mathcal{E}+mc^2 )
  \dfrac{(\mathcal{E}-V_0)- m c^2}{(\mathcal{E}-V_0) +mc^2}\right]+2n (k_x k_x' - k_z^2)\nonumber\\
  &=& \dfrac{1}{\hbar^2 c^2}\,(\mathcal{E}+ m c^2) \times \left[\dfrac{(\mathcal{E}- m c^2)[(\mathcal{E}-V_0) +mc^2]+(\mathcal{E}+mc^2 )[(\mathcal{E}-V_0)- m c^2]}{(\mathcal{E}-V_0) +mc^2}\right]+2n (k_x k_x' - k_z^2)\nonumber\\
 &=& \dfrac{1}{\hbar^2 c^2}\, (\mathcal{E}+ m c^2) \times \left[\dfrac{2 \mathcal{E}(\mathcal{E}-V_0)-2 (mc^2)^2}{(\mathcal{E}-V_0) +mc^2}\right]+2n (k_x k_x' - k_z^2)\nonumber\\
  &=& 2n \dfrac{1}{\hbar^2 c^2}\,\left[\mathcal{E}(\mathcal{E}-V_0)- m^2 c^4\right]+2n (k_x k_x' - k_z^2)\nonumber\\
  &=& 2n \left[\dfrac{\mathcal{E}^2-m^2 c^4}{\hbar^2 c^2}-\dfrac{\mathcal{E}V_0}{\hbar^2 c^2}\right]+2n (k_x k_x' - k_z^2)\nonumber\\
  &=& 2n \left[k_x^2+k_z^2-\dfrac{\mathcal{E}V_0}{\hbar^2 c^2}\right]+2n (k_x k_x' - k_z^2)\nonumber\\
  &=& 2n \left[k_x^2+k_x k_x'-\dfrac{\mathcal{E}V_0}{\hbar^2 c^2}\right]\nonumber\\
  &=& \dfrac{1}{\hbar^2 c^2}\, 2n  \left[\hbar^2 c^2(k_x^2+k_x k_x')-\mathcal{E}V_0\right]\nonumber\\
  &=& \dfrac{1}{\hbar^2 c^2}\, (\mathcal{E}+mc^2 )\times \left[\dfrac{2\left[\hbar^2 c^2(k_x^2+k_x k_x')-\mathcal{E}V_0\right]}{\mathcal{E}-V_0 +mc^2}\right],
 \end{eqnarray}
which yield
  \begin{eqnarray}
   \Delta z_{\rm t}
    &=& - \tau_y (n-1) \dfrac{ (k_x +n k_x')+\dfrac{k^2_z}{k_x k_x'} \left( k_x'+n k_x\right)}{\left(k_x +n k_x'\right)^2+\left[k_z (1-n)\right]^2}\nonumber\\
&=& - \tau_y (n-1) \dfrac{ \dfrac{\mathcal{E}- m c^2}{k_x}+  \dfrac{(\mathcal{E}-V_0)- m c^2}{k_x'}}{\dfrac{2\left[\hbar^2 c^2(k_x^2+k_x k_x')-\mathcal{E}V_0\right]}{\mathcal{E}-V_0 +mc^2}}\nonumber\\
&=& - \tau_y \dfrac{V_0}{\mathcal{E}-V_0 +mc^2} \dfrac{ \dfrac{\mathcal{E}- m c^2}{k_x}+  \dfrac{(\mathcal{E}-V_0)- m c^2}{k_x'}}{\dfrac{2\left[\hbar^2 c^2(k_x^2+k_x k_x')-\mathcal{E}V_0\right]}{\mathcal{E}-V_0 +mc^2}}\nonumber\\
&=& - \tau_y \dfrac{V_0}{2} \dfrac{ \dfrac{\mathcal{E}- m c^2}{k_x}+  \dfrac{(\mathcal{E}-V_0)- m c^2}{k_x'}}{\hbar^2 c^2 k_x(k_x+ k_x')-\mathcal{E}V_0}\nonumber\\
&=& - \tau_y \dfrac{V_0}{2} \hbar c \times  \dfrac{ \dfrac{\mathcal{E}- m c^2}{\sqrt{ \mathcal{E}^2- m^2 c^4} \cos\varphi}+  \dfrac{(\mathcal{E}-V_0)- m c^2}{\sqrt{ V_0(V_0-2\mathcal{E})+ (\mathcal{E}^2- m^2 c^4)\cos^2\varphi }}}{\sqrt{ \mathcal{E}^2- m^2 c^4} \cos\varphi \left[\sqrt{ \mathcal{E}^2- m^2 c^4} \cos\varphi+ \sqrt{ V_0(V_0-2\mathcal{E})+ (\mathcal{E}^2- m^2 c^4)\cos^2\varphi }\right]-\mathcal{E}V_0}.
  \end{eqnarray}
If we introduce the notation
\begin{eqnarray}
\mu_{\rm V} = \dfrac{mc^2}{V_0},
\end{eqnarray}
and recall $\mu_{\rm E} = \dfrac{mc^2}{\mathcal{E}}$, we then have
  \begin{eqnarray}
   \Delta z_{\rm t}
&=& - \tau_y \dfrac{V_0}{2} \hbar c \times  \dfrac{ \dfrac{\mathcal{E}- m c^2}{\sqrt{ \mathcal{E}^2- m^2 c^4} \cos\varphi}+  \dfrac{(\mathcal{E}-V_0)- m c^2}{\sqrt{ V_0(V_0-2\mathcal{E})+ (\mathcal{E}^2- m^2 c^4)\cos^2\varphi }}}{\sqrt{ \mathcal{E}^2- m^2 c^4} \cos\varphi \left[\sqrt{ \mathcal{E}^2- m^2 c^4} \cos\varphi+ \sqrt{ V_0(V_0-2\mathcal{E})+ (\mathcal{E}^2- m^2 c^4)\cos^2\varphi }\right]-\mathcal{E}V_0}\nonumber\\
&=&  - \tau_y \dfrac{V_0}{2} \hbar c \times \dfrac{1}{(mc^2)^2} \dfrac{ \dfrac{\sqrt{1-\mu_{\rm E} }}{\sqrt{ 1+\mu_{\rm E} }}\dfrac{1}{ \cos\varphi}+  \dfrac{\dfrac{1}{\mu_{\rm E}}-\dfrac{1}{\mu_{\rm V}}- 1}{\sqrt{ \dfrac{1}{\mu_{\rm V}}\left(\dfrac{1}{\mu_{\rm V}}-2\dfrac{1}{\mu_{\rm E}}\right)+ \left(\dfrac{1}{\mu^2_{\rm E}}- 1\right)\cos^2\varphi }}}{\sqrt{ \dfrac{1}{\mu^2_{\rm E}}- 1} \cos\varphi \left[\sqrt{ \dfrac{1}{\mu^2_{\rm E}}- 1} \cos\varphi+ \sqrt{ \dfrac{1}{\mu_{\rm V}}(\dfrac{1}{\mu_{\rm V}}-2\dfrac{1}{\mu_{\rm E}})+ (\dfrac{1}{\mu^2_{\rm E}}- 1)\cos^2\varphi }\,\right]-\dfrac{1}{\mu_{\rm E}}\dfrac{1}{\mu_{\rm V}}}\nonumber\\
&=&  - \frac{\hbar}{mc}\times \dfrac{\tau_y}{2} \dfrac{1}{\mu_{\rm V}}  \dfrac{ \dfrac{\sqrt{1-\mu_{\rm E} }}{\sqrt{ 1+\mu_{\rm E} }}\dfrac{1}{ \cos\varphi}+  \dfrac{\dfrac{1}{\mu_{\rm E}}-\dfrac{1}{\mu_{\rm V}}- 1}{\sqrt{ \dfrac{1}{\mu_{\rm V}}\left(\dfrac{1}{\mu_{\rm V}}-2\dfrac{1}{\mu_{\rm E}}\right)+ \left(\dfrac{1}{\mu^2_{\rm E}}- 1\right)\cos^2\varphi }}}{\sqrt{ \dfrac{1}{\mu^2_{\rm E}}- 1} \cos\varphi \left[\sqrt{ \dfrac{1}{\mu^2_{\rm E}}- 1} \cos\varphi+ \sqrt{ \dfrac{1}{\mu_{\rm V}}(\dfrac{1}{\mu_{\rm V}}-2\dfrac{1}{\mu_{\rm E}})+ (\dfrac{1}{\mu^2_{\rm E}}- 1)\cos^2\varphi }\,\right]-\dfrac{1}{\mu_{\rm E}}\dfrac{1}{\mu_{\rm V}}}.
  \end{eqnarray}
Note that the spin GH shift $\Delta {z}_{\rm t}$ is also in the order of Compton wavelength, i.e.,
\begin{eqnarray}
    \Delta {z}_{\rm t} \sim \dfrac{h}{mc}.
\end{eqnarray}

  \item \textcolor{blue}{Examples}. Here we provide some examples. Note that in our work, $\mathcal{E}$, $V_0$, $mc^2$, and $\varphi$ satisfy the conditions as shown in Eq. (\ref{eq:con-2}),
      \begin{eqnarray}\label{eq:con-2b}
&& (\mathcal{E}-V_0)^2- m^2c^4>0, \;\;\;\; \mathcal{E}^2 -  m^2 c^4>0, \;\;\;\; \mathcal{E}>0, \;\;\;\;
V_0>0, \;\;\;\; \mathcal{E}-V_0>0, \nonumber\\
&& 0\leq \dfrac{(\mathcal{E}-V_0)^2 -  m^2 c^4}{\mathcal{E}^2 -  m^2 c^4} <1, \;\;\;\; 0\leq \varphi < \varphi_{\rm cr1},
\end{eqnarray}
where $\varphi \in[0, \varphi_{\rm cr1})$ because we consider only the non-total reflection phenomenon. [Note: Seemingly, the ETSG effect is a Goos-H{\"a}nchen-like shift \cite{1947Goos}. However, they are essentially different, the latter occurs in the region of total reflection and the notion of ``spin'' is unnecessary, while the former can occur at the region of non-total reflection and the property of ``spin'' is required. In the literature, the GH shifts for optic waves and matter waves occur in the region of $\varphi \geq \varphi_{\rm cr1}$. However, the ETSG effect is an extraordinary spin effect To essentially distinguish the ETSG effect from the GH effect (or the GH shift), we restrict our study in the region $ \varphi_{\rm cr2} \in \left[0, \varphi_{\rm cr1} \right]$.

  $\;\;$ We may choose
\begin{eqnarray}
\mathcal{E} = 3 mc^2,
\end{eqnarray}
i.e.,
\begin{eqnarray}
\mu_{\rm E} = \dfrac{mc^2}{\mathcal{E}}=\frac{1}{3},
\end{eqnarray}
i.e.,
\begin{eqnarray}
\frac{1}{\sqrt{1-\dfrac{v^2}{c^2}}}=\frac{1}{\mu_{\rm E}}=3.
\end{eqnarray}
This means that in this situation the velocity of the Dirac particle reaches
\begin{eqnarray}
v =c \sqrt{1-\mu_{\rm E}^2}=\sqrt{\frac{8}{9}}\, c \approx 0.94 \,c,
\end{eqnarray}
where $c$ is the speed of light in vacuum. Besides, we may choose
\begin{eqnarray}
V_0 = \frac{1}{4} mc^2,
\end{eqnarray}
i.e.,
\begin{eqnarray}
\mu_{\rm V} = \dfrac{mc^2}{V_0} = 4.
\end{eqnarray}
In this case, the first critical angle is
\begin{eqnarray}
    \varphi_{\rm cr1}&=& \arcsin \left[\sqrt{\dfrac{(\mathcal{E}-V_0)^2 -  m^2 c^4}{\mathcal{E}^2 -  m^2 c^4}}\right]= \arcsin \left[\sqrt{\dfrac{(3-\frac{1}{4})^2 -  1}{3^2 - 1}}\right]\nonumber\\
    &=& \arcsin \left[\sqrt{\frac{105}{128}}\right] \approx \arcsin(0. 9057) \approx 1.1331\nonumber\\
    &=&  1.1331 \times \frac{180^\circ}{\pi} \approx 64.92^\circ,
\end{eqnarray}
and the second critical angle is
\begin{eqnarray}
\varphi_{\rm cr2} &=& \arctan  \sqrt{\dfrac{mc^2}{\mathcal{E}}}=\arctan \sqrt{\frac{1}{3}}= 30^\circ.
\end{eqnarray}

\emph{Example 1.---} Let the incident angle $\varphi_{\rm in}=\varphi$, with
\begin{eqnarray}
\varphi= \varphi_{\rm cr2} = 30^\circ,
\end{eqnarray}
 then
 \begin{eqnarray}
    \Delta {z}_{\rm r}=0,
\end{eqnarray}
namely, there is no the spatial shift for the reflected wave. The spatial shift for the transmitted wave is given by
  \begin{eqnarray}
   \Delta z_{\rm t}
&=&  - \frac{\hbar}{mc}\times \dfrac{\tau_y}{2} \dfrac{1}{\mu_{\rm V}}  \dfrac{ \dfrac{\sqrt{1-\mu_{\rm E} }}{\sqrt{ 1+\mu_{\rm E} }}\dfrac{1}{ \cos\varphi}+  \dfrac{\dfrac{1}{\mu_{\rm E}}-\dfrac{1}{\mu_{\rm V}}- 1}{\sqrt{ \dfrac{1}{\mu_{\rm V}}\left(\dfrac{1}{\mu_{\rm V}}-2\dfrac{1}{\mu_{\rm E}}\right)+ \left(\dfrac{1}{\mu^2_{\rm E}}- 1\right)\cos^2\varphi }}}{\sqrt{ \dfrac{1}{\mu^2_{\rm E}}- 1} \cos\varphi \left[\sqrt{ \dfrac{1}{\mu^2_{\rm E}}- 1} \cos\varphi+ \sqrt{ \dfrac{1}{\mu_{\rm V}}(\dfrac{1}{\mu_{\rm V}}-2\dfrac{1}{\mu_{\rm E}})+ (\dfrac{1}{\mu^2_{\rm E}}- 1)\cos^2\varphi }\,\right]-\dfrac{1}{\mu_{\rm E}}\dfrac{1}{\mu_{\rm V}}}\nonumber\\
&=&  - \frac{\hbar}{mc}\times \dfrac{\tau_y}{2} \times 4\times   \dfrac{ \dfrac{\sqrt{1-\dfrac{1}{3} }}{\sqrt{ 1+\dfrac{1}{3} }}\dfrac{1}{ \cos\varphi}+  \dfrac{3-\dfrac{1}{4}- 1}{\sqrt{ \dfrac{1}{4}\left(\dfrac{1}{4}-2\times 3\right)+ \left(3^2- 1\right)\cos^2\varphi }}}
{\sqrt{ 3^2- 1} \cos\varphi \left[\sqrt{ 3^2- 1} \cos\varphi+ \sqrt{ \dfrac{1}{4}(\dfrac{1}{4}-2\times 3)+ (3^2- 1)\cos^2\varphi }\,\right]-3\times \dfrac{1}{4}}\nonumber\\
&=&  - \frac{\hbar}{mc}\times 2 \tau_y \times   \dfrac{\dfrac{1}{\sqrt{2}}\dfrac{1}{ \cos\varphi}+  \dfrac{ \dfrac{7}{4}}
{\sqrt{ -\dfrac{23}{16}+ 8\cos^2\varphi }}}
{\sqrt{8} \cos\varphi \left[\sqrt{8} \cos\varphi+ \sqrt{ -\dfrac{23}{16}+ 8\cos^2\varphi }\,\right]-\dfrac{3}{4}}\nonumber\\
&=&  - \frac{\hbar}{mc}\times 2 \tau_y \times   \dfrac{\dfrac{1}{\sqrt{2}}\dfrac{2}{\sqrt{3}}+  \dfrac{ \dfrac{7}{4}}
{\sqrt{ -\dfrac{23}{16}+ 8 \times \dfrac{3}{2} }}}
{\sqrt{8} \dfrac{\sqrt{3}}{2} \left[\sqrt{8} \dfrac{\sqrt{3}}{2}+ \sqrt{ -\dfrac{23}{16}+ 8\times \dfrac{3}{2} }\,\right]-\dfrac{3}{4}}\nonumber\\
&=&  - \frac{\hbar}{mc}\times 2 \tau_y \times   \dfrac{\dfrac{2}{\sqrt{6}}+  \dfrac{ \dfrac{7}{4}}
{\dfrac{13}{4}}}
{\sqrt{6}  \left[\sqrt{6} + \dfrac{13}{4} \right]-\dfrac{3}{4}}=- \frac{\hbar}{mc} \times  2 \tau_y \times   \dfrac{\dfrac{2}{\sqrt{6}}+  \dfrac{7}{13}}
{ \dfrac{13\sqrt{6}}{4}+\dfrac{21}{4}}= - \frac{\hbar}{mc} \times  8 \tau_y \times   \dfrac{\dfrac{2}{\sqrt{6}}+  \dfrac{7}{13}}
{ 21+13\sqrt{6}}\nonumber\\
&=&- \frac{\hbar}{mc} \times  8 \tau_y \times   \dfrac{1}{39}=- \frac{h}{mc} \dfrac{1}{2\pi}\times  8 \tau_y \times   \dfrac{1}{39} \nonumber\\
&=&- \frac{h}{mc} \times  \dfrac{4 \tau_y}{39\pi}.
  \end{eqnarray}
We choose
\begin{eqnarray}
\tau_y \approx 0.92,
\end{eqnarray}
then the spatial shift for the transmitted wave is
  \begin{eqnarray}
   \Delta z_{\rm t}&=&- \frac{h}{mc} \times  \dfrac{4 \tau_y}{39\pi} \approx - 0.030 \times \frac{h}{mc},
  \end{eqnarray}
 i.e., the shift is a negative shift and is about $0.030$ Compton wavelength.

$\;\;$ Let us denote the reflected angle as $\varphi_{\rm r}$, due to Eq. (\ref{eq:momrf}), we easily know that
\begin{eqnarray}\label{eq:momrf-1}
\varphi_{\rm r}=\varphi=30^\circ.
\end{eqnarray}
Similarly, let us denote the transmitted angle as $\varphi_{\rm t}$, due to Eq. (\ref{eq:momtr}), we can have
  \begin{eqnarray}\label{eq:momtr-1}
\tan \varphi_{\rm t}&=&\dfrac{k_z}{k'_x}=\dfrac{\sqrt{ \mathcal{E}^2- m^2 c^4} \sin\varphi}{\sqrt{ (\mathcal{E}-V_0)^2- m^2 c^4 -\hbar^2 c^2 k_z^2}}= \dfrac{\sqrt{ \mathcal{E}^2- m^2 c^4} \sin\varphi}{\sqrt{ (\mathcal{E}-V_0)^2- m^2 c^4 -(\mathcal{E}^2- m^2 c^4)\sin^2\varphi}}\nonumber\\
&=& \dfrac{\sqrt{ 3^2-1} \sin\varphi}{\sqrt{ \left(3-\dfrac{1}{4}\right)^2- 1 -(3^2-1)\sin^2\varphi}}
=\dfrac{\sqrt{ 8} \sin\varphi}{\sqrt{ \dfrac{105}{16} -8\sin^2\varphi}}=\dfrac{\sqrt{ 8} \times \dfrac{1}{2}}{\sqrt{ \dfrac{105}{16} -8\times \dfrac{1}{4}}}=\sqrt{\dfrac{32}{73}},
\end{eqnarray}
thus
\begin{eqnarray}
 \varphi_{\rm t}&=& \arctan \left[\sqrt{\dfrac{32}{73}}\right] \approx 0.5848 \approx 0.5848 \times \frac{180^\circ}{\pi} \approx 33.51^\circ.
\end{eqnarray}
One may observe that $\varphi_{\rm t}$ is greater than the incident angle, i.e.,
\begin{eqnarray}
 \varphi_{\rm t} > \varphi_{\rm in}=30^\circ.
\end{eqnarray}
In Fig. \ref{fig:SGH4a}, we have illustrated the incident wave, the reflected wave and the transmitted wave.

\begin{figure}[t]
            \centering
            \includegraphics[width=75mm]{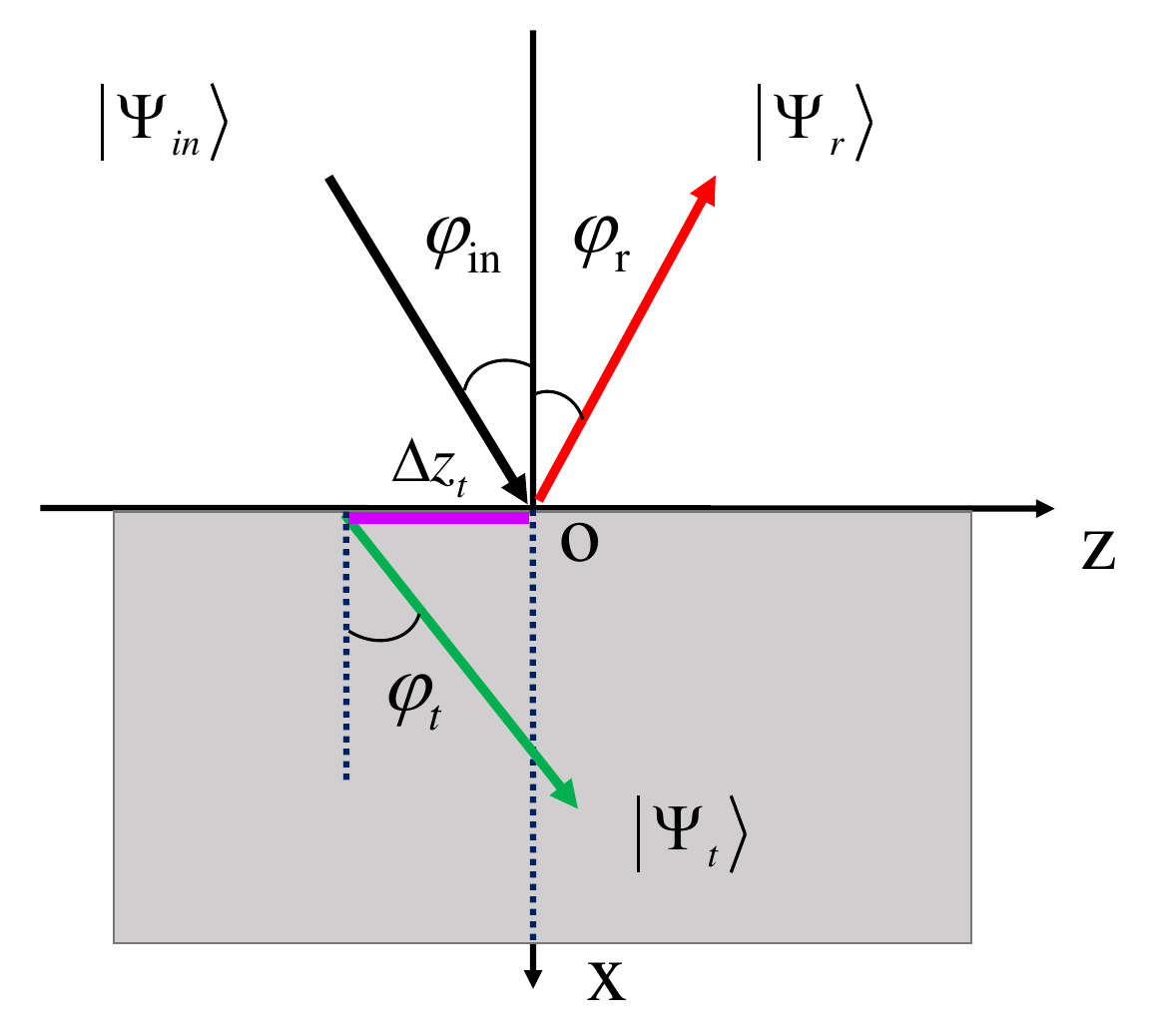}\hspace{5mm}
            \caption{Illustration of the incident wave (the black solid arrow), the reflected wave (the red solid arrow) and the transmitted wave (the green solid arrow). Here we have chosen $\mathcal{E} = 3 mc^2$, $V_0 = (1/4) mc^2$, which leads to the critical angle $\varphi_{\rm cr2}=30^\circ$. When the incident angle $\varphi_{\rm in}=\varphi_{\rm cr2}=30^\circ$, the spatial shift of the reflected wave is zero, i.e., $ \Delta z_{\rm r}$,  and the spatial shift of the transmitted wave (the purple solid line) is a negative shift and is about $0.030$ Compton wavelength, i.e., $ \Delta z_{\rm t}\approx-0.030 (h/mc)$. Note that in this case, $\varphi_{\rm r}=\varphi_{\rm in}$ and $\varphi_{\rm t} \approx 33.51^\circ > \varphi_{\rm in}$.
            }\label{fig:SGH4a}
\end{figure}

\emph{Example 2.---} Let the incident angle $\varphi_{\rm in}=\varphi$, with
\begin{eqnarray}
\varphi= \varphi_{\rm cr2} = 15^\circ,
\end{eqnarray}
 then
\begin{eqnarray}
    \Delta {z}_{\rm r}
&=& \dfrac{\hbar}{mc} \times \tau_y \, \dfrac{\sqrt{1-\mu_{\rm E} } }{\sqrt{1+\mu_{\rm E}}} \times
\dfrac{ \mu_{\rm E} \cos \varphi}{ \sin^2 \varphi + \mu_{\rm E}^2 \cos^2 \varphi }
 \left[\tan^2 \varphi -   \mu_{\rm E} \right]\nonumber\\
 &=& \dfrac{\hbar}{mc} \times \tau_y \, \dfrac{\sqrt{1-\dfrac{1}{3} } }{\sqrt{1+\dfrac{1}{3}}} \times
\dfrac{ \dfrac{1}{3} \cos \varphi}{ \sin^2 \varphi + \left(\dfrac{1}{3}\right)^2 \cos^2 \varphi }
 \left[\tan^2 \varphi -   \dfrac{1}{3} \right]\nonumber\\
 &=& \dfrac{\hbar}{mc} \times \frac{\tau_y}{\sqrt{2}} \,
\dfrac{ 3 \cos \varphi}{ 9 \sin^2 \varphi + \cos^2 \varphi }
 \left[\tan^2 \varphi -   \dfrac{1}{3} \right]\nonumber\\
 &=& \dfrac{\hbar}{mc} \times \frac{\tau_y}{\sqrt{2}} \,
\dfrac{  \cos \varphi}{ 8 \sin^2 \varphi + 1 }
 \left[3 \tan^2 \varphi -   1 \right]\nonumber\\
 &=& \dfrac{\hbar}{mc} \times \frac{\tau_y}{\sqrt{2}} \,
\dfrac{  \cos 15^\circ}{ 8 \sin^2 15^\circ + 1 }
 \left[3 \tan^2 15^\circ -   1 \right]\nonumber\\
 &=& \dfrac{h}{mc} \times \frac{\tau_y}{2\sqrt{2}\pi} \,
\dfrac{  \cos 15^\circ}{ 8 \sin^2 15^\circ + 1 }
 \left[3 \tan^2 15^\circ -   1 \right].
\end{eqnarray}
By taking $\tau_y \approx 0.92$, we have
\begin{eqnarray}
    \Delta {z}_{\rm r} \approx - 0.051 \times \dfrac{h}{mc}.
\end{eqnarray}
Namely, the spatial shift of the reflected wave is a negative shift and is about 0.051 Compton wavelength.

$\;\;$ Similarly, the spatial shift of the transmitted wave reads
  \begin{eqnarray}
   \Delta z_{\rm t}
&=&  - \frac{\hbar}{mc}\times 2 \tau_y \times   \dfrac{\dfrac{1}{\sqrt{2}}\dfrac{1}{ \cos\varphi}+  \dfrac{ \dfrac{7}{4}}
{\sqrt{ -\dfrac{23}{16}+ 8\cos^2\varphi }}}
{\sqrt{8} \cos\varphi \left[\sqrt{8} \cos\varphi+ \sqrt{ -\dfrac{23}{16}+ 8\cos^2\varphi }\,\right]-\dfrac{3}{4}}\nonumber\\
&=&  - \frac{h}{mc}\times  \frac{\tau_y}{\pi} \times   \dfrac{\dfrac{1}{\sqrt{2}}\dfrac{1}{ \cos15^\circ}+  \dfrac{ \dfrac{7}{4}}
{\sqrt{ -\dfrac{23}{16}+ 8\cos^215^\circ }}}
{\sqrt{8} \cos15^\circ \left[\sqrt{8} \cos 15^\circ+ \sqrt{ -\dfrac{23}{16}+ 8\cos^215^\circ }\,\right]-\dfrac{3}{4}}.
  \end{eqnarray}
By taking $\tau_y \approx 0.92$, we have
\begin{eqnarray}
    \Delta {z}_{\rm t} \approx -  0.034 \times \dfrac{h}{mc}.
\end{eqnarray}
Namely, the spatial shift of the transmitted wave is a negative shift and is about 0.034 Compton wavelength.

$\;\;$ Let us denote the reflected angle as $\varphi_{\rm r}$, due to Eq. (\ref{eq:momrf}), we easily know that
\begin{eqnarray}
\varphi_{\rm r}=\varphi=15^\circ.
\end{eqnarray}
Similarly, let us denote the transmitted angle as $\varphi_{\rm t}$, due to Eq. (\ref{eq:momtr}), we can have
  \begin{eqnarray}
\tan \varphi_{\rm t}&=&\dfrac{k_z}{k'_x}=\dfrac{\sqrt{ 8} \sin\varphi}{\sqrt{ \dfrac{105}{16} -8\sin^2\varphi}}=\dfrac{\sqrt{ 8} \sin15^\circ}{\sqrt{ \dfrac{105}{16} -8\sin^2 15^\circ}} \approx 0.2982,
\end{eqnarray}
thus
\begin{eqnarray}
 \varphi_{\rm t}&=& \arctan \left(0.2982\right) \approx 0.2898 \approx 0.2898 \times \frac{180^\circ}{\pi} \approx 16.60^\circ.
\end{eqnarray}
One may observe that $\varphi_{\rm t}$ is greater than the incident angle, i.e.,
\begin{eqnarray}
 \varphi_{\rm t} > \varphi_{\rm in}=15^\circ.
\end{eqnarray}
In Fig. \ref{fig:SGH5a}, we have illustrated the incident wave, the reflected wave and the transmitted wave.

\begin{figure}[t]
            \centering
            \includegraphics[width=75mm]{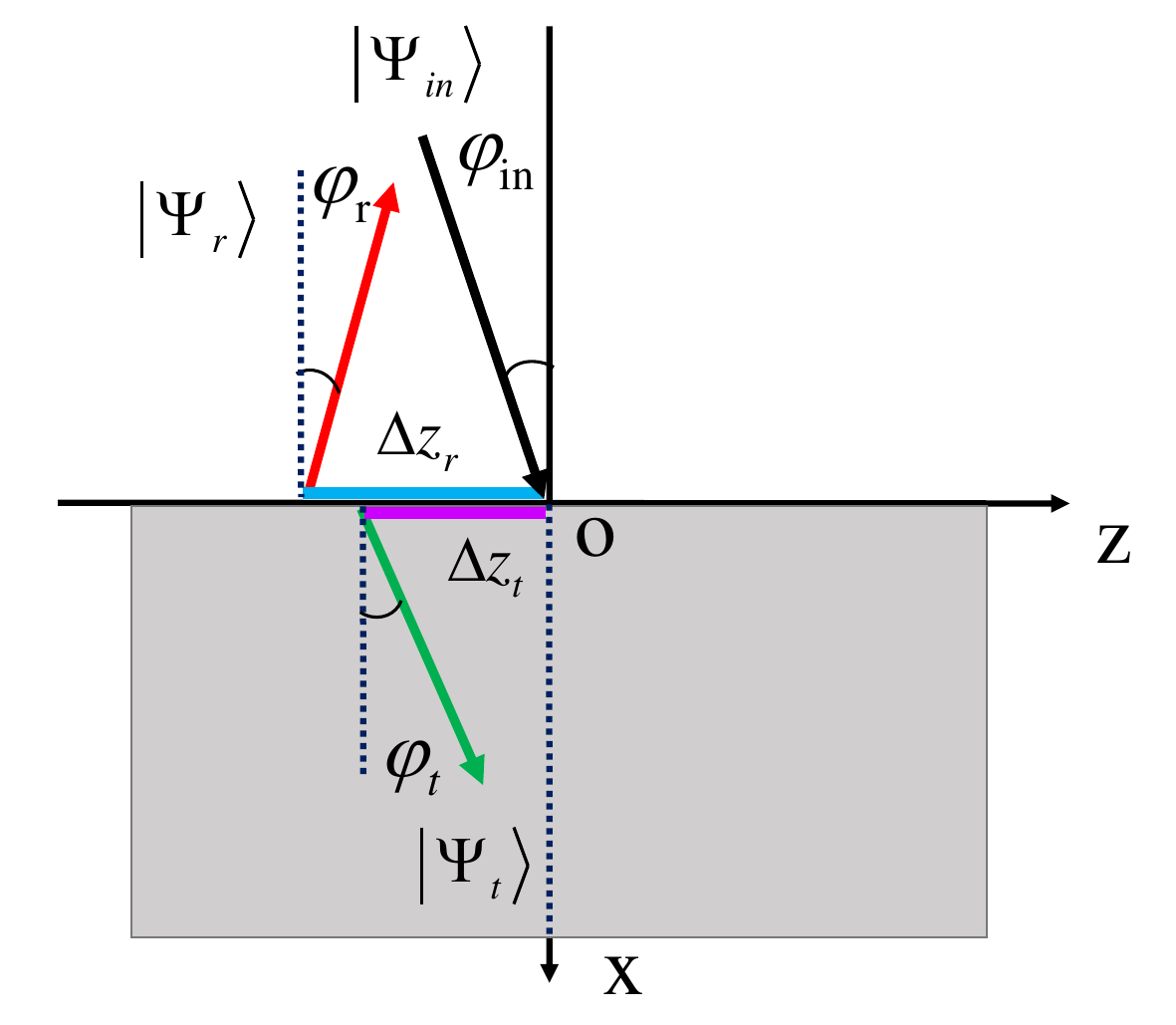}\hspace{5mm}
            \caption{Illustration of the incident wave (the black solid arrow), the reflected wave (the red solid arrow) and the transmitted wave (the green solid arrow). Here we have chosen $\mathcal{E} = 3 mc^2$, $V_0 = (1/4) mc^2$, which leads to the critical angle $\varphi_{\rm cr2}=30^\circ$. When the incident angle $\varphi_{\rm in}=15^\circ$, the spatial shift of the reflected wave (the blue solid line) is a negative shift and $ \Delta z_{\rm r}\approx - 0.051 (h/mc)$, and the spatial shift of the transmitted wave (the purple solid line) is a negative shift and $ \Delta z_{\rm t} \approx -  0.034 (h/mc)$. Note that in this case, $\varphi_{\rm r}=\varphi_{\rm in}=15^\circ$ and $\varphi_{\rm t} \approx 16.60^\circ  > \varphi_{\rm in}$.
            }\label{fig:SGH5a}
\end{figure}


\emph{Example 3.---} Let the incident angle $\varphi_{\rm in}=\varphi$, with
\begin{eqnarray}
\varphi= \varphi_{\rm cr2} = 60^\circ,
\end{eqnarray}
 then
\begin{eqnarray}
    \Delta {z}_{\rm r}
&=& \dfrac{\hbar}{mc} \times \tau_y \, \dfrac{\sqrt{1-\mu_{\rm E} } }{\sqrt{1+\mu_{\rm E}}} \times
\dfrac{ \mu_{\rm E} \cos \varphi}{ \sin^2 \varphi + \mu_{\rm E}^2 \cos^2 \varphi }
 \left[\tan^2 \varphi -   \mu_{\rm E} \right]\nonumber\\
 &=& \dfrac{\hbar}{mc} \times \frac{\tau_y}{\sqrt{2}} \,
\dfrac{  \cos \varphi}{ 8 \sin^2 \varphi + 1 }
 \left[3 \tan^2 \varphi -   1 \right]\nonumber\\
 &=& \dfrac{\hbar}{mc} \times \frac{\tau_y}{\sqrt{2}} \,
\dfrac{  \cos 60^\circ}{ 8 \sin^2 60^\circ + 1 }
 \left[3 \tan^2 60^\circ -   1 \right]\nonumber\\
 &=& \dfrac{h}{mc} \times \frac{\tau_y}{2\sqrt{2}\pi} \,
\dfrac{  \dfrac{1}{2}}{ 8 \times \dfrac{3}{4} + 1 }
 \left[3 \times 3 -   1 \right]\nonumber\\
 &=& \dfrac{h}{mc} \times \frac{\sqrt{2}\tau_y}{7\pi},
\end{eqnarray}
By taking $\tau_y \approx 0.92$, we have
\begin{eqnarray}
    \Delta {z}_{\rm r} \approx 0.059 \times \dfrac{h}{mc}.
\end{eqnarray}
Namely, the spatial shift of the reflected wave is a positive shift and is about 0.059 Compton wavelength.

$\;\;$ Similarly, the spatial shift of the transmitted wave reads
  \begin{eqnarray}
   \Delta z_{\rm t}
&=&  - \frac{\hbar}{mc}\times 2 \tau_y \times   \dfrac{\dfrac{1}{\sqrt{2}}\dfrac{1}{ \cos\varphi}+  \dfrac{ \dfrac{7}{4}}
{\sqrt{ -\dfrac{23}{16}+ 8\cos^2\varphi }}}
{\sqrt{8} \cos\varphi \left[\sqrt{8} \cos\varphi+ \sqrt{ -\dfrac{23}{16}+ 8\cos^2\varphi }\,\right]-\dfrac{3}{4}}\nonumber\\
&=&  - \frac{h}{mc}\times  \frac{\tau_y}{\pi} \times   \dfrac{\dfrac{1}{\sqrt{2}}\dfrac{1}{ \cos 60^\circ}+  \dfrac{ \dfrac{7}{4}}
{\sqrt{ -\dfrac{23}{16}+ 8\cos^2 60^\circ }}}
{\sqrt{8} \cos 60^\circ \left[\sqrt{8} \cos 60^\circ+ \sqrt{ -\dfrac{23}{16}+ 8\cos^2 60^\circ }\,\right]-\dfrac{3}{4}}.
  \end{eqnarray}
By taking $\tau_y \approx 0.92$, we have
\begin{eqnarray}
    \Delta {z}_{\rm t} \approx -0.8373 \times \dfrac{h}{mc}.
\end{eqnarray}
Namely, the spatial shift of the transmitted wave is a negative shift and is about $-0.8373$ Compton wavelength.

\begin{figure}[t]
            \centering
            \includegraphics[width=75mm]{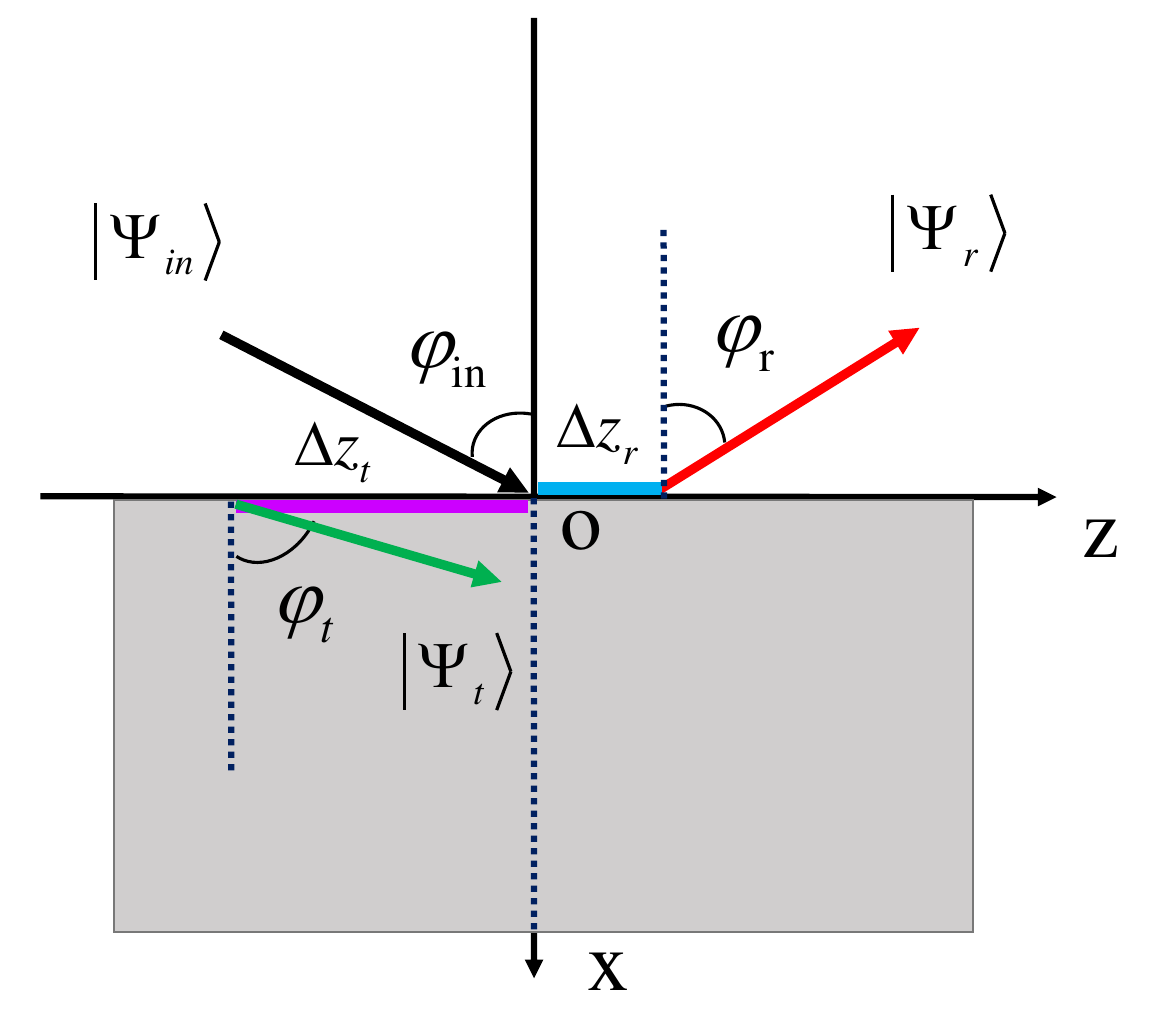}\hspace{5mm}
            \caption{Illustration of the incident wave (the black solid arrow), the reflected wave (the red solid arrow) and the transmitted wave (the green solid arrow). Here we have chosen $\mathcal{E} = 3 mc^2$, $V_0 = (1/4) mc^2$, which leads to the critical angle $\varphi_{\rm cr2}=30^\circ$. When the incident angle $\varphi_{\rm in}=60^\circ$, the spatial shift of the reflected wave is a positive shift (the blue solid line) and $ \Delta z_{\rm r}\approx 0.059 (h/mc)$, and the spatial shift of the transmitted wave (the purple solid line) is a negative shift and $ \Delta z_{\rm t} \approx -0.8373 (h/mc)$. Note that in this case, $\varphi_{\rm r}=\varphi_{\rm in}=60^\circ$ and $\varphi_{\rm t} \approx 72.98^\circ > \varphi_{\rm in} $.
            }\label{fig:SGH6a}
\end{figure}

$\;\;$ Let us denote the reflected angle as $\varphi_{\rm r}$, due to Eq. (\ref{eq:momrf}), we easily know that
\begin{eqnarray}
\varphi_{\rm r}=\varphi=60^\circ.
\end{eqnarray}
Similarly, let us denote the transmitted angle as $\varphi_{\rm t}$, due to Eq. (\ref{eq:momtr}), we can have
  \begin{eqnarray}
\tan \varphi_{\rm t}&=&\dfrac{k_z}{k'_x}=\dfrac{\sqrt{ 8} \sin\varphi}{\sqrt{ \dfrac{105}{16} -8\sin^2\varphi}}=\dfrac{\sqrt{ 8} \sin 60^\circ}{\sqrt{ \dfrac{105}{16} -8\sin^2 60^\circ}} \approx 3.2660,
\end{eqnarray}
thus
\begin{eqnarray}
 \varphi_{\rm t}&=& \arctan \left(3.2660\right) \approx 1.2737 \approx 1.2737 \times \frac{180^\circ}{\pi} \approx 72.98^\circ.
\end{eqnarray}
One may observe that $\varphi_{\rm t}$ is greater than the incident angle, i.e.,
\begin{eqnarray}
 \varphi_{\rm t} > \varphi_{\rm in}=60^\circ.
\end{eqnarray}
In Fig. \ref{fig:SGH6a}, we have illustrated the incident wave, the reflected wave and the transmitted wave.

\emph{Example 4.---} We still choose $\mathcal{E} = 3 mc^2$, $V_0 = (1/4) mc^2$, $\tau_y\approx 0.92$. Let the incident angle $\varphi$ runs from $0^\circ$ to the critical angle $\varphi_{\rm cr1}\approx 64.92^\circ$, in Table \ref{tab:varia} we list the variations of the reflected angle $\varphi_{\rm r}$, the transmitted angle $\varphi_{\rm t}$, the shift $\Delta {z}_{\rm r} $, and the shift $\Delta {z}_{\rm t} $ with respect to the incident angle. The Table \ref{tab:varia} is based on the following relations
\begin{eqnarray}
&& \varphi_{\rm r}=\varphi_{\rm in}=\varphi, \nonumber\\
&& \tan \varphi_{\rm t}=\dfrac{\sqrt{ 8} \sin\varphi}{\sqrt{ \dfrac{105}{16} -8\sin^2\varphi}}, \nonumber\\
&&     \Delta {z}_{\rm r} = \dfrac{h}{mc} \times \frac{\tau_y}{2\sqrt{2}\pi} \,
\dfrac{  \cos \varphi}{ 8 \sin^2 \varphi + 1 }
 \left[3 \tan^2 \varphi -   1 \right], \nonumber\\
&&    \Delta z_{\rm t} = - \frac{h}{mc}\times  \frac{\tau_y}{\pi} \times   \dfrac{\dfrac{1}{\sqrt{2}}\dfrac{1}{ \cos \varphi}+  \dfrac{ \dfrac{7}{4}}
{\sqrt{ -\dfrac{23}{16}+ 8\cos^2 \varphi }}}
{\sqrt{8} \cos \varphi \left[\sqrt{8} \cos \varphi+ \sqrt{ -\dfrac{23}{16}+ 8\cos^2 \varphi }\,\right]-\dfrac{3}{4}}.
\end{eqnarray}

\begin{table}[t]
\centering
\caption{The variations of the reflected angle $\varphi_{\rm r}$, the transmitted angle $\varphi_{\rm t}$, the shift $\Delta {z}_{\rm r} $, and the shift $\Delta {z}_{\rm t} $ when the incident angle $\varphi$ runs from $0^\circ$ to the critical angle $\varphi_{\rm cr1}\approx 64.92^\circ$. Here $\mu_{\rm E} = 1/3$, $\mu_{\rm V} = 4$, $\tau_y\approx 0.92$, $\varphi_{\rm r}=\varphi_{\rm in}$, and $\varphi_{\rm t}$ satisfies $\tan \varphi_{\rm t}={k_z}/{k'_x}$ with $  k_x'^2= \left[(\mathcal{E}-V_0)^2 - \left(\hbar^2 k_z^2 c^2 +m^2 c^4\right)\right]/(\hbar c)^2$. We have taken the unit of $\Delta {z}_{\rm r} $ and $\Delta {z}_{\rm t} $ as $h/mc$. }
\begin{tabular}{cccccccccccccc}
\hline\hline\hline
$\varphi_{\rm in}$ & $0^\circ$  &  $2^\circ$  &  $4^\circ$& $6^\circ$& $8^\circ$& $10^\circ$ & $12^\circ$ & $14^\circ$ & $16^\circ$ & $18^\circ$
& $20^\circ$ & $22^\circ$  \\
  \hline
$\varphi_{\rm r}$  & $0^\circ$  &  $2^\circ$  & $4^\circ$& $6^\circ$& $8^\circ$ &$10^\circ$ & $12^\circ$ & $14^\circ$ & $16^\circ$ & $18^\circ$
& $20^\circ$ & $22^\circ$  \\
  \hline
$\varphi_{\rm t}$ & $0^\circ$  &  $2.21^\circ$  &  $4.42^\circ$&  $6.62^\circ$ & $8.84^\circ$ & $11.05^\circ$& $13.27^\circ$ & $15.49^\circ$ & $17.72^\circ$ & $19.95^\circ$
& $22.19^\circ$ & $24.43^\circ$ \\
  \hline
$\Delta {z}_{\rm r}$ & $-0.1035$  & $-0.1021$  &  $-0.0980$&  $-0.0916$ & $-0.0835$& $-0.0745$ & $-0.0650$ & $-0.0556$ & $-0.0466$ & $-0.0381$
& $-0.0302$ & $-0.0231$\\
  \hline
$\Delta {z}_{\rm t}$ & $-0.0302$  &  $-0.0302$  &  $-0.0304$&  $-0.0308$&  $-0.0312$ & $-0.0318$& $-0.0326$ & $-0.0335$ & $-0.0347$ & $-0.0360$
& $-0.0376$ & $-0.0394$ \\
 \hline\hline
$\varphi_{\rm in}$& $24^\circ$ & $26^\circ$  &  $28^\circ$  &  $30^\circ$& $32^\circ$& $34^\circ$& $36^\circ$ & $38^\circ$ & $40^\circ$ & $42^\circ$ & $44^\circ$ & $46^\circ$ \\
  \hline
$\varphi_{\rm r}$& $24^\circ$ & $26^\circ$  &  $28^\circ$  &  $30^\circ$& $32^\circ$& $34^\circ$& $36^\circ$ & $38^\circ$ & $40^\circ$ & $42^\circ$ & $44^\circ$ & $46^\circ$ \\
  \hline
$\varphi_{\rm t}$ & $26.68^\circ$ & $28.95^\circ$  &  $31.22^\circ$  &  $33.51^\circ$& $35.81^\circ$& $38.13^\circ$& $40.46^\circ$ & $42.82^\circ$ & $45.21^\circ$ & $47.63^\circ$ & $50.08^\circ$ & $52.58^\circ$ \\
  \hline
$\Delta {z}_{\rm r}$ & $-0.0165$  & $-0.0105$  &  $-0.0050$  &  $0$& $0.0046$& $0.0089$& $0.0130$ & $0.0168^\circ$ & $0.0205^\circ$ & $0.0241^\circ$ & $0.0275^\circ$ & $0.0310^\circ$ \\
  \hline
$\Delta {z}_{\rm t}$ & $-0.0416$ & $-0.0441^\circ$  &  $-0.0470^\circ$  &  $-0.0505^\circ$& $-0.0546^\circ$& $-0.0595^\circ$& $-0.0653^\circ$ & $-0.0723^\circ$ & $0.0809^\circ$ & $-0.0914^\circ$ & $-0.1046^\circ$ & $-0.1214^\circ$ \\
 \hline\hline
$\varphi_{\rm in}$ & $48^\circ$ & $50^\circ$  &  $52^\circ$  &  $54^\circ$& $56^\circ$& $58^\circ$& $59^\circ$ & $60^\circ$ & $61^\circ$ & $62^\circ$ & $63^\circ$ & $64^\circ$ \\
  \hline
$\varphi_{\rm r}$& $48^\circ$ & $50^\circ$  &  $52^\circ$  &  $54^\circ$& $56^\circ$& $58^\circ$& $59^\circ$ & $60^\circ$ & $61^\circ$ & $62^\circ$ & $63^\circ$ & $64^\circ$ \\
  \hline
$\varphi_{\rm t}$ & $55.14^\circ$ & $57.76^\circ$  &  $60.46^\circ$  &  $63.28^\circ$& $66.25^\circ$& $69.44^\circ$& $71.16^\circ$ & $72.98^\circ$ & $74.94^\circ$ & $77.13^\circ$ & $79.66^\circ$ & $82.91^\circ$\\
  \hline
$\Delta {z}_{\rm r}$& $0.0345$ & $0.0381$  &  $0.0418$  &  $0.0457$& $0.0498$& $0.0543$& $0.0567$ & $0.0592$ & $0.0617$ & $0.0646$ & $0.0675$ & $0.0706$ \\
  \hline
$\Delta {z}_{\rm t}$& $-0.1433$ & $-0.1725$  &  $-0.2132$  &  $-0.2722$& $-0.3639$& $-0.5207$& $-0.6475$ & $-0.8373$ & $-1.1481$ & $-1.7366$ & $-3.2110$ & $-11.6888$ \\
 \hline\hline\hline
\end{tabular}\label{tab:varia}
\end{table}

$\;\;$ From Table \ref{tab:varia}, where the parameters $\mathcal{E} = 3 mc^2$, $V_0 = (1/4) mc^2$ and $\tau_y\approx 0.92$ are fixed, one may observe that: (i) when the incident angle is close to $0^\circ$, $\Delta {z}_{\rm r}$ has a negative shift as $-0.10 (h/mc)$, and $\Delta {z}_{\rm t}$ has a negative shift as $-0.03(h/mc)$; (ii) when the incident angle is close to the critical angle $\varphi_{\rm cr1}\approx 64.92^\circ$, $\Delta {z}_{\rm r}$ has a positive shift as $0.07 (h/mc)$, and $\Delta {z}_{\rm t}$ has a negative shift as $-11.7(h/mc)$.

\end{enumerate}

\newpage
\newpage

\section{Some Other Calculations}

\subsection{Spatial Shift in the Case that the Finite-Width Incident Wave with Finite Dimensions along both the $z$ Direction and $y$ Direction}

\subsubsection{The Finite-Width Incident Wave}

Now, for the finite-width incident wave, it can still be regarded as the composition of infinite-width waves, which is the solution of Dirac equation. The wavefront of the incident wave is
\begin{equation}
    \ket{\Psi_{\rm in}(0, y, z)}= \begin{cases}
        |\Psi_0\rangle {\rm e}^{{\rm i} \left[k_{y_0} y+k_{z_0} z- \frac{1}{\hbar} \mathcal{E}t\right]}, & \;\;\;\; |y| \leq b\,, |z| \leq a  \\
        0, &\;\;\;\; {\rm other\; case },
    \end{cases}
\end{equation}
with
\begin{equation}\label{eq:Psi0-b}
    |\Psi_0\rangle= \begin{pmatrix}
        \ell_1 (\mathcal{E} + mc^2) \\
        \ell_2(\mathcal{E}+ mc^2) \\
        [\ell_1k_{z_0} + \ell_2\,(k_{x_0}- \i k_{y_0})]c\hbar  \\
        [\ell_1(k_{x_0}+\i k_{y_0}) - \ell_2\,k_{z_0}]c\hbar
    \end{pmatrix}.
\end{equation}
Here we also define the average momentum of the beam as $ \vec{p}_0 = p_{x_0} \,\vec{e}_x +p_{y_0} \,\vec{e}_y
+p_{z_0} \,\vec{e}_z= \hbar(k_{x_0} \,\vec{e}_x+k_{y_0} \,\vec{e}_y
+k_{z_0} \,\vec{e}_z) $, and the subscripts $x_0$ , $y_0$ and $z_0$ correspond the $x$-component, $y$-component and $z$-component of the average momentum, respectively. The probability of Dirac's electron appearing outside the beam is negligible.

We consider a finite-width matter wave with finite dimensions along both the $z$ direction and $y$ direction, i.e.,
    \begin{align}\label{incident wave fouriered'}
         \ket{\Psi_{\rm in}(x,y,z)} &= \frac{1}{{2 \pi}} \int_{-\infty}^{\infty} \int_{-\infty}^{\infty}\mathrm{\Theta}(k^2 -k_y^2- k_z^2) \ket{\psi_k(k_y,k_z,x,y,z)}  \d k_y \d ( k_z) \notag \\
        &= \frac{1}{{2 \pi}} \int_{-\infty}^{\infty}\int_{-\infty}^{\infty}
        \mathrm{\Theta}(k^2 - k_y^2-k_z^2) |\psi_k( k_y,k_z)\rangle
         \exp\bigg[{\rm i} \big( k_y y+ k_z z + x \sqrt{k^2 -k_y^2- k_z^2}\big)\bigg] d k_y \d k_z,
        \end{align}
       where
         \begin{eqnarray}
    \ket{\psi_{{k}}(k_y, k_z,x,y,z)}=|\psi_{{k}}(k_y, k_z)\rangle \exp\bigg[{\rm i}\big(k_y y +k_z z + x \sqrt{k^2 -k_y^2- k_z^2}\big)\bigg],
      \end{eqnarray}
   and
           \begin{eqnarray}
                \mathrm{\Theta}(k^2 -k_z^2)=\begin{cases}
                    1,\quad k^2 -k_y^2-k_z^2\ge 0, \\
                    0,\quad k^2 -k_y^2-k_z^2 < 0.
                \end{cases}
            \end{eqnarray}
At $x = 0$, one has
            \begin{equation}
             |\psi_k(k_y, k_z)\rangle = \frac{1}{ { {2 \pi}}} \iint_{-\infty}^{\infty}
                 \ket{\Psi_{\rm in}(0,y,z)}\exp\left[-{\rm i} (k_y\,y+ k_z\,z)\right] \d y \, \d z.
             \end{equation}
     It follows that
          \begin{align}
             \ket{\psi_k(k_y , k_z)}=& \frac{1}{{  {2 \pi}}}\iint_{-\infty}^{\infty}
             \ket{\Psi_{\rm in}(0,y,z)}\exp\left[-{\rm i}(k_y y +k_z\,z)
                 \right]  \d y {\rm d}z \notag \\
             =& \frac{1}{ {{2 \pi}}}
             \ket{\Psi_0}\int_{-a}^{a} \exp\left[
                 {\rm i} (k_{z0} -k_z)z\right]{\rm d}z \int_{-b}^{b} \exp\left[
                    {\rm i} (k_{y0} -k_y)y\right]{\rm d}y \notag \\
             =& \ket{\Psi_0}\sqrt{\frac{2}{\pi}}\,\dfrac{\sin\left[(k_{z0} - k_z)a\right]}{(k_{z_0} - k_z)} \sqrt{\frac{2}{\pi}}\,\dfrac{\sin\left[(k_{y0} - k_y)b\right]}{(k_{y_0} - k_y)} \notag\\
             =&F(k_y)F(k_z)\ket{\Psi_0},
             \end{align}
     where
     \begin{eqnarray}
        F(k_y)=\sqrt{\frac{2}{\pi}}\,\dfrac{\sin\left[(k_{y0} - k_y)b\right]}{(k_{y_0} - k_y)}, \;\;\;\;\;
            F(k_z)=\sqrt{\frac{2}{\pi}}\,\dfrac{\sin\left[(k_{z0} - k_z)a\right]}{(k_{z_0} - k_z)}.
    \end{eqnarray}

     Obviously, $\ket{\psi_k(k_y, k_z )}$ cannot be the solution of Dirac's equation. However, the function $F(k_y)$ and $F(k_z)$ indicates that only the $ \ket{\psi_k(k_y, k_z )}$, with $k_y $ very near $k_{y_0}$  and  $k_z $ very near $k_{z_0}$, can have influence. When  $k_{y_0} b$ and $k_{z_0} a$ is large, we only need to consider $k_y$ near $k_{y_0}$ and $k_z$ near $k_{z_0}$. Let $k_y=k_{y0}+\delta k_y$ and $k_z=k_{z0}+\delta k_z$, by considering $\delta k_y \rightarrow 0$ and $\delta k_z \rightarrow 0$, one then has
\begin{eqnarray}
    \ket{\Psi_0} \approx \ket{\Psi},
\end{eqnarray}
with
\begin{equation}\label{eq:Psi0-c}
    |\Psi\rangle = \begin{pmatrix}
        \ell_1 (\mathcal{E} + mc^2) \\
        \ell_2(\mathcal{E}+ mc^2) \\
        [\ell_1k_{z} + \ell_2\,(k_{x}- \i k_{y})]c\hbar  \\
        [\ell_1(k_{x}+\i k_{y}) - \ell_2\,k_{z}]c\hbar
    \end{pmatrix}.
\end{equation}
Then one obtains
\begin{eqnarray}
    \ket{\psi_k(k_y, k_z)}=F(k_y) F(k_z)\ket{\Psi_0} \approx F(k_y) F(k_z)\ket{\Psi}.
\end{eqnarray}
Eventually the incident wave is equal to
\begin{eqnarray}
    \ket{\Psi_{\rm in}(x, y, z)} &=& \frac{1}{{2 \pi}}\int_{k_{y_0}-\delta k_y}^{k_{y_0}+\delta k_y} \int_{k_{z_0}-\delta k_z}^{k_{z_0}+\delta k_z} \ket{\psi_k( k_y, k_z, {x, y, z})}  \d k_y \d ( k_z).
\end{eqnarray}
Here $ \ket{\Psi_{\rm in}(x, y, z)}$ can also be regarded as the composition of the solution of Dirac's equation approximately.

\subsubsection{The Spatial Shifts}

For the infinite-width wave, one can attain
\begin{subequations}
    \begin{align}
 &A=\frac{\ell_1\left[k_x^2-(k_y^2+k_z^2)(n-1)^2-k_x'^2 n^2\right] +2 k_x (-i k_y \ell_1+k_z \ell_2) (n-1)}
 {k_x^2+(k_y^2+k_z^2)(n-1)^2+2 k_x k_x' n +k_x'^2 n^2}, \\
 &B= \frac{\ell_2\left[k_x^2-(k_y^2+k_z^2)(n-1)^2-k_x'^2 n^2\right] -2 k_x (k_z \ell_1-i k_y \ell_2) (n-1)}
 {k_x^2+(k_y^2+k_z^2)(n-1)^2+2 k_x k_x' n +k_x'^2 n^2}, \\
 &C=\dfrac{\sqrt{(\mathcal{E}-V_0)(\mathcal{E} -V_0+mc^2)}}{\sqrt{\mathcal{E}(\mathcal{E} +mc^2)}}\frac{2  n \ell_1 k_x(k_x+{n}k_x') +2n k_x k_z \ell_2(n-1)-2 \i \, n k_x k_y \ell_1 (n-1)}
 {k_x^2+(k_y^2+k_z^2)(n-1)^2+2 k_x k_x' n +k_x'^2 n^2},\\
 & D=\dfrac{\sqrt{(\mathcal{E}-V_0)(\mathcal{E} -V_0+mc^2)}}{\sqrt{\mathcal{E}(\mathcal{E} +mc^2)}}\frac{2  n \ell_2k_x (k_x +{n} k_x')+  2 \i\, n k_x k_y \ell_2 (n-1)-2nk_x  k_z \ell_1(n-1)}
 {k_x^2+(k_y^2+k_z^2)(n-1)^2+2 k_x k_x' n +k_x'^2 n^2},
     \end{align}
 \end{subequations}
which are just Eqs. (\ref{eq:2-A})-(\ref{eq:2-D}) without setting $k_y=0$.

\begin{remark}
\textcolor{blue}{Calculating the Spatial Shifts in the $z$-Basis}. In the following, let us calculate the spatial shift for the reflected wave in the $z$-basis. In this situation, the reflected wave becomes
 \begin{eqnarray}
       \ket{\Psi_{\rm r}} &\approx& \dfrac{1}{2 \pi \sqrt{2\mathcal{E}(\mathcal{E} +mc^2)}}\int_{k_{y_0}-\delta k_y}^{k_{y_0}+\delta k_y} \int_{k_{z_0}-\delta k_z}^{k_{z_0}+\delta k_z}  F(k_y) F(k_z)\left\{A \begin{pmatrix}
                \mathcal{E} + mc^2 \\
                0 \\
                k_z c {\hbar}\\
                (-k_x+ \i k_y) c {\hbar}
            \end{pmatrix}
            + B \begin{pmatrix}
                0 \\
                \mathcal{E} + mc^2 \\
                -(k_x+ \i k_y) c {\hbar} \\
                -k_z c {\hbar}
            \end{pmatrix}\right\}\notag\\
            &&{\rm exp} \left\{\frac{\i}{\hbar} [{\hbar}(-k_x x+k_y y+ k_z z) - \mathcal{E} t]\right\} \d k_x \d k_y,
\end{eqnarray}
i.e.,
\begin{eqnarray}
        \ket{\Psi_{\rm r}} &\approx& \dfrac{{\ell_1}}{2 \pi {|\ell_1|}\sqrt{2\mathcal{E}(\mathcal{E} +mc^2)}}\int_{k_{y_0}-\delta k_y}^{k_{y_0}+\delta k_y} \int_{k_{z_0}-\delta k_z}^{k_{z_0}+\delta k_z}  F(k_y) F(k_z){\rm exp}\left\{{{\rm i} \left[(-k_x x+k_y y+ k_z z) - \frac{1}{\hbar} \mathcal{E} t\right]}\right\} \notag\\
   && \Biggl\{|A| \, {\rm exp}\left[-{\rm i}\left(\theta_{\rm ra\,0} +   \left. \frac{\partial \theta_{\rm ra}}{\partial k_y} \right|_{\vec{k}=\vec{k}_0} \delta k_y +\left. \frac{\partial \theta_{\rm ra}}{\partial k_z} \right|_{\vec{k}=\vec{k}_0} \delta k_z \right)\right] \begin{pmatrix}
                \mathcal{E} + mc^2 \\
                0 \\
                k_z c {\hbar}\\
                (-k_x+ \i k_y) c {\hbar}
            \end{pmatrix}\notag\\
            &&+  |B| \, {\rm exp}\left[-{\rm i} \left(\theta_{\rm rb\,0} +   \left. \frac{\partial \theta_{\rm rb}}{\partial k_y} \right|_{\vec{k}=\vec{k}_0} \delta k_y +\left. \frac{\partial \theta_{\rm rb}}{\partial k_z} \right|_{\vec{k}=\vec{k}_0} \delta k_z \right)\right] \begin{pmatrix}
                0 \\
                \mathcal{E} + mc^2 \\
                -(k_x+ \i k_y) c {\hbar} \\
                -k_z c {\hbar}
            \end{pmatrix}\Biggr\} \d k_y \d k_z.
\end{eqnarray}
Let us denote
\begin{eqnarray}
    &&\ell_1 = \cos \frac{\theta}{2 }, \;\;\;\; \ell_2  = \sin \frac{\theta}{2 } \e^{\i \phi}, \nonumber\\
    && \frac{\ell_2}{\ell_1} = G_1 + \i G_2 = \tan \frac{\theta}{2} \e^{\i \phi}, \nonumber\\
    && |\ell_1|^2 ( 1+ G_1^2 +G_2^2)=1,
\end{eqnarray}
and
    \begin{eqnarray}
    &&    \Omega_1 = k_x^2-k_y^2(n-1)^2-k_z^2(n-1)^2-k_x'^2 n^2,\nonumber\\
    &&    \Omega_2 = 2 k_x k_y  (n-1),\nonumber\\
    &&    \Omega_3 = 2 k_x k_z  (n-1),
    \end{eqnarray}
we then have
    \begin{eqnarray}
     &&   A = \ell_1 \dfrac{\Omega_1 - \i \Omega_2 + (G_1 + \i G_2)\Omega_3}{(n k_x' + k_x)^2 + (1 - n)^2 (k_y^2 + k_z^2)}= \frac{\ell_1}{|\ell_1|}|A| \e^{-\i \theta_{\rm ra}}  , \nonumber\\
     &&   B = \ell_1\dfrac{-\Omega_3 + (G_1+\i G_2) (\Omega_1 +\i \Omega_2)}{(n k_x' + k_x)^2 + (1 - n)^2 (k_y^2 + k_z^2)}=\frac{\ell_1}{|\ell_1|}|B| \e^{-\i \theta_{\rm rb}},
    \end{eqnarray}
\begin{eqnarray}
    &&|A|^2 = |\ell_1|^2 \dfrac{(\Omega_1 + G_1 \Omega_3)^2 +  ( G_2\Omega_3-\Omega_2  )^2}{[(n k_x' + k_x)^2 + (1 - n)^2 (k_y^2 + k_z^2)]^2}\notag\\
    &&|B|^2 = |\ell_1|^2 \dfrac{(G_1\Omega_1 - G_2 \Omega_2 -\Omega_3)^2 +  ( G_2\Omega_1+ G_1\Omega_2  )^2}{[(n k_x' + k_x)^2 + (1 - n)^2 (k_y^2 + k_z^2)]^2}
\end{eqnarray}
\begin{eqnarray}
    \tan \theta_{\rm ra} = - \frac{G_2 \Omega_3- \Omega_2}{\Omega_1+G_1 \Omega_3},  \;\;\;\;\;\;\;    \tan \theta_{\rm rb} = - \frac{G_2 \Omega_1+ G_1 \Omega_2}{G_1\Omega_1 - G_2 \Omega_2 -\Omega_3}.
\end{eqnarray}
By the way, for the above phases $\theta_{\rm ra}$ and $\theta_{\rm rb}$, in generally, one does not have $\theta_{\rm ra}=\theta_{\rm rb}$ or $\theta_{\rm ra}=-\theta_{\rm rb}$, hence he does not the following relation.
\begin{eqnarray}
    \left. \frac{\partial \theta_{\rm ra}}{\partial k_y} \right|_{\vec{k}=\vec{k}_0} \ne  {\pm}   \left. \frac{\partial \theta_{\rm rb}}{\partial k_y} \right|_{\vec{k}=\vec{k}_0}, \;\;\;\;\;\;\;\;\;
   \left. \frac{\partial \theta_{\rm ra}}{\partial k_z} \right|_{\vec{k}=\vec{k}_0} \ne {\pm}    \left. \frac{\partial \theta_{\rm rb}}{\partial k_z} \right|_{\vec{k}=\vec{k}_0}.
\end{eqnarray}
Nota that $\ell_1$ and $\ell_2$ do not depend on the momentum $\vec{k}$, thus their phases do not have contributions to the spatial shifts.
As one can expect, the total spatial shift for the reflected wave will occur along both the $z$ direction and $y$ direction (i.e., the $yz$-plane), which is given by
\begin{eqnarray}
    \Delta \vec{l}_{\rm r} = \Delta y_{\rm r} \; \vec{e}_y + \Delta z_{\rm r} \; \vec{e}_z,
\end{eqnarray}
where
\begin{eqnarray}
    \Delta y_{\rm r}  = \frac{|A|^2}{|A|^2 + |B|^2 } \Delta y_{\rm ra} + \frac{|B|^2}{|A|^2 + |B|^2 } \Delta y_{\rm rb},
\end{eqnarray}
\begin{eqnarray}
    \Delta z_{\rm r}  = \frac{|A|^2}{|A|^2 + |B|^2 } \Delta z_{\rm ra} + \frac{|B|^2}{|A|^2 + |B|^2 } \Delta z_{\rm rb}.
\end{eqnarray}
with
\begin{eqnarray}
    \Delta y_{\rm ra} = \frac{\partial \theta_{\rm ra}}{\partial k_y}, \;\;\;\;\;\;\;  \Delta y_{\rm rb} = \frac{\partial \theta_{\rm rb}}{\partial k_y},
\end{eqnarray}
\begin{eqnarray}
    \Delta z_{\rm ra} = \frac{\partial \theta_{\rm ra}}{\partial k_z}, \;\;\;\;\;\;\;  \Delta z_{\rm rb} = \frac{\partial \theta_{\rm rb}}{\partial k_z}.
\end{eqnarray}
Note that $\Delta \vec{l}_{\rm r}$ is a Imbert-Fedorov-like shift \cite{1955Fedorov,1972Imbert}.
Now, we come to calculate the spatial shift in the $z$-basis. For convenience, in the following calculation, we do not distinguish much $\vec{k}$ from $\vec{k}_0$.

Because
  \begin{eqnarray}
    &&\frac{(n k_x' + k_x)^2 + (1 - n)^2 (k_y^2 + k_z^2)} {|\ell_1|^2} |A|^2    \Delta y_{\rm ra} \notag\\
&=&-\left[G_2 \frac{\partial \Omega_3}{\partial k_y}- \frac{\partial \Omega_2}{\partial k_y}\right](\Omega_1 + G_1 \Omega_3)  +(G_2 \Omega_3 - \Omega_2) \left[ \frac{\partial \Omega_1}{\partial k_y} + G_1\frac{\partial \Omega_3}{\partial k_y}\right]\notag\\
&=&G_2 \left(\Omega_3 \frac{\partial \Omega_1}{\partial k_y} -\Omega_1 \frac{\partial \Omega_3}{\partial k_y} \right)  - \left(\Omega_2 \frac{\partial \Omega_1}{\partial k_y} -\Omega_1 \frac{\partial \Omega_2}{\partial k_y} \right) +G_1 \left(\Omega_3 \frac{\partial \Omega_2}{\partial k_y} -\Omega_2 \frac{\partial \Omega_3}{\partial k_y} \right),
  \end{eqnarray}
\begin{eqnarray}
    &&\frac{(n k_x' + k_x)^2 + (1 - n)^2 (k_y^2 + k_z^2)} {|\ell_1|^2} |B|^2    \Delta y_{\rm rb} \notag\\
    &=& -\left( G_2 \frac{\partial \Omega_1}{\partial k_y} + G_1\frac{\partial \Omega_2}{\partial k_y} \right) (G_1\Omega_1 - G_2 \Omega_2 -\Omega_3 ) + (G_2 \Omega_1+ G_1 \Omega_2 ) \left(G_1 \frac{\partial \Omega_1}{\partial k_y} - G_2 \frac{\partial \Omega_2}{\partial k_y} - \frac{\partial \Omega_3}{\partial k_y}\right)\notag\\
    &=& (G_1^2 +G_2^2 ) \left(\Omega_2 \frac{\partial \Omega_1}{\partial k_y} -\Omega_1 \frac{\partial \Omega_2}{\partial k_y} \right)  + G_2  \left(\Omega_3 \frac{\partial \Omega_1}{\partial k_y} -\Omega_1 \frac{\partial \Omega_3}{\partial k_y} \right) +G_1  \left(\Omega_3 \frac{\partial \Omega_2}{\partial k_y} -\Omega_2 \frac{\partial \Omega_3}{\partial k_y} \right),
\end{eqnarray}
then we have
\begin{eqnarray}
    &&\frac{(n k_x' + k_x)^2 + (1 - n)^2 (k_y^2 + k_z^2)} {|\ell_1|^2} (|A|^2 +  |B|^2)\nonumber\\
    &=& (\Omega_1 + G_1 \Omega_3)^2 +  ( G_2\Omega_3-\Omega_2  )^2+(G_1\Omega_1 - G_2 \Omega_2 -\Omega_3)^2 +  ( G_2\Omega_1+ G_1\Omega_2  )^2\notag\\
    &=&\Omega_1^2 + G_1^2 \Omega_3^2 + 2 G_1 \Omega_1\Omega_3 + \Omega_2^2 + G_2^2 \Omega_3^2 - 2 G_2 \Omega_2\Omega_3 + G_1^2 \Omega_1^2 + G_2^2 \Omega_2^2 + \Omega_3^2 -2 G_1 G_2\Omega_1\Omega_2   \notag\\
    &&- 2 G_1 \Omega_1\Omega_3 + 2  G_2\Omega_2\Omega_3 +G_2^2\Omega_1^2 + G_1^2 \Omega_2^2 + 2 G_1 G_2\Omega_1\Omega_2 \notag\\
    &=&(1+ G_1^2 + G_2^2 ) (\Omega_1^2 +\Omega_2^2 +\Omega_3^2  ),
\end{eqnarray}
thus
\begin{eqnarray}
    \Delta y_{\rm r} &= & \frac{|A|^2}{|A|^2 + |B|^2 } \Delta y_{\rm ra} + \frac{|B|^2}{|A|^2 + |B|^2 } \Delta y_{\rm rb} \nonumber\\
    &=&\frac{2 G_2 \left(\frac{\partial\Omega_1  }{\partial k_y}\Omega_3 - \Omega_1 \frac{\partial\Omega_3 }{\partial k_y}\right)
    + {(G_1^2 +G_2^2-1 )}\left(\frac{\partial\Omega_1  }{\partial k_y} \Omega_2 - \Omega_1 \frac{\partial\Omega_2  }{\partial k_y}\right)
    +2 G_1 \left(\frac{\partial\Omega_2  }{\partial k_y} \Omega_3 - \Omega_2 \frac{\partial\Omega_3 }{\partial k_y}\right)]}
    {(1+ G_1^2 + G_2^2 ) (\Omega_1^2 +\Omega_2^2 +\Omega_3^2  )}.
\end{eqnarray}
Similarly, for $ \Delta z_r$ we have
\begin{eqnarray}
    \Delta z_{\rm r}&= & \frac{|A|^2}{|A|^2 + |B|^2 } \Delta z_{\rm ra} + \frac{|B|^2}{|A|^2 + |B|^2 } \Delta z_{\rm rb} \nonumber\\
    &=&\frac{2 G_2 \left(\frac{\partial\Omega_1  }{\partial k_z}\Omega_3 - \Omega_1 \frac{\partial\Omega_3 }{\partial k_z}\right)
    + {(G_1^2 +G_2^2-1 )}\left(\frac{\partial\Omega_1  }{\partial k_z} \Omega_2 - \Omega_1 \frac{\partial\Omega_2  }{\partial k_z}\right)
    +2 G_1 \left(\frac{\partial\Omega_2  }{\partial k_z} \Omega_3 - \Omega_2 \frac{\partial\Omega_3 }{\partial k_z}\right)]}
    {(1+ G_1^2 + G_2^2 ) (\Omega_1^2 +\Omega_2^2 +\Omega_3^2  )}.
\end{eqnarray}
$\blacksquare$
\end{remark}

\begin{remark}
\textcolor{blue}{Calculating the Spatial Shifts in the General Basis}. It is important that no matter in which basis, the expectant value of the spatial shift  $\Delta \vec{l}_{\rm r}$ is unique. Namely, one has
\begin{eqnarray}
    \Delta \vec{l}_{\rm r}  = \Delta y_{\rm r}\, \vec{e}_y+ \Delta z_{\rm r} \, \vec{e}_z,
\end{eqnarray}
where
\begin{eqnarray}
    \Delta y_{\rm r}  &=&\frac{2 G_2 \left(\frac{\partial\Omega_1  }{\partial k_y}\Omega_3 - \Omega_1 \frac{\partial\Omega_3 }{\partial k_y}\right)
    + {(G_1^2 +G_2^2-1 )}\left(\frac{\partial\Omega_1  }{\partial k_y} \Omega_2 - \Omega_1 \frac{\partial\Omega_2  }{\partial k_y}\right)
    +2 G_1 \left(\frac{\partial\Omega_2  }{\partial k_y} \Omega_3 - \Omega_2 \frac{\partial\Omega_3 }{\partial k_y}\right)]}
    {(1+ G_1^2 + G_2^2 ) (\Omega_1^2 +\Omega_2^2 +\Omega_3^2  )},
\end{eqnarray}
\begin{eqnarray}
    \Delta z_{\rm r} &=&\frac{2 G_2 \left(\frac{\partial\Omega_1  }{\partial k_z}\Omega_3 - \Omega_1 \frac{\partial\Omega_3 }{\partial k_z}\right)
    + {(G_1^2 +G_2^2-1 )}\left(\frac{\partial\Omega_1  }{\partial k_z} \Omega_2 - \Omega_1 \frac{\partial\Omega_2  }{\partial k_z}\right)
    +2 G_1 \left(\frac{\partial\Omega_2  }{\partial k_z} \Omega_3 - \Omega_2 \frac{\partial\Omega_3 }{\partial k_z}\right)]}
    {(1+ G_1^2 + G_2^2 ) (\Omega_1^2 +\Omega_2^2 +\Omega_3^2  )}.
\end{eqnarray}
Here we would like to provide a proof.

\begin{proof}
According to Remark \ref{r10}, after expanding the reflected wave in the general basis, one can have
\begin{eqnarray}
    \begin{pmatrix}
        A\\
        B
    \end{pmatrix} = A' \begin{pmatrix}
         \cos \frac{\theta}{2}\\
         \sin \frac{\theta}{2}  \e^{\i \phi}
    \end{pmatrix}+ B' \begin{pmatrix}
         \sin \frac{\theta}{2}\\
         -\cos  \frac{\theta}{2}  \e^{\i \phi}
    \end{pmatrix},
\end{eqnarray}
and
\begin{eqnarray}
    A' = \frac{A \cos \frac{\theta}{2} \e^{\i \phi}+ B   \sin \frac{\theta}{2} }{ \cos^2 \frac{\theta}{2} \e^{\i \phi}+ \sin^2 \frac{\theta}{2} \e^{\i \phi}}= \left[A \cos \frac{\theta}{2} \e^{\i \phi}+ B   \sin \frac{\theta}{2}  \right]\e^{-\i \phi},
\end{eqnarray}
\begin{eqnarray}
    B' = \frac{A \sin \frac{\theta}{2} \e^{\i \phi}- B   \cos \frac{\theta}{2} }{ \cos^2 \frac{\theta}{2} \e^{\i \phi}+ \sin^2 \frac{\theta}{2} \e^{\i \phi}}= \left[A \sin \frac{\theta}{2} \e^{\i \phi}- B   \cos \frac{\theta}{2}  \right]\e^{-\i \phi}.
\end{eqnarray}
Then we have
\begin{eqnarray}
A'& =&  \frac{\ell_1 {\rm e}^{-{\rm i} \phi}}{(n k_x' + k_x)^2 + (1 - n)^2 (k_y^2 + k_z^2)} \times \nonumber\\
&& \biggr\{ \left[(\Omega_1 + G_1 \Omega_3) \cos \frac{\theta}{2} \cos \phi -(G_2\Omega_3-\Omega_2 ) \cos \frac{\theta}{2} \sin \phi + (G_1\Omega_1 - G_2 \Omega_2 -\Omega_3)\sin \frac{\theta}{2}\right]  \notag\\
&&+ {\rm i} \left[( G_2\Omega_3-\Omega_2  ) \cos \frac{\theta}{2} \cos \phi + (\Omega_1 + G_1 \Omega_3)\cos \frac{\theta}{2} \sin \phi + ( G_2\Omega_1+ G_1\Omega_2  ) \sin \frac{\theta}{2}\right] \biggr\},
\end{eqnarray}
\begin{eqnarray}
B'& =&  \frac{\ell_1 {\rm e}^{-{\rm i} \phi}}{(n k_x' + k_x)^2 + (1 - n)^2 (k_y^2 + k_z^2)} \times \nonumber\\
&& \biggr\{ \left[(\Omega_1 + G_1 \Omega_3) \sin \frac{\theta}{2} \cos \phi - ( G_2\Omega_3-\Omega_2  )\sin \frac{\theta}{2} \sin \phi- (G_1\Omega_1 - G_2 \Omega_2 -\Omega_3) \cos \frac{\theta}{2}\right]  \notag\\
&&+ {\rm i} \left[( G_2\Omega_3-\Omega_2  )\sin \frac{\theta}{2} \cos \phi + (\Omega_1 + G_1 \Omega_3)\sin \frac{\theta}{2} \sin \phi -( G_2\Omega_1+ G_1\Omega_2  ) \cos \frac{\theta}{2}\right]\biggr\}.
\end{eqnarray}
Let us denote
\begin{eqnarray}
    M_{A'}^2 &=& \left[(\Omega_1 + G_1 \Omega_3) \cos \frac{\theta}{2} \cos \phi -(G_2\Omega_3-\Omega_2 ) \cos \frac{\theta}{2} \sin \phi
    + (G_1\Omega_1 - G_2 \Omega_2 -\Omega_3)\sin \frac{\theta}{2}\right]^2\notag\\
    &&+ \left[( G_2\Omega_3-\Omega_2  ) \cos \frac{\theta}{2} \cos \phi + (\Omega_1 + G_1 \Omega_3)\cos \frac{\theta}{2} \sin \phi + ( G_2\Omega_1+ G_1\Omega_2  ) \sin \frac{\theta}{2}\right]^2,
\end{eqnarray}

\begin{eqnarray}
    M_{B'}^2 &=& \left[(\Omega_1 + G_1 \Omega_3) \sin \frac{\theta}{2} \cos \phi - ( G_2\Omega_3-\Omega_2  )\sin \frac{\theta}{2} \sin \phi- (G_1\Omega_1 - G_2 \Omega_2 -\Omega_3) \cos \frac{\theta}{2}\right] ^2\notag\\
    &&+ \left[( G_2\Omega_3-\Omega_2  )\sin \frac{\theta}{2} \cos \phi + (\Omega_1 + G_1 \Omega_3)\sin \frac{\theta}{2} \sin \phi -( G_2\Omega_1+ G_1\Omega_2  ) \cos \frac{\theta}{2}\right]^2,
\end{eqnarray}

\begin{eqnarray}
    \tan \theta_{A'} = \frac{ \left[( G_2\Omega_3-\Omega_2  ) \cos \frac{\theta}{2} \cos \phi + (\Omega_1 + G_1 \Omega_3)\cos \frac{\theta}{2} \sin \phi + ( G_2\Omega_1+ G_1\Omega_2  ) \sin \frac{\theta}{2}\right] }{ \left[(\Omega_1 + G_1 \Omega_3) \cos \frac{\theta}{2} \cos \phi -(G_2\Omega_3-\Omega_2 ) \cos \frac{\theta}{2} \sin \phi + (G_1\Omega_1 - G_2 \Omega_2 -\Omega_3)\sin \frac{\theta}{2}\right]},
\end{eqnarray}
\begin{eqnarray}
    \tan \theta_{B'} =\frac{\left[( G_2\Omega_3-\Omega_2  )\sin \frac{\theta}{2} \cos \phi + (\Omega_1 + G_1 \Omega_3)\sin \frac{\theta}{2} \sin \phi -( G_2\Omega_1+ G_1\Omega_2  ) \cos \frac{\theta}{2}\right]}{\left[(\Omega_1 + G_1 \Omega_3) \sin \frac{\theta}{2} \cos \phi - ( G_2\Omega_3-\Omega_2  )\sin \frac{\theta}{2} \sin \phi- (G_1\Omega_1 - G_2 \Omega_2 -\Omega_3) \cos \frac{\theta}{2}\right]  },
\end{eqnarray}
then we have
\begin{eqnarray}
    A' =\frac{\ell_1 M_{A'}}{(n k_x' + k_x)^2 + (1 - n)^2 (k_y^2 + k_z^2)}{\rm e}^{-{\rm i} \phi + \i \theta_{A'}},
\end{eqnarray}
\begin{eqnarray}
    B' =\frac{\ell_1 M_{B'}}{(n k_x' + k_x)^2 + (1 - n)^2 (k_y^2 + k_z^2)}{\rm e}^{-{\rm i} \phi + \i \theta_{B'}}.
\end{eqnarray}

In this case, the spatial shifts for the reflected wave become
\begin{eqnarray}
    \Delta y_{\rm r}  = \frac{|A'|^2}{|A'|^2 + |B'|^2 } \Delta y_{\rm ra'} + \frac{|B'|^2}{|A'|^2 + |B'|^2 } \Delta y_{\rm rb'},
\end{eqnarray}
\begin{eqnarray}
    \Delta z_{\rm r}  = \frac{|A'|^2}{|A'|^2 + |B'|^2 } \Delta z_{\rm ra'} + \frac{|B'|^2}{|A'|^2 + |B'|^2 } \Delta z_{\rm rb'}.
\end{eqnarray}
For the $ \Delta y_{\rm r}$, for simplicity, we designate
\begin{eqnarray}
    \frac{\partial \Omega_1  }{\partial k_y} = \Omega_1', \;\;\;\;\; \frac{\partial \Omega_2  }{\partial k_y} = \Omega_2', \;\;\;\;\;
    \frac{\partial \Omega_3  }{\partial k_y} = \Omega_3',
\end{eqnarray}
then we have
\begin{eqnarray}
    &&\frac{(n k_x' + k_x)^2 + (1 - n)^2 (k_y^2 + k_z^2)} {|\ell_1|^2} |A'|^2    \Delta y_{\rm ra'} \notag\\
    &=&\left[( G_2\Omega'_3-\Omega'_2  ) \cos \frac{\theta}{2} \cos \phi + (\Omega'_1 + G_1 \Omega'_3)\cos \frac{\theta}{2} \sin \phi + ( G_2\Omega'_1+ G_1\Omega'_2  ) \sin \frac{\theta}{2}\right]\times \notag\\
    &&{ \left[(\Omega_1 + G_1 \Omega_3) \cos \frac{\theta}{2} \cos \phi -(G_2\Omega_3-\Omega_2 ) \cos \frac{\theta}{2} \sin \phi + (G_1\Omega_1 - G_2 \Omega_2 -\Omega_3)\sin \frac{\theta}{2}\right]} \notag\\
    &&-\left[( G_2\Omega_3-\Omega_2  ) \cos \frac{\theta}{2} \cos \phi + (\Omega_1 + G_1 \Omega_3)\cos \frac{\theta}{2} \sin \phi + ( G_2\Omega_1+ G_1\Omega_2  ) \sin \frac{\theta}{2}\right]\times \notag\\
    &&{ \left[(\Omega'_1 + G_1 \Omega'_3) \cos \frac{\theta}{2} \cos \phi -(G_2\Omega'_3-\Omega'_2 ) \cos \frac{\theta}{2} \sin \phi + (G_1\Omega'_1 - G_2 \Omega'_2 -\Omega'_3)\sin \frac{\theta}{2}\right]}
\end{eqnarray}
\begin{eqnarray}
    &&\frac{(n k_x' + k_x)^2 + (1 - n)^2 (k_y^2 + k_z^2)} {|\ell_1|^2} |B'|^2    \Delta y_{\rm rb'} \notag\\
    &=&{\left[( G_2\Omega'_3-\Omega'_2  )\sin \frac{\theta}{2} \cos \phi + (\Omega'_1 + G_1 \Omega'_3)\sin \frac{\theta}{2} \sin \phi -( G_2\Omega'_1+ G_1\Omega'_2  ) \cos \frac{\theta}{2}\right]} \times \notag\\
    &&{\left[(\Omega_1 + G_1 \Omega_3) \sin \frac{\theta}{2} \cos \phi - ( G_2\Omega_3-\Omega_2  )\sin \frac{\theta}{2} \sin \phi- (G_1\Omega_1 - G_2 \Omega_2 -\Omega_3) \cos \frac{\theta}{2}\right]  }\notag\\
    &&-{\left[( G_2\Omega_3-\Omega_2  )\sin \frac{\theta}{2} \cos \phi + (\Omega_1 + G_1 \Omega_3)\sin \frac{\theta}{2} \sin \phi -( G_2\Omega_1+ G_1\Omega_2  ) \cos \frac{\theta}{2}\right]} \times  \notag\\
    &&{\left[(\Omega'_1 + G_1 \Omega'_3) \sin \frac{\theta}{2} \cos \phi - ( G_2\Omega'_3-\Omega'_2  )\sin \frac{\theta}{2} \sin \phi- (G_1\Omega'_1 - G_2 \Omega'_2 -\Omega'_3) \cos \frac{\theta}{2}\right]  }.
\end{eqnarray}
Because
\begin{eqnarray}
    && -( G_2\Omega'_3-\Omega'_2  )(\Omega_1 + G_1 \Omega_3) \cos \frac{\theta}{2} \cos \phi \cos \frac{\theta}{2} \cos \phi + ( G_2\Omega_3-\Omega_2  )(\Omega'_1 + G_1 \Omega'_3)\cos \frac{\theta}{2} \cos \phi \cos \frac{\theta}{2} \cos \phi \notag\\
    &=&[ G_2 (\Omega'_1 \Omega_3 - \Omega_1 \Omega'_3)- (\Omega'_1 \Omega_2 - \Omega_1 \Omega'_2)+G_1 (\Omega'_2 \Omega_3 - \Omega_2 \Omega'_3)]\cos^2 \frac{\theta}{2} \cos^2 \phi,
\end{eqnarray}
\begin{eqnarray}
   && (\Omega'_1 + G_1 \Omega'_3)\cos \frac{\theta}{2} \sin \phi  (G_2\Omega_3-\Omega_2 ) \cos \frac{\theta}{2} \sin \phi -(\Omega_1 + G_1 \Omega_3)\cos \frac{\theta}{2} \sin \phi  (G_2\Omega'_3-\Omega'_2 ) \cos \frac{\theta}{2} \sin \phi \notag\\
   &=& [ G_2 (\Omega'_1 \Omega_3 - \Omega_1 \Omega'_3)- (\Omega'_1 \Omega_2 - \Omega_1 \Omega'_2)+G_1 (\Omega'_2 \Omega_3 - \Omega_2 \Omega'_3)]\cos^2 \frac{\theta}{2} \sin^2  \phi,
\end{eqnarray}
\begin{eqnarray}
    && - ( G_2\Omega'_1+ G_1\Omega'_2  ) \sin \frac{\theta}{2}(G_1\Omega_1 - G_2 \Omega_2 -\Omega_3) \sin \frac{\theta}{2}
     +( G_2\Omega_1+ G_1\Omega_2  ) \sin \frac{\theta}{2}(G_1\Omega'_1 + G_2 \Omega'_2 -\Omega'_3) \sin \frac{\theta}{2} \notag\\
     &=& [ G_2 (\Omega'_1 \Omega_3 - \Omega_1 \Omega'_3)+(G_1^2 +G_2^2) (\Omega'_1 \Omega_2 - \Omega_1 \Omega'_2)+G_1 (\Omega'_2 \Omega_3 - \Omega_2 \Omega'_3)] \sin^2 \frac{\theta}{2},
\end{eqnarray}
\begin{eqnarray}
   && ( G_2\Omega'_3-\Omega'_2  )\cos \frac{\theta}{2} \cos \phi (G_2\Omega_3-\Omega_2 ) \cos \frac{\theta}{2} \sin \phi -(\Omega'_1 + G_1 \Omega'_3)\cos \frac{\theta}{2} \sin \phi (\Omega_1 + G_1 \Omega_3) \cos \frac{\theta}{2} \cos \phi \notag\\
    &&-  ( G_2\Omega_3-\Omega_2  )\cos \frac{\theta}{2} \cos \phi (G_2\Omega'_3-\Omega'_2 ) \cos \frac{\theta}{2} \sin \phi +(\Omega_1 + G_1 \Omega_3)\cos \frac{\theta}{2} \sin \phi (\Omega'_1 + G_1 \Omega'_3) \cos \frac{\theta}{2} \cos \phi\notag\\
    &=&0,
\end{eqnarray}
\begin{eqnarray}
    &&( G_2\Omega'_3-\Omega'_2  ) \cos \frac{\theta}{2} \cos \phi (G_1\Omega_1 - G_2 \Omega_2 -\Omega_3) \sin \frac{\theta}{2} +( G_2\Omega'_1+ G_1\Omega'_2  ) \sin \frac{\theta}{2}(\Omega_1 + G_1 \Omega_3) \cos \frac{\theta}{2} \cos \phi\notag\\
&&-( G_2\Omega_3-\Omega_2  ) \cos \frac{\theta}{2} \cos \phi (G_1\Omega'_1 - G_2 \Omega'_2 -\Omega'_3) \sin \frac{\theta}{2} -( G_2\Omega_1+ G_1\Omega_2  ) \sin \frac{\theta}{2}(\Omega'_1 + G_1 \Omega'_3) \cos \frac{\theta}{2} \cos \phi\notag\\
&=&[(G_2^2+G_1^2+1)(\Omega'_2 \Omega_3 - \Omega_2 \Omega'_3) ]\cos \frac{\theta}{2}  \sin \frac{\theta}{2}\cos \phi,
\end{eqnarray}
\begin{eqnarray}
    &&-(\Omega'_1 + G_1 \Omega'_3)\cos \frac{\theta}{2} \sin \phi(G_1\Omega_1 - G_2 \Omega_2 -\Omega_3) \sin \frac{\theta}{2}+( G_2\Omega'_1+ G_1\Omega'_2  ) \sin \frac{\theta}{2}(G_2\Omega_3-\Omega_2 )  \cos \frac{\theta}{2} \sin \phi\notag\\
    &&+(\Omega_1 + G_1 \Omega_3)\cos \frac{\theta}{2} \sin \phi(G_1\Omega'_1 - G_2 \Omega'_2 -\Omega'_3) \sin \frac{\theta}{2}-( G_2\Omega_1+ G_1\Omega_2  ) \sin \frac{\theta}{2}(G_2\Omega'_3-\Omega'_2 )  \cos \frac{\theta}{2} \sin \phi\notag\\
    &=&(1+G^2_1 +G^2_2)(\Omega'_1 \Omega_3 -\Omega_1 \Omega'_3)\cos \frac{\theta}{2} \sin \frac{\theta}{2} \sin \phi,
\end{eqnarray}
therefore one attains
\begin{eqnarray}
    &&\frac{(n k_x' + k_x)^2 + (1 - n)^2 (k_y^2 + k_z^2)} {|\ell_1|^2} |A'|^2    \Delta y_{\rm ra'}  \notag\\
    &=&[ G_2 (\Omega'_1 \Omega_3 - \Omega_1 \Omega'_3)+G_1 (\Omega'_2 \Omega_3 - \Omega_2 \Omega'_3)]+[(G_2^2+G_1^2+1)(\Omega'_2 \Omega_3 - \Omega_2 \Omega'_3) ]\sin \frac{\theta}{2}  \sin \frac{\theta}{2}\cos \phi\notag\\
   &&-(1+G^2_1 +G^2_2)(\Omega'_1 \Omega_3 -\Omega_1 \Omega'_3)\cos \frac{\theta}{2} \sin \frac{\theta}{2} \sin \phi + (\Omega'_1 \Omega_2 - \Omega_1 \Omega'_2) \left[(G_1^2+G_2^2)\sin^2 \frac{\theta}{2}-\cos^2 \frac{\theta}{2}\right].
\end{eqnarray}

Similarly, because
\begin{eqnarray}
    && -( G_2\Omega'_3-\Omega'_2  )(\Omega_1 + G_1 \Omega_3) \sin \frac{\theta}{2} \cos \phi \sin \frac{\theta}{2} \cos \phi + ( G_2\Omega_3-\Omega_2  )(\Omega'_1 + G_1 \Omega'_3)\sin \frac{\theta}{2} \cos \phi \sin \frac{\theta}{2} \cos \phi \notag\\
    &=&[ G_2 (\Omega'_1 \Omega_3 - \Omega_1 \Omega'_3)- (\Omega'_1 \Omega_2 - \Omega_1 \Omega'_2)+G_1 (\Omega'_2 \Omega_3 - \Omega_2 \Omega'_3)]\sin^2 \frac{\theta}{2} \cos^2 \phi,
\end{eqnarray}
\begin{eqnarray}
   && (\Omega'_1 + G_1 \Omega'_3)\sin \frac{\theta}{2} \sin \phi  (G_2\Omega_3-\Omega_2 ) \sin \frac{\theta}{2} \sin \phi -(\Omega_1 + G_1 \Omega_3)\sin \frac{\theta}{2} \sin \phi  (G_2\Omega'_3-\Omega'_2 ) \sin \frac{\theta}{2} \sin \phi \notag\\
   &=& [ G_2 (\Omega'_1 \Omega_3 - \Omega_1 \Omega'_3)- (\Omega'_1 \Omega_2 - \Omega_1 \Omega'_2)+G_1 (\Omega'_2 \Omega_3 - \Omega_2 \Omega'_3)]\sin^2 \frac{\theta}{2} \sin^2  \phi,
\end{eqnarray}
\begin{eqnarray}
    && - ( G_2\Omega'_1+ G_1\Omega'_2  ) \cos \frac{\theta}{2}(G_1\Omega_1 - G_2 \Omega_2 -\Omega_3) \cos \frac{\theta}{2}
     +( G_2\Omega_1+ G_1\Omega_2  ) \cos \frac{\theta}{2}(G_1\Omega'_1 + G_2 \Omega'_2 -\Omega'_3) \cos \frac{\theta}{2} \notag\\
     &=& [ G_2 (\Omega'_1 \Omega_3 - \Omega_1 \Omega'_3)+(G_1^2 +G_2^2) (\Omega'_1 \Omega_2 - \Omega_1 \Omega'_2)+G_1 (\Omega'_2 \Omega_3 - \Omega_2 \Omega'_3)] \cos^2 \frac{\theta}{2},
\end{eqnarray}
\begin{eqnarray}
   && ( G_2\Omega'_3-\Omega'_2  )\sin \frac{\theta}{2} \cos \phi (G_2\Omega_3-\Omega_2 ) \sin \frac{\theta}{2} \sin \phi -(\Omega'_1 + G_1 \Omega'_3)\sin \frac{\theta}{2} \sin \phi (\Omega_1 + G_1 \Omega_3) \sin \frac{\theta}{2} \cos \phi \notag\\
    &&-  ( G_2\Omega_3-\Omega_2  )\sin \frac{\theta}{2} \cos \phi (G_2\Omega'_3-\Omega'_2 ) \sin \frac{\theta}{2} \sin \phi +(\Omega_1 + G_1 \Omega_3)\sin \frac{\theta}{2} \sin \phi (\Omega'_1 + G_1 \Omega'_3) \sin \frac{\theta}{2} \cos \phi\notag\\
    &=& 0,
\end{eqnarray}
\begin{eqnarray}
    &&-( G_2\Omega'_3-\Omega'_2  ) \sin \frac{\theta}{2} \cos \phi (G_1\Omega_1 - G_2 \Omega_2 -\Omega_3) \cos \frac{\theta}{2} -( G_2\Omega'_1+ G_1\Omega'_2  ) \sin \frac{\theta}{2}(\Omega_1 + G_1 \Omega_3) \cos \frac{\theta}{2} \cos \phi\notag\\
&&+( G_2\Omega_3-\Omega_2  ) \sin \frac{\theta}{2} \cos \phi (G_1\Omega'_1 - G_2 \Omega'_2 -\Omega'_3) \cos \frac{\theta}{2} +( G_2\Omega_1+ G_1\Omega_2  ) \sin \frac{\theta}{2}(\Omega'_1 + G_1 \Omega'_3) \cos \frac{\theta}{2} \cos \phi\notag\\
&=& -[(G_2^2+G_1^2+1)(\Omega'_2 \Omega_3 - \Omega_2 \Omega'_3) ]\sin \frac{\theta}{2}  \cos \frac{\theta}{2}\cos \phi,
\end{eqnarray}
\begin{eqnarray}
    &&(\Omega'_1 + G_1 \Omega'_3)\sin \frac{\theta}{2} \sin \phi(G_1\Omega_1 - G_2 \Omega_2 -\Omega_3) \cos \frac{\theta}{2}-( G_2\Omega'_1+ G_1\Omega'_2  ) \sin \frac{\theta}{2}(G_2\Omega_3-\Omega_2 )  \cos \frac{\theta}{2} \sin \phi\notag\\
    &&-(\Omega_1 + G_1 \Omega_3)\sin \frac{\theta}{2} \sin \phi(G_1\Omega'_1 - G_2 \Omega'_2 -\Omega'_3) \cos \frac{\theta}{2}+( G_2\Omega_1+ G_1\Omega_2  ) \sin \frac{\theta}{2}(G_2\Omega'_3-\Omega'_2 )  \cos \frac{\theta}{2} \sin \phi\notag\\
    &=&-(1+G^2_1 +G^2_2)(\Omega'_1 \Omega_3 -\Omega_1 \Omega'_3)\sin \frac{\theta}{2} \cos \frac{\theta}{2} \sin \phi,
\end{eqnarray}
so one  attains
\begin{eqnarray}
    &&\frac{(n k_x' + k_x)^2 + (1 - n)^2 (k_y^2 + k_z^2)} {|\ell_1|^2} |B'|^2    \Delta y_{\rm rb'}  \notag\\
    &=&[ G_2 (\Omega'_1 \Omega_3 - \Omega_1 \Omega'_3)+G_1 (\Omega'_2 \Omega_3 - \Omega_2 \Omega'_3)]-[(G_2^2+G_1^2+1)(\Omega'_2 \Omega_3 - \Omega_2 \Omega'_3) ]\sin \frac{\theta}{2}  \sin \frac{\theta}{2}\cos \phi\notag\\
    &&+(1+G^2_1 +G^2_2)(\Omega'_1 \Omega_3 -\Omega_1 \Omega'_3)\cos \frac{\theta}{2} \sin \frac{\theta}{2} \sin \phi+(\Omega'_1 \Omega_2 - \Omega_1 \Omega'_2) [(G_1^2+G_2^2)\cos^2 \frac{\theta}{2}-\sin^2 \frac{\theta}{2}] .
\end{eqnarray}
Based on the above results, we finally have
\begin{eqnarray}
    &&\frac{(n k_x' + k_x)^2 + (1 - n)^2 (k_y^2 + k_z^2)} {|\ell_1|^2} [|A'|^2    \Delta y_{ra'}+|B'|^2    \Delta y_{\rm rb'}]\notag\\
&=&2[ G_2 (\Omega'_1 \Omega_3 - \Omega_1 \Omega'_3)+G_1 (\Omega'_2 \Omega_3 - \Omega_2 \Omega'_3)]+ [(G_1^2+G_2^2)-1] (\Omega'_1 \Omega_2 - \Omega_1 \Omega'_2).
\end{eqnarray}

Due to
\begin{eqnarray}
    |A'|^2 + |B'|^2 =  |A|^2 + |B|^2 = \frac{|\ell_1|^2(1+ G_1^2 + G_2^2 ) (\Omega_1^2 +\Omega_2^2 +\Omega_3^2  )}{(n k_x' + k_x)^2 + (1 - n)^2 (k_y^2 + k_z^2)}  ,
\end{eqnarray}
\begin{eqnarray}
    \Delta y_{\rm r} &= &\frac{2[ G_2 (\Omega'_1 \Omega_3 - \Omega_1 \Omega'_3)+G_1 (\Omega'_2 \Omega_3 - \Omega_2 \Omega'_3)]+ [(G_1^2+G_2^2)-1] (\Omega'_1 \Omega_2 - \Omega_1 \Omega'_2)}{(1+ G_1^2 + G_2^2 ) (\Omega_1^2 +\Omega_2^2 +\Omega_3^2  )}\notag\\
    &=&\frac{2[ G_2 (\frac{\partial\Omega_1  }{\partial k_y}\Omega_3 - \Omega_1 \frac{\partial\Omega_3 }{\partial k_y})+G_1 (\frac{\partial\Omega_2  }{\partial k_y} \Omega_3 - \Omega_2 \frac{\partial\Omega_3 }{\partial k_y})]+(G_1^2+G_2^2-1)(\frac{\partial\Omega_1  }{\partial k_y} \Omega_2 - \Omega_1 \frac{\partial\Omega_2  }{\partial k_y})}{(1+ G_1^2 + G_2^2 ) (\Omega_1^2 +\Omega_2^2 +\Omega_3^2  )}
\end{eqnarray}
Thus we prove that the result of $\Delta y_{\rm r}$ is unique. For $ \Delta z_{\rm r}$, similarly it is also unique and
\begin{eqnarray}
    \Delta z_r
    &=&\frac{2[ G_2 (\frac{\partial\Omega_1  }{\partial k_z}\Omega_3 - \Omega_1 \frac{\partial\Omega_3 }{\partial k_z})+G_1 (\frac{\partial\Omega_2  }{\partial k_z} \Omega_3 - \Omega_2 \frac{\partial\Omega_3 }{\partial k_z})]+(G_1^2+G_2^2-1) (\frac{\partial\Omega_1  }{\partial k_z} \Omega_2 - \Omega_1 \frac{\partial\Omega_2  }{\partial k_z})}{(1+ G_1^2 + G_2^2 ) (\Omega_1^2 +\Omega_2^2 +\Omega_3^2  )}.
\end{eqnarray}
This ends the proof for the reflected wave.
\end{proof}

For the transmitted wave, one  has
\begin{subequations}
\begin{eqnarray}
    C&=&\dfrac{\sqrt{(\mathcal{E}-V_0)(\mathcal{E} -V_0+mc^2)}}{\sqrt{\mathcal{E}(\mathcal{E} +mc^2)}}\frac{2  n \ell_1 k_x(k_x+{n}k_x') +2n k_xk_z \ell_2(n-1)-2 \i \, n k_x k_y \ell_1 (n-1)}
{k_x^2+(k_y^2+k_z^2)(n-1)^2+2 k_x k_x' n +k_x'^2 n^2},
 \end{eqnarray}
 \begin{eqnarray}
    D&=&\dfrac{\sqrt{(\mathcal{E}-V_0)(\mathcal{E} -V_0+mc^2)}}{\sqrt{\mathcal{E}(\mathcal{E} +mc^2)}}\frac{2  n \ell_2k_x (k_x +{n}k_x')+  2 \i\, n k_x k_y \ell_2 (n-1)-2nk_x  k_z \ell_1(n-1)}
{k_x^2+(k_y^2+k_z^2)(n-1)^2+2 k_x k_x' n +k_x'^2 n^2}.
 \end{eqnarray}
\end{subequations}
We designate
\begin{eqnarray}
   \Omega_{k1}= k_x +{n}k_x', \;\;\;\;\;  \Omega_{k2} =(n-1)k_y , \;\;\;\;\;  \Omega_{k3}=(n-1)k_z,
\end{eqnarray}
then
\begin{subequations}
    \begin{eqnarray}
        C= \dfrac{\sqrt{(\mathcal{E}-V_0)(\mathcal{E} -V_0+mc^2)}}{\sqrt{\mathcal{E}(\mathcal{E} +mc^2)}}\frac{2  n \ell_1 k_x [\Omega_{k1} - \i \Omega_{k2} + (G_1 +\i G_2 ) \Omega_{k3}]}
        {k_x^2+(k_y^2+k_z^2)(n-1)^2+2 k_x k_x' n +k_x'^2 n^2} = \frac{\ell_1}{|\ell_1|} |C| \e^{-\i \theta_{\rm tc}},
    \end{eqnarray}
    \begin{eqnarray}
        D= \dfrac{\sqrt{(\mathcal{E}-V_0)(\mathcal{E} -V_0+mc^2)}}{\sqrt{\mathcal{E}(\mathcal{E} +mc^2)}}\frac{2  n \ell_1 k_x [-\Omega_{k3}  + (G_1 +\i G_2 ) (\Omega_{k1}+\i \Omega_{k2})]}
        {k_x^2+(k_y^2+k_z^2)(n-1)^2+2 k_x k_x' n +k_x'^2 n^2}=\frac{\ell_1}{|\ell_1|} |D| \e^{-\i \theta_{\rm td}},
    \end{eqnarray}
\end{subequations}
where
\begin{eqnarray}
    \tan \theta_{\rm tc} = - \frac{G_2 \Omega_{k3}- \Omega_{k2}}{\Omega_{k1}}, \;\;\;\;\;\;    \tan \theta_{\rm rb} = - \frac{G_2 \Omega_{k1}}{ - G_2 \Omega_{k2} -\Omega_{k3}}.
\end{eqnarray}

The spatial shift for the transmitted wave is
\begin{eqnarray}
    \Delta \vec{l}_{\rm t} = \Delta y_{\rm t} \, \vec{e}_y +\Delta z_{\rm t}\,  \vec{e}_z,
\end{eqnarray}
where
\begin{eqnarray}
    \Delta y_{\rm t}  = \frac{|C|^2}{|C|^2 + |D|^2 } \Delta y_{\rm tc} + \frac{|D|^2}{|C|^2 + |D|^2 } \Delta y_{\rm td}
\end{eqnarray}
\begin{eqnarray}
    \Delta z_{\rm t}  = \frac{|C|^2}{|C|^2 + |D|^2 } \Delta z_{\rm tc} + \frac{|D|^2}{|C|^2 + |D|^2 } \Delta z_{\rm td} .
\end{eqnarray}
with
\begin{eqnarray}
    \Delta y_{\rm tc} = \frac{\partial \theta_{\rm tc}}{\partial k_y}, \;\;\;\;\;\;  \Delta y_{\rm td} = \frac{\partial \theta_{\rm td}}{\partial k_y},
\end{eqnarray}
\begin{eqnarray}
    \Delta z_{\rm tc} = \frac{\partial \theta_{\rm ra}}{\partial k_z}, \;\;\;\;\;\;  \Delta z_{\rm td} = \frac{\partial \theta_{\rm rb}}{\partial k_z}.
\end{eqnarray}

Similarly, the expectant value of the spatial shift for the transmitted wave is also unique, which is
\begin{eqnarray}
    \Delta \vec{l}_{\rm t} = \Delta y_{\rm t} \, \vec{e}_y +\Delta z_{\rm t}\,  \vec{e}_z,
\end{eqnarray}
\begin{eqnarray}
    \Delta y_{\rm t}
    &=&\frac{2\left[ G_2 \left(\frac{\partial\Omega_{k1}  }{\partial k_y}\Omega_{k3} - \Omega_{k1} \frac{\partial\Omega_{k3} }{\partial k_y}\right)+G_1 \left(\frac{\partial \Omega_{k2}  }{\partial k_y} \Omega_{k3} - \Omega_{k2} \frac{\partial \Omega_{k3} }{\partial k_y}\right)\right]+(G_1^2+G_2^2 -1) \left(\frac{\partial\Omega_{k1}  }{\partial k_y} \Omega_{k2} - \Omega_{k1} \frac{\partial \Omega_{k2}  }{\partial k_y}\right)}{(1+ G_1^2 + G_2^2 ) ({\Omega_{k1}}^2 +\Omega_{k2}^2 +\Omega_{k3}^2  )},
\end{eqnarray}
\begin{eqnarray}
    \Delta z_{\rm t}
    &=&\frac{2 \left[ G_2 \left(\frac{\partial\Omega_{k1}  }{\partial k_z}\Omega_{k3} - \Omega_{k1} \frac{\partial\Omega_{k3} }{\partial k_z}\right)+G_1 \left(\frac{\partial \Omega_{k2}  }{\partial k_z} \Omega_{k3} - \Omega_{k2} \frac{\partial \Omega_{k3} }{\partial k_z}\right)\right]+(G_1^2+G_2^2 -1) \left(\frac{\partial\Omega_{k1}  }{\partial k_z} \Omega_{k2} - \Omega_{k1} \frac{\partial \Omega_{k2}  }{\partial k_z} \right)}{(1+ G_1^2 + G_2^2 ) ({\Omega_{k1}}^2 +\Omega_{k2}^2 +\Omega_{k3}^2  )}.
\end{eqnarray}
$\blacksquare$
\end{remark}

\begin{remark}
\textcolor{blue}{The Special Spin Direction for $k_y\neq 0$}. In Sec. \ref{s11}, we have considered the incident plane of the incident wave is just the $xz$-plane, i.e.,
\begin{eqnarray}
\vec{p}=\hbar \vec{k}= \hbar(k_x, 0, k_z),
\end{eqnarray}
which means the $y$-component of the momentum is zero. In this case, there is a special spin direction, i.e., the $y$-axis, for which the incident wave, the reflected wave, and the transmitted wave share the same spin direction. The spin direction of the incident wave is characterized by
$\begin{pmatrix}
    \ell_1\\
    \ell_2
\end{pmatrix}$. Similarly,  The spin direction of the reflected wave and the transmitted wave are characterized by
$\begin{pmatrix}
    A\\
    B
\end{pmatrix}$ and
$\begin{pmatrix}
    C\\
    D
\end{pmatrix}$, respectively. If their spin directions are the same, then the condition is
\begin{eqnarray}\label{eq:condition-1}
    \frac{B}{A}=\frac{D}{C}= {\frac{\ell_2}{\ell_1}}.
\end{eqnarray}
For the case of $k_y=0$, in Sec. \ref{s11} we have known that the condition is
\begin{eqnarray}
   \frac{B}{A}=\frac{D}{C}= {\frac{\ell_2}{\ell_1}}=\pm {\rm i},
\end{eqnarray}
which corresponds to the $y$-axis (i.e., $\ell_2/\ell_1={\rm i}$ corresponds to the positive $y$ direction, and $\ell_2/\ell_1=-{\rm i}$ corresponds to the negative $y$ direction).

Now, it gives rise to a natural question: What is the special spin direction if $k_y\neq 0$? To reach this purpose, we need to solve the condition (\ref{eq:condition-1}). The following is the answer.

\emph{Answer.---}We designate the momentum of the incident wave as
\begin{eqnarray}
\vec{p} = \hbar \vec{k}=\hbar k \left( \cos \phi_k' , \sin \phi_k' \cos \phi_k'', \sin \phi_k' \sin \phi_k'' \right),
\end{eqnarray}
then the answer is as follows: the special spin direction is given by
\begin{eqnarray}\label{eq:ss-1}
\vec{\tau} = (\sin \theta_k \cos \phi_k , \sin \theta_k \sin \phi_k , \cos \theta_k)= ( 0,\sin \phi_k'', -\cos \phi_k'').
\end{eqnarray}

We now come to verify this point. Eq. (\ref{eq:ss-1}) implies that
\begin{eqnarray}
\phi_k =\frac{\pi}{2}, \;\;\;\; \theta_k = \pi -\phi_k'',
\end{eqnarray}
which corresponds to the following coefficients
\begin{eqnarray}\label{eq:ll-1}
    \ell_1 = \sin \frac{\phi_k''}{2},\;\;\;\;\;\; \ell_2 = {\rm i} \cos \frac{\phi_k''}{2},
\end{eqnarray}
and
\begin{eqnarray}
\frac{\ell_2}{\ell_1}=\frac{{\rm i} \cos \frac{\phi_k''}{2}}{\sin \frac{\phi_k''}{2}}={\rm i} \cot \frac{\phi_k''}{2}.
\end{eqnarray}
When $\phi_k''=\pi/2$ and $3\pi/2$, $k_y$ becomes zero, and accordingly, $\ell_2/\ell_1$ reduces to ${\rm i}$ and $-{\rm i}$.

Based on Eq. (\ref{eq:ll-1}), we can have
    \begin{eqnarray}
 A&=&\frac{\ell_1\left[k_x^2-(k_y^2+k_z^2)(n-1)^2-k_x'^2 n^2\right] +2 k_x (-i k_y \ell_1+k_z \ell_2) (n-1)}
 {k_x^2+(k_y^2+k_z^2)(n-1)^2+2 k_x k_x' n +k_x'^2 n^2} \notag\\
 &=&\ell_1\frac{\left[k_x^2-(k_y^2+k_z^2)(n-1)^2-k_x'^2 n^2\right] +2 k_x \left(-i k \sin \phi_k' \cos \phi_k''+\i k \sin \phi_k' \sin \phi_k'' \cot \frac{\phi_k''}{2} \right) (n-1)}
 {k_x^2+(k_y^2+k_z^2)(n-1)^2+2 k_x k_x' n +k_x'^2 n^2} \notag \\
 &=&\ell_1\frac{\left[k_x^2-(k_y^2+k_z^2)(n-1)^2-k_x'^2 n^2\right] +2 \i  k_x k \sin \phi_k' (n-1)}
 {k_x^2+(k_y^2+k_z^2)(n-1)^2+2 k_x k_x' n +k_x'^2 n^2},
     \end{eqnarray}

     \begin{eqnarray}
        B&=& \frac{\ell_2\left[k_x^2-(k_y^2+k_z^2)(n-1)^2-k_x'^2 n^2\right] -2 k_x (k_z \ell_1-i k_y \ell_2) (n-1)}
 {k_x^2+(k_y^2+k_z^2)(n-1)^2+2 k_x k_x' n +k_x'^2 n^2} \notag \\
 &=& \ell_2\frac{\left[k_x^2-(k_y^2+k_z^2)(n-1)^2-k_x'^2 n^2\right] -2 k_x \left(- \i k \sin \phi_k' \sin \phi_k''  \tan \frac{\phi_k''}{2}-\i k \sin \phi_k' \cos \phi_k'' \right) (n-1)}
 {k_x^2+(k_y^2+k_z^2)(n-1)^2+2 k_x k_x' n +k_x'^2 n^2} \notag \\
 &=& \ell_2\frac{\left[k_x^2-(k_y^2+k_z^2)(n-1)^2-k_x'^2 n^2\right] +2 \i  k_x k \sin \phi_k' (n-1)}
 {k_x^2+(k_y^2+k_z^2)(n-1)^2+2 k_x k_x' n +k_x'^2 n^2}=A\, \dfrac{\ell_2}{\ell_1},
     \end{eqnarray}

     \begin{eqnarray}
        C&=&\dfrac{\sqrt{(\mathcal{E}-V_0)(\mathcal{E} -V_0+mc^2)}}{\sqrt{\mathcal{E}(\mathcal{E} +mc^2)}}\frac{2  n \ell_1 k_x(k_x+{n}k_x') +2n k_x k_z \ell_2(n-1)-2 \i \, n k_x k_y \ell_1 (n-1)}
 {k_x^2+(k_y^2+k_z^2)(n-1)^2+2 k_x k_x' n +k_x'^2 n^2} \notag\\
 &=&  \dfrac{\sqrt{(\mathcal{E}-V_0)(\mathcal{E} -V_0+mc^2)}}{\sqrt{\mathcal{E}(\mathcal{E} +mc^2)}}\frac{2  n \ell_1 k_x(k_x+{n}k_x') +2n k_x (n-1) [k_z \ell_2- \i \,  k_y \ell_1] }
 {k_x^2+(k_y^2+k_z^2)(n-1)^2+2 k_x k_x' n +k_x'^2 n^2} \notag\\
 &=& \ell_1\dfrac{\sqrt{(\mathcal{E}-V_0)(\mathcal{E} -V_0+mc^2)}}{\sqrt{\mathcal{E}(\mathcal{E} +mc^2)}}\frac{2  n k_x(k_x+{n}k_x') +2 \i n k_x (n-1) k \sin \phi_k'  }
 {k_x^2+(k_y^2+k_z^2)(n-1)^2+2 k_x k_x' n +k_x'^2 n^2},
     \end{eqnarray}

     \begin{eqnarray}
        D&=&\dfrac{\sqrt{(\mathcal{E}-V_0)(\mathcal{E} -V_0+mc^2)}}{\sqrt{\mathcal{E}(\mathcal{E} +mc^2)}}\frac{2  n \ell_2k_x (k_x +{n}k_x')+  2 \i\, n k_x k_y \ell_2 (n-1)-2nk_x  k_z \ell_1(n-1)}
 {k_x^2+(k_y^2+k_z^2)(n-1)^2+2 k_x k_x' n +k_x'^2 n^2}\notag\\
 &=&\dfrac{\sqrt{(\mathcal{E}-V_0)(\mathcal{E} -V_0+mc^2)}}{\sqrt{\mathcal{E}(\mathcal{E} +mc^2)}}\frac{2  n \ell_2k_x (k_x +{n}k_x')-  2  n(n-1)  k_x [-\i k_y \ell_2 + k_z \ell_1]}
 {k_x^2+(k_y^2+k_z^2)(n-1)^2+2 k_x k_x' n +k_x'^2 n^2}\notag\\
 &=&\ell_2\dfrac{\sqrt{(\mathcal{E}-V_0)(\mathcal{E} -V_0+mc^2)}}{\sqrt{\mathcal{E}(\mathcal{E} +mc^2)}}\frac{2  n k_x(k_x+{n}k_x') +2 \i n k_x (n-1) k \sin \phi_k'  }
 {k_x^2+(k_y^2+k_z^2)(n-1)^2+2 k_x k_x' n +k_x'^2 n^2}=C\, \dfrac{\ell_2}{\ell_1}.
     \end{eqnarray}
Therefore, one can easily verify that the condition (\ref{eq:condition-1}) is valid.
$\blacksquare$
\end{remark}

\begin{remark}
\textcolor{blue}{Calculating the Spatial Shifts in the Special Basis}.
From above, we have known that the special spin direction is
\begin{eqnarray}
\vec{\tau} = ( 0,\sin \phi_k'', -\cos \phi_k''),
\end{eqnarray}
if the momentum of the incident wave is denoted as
$\vec{p} = \hbar \vec{k}=\hbar k \left( \cos \phi_k' , \sin \phi_k' \cos \phi_k'', \sin \phi_k' \sin \phi_k'' \right)$.
For convenience, let us denote the special spin direction as the $\hat{w}$-axis, i.e.,
\begin{eqnarray}
\hat{w}=( 0,\sin \phi_k'', -\cos \phi_k''),
\end{eqnarray}
which corresponds to the following state
\begin{eqnarray}
|\uparrow\rangle_{\rm w}
=  \begin{pmatrix}
    \sin \frac{\phi_k''}{2}\\
    {\rm i} \cos \frac{\phi_k''}{2}
\end{pmatrix}.
\end{eqnarray}
The state orthogonal to $|\uparrow\rangle_{\rm w}$ can be worked out as
\begin{eqnarray}
|\downarrow\rangle_{\rm w} =
\begin{pmatrix}
    \cos \frac{\phi_k''}{2}\\
    -{\rm i} \sin \frac{\phi_k''}{2}
\end{pmatrix},
\end{eqnarray}
which corresponds to the negative $\hat{w}$ direction. Obviously, $\{|\uparrow\rangle_{\rm w}, |\downarrow\rangle_{\rm w}\}$ forms a special basis. In the following, we come to calculate the spatial shift in this $\hat{w}$-basis.

Very similar to Eq. (\ref{eq:r-1a}), in the $z$-basis, the reflected wave $\ket{\Psi_{\rm r}}$ with $k_y\neq 0$ is given by
\begin{eqnarray}\label{eq:r-2a}
    &&    \ket{\Psi_{\rm r}} =\left\{A\, \ket{{\tilde{s}}_z = +\frac{\hbar}{2}}' + B\,  \ket{{\tilde{s}}_z = -\frac{\hbar}{2}}'\right\} {\rm e}^{\frac{\i}{\hbar} [{\hbar}(-k_x x+k_y y + k_z z) - \mathcal{E} t]},
\end{eqnarray}
where
\begin{eqnarray}
    &&    \ket{{\tilde{s}}_z = +\frac{\hbar}{2}}'= \frac{1}{\sqrt{2\mathcal{E}(\mathcal{E} +mc^2)}}  \begin{pmatrix}
        (\mathcal{E}+mc^2) |\uparrow\rangle_z\\
        c\hbar (\vec{\sigma} \cdot \vec{k}_{\rm r}) |\uparrow\rangle_z
    \end{pmatrix}, \nonumber\\
    &&  \ket{{\tilde{s}}_z = -\frac{\hbar}{2}}'= \frac{1}{\sqrt{2\mathcal{E}(\mathcal{E} +mc^2)}}  \begin{pmatrix}
        (\mathcal{E}+mc^2) |\downarrow\rangle_z\\
        c\hbar (\vec{\sigma} \cdot \vec{k}_{\rm r}) |\downarrow\rangle_z
    \end{pmatrix}.
\end{eqnarray}

Certainly, the reflected wave $\ket{\Psi_{\rm r}}$ can be also expanded in the $\hat{w}$-basis. We can have
\begin{eqnarray}\label{eq:r-2b}
    &&    \ket{\Psi_{\rm r}} =\left\{A'\, \ket{{\tilde{s}}_w = +\frac{\hbar}{2}} + B'\,  \ket{{\tilde{s}}_w = -\frac{\hbar}{2}}\right\} {\rm e}^{\frac{\i}{\hbar} [{\hbar}(-k_x x+k_y y + k_z z) - \mathcal{E} t]},
\end{eqnarray}
where
\begin{eqnarray}
    &&    \ket{{\tilde{s}}_{\rm w} = +\frac{\hbar}{2}}= \frac{1}{\sqrt{2\mathcal{E}(\mathcal{E} +mc^2)}}  \begin{pmatrix}
        (\mathcal{E}+mc^2) |\uparrow\rangle_{\rm w}\\
        c\hbar (\vec{\sigma} \cdot \vec{k}_{\rm r}) |\uparrow\rangle_{\rm w}
    \end{pmatrix}, \nonumber\\
    &&  \ket{{\tilde{s}}_{\rm w} = -\frac{\hbar}{2}}= \frac{1}{\sqrt{2\mathcal{E}(\mathcal{E} +mc^2)}}  \begin{pmatrix}
        (\mathcal{E}+mc^2) |\downarrow\rangle_{\rm w}\\
        c\hbar (\vec{\sigma} \cdot \vec{k}_{\rm r}) |\downarrow\rangle_{\rm w}
    \end{pmatrix}.
\end{eqnarray}
Then we have
\begin{eqnarray}
    A |\uparrow\rangle_z + B |\downarrow\rangle_z= A' |\uparrow\rangle_{\rm w}+B' |\downarrow\rangle_{\rm w},
\end{eqnarray}
i.e.,
\begin{eqnarray}
    \begin{pmatrix}
        A\\
        B
    \end{pmatrix} = A' \begin{pmatrix}
    \sin \frac{\phi_k''}{2}\\
    {\rm i} \cos \frac{\phi_k''}{2}
\end{pmatrix}+ B' \begin{pmatrix}
    \cos \frac{\phi_k''}{2}\\
    -{\rm i} \sin \frac{\phi_k''}{2}
\end{pmatrix},
\end{eqnarray}
and
\begin{eqnarray}
 &&   A' = A \sin \frac{\phi_k''}{2} -{\rm i} B   \cos \frac{\phi_k''}{2}, \nonumber\\
 &&   B' = A \cos \frac{\phi_k''}{2} +{\rm i} B   \sin \frac{\phi_k''}{2}.
\end{eqnarray}
Due to
\begin{eqnarray}
& &A=\frac{\ell_1\left[k_x^2-(k_y^2+k_z^2)(n-1)^2-k_x'^2 n^2\right] +2 k_x (- {\rm i} k_y \ell_1+k_z \ell_2) (n-1)}
 {k_x^2+(k_y^2+k_z^2)(n-1)^2+2 k_x k_x' n +k_x'^2 n^2}, \nonumber\\
& &B= \frac{\ell_2\left[k_x^2-(k_y^2+k_z^2)(n-1)^2-k_x'^2 n^2\right] -2 k_x (k_z \ell_1- {\rm i} k_y \ell_2) (n-1)}
 {k_x^2+(k_y^2+k_z^2)(n-1)^2+2 k_x k_x' n +k_x'^2 n^2},
\end{eqnarray}
or they are written as
    \begin{eqnarray}
     &&   A =  \dfrac{ \ell_1 (\Omega_1 - \i \Omega_2) + \ell_2 \Omega_3}{(n k_x' + k_x)^2 + (1 - n)^2 (k_y^2 + k_z^2)}, \nonumber\\
     &&   B = \dfrac{- \ell_1 \Omega_3 + \ell_2 (\Omega_1 +\i \Omega_2)}{(n k_x' + k_x)^2 + (1 - n)^2 (k_y^2 + k_z^2)},
    \end{eqnarray}
with
    \begin{eqnarray}
    &&    \Omega_1 = k_x^2-k_y^2(n-1)^2-k_z^2(n-1)^2-k_x'^2 n^2,\nonumber\\
    &&    \Omega_2 = 2 (n-1) k_x k_y,\nonumber\\
    &&    \Omega_3 = 2 (n-1) k_x k_z, \nonumber\\
    \end{eqnarray}
and
\begin{eqnarray}
&& \ell_1=\cos\frac{\theta}{2}, \;\;\;\; \ell_2=\sin\frac{\theta}{2} {\rm e}^{{\rm i}\phi}, \nonumber\\
&& \frac{\ell_2}{\ell_1}= G_1+ {\rm i} G_2=\tan\frac{\theta}{2} {\rm e}^{{\rm i}\phi},\nonumber\\
&& \vec{k}= k \left( \cos \phi_k' , \sin \phi_k' \cos \phi_k'', \sin \phi_k' \sin \phi_k'' \right),
\end{eqnarray}
we then have
\begin{eqnarray}
   A' &=& A \sin \frac{\phi_k''}{2} -{\rm i} B   \cos \frac{\phi_k''}{2} \nonumber\\
   &=&  \dfrac{ \ell_1 (\Omega_1 - \i \Omega_2) + \ell_2 \Omega_3}{(n k_x' + k_x)^2 + (1 - n)^2 (k_y^2 + k_z^2)} \sin \frac{\phi_k''}{2}
   -{\rm i}  \dfrac{- \ell_1 \Omega_3 + \ell_2 (\Omega_1 +\i \Omega_2)}{(n k_x' + k_x)^2 + (1 - n)^2 (k_y^2 + k_z^2)} \cos \frac{\phi_k''}{2} \nonumber\\
   &=&  \dfrac{ (\ell_1 \Omega_1+\ell_2 \Omega_3 ) - \i \ell_1 \Omega_2}{(n k_x' + k_x)^2 + (1 - n)^2 (k_y^2 + k_z^2)} \sin \frac{\phi_k''}{2}
   -{\rm i}  \dfrac{ (- \ell_1 \Omega_3 + \ell_2 \Omega_1)+ {\rm i}\ell_2 \Omega_2}{(n k_x' + k_x)^2 + (1 - n)^2 (k_y^2 + k_z^2)} \cos \frac{\phi_k''}{2} \nonumber\\
   &=&  \dfrac{ (\ell_1 \Omega_1+\ell_2 \Omega_3 ) - \i \ell_1 \Omega_2}{(n k_x' + k_x)^2 + (1 - n)^2 (k_y^2 + k_z^2)} \sin \frac{\phi_k''}{2}
   +  \dfrac{ {\rm i}(\ell_1 \Omega_3 - \ell_2 \Omega_1)+ \ell_2 \Omega_2}{(n k_x' + k_x)^2 + (1 - n)^2 (k_y^2 + k_z^2)} \cos \frac{\phi_k''}{2} \nonumber\\
   &=&  \dfrac{ \left[(\ell_1 \Omega_1+\ell_2 \Omega_3 )\sin \frac{\phi_k''}{2}+ \ell_2 \Omega_2 \cos \frac{\phi_k''}{2}\right] + {\rm i}\left[ -\ell_1 \Omega_2 \sin \frac{\phi_k''}{2}+ (\ell_1 \Omega_3 - \ell_2 \Omega_1) \cos \frac{\phi_k''}{2}\right]}{(n k_x' + k_x)^2 + (1 - n)^2 (k_y^2 + k_z^2)},
\end{eqnarray}
\begin{eqnarray}
   B' &=& A \cos \frac{\phi_k''}{2} +{\rm i} B   \sin \frac{\phi_k''}{2} \nonumber\\
   &=&  \dfrac{ \ell_1 (\Omega_1 - \i \Omega_2) + \ell_2 \Omega_3}{(n k_x' + k_x)^2 + (1 - n)^2 (k_y^2 + k_z^2)} \cos \frac{\phi_k''}{2}
   +{\rm i} \dfrac{- \ell_1 \Omega_3 + \ell_2 (\Omega_1 +\i \Omega_2)}{(n k_x' + k_x)^2 + (1 - n)^2 (k_y^2 + k_z^2)}
   \sin \frac{\phi_k''}{2} \nonumber\\
   &=&  \dfrac{ (\ell_1 \Omega_1+\ell_2 \Omega_3 ) - \i \ell_1 \Omega_2}{(n k_x' + k_x)^2 + (1 - n)^2 (k_y^2 + k_z^2)} \cos \frac{\phi_k''}{2}
   +{\rm i}  \dfrac{ (- \ell_1 \Omega_3 + \ell_2 \Omega_1)+ {\rm i}\ell_2 \Omega_2}{(n k_x' + k_x)^2 + (1 - n)^2 (k_y^2 + k_z^2)} \sin \frac{\phi_k''}{2} \nonumber\\
   &=&  \dfrac{ (\ell_1 \Omega_1+\ell_2 \Omega_3 ) - \i \ell_1 \Omega_2}{(n k_x' + k_x)^2 + (1 - n)^2 (k_y^2 + k_z^2)} \cos \frac{\phi_k''}{2}
     -\dfrac{ {\rm i}(\ell_1 \Omega_3 - \ell_2 \Omega_1)+ \ell_2 \Omega_2}{(n k_x' + k_x)^2 + (1 - n)^2 (k_y^2 + k_z^2)} \sin \frac{\phi_k''}{2} \nonumber\\
   &=&  \dfrac{ \left[(\ell_1 \Omega_1+\ell_2 \Omega_3 )\cos \frac{\phi_k''}{2}-\ell_2 \Omega_2 \sin \frac{\phi_k''}{2}\right] + {\rm i}\left[ -\ell_1 \Omega_2 \cos \frac{\phi_k''}{2}- (\ell_1 \Omega_3 - \ell_2 \Omega_1) \sin \frac{\phi_k''}{2}\right]}{(n k_x' + k_x)^2 + (1 - n)^2 (k_y^2 + k_z^2)}.
\end{eqnarray}

\emph{Discussion.---}When $k_y=0$, we can have $\Omega_2=2 (n-1) k_x k_y=0$, and $\phi_k''=\frac{\pi}{2}$, $\cos \frac{\phi_k''}{2}=\sin \frac{\phi_k''}{2}=\frac{1}{\sqrt{2}}$, then $A'$ and $B'$ reduces to
\begin{eqnarray}
   A' &=&  \dfrac{ \left[(\ell_1 \Omega_1+\ell_2 \Omega_3 )\sin \frac{\phi_k''}{2}+ \ell_2 \Omega_2 \cos \frac{\phi_k''}{2}\right] + {\rm i}\left[ -\ell_1 \Omega_2 \sin \frac{\phi_k''}{2}+ (\ell_1 \Omega_3 - \ell_2 \Omega_1) \cos \frac{\phi_k''}{2}\right]}{(n k_x' + k_x)^2 + (1 - n)^2 (k_y^2 + k_z^2)}\nonumber\\
&=&  \frac{1}{\sqrt{2}}\dfrac{ (\ell_1 \Omega_1+\ell_2 \Omega_3 )  + {\rm i} (\ell_1 \Omega_3 - \ell_2 \Omega_1)}{(n k_x' + k_x)^2 + (1 - n)^2  k_z^2}= \frac{1}{\sqrt{2}}\dfrac{ \ell_1 (\Omega_1+{\rm i} \Omega_3 ) + \ell_2 (\Omega_3 -{\rm i} \Omega_1)}{(n k_x' + k_x)^2 + (1 - n)^2  k_z^2}\nonumber\\
&=&  \frac{1}{\sqrt{2}}\dfrac{ \ell_1 (\Omega_1+{\rm i} \Omega_3 ) -{\rm i} \ell_2 (\Omega_1+ {\rm i}\Omega_3)}{(n k_x' + k_x)^2 + (1 - n)^2  k_z^2}=\frac{\ell_1-{\rm i} \ell_2 }{\sqrt{2}} \dfrac{ (\Omega_1+{\rm i} \Omega_3 ) }{(n k_x' + k_x)^2 + (1 - n)^2  k_z^2},
\end{eqnarray}
\begin{eqnarray}
   B'  &=&  \dfrac{ \left[(\ell_1 \Omega_1+\ell_2 \Omega_3 )\cos \frac{\phi_k''}{2}-\ell_2 \Omega_2 \sin \frac{\phi_k''}{2}\right] + {\rm i}\left[ -\ell_1 \Omega_2 \cos \frac{\phi_k''}{2}- (\ell_1 \Omega_3 - \ell_2 \Omega_1) \sin \frac{\phi_k''}{2}\right]}{(n k_x' + k_x)^2 + (1 - n)^2 (k_y^2 + k_z^2)}\nonumber\\
  &=&  \dfrac{1}{\sqrt{2}}\dfrac{ (\ell_1 \Omega_1+\ell_2 \Omega_3 ) - {\rm i} (\ell_1 \Omega_3 - \ell_2 \Omega_1) }{(n k_x' + k_x)^2 + (1 - n)^2 k_z^2}=  \frac{\ell_1+{\rm i} \ell_2 }{\sqrt{2}}\dfrac{ (\Omega_1-{\rm i} \Omega_3 ) }{(n k_x' + k_x)^2 + (1 - n)^2  k_z^2},\nonumber\\
\end{eqnarray}
i.e.,
\begin{eqnarray}\label{eq:simple}
 &&   A' =  \frac{{(\ell_1-{\rm i} \ell_2)}}{\sqrt{2}}  \biggr|\dfrac{ (\Omega_1+{\rm i} \Omega_3 ) }{(n k_x' + k_x)^2 + (1 - n)^2  k_z^2}\biggr|\, {\rm e}^{-{\rm i}\theta_{\rm ra}}, \nonumber\\
&&   B' =  \frac{(\ell_1+{\rm i} \ell_2)}{\sqrt{2}}  \biggr|\dfrac{ (\Omega_1-{\rm i} \Omega_3 ) }{(n k_x' + k_x)^2 + (1 - n)^2  k_z^2}\biggr|\, {\rm e}^{-{\rm i}\theta_{\rm rb}},
\end{eqnarray}
with
\begin{eqnarray}
&&\tan \theta_{\rm ra}=-\frac{\Omega_3}{\Omega_1}, \;\;\;\;\; \theta_{\rm rb}=-\theta_{\rm ra}.
\end{eqnarray}
In the $\hat{w}$-basis, the reflected wave reads
\begin{eqnarray}
\ket{\Psi_{\rm r}} & = & \left\{ |A'|\, { {\rm e}^{{\rm i}{\rm arg}[\ell_1-{\rm i} \ell_2]}} {\rm e}^{-{\rm i}\theta_{\rm ra}} \ket{{\tilde{s}}_{\rm w} = +\frac{\hbar}{2}} + |B'|\, {{\rm e}^{{\rm i}{\rm arg}[\ell_1+{\rm i} \ell_2]}}{\rm e}^{{\rm i}\theta_{\rm ra}} \ket{{\tilde{s}}_{\rm w} = -\frac{\hbar}{2}} \right\} {\rm e}^{\frac{\i}{\hbar} [{\hbar}(-k_x x+ k_z z) - \mathcal{E} t]}.
\end{eqnarray}
When calculate the spatial shift, the phases of $\ell_1\pm {\rm i} \ell_2$ have no contributions, because $\ell_1$ and $\ell_2$ do not depend on the momentum. Since $\theta_{\rm rb}=-\theta_{\rm ra}$, thus the shift caused by $\theta_{\rm rb}$ will be opposite to that of $\theta_{\rm ra}$.

However, when $k_y \neq 0$, since the angle $\phi_k''$ is related to the momentum, one does not have the above simple relation as shown in Eq. (\ref{eq:simple}). Thus in comparison to the $z$-basis, it is not more convenient to calculate the spatial shifts in the special $\hat{w}$-basis.
$\blacksquare$
\end{remark}

\begin{remark}
\textcolor{blue}{Explicit Expression of the Spatial Shift for the Reflected Wave}.
From previous calculation, we have known that the spatial shift $\Delta \vec{l}_{\rm r}$ is unique, which is
\begin{eqnarray}
    \Delta \vec{l}_{\rm r}  = \Delta y_{\rm r} \vec{e}_y+ \Delta z_{\rm r} \vec{e}_z,
\end{eqnarray}
where
\begin{eqnarray}
 &&   \Delta y_{\rm r}  =\frac{2 G_2 \left(\frac{\partial\Omega_1  }{\partial k_y}\Omega_3 - \Omega_1 \frac{\partial\Omega_3 }{\partial k_y}\right)
    + {(G_1^2 +G_2^2-1 )}\left(\frac{\partial\Omega_1  }{\partial k_y} \Omega_2 - \Omega_1 \frac{\partial\Omega_2  }{\partial k_y}\right)
    +2 G_1 \left(\frac{\partial\Omega_2  }{\partial k_y} \Omega_3 - \Omega_2 \frac{\partial\Omega_3 }{\partial k_y}\right)]}
    {(1+ G_1^2 + G_2^2 ) (\Omega_1^2 +\Omega_2^2 +\Omega_3^2  )}, \nonumber\\
 &&    \Delta z_{\rm r} =\frac{2 G_2 \left(\frac{\partial\Omega_1  }{\partial k_z}\Omega_3 - \Omega_1 \frac{\partial\Omega_3 }{\partial k_z}\right)
    + {(G_1^2 +G_2^2-1 )}\left(\frac{\partial\Omega_1  }{\partial k_z} \Omega_2 - \Omega_1 \frac{\partial\Omega_2  }{\partial k_z}\right)
    +2 G_1 \left(\frac{\partial\Omega_2  }{\partial k_z} \Omega_3 - \Omega_2 \frac{\partial\Omega_3 }{\partial k_z}\right)]}
    {(1+ G_1^2 + G_2^2 ) (\Omega_1^2 +\Omega_2^2 +\Omega_3^2  )}.
\end{eqnarray}
With the help of
    \begin{eqnarray}
    &&    \Omega_1 = k_x^2-k_y^2(n-1)^2-k_z^2(n-1)^2-k_x'^2 n^2, \;\;\;\; \Omega_2 = 2 (n-1) k_x k_y,\;\;\;\;\;  \Omega_3 = 2 (n-1) k_x k_z, \nonumber\\
    && \ell_1=\cos\frac{\theta}{2}, \;\;\;\; \ell_2=\sin\frac{\theta}{2} {\rm e}^{{\rm i}\phi}, \;\;\;\;\; \frac{\ell_2}{\ell_1}= G_1+ {\rm i} G_2=\tan\frac{\theta}{2} {\rm e}^{{\rm i}\phi},
    \end{eqnarray}
here we would like to work out its explicit expression. We can have
\begin{eqnarray}
 \Omega_1 &=& k_x^2-k_y^2(n-1)^2-k_z^2(n-1)^2-k_x'^2 n^2\nonumber\\
&=& k_x^2-n^2{k_x'}^2 - {(k_y^2+k_z^2)}(1 - 2n+n^2) \nonumber\\
&=& k_x^2+ (2n-1)  ({k_y^2}+k_z^2)-n^2 ({k_x'}^2+{k_y^2}+k_z^2)\nonumber\\
&=&\frac{{{\mathcal{E}}^2-m^2c^4 -c^2 \hbar^2\, ({k_y^2}+k_z^2)}}{c^2\hbar^2 }+ (2n-1)  ({k_y^2}+k_z^2)
-n^2 \frac{\left(\mathcal{E}-V_0\right)^2-m^2c^4}{c^2\hbar^2 }\nonumber\\
&=&\frac{{\mathcal{E}}^2-m^2c^4 }{c^2\hbar^2 }+ 2(n-1)  ({k_y^2}+k_z^2)
-n^2 \frac{\left(\mathcal{E}-V_0\right)^2-m^2c^4}{c^2\hbar^2 }\nonumber\\
&=&\frac{{\mathcal{E}}^2-m^2c^4 }{c^2\hbar^2 }+ 2(n-1)  ({k_y^2}+k_z^2)
-\frac{(\mathcal{E}+mc^2)^2 }{(\mathcal{E}-V_0 +mc^2)^2} \frac{\left(\mathcal{E}-V_0\right)^2-m^2c^4}{c^2\hbar^2 }\nonumber\\
&=&\frac{{\mathcal{E}}^2-m^2c^4 }{c^2\hbar^2 }+ 2(n-1)  ({k_y^2}+k_z^2)
-\frac{(\mathcal{E}+mc^2)^2 }{(\mathcal{E}-V_0 +mc^2)} \frac{\left(\mathcal{E}-V_0\right)-mc^2}{c^2\hbar^2 }\nonumber\\
&=&2(n-1)  ({k_y^2}+k_z^2)+
\frac{({\mathcal{E}}^2-m^2c^4)(\mathcal{E}-V_0 +mc^2)-(\mathcal{E}+mc^2)^2\left(\mathcal{E}-V_0-mc^2\right) }{(\mathcal{E}-V_0 +mc^2)c^2\hbar^2 } \nonumber\\
&=&2(n-1)  ({k_y^2}+k_z^2)+(\mathcal{E}+mc^2) \times
\frac{({\mathcal{E}}-mc^2)(\mathcal{E}-V_0 +mc^2)-(\mathcal{E}+mc^2)\left(\mathcal{E}-V_0-mc^2\right) }{(\mathcal{E}-V_0 +mc^2)c^2\hbar^2 } \nonumber\\
&=&2(n-1)  ({k_y^2}+k_z^2)+(\mathcal{E}+mc^2) \times
\frac{{\mathcal{E}} \times 2mc^2 -mc^2 \times  2(\mathcal{E}-V_0)  }{(\mathcal{E}-V_0 +mc^2)c^2\hbar^2 } \nonumber\\
&=&2\left(\frac{\mathcal{E}+mc^2 }{\mathcal{E}-V_0 +mc^2}-1\right)  ({k_y^2}+k_z^2)+(\mathcal{E}+mc^2) \times
\frac{2 V_0 mc^2 }{(\mathcal{E}-V_0 +mc^2)c^2\hbar^2 } \nonumber\\
&=&\frac{2 V_0 }{\mathcal{E}-V_0 +mc^2}  ({k_y^2}+k_z^2)+(\mathcal{E}+mc^2) \times
\frac{2 V_0 mc^2 }{(\mathcal{E}-V_0 +mc^2)c^2\hbar^2 } \nonumber\\
&=&\frac{2 V_0 }{\mathcal{E}-V_0 +mc^2} \left[ ({k_y^2}+k_z^2)+
\frac{(\mathcal{E}+mc^2)  mc^2 }{c^2\hbar^2 }\right]\nonumber\\
&=& 2(n-1) \left[ ({k_y^2}+k_z^2)+
\frac{(\mathcal{E}+mc^2)  mc^2 }{c^2\hbar^2 }\right],
\end{eqnarray}
\begin{eqnarray}
    \frac{\partial \Omega_1}{\partial k_y}= 4(n-1)k_y, \;\;\;\;\;\;\; \frac{\partial \Omega_1}{\partial k_z}= 4(n-1)k_z,
\end{eqnarray}
and
\begin{eqnarray}
    \frac{\partial \Omega_2}{\partial k_y}= 2 (n-1) \left[- \frac{k_y^2}{k_x} + k_x \right], \;\;\;\;\;\;\;
    \frac{\partial \Omega_2}{\partial k_z}= -2 (n-1) \frac{k_z k_y }{k_x},
\end{eqnarray}
\begin{eqnarray}
    \frac{\partial \Omega_3}{\partial k_y}= -2 (n-1) \frac{k_y k_z }{k_x}, \;\;\;\;\;\;\;
    \frac{\partial \Omega_3}{\partial k_z}= 2 (n-1) \left[-\frac{k_z^2 }{k_x}+k_x \right],
\end{eqnarray}
which lead to
\begin{eqnarray}
    \frac{\partial\Omega_1  }{\partial k_y}\Omega_3 - \Omega_1 \frac{\partial\Omega_3 }{\partial k_y} &=  & 4(n-1)^2 \left\{2k_yk_x k_z +\left[{k_y^2}+k_z^2
    +\frac{(\mathcal{E}+mc^2)  mc^2 }{c^2\hbar^2 }\right]\left(\frac{k_y k_z }{k_x}\right)  \right\} \notag\\
    &=&4(n-1)^2 \left(\frac{k_y k_z }{k_x}\right)\left\{2k_x^2  +{k_y^2}+k_z^2
    +\frac{(\mathcal{E}+mc^2)  mc^2 }{c^2\hbar^2 } \right\} \notag\\
    &=& 4(n-1)^2 \left(\frac{k_y k_z }{k_x}\right)\left\{\frac{2(\mathcal{E}^2 -m^2c^4)}{c^2\hbar^2 }  -{k_y^2}-k_z^2
    +\frac{(\mathcal{E}+mc^2)  mc^2 }{c^2\hbar^2 } \right\} \notag\\
    &=& 4(n-1)^2 \left(\frac{k_y k_z }{k_x}\right)\left\{\frac{(2\mathcal{E} -mc^2)(\mathcal{E} + mc^2 )}{c^2\hbar^2 }  -{k_y^2}-k_z^2
     \right\},
\end{eqnarray}
\begin{eqnarray}
    \frac{\partial\Omega_1  }{\partial k_y}\Omega_2 - \Omega_1 \frac{\partial\Omega_2 }{\partial k_y} =4(n-1)^2 \left\{2k_y^2 k_x  +\left[{k_y^2}+k_z^2
    +\frac{(\mathcal{E}+mc^2)  mc^2 }{c^2\hbar^2 }\right]\left(\frac{k_y^2  }{k_x}-k_x\right)  \right\},
\end{eqnarray}

\begin{eqnarray}
    \frac{\partial\Omega_2  }{\partial k_y}\Omega_3 - \Omega_2 \frac{\partial\Omega_3 }{\partial k_y}
    =4(n-1)^2 \left\{-\left(\frac{k_y^2 }{k_x}-k_x\right)k_x k_z   +k_x k_y\left(\frac{k_y k_z }{k_x}\right)  \right\}
    = 4(n-1)^2 k_x^2 k_z.
\end{eqnarray}

Furthermore, because
\begin{eqnarray}
\vec{k} =k \left( \cos \phi_k' , \sin \phi_k' \cos \phi_k'', \sin \phi_k' \sin \phi_k'' \right), \;\;\;\;\;
k =\frac{\sqrt{\mathcal{E}^2 - m^2c^4}}{{c\hbar}},
    \end{eqnarray}
one can have
    \begin{eqnarray}
        \frac{\partial\Omega_1  }{\partial k_y}\Omega_3 - \Omega_1 \frac{\partial\Omega_3 }{\partial k_y}
        &=& (n-1)^2 k\sin (2 \phi_k'') \sin \phi_k' \tan \phi_k'
        \left[\frac{(3\mathcal{E} -mc^2)(\mathcal{E} + mc^2 )}{c^2\hbar^2 } + k^2 \cos (2\phi_k')\right],
    \end{eqnarray}
    \begin{eqnarray}
        \frac{\partial\Omega_1  }{\partial k_y}\Omega_2 - \Omega_1 \frac{\partial\Omega_2 }{\partial k_y} &=&
        \frac{1}{2}(n-1)^2 \frac{k}{\cos \phi_k'}   \biggl\{4 k^2 \sin^2 (2\phi_k') \cos^2 \phi_k'' \notag\\
       &&   - \frac{\mathcal{E} +mc^2}{c^2 \hbar^2} [\mathcal{E} +mc^2- (\mathcal{E} -mc^2) \cos (2 \phi_k') ] [1+3\cos (2 \phi_k')- 2\cos (2 \phi_k'') \sin^2 \phi_k' ]    \biggr\},
    \end{eqnarray}
    \begin{eqnarray}
        \frac{\partial\Omega_2  }{\partial k_y}\Omega_3 - \Omega_2 \frac{\partial\Omega_3 }{\partial k_y} =4(n-1)^2k^{3} \sin \phi_k' \cos^2 \phi_k'\sin \phi_k'' ,
    \end{eqnarray}
\begin{eqnarray}
    (\Omega_1^2 +\Omega_2^2 +\Omega_3^2  )= \frac{2(n-1)^2}{c^4 \hbar^4}(\mathcal{E} + mc^2 )^2 [\mathcal{E}^2 + m^2c^4 -( \mathcal{E}^2 - m^2c^4) \cos (2 \phi_k')].
\end{eqnarray}
In addition, one has
\begin{eqnarray}
  &&  \frac{2G_1}{1+ G_1^2 + G_2^2 } = \frac{2\tan \frac{\theta}{2}  \cos \phi}{1+\tan^2 \frac{\theta}{2}}=\sin {\theta} \cos \phi, \nonumber\\    &&\frac{2G_2}{1+ G_1^2 + G_2^2 } = \frac{2\tan \frac{\theta}{2}  \sin \phi}{1+\tan^2 \frac{\theta}{2}}=\sin {\theta} \sin \phi,\nonumber\\
  && \frac{G_1^2 + G_2^2 -1}{1+ G_1^2 + G_2^2 } = \frac{\tan^2 \frac{\theta}{2}-1}{1+\tan^2 \frac{\theta}{2}}= -\cos {\theta},
\end{eqnarray}
so
\begin{eqnarray}
    \Delta y_{\rm r} & = & \frac{ \hbar^2 c^2 k\sin {\theta} \sin \phi \sin (2 \phi_k'') \sin \phi_k' \tan \phi_k' [{(3\mathcal{E} -mc^2)(\mathcal{E} + mc^2 )}+{c^2\hbar^2 }  k^2 \cos (2\phi_k')]}{2(\mathcal{E} + mc^2 )^2 [\mathcal{E}^2 + m^2c^4 -( \mathcal{E}^2 - m^2c^4) \cos (2 \phi_k')]}\notag\\
    &&+\frac{ 2\hbar^4 c^4 k^{3}\sin {\theta} \cos \phi  \sin \phi_k' \cos^2 \phi_k'\sin \phi_k''}{(\mathcal{E} + mc^2 )^2 [\mathcal{E}^2 + m^2c^4 -( \mathcal{E}^2 - m^2c^4) \cos (2 \phi_k')]}{-}\cos \theta {c^2\hbar^2 }k  \times \notag\\
    &&\frac{  \biggl[4  {\hbar^2 c^2}k^2 \sin^2 (2\phi_k') \cos^2 \phi_k'' - ({\mathcal{E} +mc^2}) [\mathcal{E} +mc^2- (\mathcal{E} -mc^2) \cos (2 \phi_k') ] [1+3\cos (2 \phi_k')- 2\cos (2 \phi_k'') \sin^2 \phi_k' ]    \biggr]}{4 \cos \phi_k'(\mathcal{E} + mc^2 )^2 [\mathcal{E}^2 + m^2c^4 -( \mathcal{E}^2 - m^2c^4) \cos (2 \phi_k')]}.\notag\\
\end{eqnarray}

For $\Delta z_{\rm r} $, one can have
\begin{eqnarray}
    \frac{\partial\Omega_1  }{\partial k_z}\Omega_3 - \Omega_1 \frac{\partial\Omega_3 }{\partial k_z} &=  & 4(n-1)^2 \left\{2k_z^2 k_x  +\left[{k_y^2}+k_z^2
    +\frac{(\mathcal{E}+mc^2)  mc^2 }{c^2\hbar^2 }\right]\left(\frac{k_z^2 }{k_x}-k_x \right)  \right\},
\end{eqnarray}

\begin{eqnarray}
    \frac{\partial\Omega_1  }{\partial k_z}\Omega_2 - \Omega_1 \frac{\partial\Omega_2 }{\partial k_z} =
    4(n-1)^2 \left(\frac{k_y k_z }{k_x}\right)\left\{\frac{(2\mathcal{E} -mc^2)(\mathcal{E} + mc^2 )}{c^2\hbar^2 }  -{k_y^2}-k_z^2
     \right\},
\end{eqnarray}

\begin{eqnarray}
    \frac{\partial\Omega_2  }{\partial k_z}\Omega_3 - \Omega_2 \frac{\partial\Omega_3 }{\partial k_z}
    =4(n-1)^2 \left\{\left(\frac{ k_z^2 }{k_x}-k_x\right)k_x k_y   -k_x k_z\left(\frac{k_y k_z }{k_x}\right)  \right\}
    =-4(n-1)^2 k_x^2 k_y.
\end{eqnarray}
which yield
\begin{eqnarray}
    \frac{\partial\Omega_1  }{\partial k_z}\Omega_3 - \Omega_1 \frac{\partial\Omega_3 }{\partial k_z} &=& \frac{1}{2}(n-1)^2 \frac{k}{\cos \phi_k'}   \biggl\{4 k^2 \sin^2 (2\phi_k') \sin^2 \phi_k'' \notag\\
    && - \frac{\mathcal{E} +mc^2}{\hbar^2 c^2} [\mathcal{E} +mc^2- (\mathcal{E} -mc^2) \cos (2 \phi_k') ] [1+3\cos (2 \phi_k')+ 2\cos (2 \phi_k'') \sin^2 \phi_k' ]    \biggr\},
\end{eqnarray}
\begin{eqnarray}
    \frac{\partial\Omega_1  }{\partial k_z}\Omega_2 - \Omega_1 \frac{\partial\Omega_2 }{\partial k_z} =
    (n-1)^2 k\sin (2 \phi_k'') \sin \phi_k' \tan \phi_k' \left[\frac{(3\mathcal{E} -mc^2)(\mathcal{E} + mc^2 )}{c^2\hbar^2 } + k^2 \cos (2\phi_k')\right],
\end{eqnarray}
\begin{eqnarray}
    \frac{\partial\Omega_2  }{\partial k_z}\Omega_3 - \Omega_2 \frac{\partial\Omega_3 }{\partial k_z} =-4(n-1)^2k^{3} \sin \phi_k' \cos^2 \phi_k'\cos \phi_k''.
\end{eqnarray}
So finally one has
\begin{eqnarray}
    \Delta z_{\rm r}  &=&  -\frac{ \hbar^2 c^2 k \cos \theta \sin (2 \phi_k'') \sin \phi_k' \tan \phi_k' [{(3\mathcal{E} -mc^2)(\mathcal{E} + mc^2 )}+{c^2\hbar^2 }  k^2 \cos (2\phi_k')]}{2(\mathcal{E} + mc^2 )^2 [\mathcal{E}^2 + m^2c^4 -( \mathcal{E}^2 - m^2c^4) \cos (2 \phi_k')]}\notag\\
    &&-\frac{ 2\hbar^4 c^4 k^{3}\sin {\theta} \cos \phi  \sin \phi_k' \cos^2 \phi_k'\cos \phi_k''}{(\mathcal{E} + mc^2 )^2 [\mathcal{E}^2 + m^2c^4 -( \mathcal{E}^2 - m^2c^4) \cos (2 \phi_k')]}+ \sin {\theta} \sin \phi {c^2\hbar^2 }k  \times \notag\\
    &&\frac{   \biggl[4  {\hbar^2 c^2}k^2 \sin^2 (2\phi_k') \sin^2 \phi_k'' - ({\mathcal{E} +mc^2}) [\mathcal{E} +mc^2- (\mathcal{E} -mc^2) \cos (2 \phi_k') ] [1+3\cos (2 \phi_k')+ 2\cos (2 \phi_k'') \sin^2 \phi_k' ]    \biggr]}{4 \cos \phi_k'(\mathcal{E} + mc^2 )^2 [\mathcal{E}^2 + m^2c^4 -( \mathcal{E}^2 - m^2c^4) \cos (2 \phi_k')]}.\notag\\
\end{eqnarray}
One may notice that $\Delta y_{\rm r}$ and $\Delta z_{\rm r}$ (hence $\Delta \vec{l}_{\rm r}$) do not depend on the potential $V_0$.

When $k_y=0$, i.e., $\cos \phi_k'' = 0$, one has
\begin{eqnarray}
    \Delta z_{\rm r} &=& \frac{  \sin {\theta} \sin \phi {c^2\hbar^2 }k   \biggl[4  {\hbar^2 c^2}k^2 \sin^2 (2\phi_k')- ({\mathcal{E} +mc^2}) [\mathcal{E} +mc^2- (\mathcal{E} -mc^2) \cos (2 \phi_k') ] [1+3\cos (2 \phi_k')- 2 \sin^2 \phi_k' ]    \biggr]}{4 \cos \phi_k'(\mathcal{E} + mc^2 )^2 [\mathcal{E}^2 + m^2c^4 -( \mathcal{E}^2 - m^2c^4) \cos (2 \phi_k')]}\notag\\
    &=&\frac{  \sin {\theta} \sin \phi {c^2\hbar^2 }k   \biggl[  (\mathcal{E}^2 -m^2c^4)\sin^2 (2\phi_k')- ({\mathcal{E} +mc^2}) [\mathcal{E} +mc^2- (\mathcal{E} -mc^2) \cos (2 \phi_k') ]\cos (2 \phi_k')    \biggr]}{2 \cos \phi_k'(\mathcal{E} + mc^2 )^2 [\mathcal{E}^2\sin^2 \phi_k' + m^2c^4  \cos^2 \phi_k']}\notag\\
    &=&\frac{  \sin {\theta} \sin \phi {c^2\hbar^2 }k   \biggl[  \mathcal{E}\sin^2 \phi_k'- mc^2\cos^2 \phi_k'   \biggr]}{\cos \phi_k'(\mathcal{E} + mc^2 )[\mathcal{E}^2\sin^2 \phi_k' + m^2c^4  \cos^2 \phi_k']}\notag\\
    &=&\frac{  \sin {\theta} \sin \phi \, {c\hbar } \dfrac{\sqrt{\mathcal{E} -mc^2 }}{\sqrt{\mathcal{E} +mc^2 }} \biggl[  \mathcal{E}\cos \phi_k'\tan^2 \phi_k'- mc^2\cos\phi_k'   \biggr]}{\mathcal{E}^2\sin^2 \phi_k' + m^2c^4  \cos^2 \phi_k'}\nonumber\\
     &=& (\sin\theta\sin\phi) \, \dfrac{\sqrt{\mathcal{E}-mc^2} }{\sqrt{{\mathcal{E}}+mc^2}}
\dfrac{c\hbar {\mathcal{E}} \cos \varphi}{\left[ {\mathcal{E}}^2 \sin^2 \varphi + m^2c^4 \cos^2 \varphi \right]}
 \left[\tan^2 \varphi -   \dfrac{mc^2}{\mathcal{E}} \right],
\end{eqnarray}
which naturally reduces to Eq. (\ref{eq:z-1d}) with $\varphi \equiv \phi_k'$, i.e., it is equal to the case where the finite-width incident wave with finite dimensions just along the $z$ direction and $k_y=0$.
$\blacksquare$
\end{remark}

\begin{remark}
\textcolor{blue}{Explicit Expression of the Spatial Shift for the Transmitted Wave}.
From previous calculation, we have known that the spatial shift  $\Delta \vec{l}_{\rm t}$ is unique, which is
\begin{eqnarray}
    \Delta \vec{l}_{\rm t} = \Delta y_{\rm t} \, \vec{e}_y +\Delta z_{\rm t}\,  \vec{e}_z,
\end{eqnarray}
\begin{eqnarray}
    \Delta y_{\rm t}
    &=&\frac{2\left[ G_2 \left(\frac{\partial\Omega_{k1}  }{\partial k_y}\Omega_{k3} - \Omega_{k1} \frac{\partial\Omega_{k3} }{\partial k_y}\right)+G_1 \left(\frac{\partial \Omega_{k2}  }{\partial k_y} \Omega_{k3} - \Omega_{k2} \frac{\partial \Omega_{k3} }{\partial k_y}\right)\right]+(G_1^2+G_2^2 -1) \left(\frac{\partial\Omega_{k1}  }{\partial k_y} \Omega_{k2} - \Omega_{k1} \frac{\partial \Omega_{k2}  }{\partial k_y}\right)}{(1+ G_1^2 + G_2^2 ) ({\Omega_{k1}}^2 +\Omega_{k2}^2 +\Omega_{k3}^2  )}, \nonumber\\
    \Delta z_{\rm t}
    &=&\frac{2 \left[ G_2 \left(\frac{\partial\Omega_{k1}  }{\partial k_z}\Omega_{k3} - \Omega_{k1} \frac{\partial\Omega_{k3} }{\partial k_z}\right)+G_1 \left(\frac{\partial \Omega_{k2}  }{\partial k_z} \Omega_{k3} - \Omega_{k2} \frac{\partial \Omega_{k3} }{\partial k_z}\right)\right]+(G_1^2+G_2^2 -1) \left(\frac{\partial\Omega_{k1}  }{\partial k_z} \Omega_{k2} - \Omega_{k1} \frac{\partial \Omega_{k2}  }{\partial k_z} \right)}{(1+ G_1^2 + G_2^2 ) ({\Omega_{k1}}^2 +\Omega_{k2}^2 +\Omega_{k3}^2  )}.
\end{eqnarray}
With the help of
\begin{eqnarray}
 &&  \Omega_{k1}= k_x +{n}k_x', \;\;\;\;\;  \Omega_{k2} =(n-1)k_y , \;\;\;\;\;  \Omega_{k3}=(n-1)k_z, \nonumber\\
&& \ell_1=\cos\frac{\theta}{2}, \;\;\;\; \ell_2=\sin\frac{\theta}{2} {\rm e}^{{\rm i}\phi}, \;\;\;\; \frac{\ell_2}{\ell_1}= G_1+ {\rm i} G_2=\tan\frac{\theta}{2} {\rm e}^{{\rm i}\phi},
\end{eqnarray}
here we would like to work out its explicit expression.

First, one has
\begin{eqnarray}
    \frac{\partial \Omega_{k1}}{\partial k_y}= - \left(\frac{1}{k_x}+\frac{{n}}{k'_x}\right)k_y,
    \;\;\;\;\; \frac{\partial \Omega_{k1}}{\partial k_z}= - \left(\frac{1}{k_x}+\frac{{n}}{k'_x}\right)k_z,
\end{eqnarray}
\begin{eqnarray}
    \frac{\partial \Omega_{k2}}{\partial k_y}= n-1, \;\;\;\;\;  \frac{\partial \Omega_{k2}}{\partial k_z}= 0,
\end{eqnarray}
\begin{eqnarray}
    \frac{\partial \Omega_{k3}}{\partial k_y}= 0, \;\;\;\;\;  \frac{\partial \Omega_{k3}}{\partial k_z}= n-1,
\end{eqnarray}
then one has
\begin{eqnarray}
    \frac{\partial\Omega_{k1}  }{\partial k_y}\Omega_{k3} - \Omega_{k1} \frac{\partial\Omega_{k3} }{\partial k_y} =  \frac{\partial\Omega_{k1}  }{\partial k_y}\Omega_{k3} = - \left(\frac{1}{k_x}+\frac{{n}}{k'_x}\right)k_y (n-1)k_z ,
\end{eqnarray}
\begin{eqnarray}
    \frac{\partial \Omega_{k2}  }{\partial k_y} \Omega_{k3} - \Omega_{k2} \frac{\partial \Omega_{k3} }{\partial k_y} =    \frac{\partial \Omega_{k2}  }{\partial k_y} \Omega_{k3} = (n-1)^2 k_z \,,\,
\end{eqnarray}
\begin{eqnarray}
    \frac{\partial\Omega_{k1}  }{\partial k_y} \Omega_{k2} - \Omega_{k1} \frac{\partial \Omega_{k2}  }{\partial k_y} = - (n-1)\left(\frac{1}{k_x}+\frac{{n}}{k'_x}\right)k_y^2 -(n-1) (k_x +{n}k_x'),
\end{eqnarray}

\begin{eqnarray}
    \frac{\partial\Omega_{k1}  }{\partial k_z}\Omega_{k3} - \Omega_{k1} \frac{\partial\Omega_{k3} }{\partial k_z} =  -(n-1) \left(\frac{1}{k_x}+\frac{{n}}{k'_x}\right)k_z^2  -(n-1) (k_x +{n}k_x'),
\end{eqnarray}
\begin{eqnarray}
    \frac{\partial \Omega_{k2}  }{\partial k_z} \Omega_{k3} - \Omega_{k2} \frac{\partial \Omega_{k3} }{\partial k_z} =   - \frac{\partial \Omega_{k3}  }{\partial k_y} \Omega_{k2} = -(n-1)^2 k_y,
\end{eqnarray}
\begin{eqnarray}
    \frac{\partial\Omega_{k1}  }{\partial k_z} \Omega_{k2} - \Omega_{k1} \frac{\partial \Omega_{k2}  }{\partial k_z} =   \frac{\partial\Omega_{k1}  }{\partial k_z} \Omega_{k2} = - \left(\frac{1}{k_x}+\frac{{n}}{k'_x}\right)k_y (n-1)k_z,
\end{eqnarray}

\begin{eqnarray}
    {\Omega_{k1}}^2 +\Omega_{k2}^2 +\Omega_{k3}^2  = (k_x +{n}k_x' )^2 + (n-1)^2 (k_y^2 +k_z^2 ).
\end{eqnarray}
Finally, one obtains the shifts as
\begin{eqnarray} \label{eq:tt-1}
    \Delta y_{\rm t}
    &=&(n-1)\frac{-\sin \theta \sin \phi   \left(\dfrac{1}{k_x}+\dfrac{{n}}{k'_x}\right)k_y k_z +  \sin \theta \cos \phi (n-1) k_z { + }\cos \theta  \left[\left(\dfrac{1}{k_x}+\dfrac{{n}}{k'_x}\right)k_y^2 +(k_x +{n}k_x')\right]}{(k_x +{n}k_x' )^2 + (n-1)^2 (k_y^2 +k_z^2 )},
\end{eqnarray}

\begin{eqnarray} \label{eq:tt-2}
    \Delta z_{\rm t}
    &=&(n-1)\frac{-\sin \theta \sin \phi   \left[\left(\dfrac{1}{k_x}+\dfrac{{n}}{k'_x}\right)k_z^2  +(k_x +{n}k_x')\right]- \sin \theta \cos \phi (n-1) k_y { + }\cos \theta \left(\dfrac{1}{k_x}+\dfrac{{n}}{k'_x}\right)k_y k_z}{(k_x +{n}k_x' )^2 + (n-1)^2 (k_y^2 +k_z^2 )}.
\end{eqnarray}
When $k_y=0$, one has
\begin{eqnarray}
    \Delta z_{\rm t}
    &=& (n-1)\frac{-\sin \theta \sin \phi   \left[\left(\dfrac{1}{k_x}+\dfrac{{n}}{k'_x}\right)k_z^2  +(k_x +{n}k_x')\right]}{(k_x +{n}k_x' )^2 + (n-1)^2 k_z^2 }\nonumber\\
        &=& (\sin\theta\sin\phi) \, (1-n) \dfrac{ (k_x +n k_x')+k^2_z \left( \dfrac{1}{ k_x}+n\dfrac{1}{k_x'}\right)}{\left(k_x +n k_x'\right)^2+\left[k_z (1-n)\right]^2},
\end{eqnarray}
which naturally reduces to Eq. (\ref{eq:z-1e}), i.e., it is equal to the case where the finite-width incident wave with finite dimensions just along the $z$ direction and $k_y = 0 $.
$\blacksquare$
\end{remark}

\begin{remark}
\textcolor{blue}{The Simple Vector-Form Expression for $\Delta \vec{l}_{\rm r}$}. From above, we have known that
\begin{eqnarray}
    \Delta \vec{l}_{\rm r} &=& \Delta y_{\rm r} \, \vec{e}_y +\Delta z_{\rm r}\,  \vec{e}_z,\nonumber\\
      \Delta y_{\rm r} & = & \frac{ \hbar^2 c^2 k\sin {\theta} \sin \phi \sin (2 \phi_k'') \sin \phi_k' \tan \phi_k' [{(3\mathcal{E} -mc^2)(\mathcal{E} + mc^2 )}+{c^2\hbar^2 }  k^2 \cos (2\phi_k')]}{2(\mathcal{E} + mc^2 )^2 [\mathcal{E}^2 + m^2c^4 -( \mathcal{E}^2 - m^2c^4) \cos (2 \phi_k')]}\notag\\
    &&+\frac{ 2\hbar^4 c^4 k^{3}\sin {\theta} \cos \phi  \sin \phi_k' \cos^2 \phi_k'\sin \phi_k''}{(\mathcal{E} + mc^2 )^2 [\mathcal{E}^2 + m^2c^4 -( \mathcal{E}^2 - m^2c^4) \cos (2 \phi_k')]}{-}\cos \theta {c^2\hbar^2 }k  \times \notag\\
    &&\frac{  \biggl[4  {\hbar^2 c^2}k^2 \sin^2 (2\phi_k') \cos^2 \phi_k'' - ({\mathcal{E} +mc^2}) [\mathcal{E} +mc^2- (\mathcal{E} -mc^2) \cos (2 \phi_k') ] [1+3\cos (2 \phi_k')- 2\cos (2 \phi_k'') \sin^2 \phi_k' ]    \biggr]}{4 \cos \phi_k'(\mathcal{E} + mc^2 )^2 [\mathcal{E}^2 + m^2c^4 -( \mathcal{E}^2 - m^2c^4) \cos (2 \phi_k')]},\notag\\
        \Delta z_{\rm r}  &=&  -\frac{ \hbar^2 c^2 k \cos \theta \sin (2 \phi_k'') \sin \phi_k' \tan \phi_k' [{(3\mathcal{E} -mc^2)(\mathcal{E} + mc^2 )}+{c^2\hbar^2 }  k^2 \cos (2\phi_k')]}{2(\mathcal{E} + mc^2 )^2 [\mathcal{E}^2 + m^2c^4 -( \mathcal{E}^2 - m^2c^4) \cos (2 \phi_k')]}\notag\\
    &&-\frac{ 2\hbar^4 c^4 k^{3}\sin {\theta} \cos \phi  \sin \phi_k' \cos^2 \phi_k'\cos \phi_k''}{(\mathcal{E} + mc^2 )^2 [\mathcal{E}^2 + m^2c^4 -( \mathcal{E}^2 - m^2c^4) \cos (2 \phi_k')]}+ \sin {\theta} \sin \phi {c^2\hbar^2 }k  \times \notag\\
    &&\frac{   \biggl[4  {\hbar^2 c^2}k^2 \sin^2 (2\phi_k') \sin^2 \phi_k'' - ({\mathcal{E} +mc^2}) [\mathcal{E} +mc^2- (\mathcal{E} -mc^2) \cos (2 \phi_k') ] [1+3\cos (2 \phi_k')+ 2\cos (2 \phi_k'') \sin^2 \phi_k' ]    \biggr]}{4 \cos \phi_k'(\mathcal{E} + mc^2 )^2 [\mathcal{E}^2 + m^2c^4 -( \mathcal{E}^2 - m^2c^4) \cos (2 \phi_k')]}.\notag\\
 \end{eqnarray}
By introducing the following vectors
\begin{eqnarray}
 &&   \vec{e}_x= (1, 0, 0), \nonumber\\
 &&   \vec{\tau} = (\tau_x, \tau_y, \tau_z)=(\sin \theta \cos \phi , \sin \theta \sin \phi , \cos \theta),\nonumber\\
 &&   \vec{k}=(k_x, k_y, k_z)=k \left( \cos \phi_k' , \sin \phi_k' \cos \phi_k'', \sin \phi_k' \sin \phi_k'' \right), \;\;\;
 k =\frac{\sqrt{\mathcal{E}^2 - m^2c^4}}{{c\hbar}},\nonumber\\
 &&   \vec{k}_{\rm r}= (-k_x, k_y, k_z),
\end{eqnarray}
we aim is to rewrite $\Delta \vec{l}_{\rm r}$ in a good-looking vector-form.

We can have
\begin{eqnarray}
      \Delta y_{\rm r} & = & \frac{ \hbar^2 c^2 k\sin {\theta} \sin \phi \sin (2 \phi_k'') \sin \phi_k' \tan \phi_k' [{(3\mathcal{E} -mc^2)(\mathcal{E} + mc^2 )}+{c^2\hbar^2 }  k^2 \cos (2\phi_k')]}{2(\mathcal{E} + mc^2 )^2 [\mathcal{E}^2 + m^2c^4 -( \mathcal{E}^2 - m^2c^4) \cos (2 \phi_k')]}\notag\\
    &&+\frac{ 2\hbar^4 c^4 k^{3}\sin {\theta} \cos \phi  \sin \phi_k' \cos^2 \phi_k'\sin \phi_k''}{(\mathcal{E} + mc^2 )^2 [\mathcal{E}^2 + m^2c^4 -( \mathcal{E}^2 - m^2c^4) \cos (2 \phi_k')]}{-}\cos \theta {c^2\hbar^2 }k  \times \notag\\
    &&\frac{  \biggl[4  {\hbar^2 c^2}k^2 \sin^2 (2\phi_k') \cos^2 \phi_k'' - ({\mathcal{E} +mc^2}) [\mathcal{E} +mc^2- (\mathcal{E} -mc^2) \cos (2 \phi_k') ] [1+3\cos (2 \phi_k')- 2\cos (2 \phi_k'') \sin^2 \phi_k' ]    \biggr]}{4 \cos \phi_k'(\mathcal{E} + mc^2 )^2 [\mathcal{E}^2 + m^2c^4 -( \mathcal{E}^2 - m^2c^4) \cos (2 \phi_k')]}, \nonumber\\
 & = & \frac{\tau_y \hbar^2 c^2 k  \sin (2 \phi_k'') \sin \phi_k' \tan \phi_k'
 2[\mathcal{E} (\mathcal{E} + mc^2 )+{c^2\hbar^2 }  k_x^2]}
 {2(\mathcal{E} + mc^2 )^2 2[\mathcal{E}^2 -{c^2\hbar^2 }  k_x^2]}\notag\\
    &&+\frac{ \tau_x 2\hbar^4 c^4 k^{3}   \sin \phi_k' \cos^2 \phi_k'\sin \phi_k''}
    {(\mathcal{E} + mc^2 )^2 2[\mathcal{E}^2 -{c^2\hbar^2 }  k_x^2]}- \tau_z {c^2\hbar^2 }k  \times \notag\\
    &&\frac{  \biggl[4  {\hbar^2 c^2}k^2 \sin^2 (2\phi_k') \cos^2 \phi_k'' -
    2[\mathcal{E} (\mathcal{E} + mc^2 )-{c^2\hbar^2 }  k_x^2]
    [1+3\cos (2 \phi_k')- 2\cos (2 \phi_k'') \sin^2 \phi_k' ]    \biggr]}
    {4 \cos \phi_k'(\mathcal{E} + mc^2 )^2 2[\mathcal{E}^2 -{c^2\hbar^2 }  k_x^2]}, \nonumber\\
 & = & \frac{\tau_y \hbar^2 c^2 k  \sin (2 \phi_k'') \sin \phi_k' \tan \phi_k'
 [\mathcal{E} (\mathcal{E} + mc^2 )+{c^2\hbar^2 }  k_x^2]}
 {2(\mathcal{E} + mc^2 )^2 [\mathcal{E}^2 -{c^2\hbar^2 }  k_x^2]}+\frac{ \tau_x \hbar^4 c^4 k^{3}   \sin \phi_k' \cos^2 \phi_k'\sin \phi_k''}
    {(\mathcal{E} + mc^2 )^2 [\mathcal{E}^2 -{c^2\hbar^2 }  k_x^2]} \nonumber\\
   && - \tau_z {c^2\hbar^2 }k  \times \frac{  \biggl[2  {\hbar^2 c^2}k^2 \sin^2 (2\phi_k') \cos^2 \phi_k'' -
    [\mathcal{E} (\mathcal{E} + mc^2 )-{c^2\hbar^2 }  k_x^2]
    [1+3\cos (2 \phi_k')- 2\cos (2 \phi_k'') \sin^2 \phi_k' ]    \biggr]}
    {2 \cos \phi_k'(\mathcal{E} + mc^2 )^2 2[\mathcal{E}^2 -{c^2\hbar^2 }  k_x^2]}, \nonumber\\
 & = & \frac{\tau_y \hbar^2 c^2 2 k_y k_z
 [\mathcal{E} (\mathcal{E} + mc^2 )+{c^2\hbar^2 }  k_x^2]}
 { k_x 2(\mathcal{E} + mc^2 )^2 [\mathcal{E}^2 -{c^2\hbar^2 }  k_x^2]}+\frac{ \tau_x \hbar^4 c^4 k_z k_x^2}
    {(\mathcal{E} + mc^2 )^2 [\mathcal{E}^2 -{c^2\hbar^2 }  k_x^2]} \nonumber\\
   && - \tau_z {c^2\hbar^2 }  \times \frac{  \biggl[2  {\hbar^2 c^2} 4 k_x^2 k_y^2 -
    k^2 [\mathcal{E} (\mathcal{E} + mc^2 )-{c^2\hbar^2 }  k_x^2]
    [1+3\cos (2 \phi_k')- 2\cos (2 \phi_k'') \sin^2 \phi_k' ]    \biggr]}
    {2 k_x(\mathcal{E} + mc^2 )^2 2[\mathcal{E}^2 -{c^2\hbar^2 }  k_x^2]}, \nonumber\\
 & = & \frac{\tau_y \hbar^2 c^2 k_y k_z
 [\mathcal{E} (\mathcal{E} + mc^2 )+{c^2\hbar^2 }  k_x^2]}
 { k_x (\mathcal{E} + mc^2 )^2 [\mathcal{E}^2 -{c^2\hbar^2 }  k_x^2]}+\frac{ \tau_x \hbar^4 c^4 k_z k_x^2}
    {(\mathcal{E} + mc^2 )^2 [\mathcal{E}^2 -{c^2\hbar^2 }  k_x^2]} \nonumber\\
   && - \tau_z {c^2\hbar^2 }  \times \frac{  \biggl[ 8  {\hbar^2 c^2}  k_x^2 k_y^2 -
    [\mathcal{E} (\mathcal{E} + mc^2 )-{c^2\hbar^2 }  k_x^2]
    [6k_x^2 -2 k^2 -2(k_y^2-k_z^2) ]    \biggr]}
    {2 k_x(\mathcal{E} + mc^2 )^2 2[\mathcal{E}^2 -{c^2\hbar^2 }  k_x^2]}, \nonumber\\
 & = & \frac{\tau_y \hbar^2 c^2 k_y k_z
 [\mathcal{E} (\mathcal{E} + mc^2 )+{c^2\hbar^2 }  k_x^2]}
 { k_x (\mathcal{E} + mc^2 )^2 [\mathcal{E}^2 -{c^2\hbar^2 }  k_x^2]}+\frac{ \tau_x \hbar^4 c^4 k_z k_x^2}
    {(\mathcal{E} + mc^2 )^2 [\mathcal{E}^2 -{c^2\hbar^2 }  k_x^2]} \nonumber\\
     && -  \frac{  \tau_z {\hbar^2 c^2 } \biggl[   {\hbar^2 c^2}  k_x^2 (k_x^2+ k_y^2) -
    [\mathcal{E} (\mathcal{E} + mc^2 )]
    [k_x^2 -k_y^2 ]    \biggr]}
    { k_x(\mathcal{E} + mc^2 )^2 [\mathcal{E}^2 -{c^2\hbar^2 }  k_x^2]},
\end{eqnarray}
and
\begin{eqnarray}
        \Delta z_{\rm r}  &=&  -\frac{ \hbar^2 c^2 k \cos \theta \sin (2 \phi_k'') \sin \phi_k' \tan \phi_k' [{(3\mathcal{E} -mc^2)(\mathcal{E} + mc^2 )}+{c^2\hbar^2 }  k^2 \cos (2\phi_k')]}{2(\mathcal{E} + mc^2 )^2 [\mathcal{E}^2 + m^2c^4 -( \mathcal{E}^2 - m^2c^4) \cos (2 \phi_k')]}\notag\\
    &&-\frac{ 2\hbar^4 c^4 k^{3}\sin {\theta} \cos \phi  \sin \phi_k' \cos^2 \phi_k'\cos \phi_k''}{(\mathcal{E} + mc^2 )^2 [\mathcal{E}^2 + m^2c^4 -( \mathcal{E}^2 - m^2c^4) \cos (2 \phi_k')]}+ \sin {\theta} \sin \phi {c^2\hbar^2 }k  \times \notag\\
    &&\frac{   \biggl[4  {\hbar^2 c^2}k^2 \sin^2 (2\phi_k') \sin^2 \phi_k'' - ({\mathcal{E} +mc^2}) [\mathcal{E} +mc^2- (\mathcal{E} -mc^2) \cos (2 \phi_k') ] [1+3\cos (2 \phi_k')+ 2\cos (2 \phi_k'') \sin^2 \phi_k' ]    \biggr]}{4 \cos \phi_k'(\mathcal{E} + mc^2 )^2 [\mathcal{E}^2 + m^2c^4 -( \mathcal{E}^2 - m^2c^4) \cos (2 \phi_k')]} \nonumber\\
 & = & -\frac{\tau_z \hbar^2 c^2 k_y k_z
 [\mathcal{E} (\mathcal{E} + mc^2 )+{c^2\hbar^2 }  k_x^2]}
 { k_x (\mathcal{E} + mc^2 )^2 [\mathcal{E}^2 -{c^2\hbar^2 }  k_x^2]}-\frac{ \tau_x \hbar^4 c^4 k_y k_x^2}
    {(\mathcal{E} + mc^2 )^2 [\mathcal{E}^2 -{c^2\hbar^2 }  k_x^2]} + \sin {\theta} \sin \phi {c^2\hbar^2 }k  \times \notag\\
    &&\frac{   \biggl[4  {\hbar^2 c^2}k^2 \sin^2 (2\phi_k') \sin^2 \phi_k'' - ({\mathcal{E} +mc^2}) [\mathcal{E} +mc^2- (\mathcal{E} -mc^2) \cos (2 \phi_k') ] [1+3\cos (2 \phi_k')+ 2\cos (2 \phi_k'') \sin^2 \phi_k' ]    \biggr]}{4 \cos \phi_k'(\mathcal{E} + mc^2 )^2 [\mathcal{E}^2 + m^2c^4 -( \mathcal{E}^2 - m^2c^4) \cos (2 \phi_k')]} \nonumber\\
     & = & -\frac{\tau_z \hbar^2 c^2 k_y k_z
 [\mathcal{E} (\mathcal{E} + mc^2 )+{c^2\hbar^2 }  k_x^2]}
 { k_x (\mathcal{E} + mc^2 )^2 [\mathcal{E}^2 -{c^2\hbar^2 }  k_x^2]}-\frac{ \tau_x \hbar^4 c^4 k_y k_x^2}
    {(\mathcal{E} + mc^2 )^2 [\mathcal{E}^2 -{c^2\hbar^2 }  k_x^2]} + \tau_y {c^2\hbar^2 }k  \times \notag\\
    &&\frac{   \biggl[4  {\hbar^2 c^2}k^2 \sin^2 (2\phi_k') \sin^2 \phi_k'' -  2[\mathcal{E} (\mathcal{E} + mc^2 )-{c^2\hbar^2 }  k_x^2] [1+3\cos (2 \phi_k')+ 2\cos (2 \phi_k'') \sin^2 \phi_k' ]    \biggr]}{4 \cos \phi_k'(\mathcal{E} + mc^2 )^2 [\mathcal{E}^2 + m^2c^4 -( \mathcal{E}^2 - m^2c^4) \cos (2 \phi_k')]} \nonumber\\
     & = & -\frac{\tau_z \hbar^2 c^2 k_y k_z
 [\mathcal{E} (\mathcal{E} + mc^2 )+{c^2\hbar^2 }  k_x^2]}
 { k_x (\mathcal{E} + mc^2 )^2 [\mathcal{E}^2 -{c^2\hbar^2 }  k_x^2]}-\frac{ \tau_x \hbar^4 c^4 k_y k_x^2}
    {(\mathcal{E} + mc^2 )^2 [\mathcal{E}^2 -{c^2\hbar^2 }  k_x^2]}  \notag\\
    && + \tau_y {c^2\hbar^2 }  \times \frac{   \biggl[2  {\hbar^2 c^2}k^4 \sin^2 (2\phi_k') \sin^2 \phi_k'' -  [\mathcal{E} (\mathcal{E} + mc^2 )-{c^2\hbar^2 }  k_x^2] [1+3\cos (2 \phi_k')+ 2\cos (2 \phi_k'') \sin^2 \phi_k' ]    \biggr]}{2 k \cos \phi_k'(\mathcal{E} + mc^2 )^2 2[\mathcal{E}^2 -{c^2\hbar^2 }  k_x^2]} \nonumber\\
     & = & -\frac{\tau_z \hbar^2 c^2 k_y k_z
 [\mathcal{E} (\mathcal{E} + mc^2 )+{c^2\hbar^2 }  k_x^2]}
 { k_x (\mathcal{E} + mc^2 )^2 [\mathcal{E}^2 -{c^2\hbar^2 }  k_x^2]}-\frac{ \tau_x \hbar^4 c^4 k_y k_x^2}
    {(\mathcal{E} + mc^2 )^2 [\mathcal{E}^2 -{c^2\hbar^2 }  k_x^2]}  \notag\\
    && + \tau_y {c^2\hbar^2 }  \times \frac{   \biggl[2  {\hbar^2 c^2}k^4 \sin^2 (2\phi_k') \sin^2 \phi_k'' -  k^2 [\mathcal{E} (\mathcal{E} + mc^2 )-{c^2\hbar^2 }  k_x^2] [1+3\cos (2 \phi_k')+ 2\cos (2 \phi_k'') \sin^2 \phi_k' ]    \biggr]}{4 k_x (\mathcal{E} + mc^2 )^2
    [\mathcal{E}^2 -{c^2\hbar^2 }  k_x^2]} \nonumber\\
     & = & -\frac{\tau_z \hbar^2 c^2 k_y k_z
 [\mathcal{E} (\mathcal{E} + mc^2 )+{c^2\hbar^2 }  k_x^2]}
 { k_x (\mathcal{E} + mc^2 )^2 [\mathcal{E}^2 -{c^2\hbar^2 }  k_x^2]}-\frac{ \tau_x \hbar^4 c^4 k_y k_x^2}
    {(\mathcal{E} + mc^2 )^2 [\mathcal{E}^2 -{c^2\hbar^2 }  k_x^2]}  \notag\\
    && + \tau_y {c^2\hbar^2 }  \times \frac{   \biggl[8  {\hbar^2 c^2} k_x^2 k_z^2 -  [\mathcal{E} (\mathcal{E} + mc^2 )-{c^2\hbar^2 }  k_x^2] [6k_x^2 -2 k^2 + 2(k_y^2-k_z^2)\biggr]}{4 k_x (\mathcal{E} + mc^2 )^2
    [\mathcal{E}^2 -{c^2\hbar^2 }  k_x^2]} \nonumber\\
     & = & -\frac{\tau_z \hbar^2 c^2 k_y k_z
 [\mathcal{E} (\mathcal{E} + mc^2 )+{c^2\hbar^2 }  k_x^2]}
 { k_x (\mathcal{E} + mc^2 )^2 [\mathcal{E}^2 -{c^2\hbar^2 }  k_x^2]}-\frac{ \tau_x \hbar^4 c^4 k_y k_x^2}
    {(\mathcal{E} + mc^2 )^2 [\mathcal{E}^2 -{c^2\hbar^2 }  k_x^2]}  \notag\\
    && + \tau_y {c^2\hbar^2 }  \times \frac{   \biggl[8  {\hbar^2 c^2} k_x^2 k_z^2 -  [\mathcal{E} (\mathcal{E} + mc^2 )-{c^2\hbar^2 }  k_x^2] 4[k_x^2 - k_z^2]\biggr]}{4 k_x (\mathcal{E} + mc^2 )^2
    [\mathcal{E}^2 -{c^2\hbar^2 }  k_x^2]} \nonumber\\
     & = & -\frac{\tau_z \hbar^2 c^2 k_y k_z
 [\mathcal{E} (\mathcal{E} + mc^2 )+{c^2\hbar^2 }  k_x^2]}
 { k_x (\mathcal{E} + mc^2 )^2 [\mathcal{E}^2 -{c^2\hbar^2 }  k_x^2]}-\frac{ \tau_x \hbar^4 c^4 k_y k_x^2}
    {(\mathcal{E} + mc^2 )^2 [\mathcal{E}^2 -{c^2\hbar^2 }  k_x^2]}  \notag\\
    && + \frac{  \tau_y {\hbar^2 c^2 } \biggl[   {\hbar^2 c^2}  k_x^2 (k_x^2+ k_z^2) -
    [\mathcal{E} (\mathcal{E} + mc^2 )]
    [k_x^2 -k_z^2 ]    \biggr]}
    { k_x(\mathcal{E} + mc^2 )^2 [\mathcal{E}^2 -{c^2\hbar^2 }  k_x^2]}.
 \end{eqnarray}
i.e.,
\begin{eqnarray}
      \Delta y_{\rm r}
& = &  \frac{\hbar^2 c^2} { k_x (\mathcal{E} + mc^2 )^2 [\mathcal{E}^2 -{c^2\hbar^2 }  k_x^2]} \times
 \biggr\{{\tau_y k_y k_z
 [\mathcal{E} (\mathcal{E} + mc^2 )+{c^2\hbar^2 }  k_x^2]}
 +{ \tau_x \hbar^2 c^2 k_z k_x^3}
   \nonumber\\
     && -  \tau_z \biggl[   {\hbar^2 c^2}  k_x^2 (k_x^2+ k_y^2) -
    [\mathcal{E} (\mathcal{E} + mc^2 )] [k_x^2 -k_y^2 ] \biggr] \biggr\}\nonumber\\
& = &  \frac{\hbar^2 c^2} { k_x (\mathcal{E} + mc^2 )^2 [\mathcal{E}^2 -{c^2\hbar^2 }  k_x^2]} \times \nonumber\\
&&  \biggr\{ \mathcal{E} (\mathcal{E} + mc^2 ) \left[\tau_y  k_y k_z +\tau_z (k_x^2 -k_y^2)\right]
+ \left[\tau_y  k_y k_z
 {c^2\hbar^2 }  k_x^2+ \tau_x \hbar^2 c^2 k_z k_x^3
 -   \tau_z   {\hbar^2 c^2}  k_x^2 (k_x^2+ k_y^2) \right]\biggr\}
   \nonumber\\
& = &  \frac{\hbar^2 c^2} { k_x (\mathcal{E} + mc^2 )^2 [\mathcal{E}^2 -{c^2\hbar^2 }  k_x^2]} \times   \biggr\{ \mathcal{E} (\mathcal{E} + mc^2 ) \left[\tau_y  k_y k_z +\tau_z (k_x^2 -k_y^2)\right]
+ {\hbar^2 c^2} k_x^2  \left[\tau_y  k_y k_z
  + \tau_x  k_z k_x - \tau_z (k_x^2+ k_y^2) \right]\biggr\}
   \nonumber\\
& = &  \frac{\hbar^2 c^2} { k_x (\mathcal{E} + mc^2 )^2 [\mathcal{E}^2 -{c^2\hbar^2 }  k_x^2]} \times   \biggr\{ \mathcal{E} (\mathcal{E} + mc^2 ) \left[(\tau_y  k_y) k_z + (k_x^2 -k_y^2) \tau_z \right]
+ {\hbar^2 c^2} k_x^2  \left[(\vec{k}\cdot \vec{\tau})k_z - (\vec{k}\cdot\vec{k}) \tau_z \right]\biggr\}
   \nonumber\\
& = &  \frac{\hbar^2 c^2} { k_x (\mathcal{E} + mc^2 )^2 [\mathcal{E}^2 -{c^2\hbar^2 }  k_x^2]} \times   \biggr\{ \mathcal{E} (\mathcal{E} + mc^2 ) \left[(\tau_y  k_y) k_z + (k_x^2 -k_y^2) \tau_z \right]
+ {\hbar^2 c^2} k_x^2  \left[(\vec{k}\cdot \vec{\tau})k_z - (\vec{k}\cdot\vec{k}) \tau_z \right]\biggr\}
   \nonumber\\
& = &  \frac{\hbar^2 c^2} { k_x (\mathcal{E} + mc^2 )^2 [\mathcal{E}^2 -{c^2\hbar^2 }  k_x^2]} \times   \biggr\{ \mathcal{E} (\mathcal{E} + mc^2 ) \left[ (\tau_y  k_z -\tau_z k_y)\tau_y + (k_x^2) \tau_z \right]
+ {\hbar^2 c^2} k_x^2  \left[(\vec{k}\cdot \vec{\tau})k_z - (\vec{k}\cdot\vec{k}) \tau_z \right]\biggr\}
   \nonumber\\
& = &  \frac{\hbar^2 c^2} { k_x (\mathcal{E} + mc^2 )^2 [\mathcal{E}^2 -{c^2\hbar^2 }  k_x^2]} \times  \biggr\{ \mathcal{E} (\mathcal{E} + mc^2 ) [(\vec{\tau}\times \vec{k})\cdot \vec{e}_x ]\tau_y +(k_x^2)\left[\mathcal{E} (\mathcal{E} + mc^2 ) - {c^2\hbar^2 }(\vec{k}\cdot\vec{k})  \right]\tau_z+ {c^2\hbar^2 }  k_x^2 (\vec{k}\cdot \vec{\tau})k_z\biggr\}
   \nonumber\\
& = &  \frac{\hbar^2 c^2} { k_x (\mathcal{E} + mc^2 )^2 [\mathcal{E}^2 -{c^2\hbar^2 }  k_x^2]} \times  \biggr\{ \mathcal{E} (\mathcal{E} + mc^2 ) [(\vec{\tau}\times \vec{k})\cdot \vec{e}_x ]\tau_y +(k_x^2)mc^2 (\mathcal{E} + mc^2 ) \tau_z+ {c^2\hbar^2 }  k_x^2 (\vec{k}\cdot \vec{\tau})k_z\biggr\},
\end{eqnarray}
and
\begin{eqnarray}
\Delta z_{\rm r}
& = &  -\frac{\hbar^2 c^2 }
 { k_x (\mathcal{E} + mc^2 )^2 [\mathcal{E}^2 -{c^2\hbar^2 }  k_x^2]}\times
 \biggr\{{\tau_z  k_y k_z
 [\mathcal{E} (\mathcal{E} + mc^2 )+{c^2\hbar^2 }  k_x^2]}
+{ \tau_x \hbar^2 c^2 k_y k_x^3}
   \notag\\
    && -  \tau_y  \biggl[   {\hbar^2 c^2}  k_x^2 (k_x^2+ k_z^2) -
    [\mathcal{E} (\mathcal{E} + mc^2 )]
    [k_x^2 -k_z^2 ]    \biggr]\biggr\}
   \nonumber\\
&=&  -\frac{\hbar^2 c^2 }
 { k_x (\mathcal{E} + mc^2 )^2 [\mathcal{E}^2 -{c^2\hbar^2 }  k_x^2]}\times \nonumber\\
&&  \biggr\{ \mathcal{E} (\mathcal{E} + mc^2 ) \left[\tau_z  k_y k_z +\tau_y (k_x^2 -k_z^2)\right]+ \left[\tau_z  k_y k_z
 {c^2\hbar^2 }  k_x^2+ \tau_x \hbar^2 c^2 k_y k_x^3
 -   \tau_y   {\hbar^2 c^2}  k_x^2 (k_x^2+ k_z^2) \right]\biggr\}
   \nonumber\\
&=&  -\frac{\hbar^2 c^2 }
 { k_x (\mathcal{E} + mc^2 )^2 [\mathcal{E}^2 -{c^2\hbar^2 }  k_x^2]}\times   \biggr\{ \mathcal{E} (\mathcal{E} + mc^2 ) \left[\tau_z  k_y k_z +\tau_y (k_x^2 -k_z^2)\right]+ {c^2\hbar^2 }  k_x^2 \left[k_y k_z \tau_z
+ k_y k_x \tau_x - (k_x^2+ k_z^2) \tau_y \right]\biggr\}
   \nonumber\\
&=&  -\frac{\hbar^2 c^2 }
 { k_x (\mathcal{E} + mc^2 )^2 [\mathcal{E}^2 -{c^2\hbar^2 }  k_x^2]}\times   \biggr\{ \mathcal{E} (\mathcal{E} + mc^2 ) \left[ (\tau_z   k_z) k_y + (k_x^2 -k_z^2) \tau_y\right]+ {c^2\hbar^2 }  k_x^2 \left[(\vec{k}\cdot \vec{\tau})k_y - (\vec{k}\cdot\vec{k}) \tau_y \right]\biggr\}\nonumber\\
&=&  -\frac{\hbar^2 c^2 }
 { k_x (\mathcal{E} + mc^2 )^2 [\mathcal{E}^2 -{c^2\hbar^2 }  k_x^2]}\times   \biggr\{ \mathcal{E} (\mathcal{E} + mc^2 )
 \left[ -(\tau_y k_z-\tau_z   k_y) k_z + (k_x^2) \tau_y\right]+ {c^2\hbar^2 }  k_x^2 \left[ (\vec{k}\cdot \vec{\tau})k_y - (\vec{k}\cdot\vec{k}) \tau_y \right]\biggr\}\nonumber\\
&=&  -\frac{\hbar^2 c^2 }
 { k_x (\mathcal{E} + mc^2 )^2 [\mathcal{E}^2 -{c^2\hbar^2 }  k_x^2]}\times  \nonumber\\
 && \biggr\{ -\mathcal{E} (\mathcal{E} + mc^2 ) [(\vec{\tau}\times \vec{k})\cdot\vec{e}_x] k_z
 + (k_x^2)\left[\mathcal{E} (\mathcal{E} + mc^2 ) - {c^2\hbar^2 }(\vec{k}\cdot\vec{k})  \right]\tau_y+ {c^2\hbar^2 }  k_x^2 (\vec{k}\cdot \vec{\tau})k_y \biggr\}\nonumber\\
&=&  -\frac{\hbar^2 c^2 }
 { k_x (\mathcal{E} + mc^2 )^2 [\mathcal{E}^2 -{c^2\hbar^2 }  k_x^2]}\times  \biggr\{ -\mathcal{E} (\mathcal{E} + mc^2 ) [(\vec{\tau}\times \vec{k})\cdot\vec{e}_x] k_z
 + (k_x^2) mc^2 (\mathcal{E} + mc^2 ) \tau_y+ {c^2\hbar^2 }  k_x^2 (\vec{k}\cdot \vec{\tau})k_y \biggr\}\nonumber\\
&=&  \frac{\hbar^2 c^2 }
 { k_x (\mathcal{E} + mc^2 )^2 [\mathcal{E}^2 -{c^2\hbar^2 }  k_x^2]}\times  \biggr\{ \mathcal{E} (\mathcal{E} + mc^2 ) [(\vec{\tau}\times \vec{k})\cdot\vec{e}_x] k_z
 - (k_x^2) mc^2 (\mathcal{E} + mc^2 ) \tau_y - {c^2\hbar^2 }  k_x^2 (\vec{k}\cdot \vec{\tau})k_y \biggr\}.
\end{eqnarray}
By using
\begin{eqnarray}
&& \tau_z= (\vec{\tau} \times \vec{e}_x)_y, \;\;\; k_z= (\vec{k} \times \vec{e}_x)_y, \nonumber\\
&& \tau_y= -(\vec{\tau} \times \vec{e}_x)_z, \;\;\; k_y= -(\vec{k} \times \vec{e}_x)_z, \nonumber\\
&& (\vec{\tau} \times \vec{e}_x)_x=0, \;\;\; (\vec{k} \times \vec{e}_x)_x=0, \nonumber\\
 &&   \vec{\eta}= k_y \vec{e}_y+k_z \vec{e}_z=(0, k_y, k_z)=\vec{k}-(\vec{k}\cdot \vec{e}_x) \vec{e}_x,
\end{eqnarray}
then we have the simple vector form as
\begin{eqnarray}
\Delta \vec{l}_{\rm r} &=& \Delta y_{\rm r} \, \vec{e}_y +\Delta z_{\rm r}\,  \vec{e}_z = \frac{\hbar^2 c^2 }
 { k_x (\mathcal{E} + mc^2 )^2 [\mathcal{E}^2 -{c^2\hbar^2 }  k_x^2]}\times\nonumber\\
&&    \biggr\{ \mathcal{E} (\mathcal{E} + mc^2 ) [(\vec{\tau}\times \vec{k})\cdot\vec{e}_x] \vec{\eta}
 + (k_x^2) mc^2 (\mathcal{E} + mc^2 ) (\vec{\tau} \times \vec{e}_x) + {c^2\hbar^2 }  k_x^2 (\vec{k}\cdot \vec{\tau}) (\vec{k} \times \vec{e}_x) \biggr\}.
\end{eqnarray}
One may observe that when the spin direction is reversed, i.e, $\vec{\tau} \mapsto -\vec{\tau}$, the spatial shift $\Delta \vec{l}_{\rm r}$ is also reversed, thus such a shift is a spin effect. $\blacksquare$
\end{remark}

\begin{remark}
\textcolor{blue}{The Simple Vector-Form Expression for $\Delta \vec{l}_{\rm t}$}. From above, we have known that
\begin{eqnarray}
 &&   \Delta \vec{l}_{\rm t} = \Delta y_{\rm t} \, \vec{e}_y +\Delta z_{\rm t}\,  \vec{e}_z,\nonumber\\
 && \Delta y_{\rm t}
    =(n-1)\frac{-\sin \theta \sin \phi   \left(\dfrac{1}{k_x}+\dfrac{{n}}{k'_x}\right)k_y k_z +  \sin \theta \cos \phi (n-1) k_z { + }\cos \theta  \left[\left(\dfrac{1}{k_x}+\dfrac{{n}}{k'_x}\right)k_y^2 +(k_x +{n}k_x')\right]}{(k_x +{n}k_x' )^2 + (n-1)^2 (k_y^2 +k_z^2 )},\nonumber\\
 &&     \Delta z_{\rm t}
    =(n-1)\frac{-\sin \theta \sin \phi   \left[\left(\dfrac{1}{k_x}+\dfrac{{n}}{k'_x}\right)k_z^2  +(k_x +{n}k_x')\right]- \sin \theta \cos \phi (n-1) k_y { + }\cos \theta \left(\dfrac{1}{k_x}+\dfrac{{n}}{k'_x}\right)k_y k_z}{(k_x +{n}k_x' )^2 + (n-1)^2 (k_y^2 +k_z^2 )}.
\end{eqnarray}
By introducing the following vectors
\begin{eqnarray}
 &&   \vec{e}_x= (1, 0, 0), \nonumber\\
 &&   \vec{\tau} = (\tau_x, \tau_y, \tau_z)=(\sin \theta \cos \phi , \sin \theta \sin \phi , \cos \theta), \nonumber\\
 &&   \vec{k}=(k_x, k_y, k_z), \nonumber\\
 &&    \vec{k}_{\rm r}= (-k_x, k_y, k_z), \;\;\; \vec{k}_{\rm t}= (k_x', k_y, k_z),
\end{eqnarray}
we aim is to rewrite $\Delta \vec{l}_{\rm t}$ in a good-looking vector-form.

One can have
\begin{eqnarray}
    \Delta y_{\rm t}
    &=&(n-1)\frac{-\sin \theta \sin \phi   \left(\dfrac{1}{k_x}+\dfrac{{n}}{k'_x}\right)k_y k_z +  \sin \theta \cos \phi (n-1) k_z { + }\cos \theta  \left[\left(\dfrac{1}{k_x}+\dfrac{{n}}{k'_x}\right)k_y^2 +(k_x +{n}k_x')\right]}{(k_x +{n}k_x' )^2 + (n-1)^2 (k_y^2 +k_z^2 )}\nonumber\\
    &=& (n-1)\frac{-\tau_y   \left(\dfrac{1}{k_x}+\dfrac{{n}}{k'_x}\right)k_y k_z +  \tau_x (n-1) k_z  + \tau_z  \left[\left(\dfrac{1}{k_x}+\dfrac{{n}}{k'_x}\right)k_y^2 +(k_x +{n}k_x')\right]}{(k_x +{n}k_x' )^2 + (n-1)^2 (k_y^2 +k_z^2 )}\nonumber\\
    &=& (n-1)\frac{ (\tau_z k_y-\tau_y k_z)   \left(\dfrac{1}{k_x}+\dfrac{{n}}{k'_x}\right) k_y  + \tau_z  (k_x +{n}k_x') +  (n-1) \tau_x k_z}
    {(k_x +{n}k_x' )^2 + (n-1)^2 (k_y^2 +k_z^2 )}\nonumber\\
    &=&(n-1)\frac{ -(\vec{\tau} \times  \vec{k})\cdot \vec{e}_x   \left(\dfrac{1}{k_x}+\dfrac{{n}}{k'_x}\right) k_y  + (\vec{\tau} \times \vec{e}_x)_y  (k_x +{n}k_x') -  (n-1) \tau_x(\vec{e}_x \times \vec{k})_y}
    {(k_x +{n}k_x' )^2 + (n-1)^2 (k_y^2 +k_z^2 )},
\end{eqnarray}

\begin{eqnarray}
    \Delta z_{\rm t}
    &=&(n-1)\frac{-\sin \theta \sin \phi   \left[\left(\dfrac{1}{k_x}+\dfrac{{n}}{k'_x}\right)k_z^2  +(k_x +{n}k_x')\right]- \sin \theta \cos \phi (n-1) k_y { + }\cos \theta \left(\dfrac{1}{k_x}+\dfrac{{n}}{k'_x}\right)k_y k_z}{(k_x +{n}k_x' )^2 + (n-1)^2 (k_y^2 +k_z^2 )}\notag\\
    &=&(n-1)\frac{-\tau_y  \left[\left(\dfrac{1}{k_x}+\dfrac{{n}}{k'_x}\right)k_z^2  +(k_x +{n}k_x')\right]- \tau_x (n-1) k_y { + }\tau_z\left(\dfrac{1}{k_x}+\dfrac{{n}}{k'_x}\right)k_y k_z}{(k_x +{n}k_x' )^2 + (n-1)^2 (k_y^2 +k_z^2 )} \notag\\
    &=&(n-1)\frac{ (\tau_z k_y-\tau_y k_z)   \left(\dfrac{1}{k_x}+\dfrac{{n}}{k'_x}\right) k_z  -\tau_y  (k_x +{n}k_x') -  (n-1) \tau_x k_y}
    {(k_x +{n}k_x' )^2 + (n-1)^2 (k_y^2 +k_z^2 )}\nonumber\\
    &=&(n-1)\frac{ -(\vec{\tau} \times  \vec{k})\cdot \vec{e}_x   \left(\dfrac{1}{k_x}+\dfrac{{n}}{k'_x}\right) k_z  + (\vec{\tau} \times \vec{e}_x)_z  (k_x +{n}k_x') -  (n-1) \tau_x(\vec{e}_x \times \vec{k})_z}
    {(k_x +{n}k_x' )^2 + (n-1)^2 (k_y^2 +k_z^2 )},
\end{eqnarray}
which lead to
\begin{eqnarray}
    \Delta \vec{l}_{\rm t}
    &=& \Delta y_{\rm t} \, \vec{e}_y +\Delta z_{\rm t}\,  \vec{e}_z,\nonumber\\
    &=& (n-1)\frac{ -(\vec{\tau} \times  \vec{k})\cdot \vec{e}_x   \left(\dfrac{1}{k_x}+\dfrac{{n}}{k'_x}\right) (k_y \vec{e}_y+k_z \vec{e}_z)  + (\vec{\tau} \times \vec{e}_x)  (k_x +{n}k_x') -  (n-1)\tau_x (\vec{e}_x \times \vec{k})}
    {(k_x +{n}k_x' )^2 + (n-1)^2 (k_y^2 +k_z^2 )}\notag\\
   & = &(n-1)\frac{ -(\vec{\tau} \times  \vec{k})\cdot \vec{e}_x   \left(\dfrac{1}{k_x}+\dfrac{{n}}{k'_x}\right) (k_y \vec{e}_y+k_z \vec{e}_z)  + (\vec{\tau} \times \vec{e}_x)  (k_x +{n}k_x') -  (n-1)(\vec{\tau} \cdot \vec{e}_x) (\vec{e}_x \times \vec{k})}
    {(k_x +{n}k_x' )^2 + (n-1)^2 (k_y^2 +k_z^2 )}.
\end{eqnarray}
By introducing further
\begin{eqnarray}
 &&   \vec{\eta}= k_y \vec{e}_y+k_z \vec{e}_z=(0, k_y, k_z)=\vec{k}-(\vec{k}\cdot \vec{e}_x) \vec{e}_x, \nonumber\\
 &&    \kappa_1=\dfrac{1}{k_x}+\dfrac{{n}}{k'_x}, \;\;\;\; \kappa_2=k_x +{n}k_x', \;\;\;\; \kappa_3=\dfrac{1}{(k_x +{n}k_x' )^2 + (n-1)^2 (k_y^2 +k_z^2 )},
\end{eqnarray}
we finally have
\begin{eqnarray}
    \Delta \vec{l}_{\rm t}
       & = & - \kappa_3(n-1) \left\{\kappa_1 \left[(\vec{\tau} \times  \vec{k})\cdot \vec{e}_x\right] \vec{\eta} - \kappa_2 (\vec{\tau} \times \vec{e}_x) + (n-1) (\vec{\tau} \cdot \vec{e}_x) (\vec{e}_x \times \vec{k})\right\}.
\end{eqnarray}
Similarly, one may observe that when the spin direction is reversed, i.e, $\vec{\tau} \mapsto -\vec{\tau}$, the spatial shift $\Delta \vec{l}_{\rm t}$ is also reversed, thus such a shift is a spin effect. In addition, the scalar potential (\ref{eq:poten}) corresponds to an electric field $\vec{\mathbb{E}}$ along the $x$-axis, if one denotes $\vec{e}_x=\vec{\mathbb{E}}/|\vec{\mathbb{E}}|$, he may observe that the above spatial shifts are related to the interaction between spin and the electric field. $\blacksquare$
\end{remark}

\newpage

\subsection{The Case in a Continuous Electric Field}

Considering the potential barrier, there are spatial shifts related to the spin of Dirac's particle.
This potential barrier can been regarded that it provide an infinity electric field at an infinitesimal intervals.
This truth inspires us whether our model can cause a similar phenomena, if we consider a uniform electric field distribution.

We consider a scalar potential as
\begin{equation}
    V(x)= q \Phi(x) = \begin{cases}
         -q \mathbb{E}_0 \, x  , & x > 0, \\
         0, & x \leq 0,
     \end{cases}
 \end{equation}
where $q=-e$ (here $e=|e|>0$) is the electric charge of the Dirac's electron, $\Phi(x)=-\mathbb{E}_0 \, x$ is the scalar potential with $\mathbb{E}_0>0$. Accordingly, the electric field is given by
\begin{eqnarray}
 \vec{\mathbb{E}} &=& -\vec{\nabla}\Phi(x)=- \frac{\partial \Phi}{\partial x} \vec{e}_x
 -\frac{\partial \Phi}{\partial y} \vec{e}_y-\frac{\partial \Phi}{\partial z} \vec{e}_z= - \frac{\partial \Phi}{\partial x} \vec{e}_x = \mathbb{E}_0 \, \vec{e}_x,
 \end{eqnarray}
which is a uniform electric field along the positive $x$-axis. By considering $-q=e>0$, $\mathbb{E}_0>0$, thus $V(x)= -q \mathbb{E}_0 \, x $ is greater than 0 when $x>0$.

When $x>0$, Dirac's equation becomes
 \begin{equation}
    \left\{ c (\vec{\alpha}\cdot\vec{p})+\beta\,m\,c^2 + \left[V(x)-\mathcal{E}\right]\,\mathbb{I}\right\}\, \ket{\Psi}=0.
\end{equation}
However, such a eigen-problem is not easy to be solved. Here we try to divide the whole space into an infinite number of infinitesimal intervals along the $x$-axis, whose width is $\d x \rightarrow 0 $. In this case, we can assume a matter plane wave goes through many infinitesimal-width potential barriers and adopt our above approach to attain the spatial shifts when the wave goes through any barrier.

For any barrier, we assume the incident wave has an energy $\mathcal{E}$ and the scalar potential of the transmitted end is $ V= -q \mathbb{E}_0 \, \d x $, which means

\begin{eqnarray}
V= -q \mathbb{E}_0 \, \d x\approx 0,
\end{eqnarray}
and further
\begin{eqnarray}
    n\approx1  \,\,,\, k_x' \approx k_x \,\,,\,\, n-1 = \frac{V}{E-V+mc^2 } \approx \frac{V}{E+mc^2 }\,\,,\,\, (n-1)^2 \approx 0.
\end{eqnarray}
Hence, for the reflected wave and the transmitted wave, we can find
\begin{eqnarray}
 &&A=\frac{\ell_1\left[k_x^2-(k_y^2+k_z^2)(n-1)^2-k_x'^2 n^2\right] +2 k_x (-i k_y \ell_1+k_z \ell_2) (n-1)}
 {k_x^2+(k_y^2+k_z^2)(n-1)^2+2 k_x k_x' n +k_x'^2 n^2} \approx 0, \nonumber\\
 &&B= \frac{\ell_2\left[k_x^2-(k_y^2+k_z^2)(n-1)^2-k_x'^2 n^2\right] -2 k_x (k_z \ell_1-i k_y \ell_2) (n-1)}
 {k_x^2+(k_y^2+k_z^2)(n-1)^2+2 k_x k_x' n +k_x'^2 n^2}\approx 0,
\end{eqnarray}
and
\begin{eqnarray}
 C &=& \dfrac{\sqrt{(\mathcal{E}-V_0)(\mathcal{E} -V_0+mc^2)}}{\sqrt{\mathcal{E}(\mathcal{E} +mc^2)}}\frac{2  n \ell_1 k_x(k_x+{n}k_x') +2n k_x k_z \ell_2(n-1)-2 \i \, n k_x k_y \ell_1 (n-1)}
 {k_x^2+(k_y^2+k_z^2)(n-1)^2+2 k_x k_x' n +k_x'^2 n^2} \nonumber\\
 &=& \dfrac{\sqrt{\mathcal{E}-V_0}}{\sqrt{\mathcal{E}}}\dfrac{1}{\sqrt{n}} \frac{2  n \ell_1 k_x(k_x+{n}k_x') +2n k_x k_z \ell_2(n-1)-2 \i \, n k_x k_y \ell_1 (n-1)}
 {k_x^2+(k_y^2+k_z^2)(n-1)^2+2 k_x k_x' n +k_x'^2 n^2} \nonumber\\
 &\approx& \dfrac{\sqrt{\mathcal{E}-V_0}}{\sqrt{\mathcal{E}}}\dfrac{1}{\sqrt{n}}
 \frac{2  n \ell_1 k_x(k_x+{n}k_x')}
 {k_x^2+2 k_x k_x' n +k_x'^2 n^2} \nonumber\\
 &\approx& \dfrac{\sqrt{\mathcal{E}}}{\sqrt{\mathcal{E}}}\dfrac{1}{\sqrt{n}}
 \frac{2  n \ell_1 k_x(k_x+{n}k_x')}
 {k_x^2+2 k_x k_x' n +k_x'^2 n^2} \approx {\ell_1}, \nonumber\\
  D &= &\dfrac{\sqrt{(\mathcal{E}-V_0)(\mathcal{E} -V_0+mc^2)}}{\sqrt{\mathcal{E}(\mathcal{E} +mc^2)}}\frac{2  n \ell_2k_x (k_x +{n}k_x')+  2 \i\, n k_x k_y \ell_2 (n-1)-2nk_x  k_z \ell_1(n-1)}
 {k_x^2+(k_y^2+k_z^2)(n-1)^2+2 k_x k_x' n +k_x'^2 n^2}\approx {\ell_2},
\end{eqnarray}
which means the reflection coefficient and the transmission coefficient are
\begin{eqnarray}
\mathcal{R}\approx 0, \;\;\;\;\;\; \mathcal{T} \approx 1.
\end{eqnarray}
Because
\begin{eqnarray}
\frac{D}{C}= \frac{\ell_2}{\ell_1},
\end{eqnarray}
thus the spin direction of the transmitted wave does not change at the whole process.

The above fact indicates there is no reflected wave and the transmitted wave will not change the spin direction. So we merely need to study the transmitted wave. From Eq. (\ref{eq:tt-1}) and Eq. (\ref{eq:tt-2}) we have known that
\begin{eqnarray} \label{eq:tt-1a}
    \Delta y_{\rm t}
    &=&(n-1)\frac{-\sin \theta \sin \phi   \left(\dfrac{1}{k_x}+\dfrac{{n}}{k'_x}\right)k_y k_z +  \sin \theta \cos \phi (n-1) k_z {+}\cos \theta  \left[\left(\dfrac{1}{k_x}+\dfrac{{n}}{k'_x}\right)k_y^2 +(k_x +{n}k_x')\right]}{(k_x +{n}k_x' )^2 + (n-1)^2 (k_y^2 +k_z^2 )}, \nonumber\\
    \Delta z_{\rm t}
    &=&(n-1)\frac{-\sin \theta \sin \phi   \left[\left(\dfrac{1}{k_x}+\dfrac{{n}}{k'_x}\right)k_z^2  +(k_x +{n}k_x')\right]- \sin \theta \cos \phi (n-1) k_y {+}\cos \theta \left(\dfrac{1}{k_x}+\dfrac{{n}}{k'_x}\right)k_y k_z}{(k_x +{n}k_x' )^2 + (n-1)^2 (k_y^2 +k_z^2 )}.
\end{eqnarray}
By considering
\begin{eqnarray}
k_x' \approx k_x,
\end{eqnarray}
[note that the momentum $k_x = k_x (x)$ is changed under the action of the electric field], we have
\begin{eqnarray}
    \Delta y_{\rm t}
    &=&(n-1)\frac{-\sin \theta \sin \phi   \dfrac{2k_y k_z}{k_x} +  \sin \theta \cos \phi (n-1) k_z { + }\cos \theta  \left[ \dfrac{2k_y^2}{k_x} +(2k_x)\right]}{(2k_x )^2 + (n-1)^2 (k_y^2 +k_z^2 )}, \nonumber\\
    \Delta z_{\rm t}
    &=&(n-1)\frac{-\sin \theta \sin \phi   \left[  \dfrac{2k_z^2}{ k_x}  +(2k_x)\right]- \sin \theta \cos \phi (n-1) k_y { + }\cos \theta \dfrac{2k_y k_z}{k_x}}{(2k_x)^2 + (n-1)^2 (k_y^2 +k_z^2 )}.
\end{eqnarray}
We keep the order of $(n-1)$ and neglect the order $(n-1)^2$, the above equation becomes
\begin{eqnarray}
    \Delta y_{\rm t}
    &=&(n-1)\frac{-\sin \theta \sin \phi   \dfrac{2k_y k_z}{k_x} +\cos \theta  \left[ \dfrac{2k_y^2}{k_x} +(2k_x)\right]}{(2k_x )^2 }, \nonumber\\
    \Delta z_{\rm t}
    &=&(n-1)\frac{-\sin \theta \sin \phi   \left[  \dfrac{2k_z^2}{k_x}  +(2k_x)\right] +\cos \theta \dfrac{2k_y k_z}{k_x}}{(2k_x)^2 },
\end{eqnarray}
i.e.,
\begin{eqnarray}
    \Delta y_{\rm t}
    &=&(n-1)\frac{-\sin \theta \sin \phi   k_y k_z  +\cos \theta  \left[ k_y^2+  k_x^2 \right]}{2 k_x^3}, \nonumber\\
    \Delta z_{\rm t}
    &=&(n-1)\frac{-\sin \theta \sin \phi   \left[k_z^2 + k_x^2 \right] +\cos \theta k_y k_z }{2 k_x^3}.
\end{eqnarray}
Due to
\begin{eqnarray}
n -1 &=& \frac{\mathcal{E}+mc^2 }{\mathcal{E}-V +mc^2}-1=\frac{V }{\mathcal{E}-V +mc^2}\approx  \frac{V }{\mathcal{E}+mc^2} \approx \frac{-q \mathbb{E}_0 \, \d x }{\mathcal{E}+mc^2},
\end{eqnarray}
we have
\begin{eqnarray}
    \Delta y_{\rm t}
    &=&\frac{q \mathbb{E}_0 \, \d x }{\mathcal{E}+mc^2} \frac{\sin \theta \sin \phi   k_y k_z  -\cos \theta  \left[ k_y^2+  k_x^2 \right]}{2 k_x^3}, \nonumber\\
    \Delta z_{\rm t}
    &=&\frac{q \mathbb{E}_0 \, \d x }{\mathcal{E}+mc^2}\frac{\sin \theta \sin \phi   \left[k_z^2 + k_x^2 \right] -\cos \theta k_y k_z }{2 k_x^3}.
\end{eqnarray}
Due to
\begin{eqnarray}
    \vec{\tau}= (\sin \theta \cos \phi, \sin \theta \sin \phi, \cos \theta),
\end{eqnarray}
then one obtains
\begin{eqnarray}
 &&  \sin \theta \cos \phi=  \tau_x, \;\;\;\;\;\; \sin \theta \sin \phi=  \tau_y, \;\;\;\;\;\; \cos \theta= \tau_z,
\end{eqnarray}
and
\begin{eqnarray}
    \Delta y_{\rm t} \approx {\rm d} y_{\rm t}
    &=& \frac{q \mathbb{E}_0 \, \d x }{\mathcal{E}+mc^2}   \frac{ \tau_y  k_y k_z  - \tau_z \left[ k_y^2+ k_x^2 \right]}{2 k_x^3}, \nonumber\\
    \Delta z_{\rm t} \approx {\rm d} z_{\rm t}
    &=& \frac{q \mathbb{E}_0 \, \d x }{\mathcal{E}+mc^2}  \frac{ \tau_y   \left[k_z^2 + k_x^2 \right] -\tau_z k_y k_z }{2k_x^3}.
\end{eqnarray}
Then we have
\begin{eqnarray}
    \Delta \vec{l}_{\rm t} & \approx &  {\rm d} \vec{l}_{\rm t}  = {\rm d} y_{\rm t} \, \vec{e}_y + {\rm d} z_{\rm t}\,  \vec{e}_z\nonumber\\
    &=&  \frac{q \mathbb{E}_0 \, \d x }{\mathcal{E}+mc^2}   \frac{ \tau_y   k_y k_z  - \tau_z \left[ k_y^2+  k_x^2 \right]}{2k_x^3} \, \vec{e}_y +\frac{q \mathbb{E}_0 \, \d x }{\mathcal{E}+mc^2}  \frac{ \tau_y   \left[k_z^2 + k_x^2 \right] -\tau_z k_y k_z }{2 k_x^3}\,  \vec{e}_z\nonumber\\
     &=& \frac{q \mathbb{E}_0 \, \d x }{2 k_x^3(\mathcal{E}+mc^2)}
     \biggr\{\left[\tau_y  k_y k_z  - \tau_z \left( k_y^2+  k_x^2 \right)\right]\, \vec{e}_y+
     \left[\tau_y   \left[k_z^2 + k_x^2 \right] -\tau_z k_y k_z \right]\,  \vec{e}_z\biggr\},
\end{eqnarray}
i.e.,
\begin{eqnarray}
    \Delta \vec{l}_{\rm t} & \approx &
      \frac{q \mathbb{E}_0 \, \d x }{2 k_x^3(\mathcal{E}+mc^2)}
     \biggr\{\left[\tau_y  k_y k_z  - \tau_z \left( k_y^2+  k_x^2 \right)\right]\, \vec{e}_y+
     \left[\tau_y   \left[k_z^2 + k_x^2 \right] -\tau_z k_y k_z \right]\,  \vec{e}_z\biggr\}\nonumber\\
& = &  \frac{q \mathbb{E}_0 \, \d x }{2 k_x^3(\mathcal{E}+mc^2)}
     \biggr\{\left[(\tau_y  k_y k_z  - \tau_z  k_y^2)- \tau_z k_x^2 \right]\, \vec{e}_y+
     \left[\tau_y k_x^2 +  ( \tau_y k_z^2-\tau_z k_y k_z) \right]\,  \vec{e}_z\biggr\}\nonumber\\
& = &  \frac{q \mathbb{E}_0 \, \d x }{2 k_x^3(\mathcal{E}+mc^2)}
     \biggr\{\left[(\tau_y  k_z  - \tau_z  k_y) k_y - \tau_z k_x^2 \right]\, \vec{e}_y+
     \left[( \tau_y k_z-\tau_z k_y )k_z+\tau_y k_x^2  \right]\,  \vec{e}_z\biggr\}\nonumber\\
& = &  \frac{q \mathbb{E}_0 \, \d x }{2 k_x^3(\mathcal{E}+mc^2)}
     \biggr\{ (\tau_y  k_z  - \tau_z  k_y) (k_y \vec{e}_y + k_z \vec{e}_z) + k_x^2 ( - \tau_z \vec{e}_y+ \tau_y \vec{e}_z)\biggr\}\nonumber\\
& = &  \frac{q \mathbb{E}_0 \, \d x }{2 k_x^3(\mathcal{E}+mc^2)}
     \biggr\{ [(\vec{\tau}\times\vec{k})\cdot\vec{e}_x] (k_y \vec{e}_y + k_z \vec{e}_z) + k_x^2 [ - (\vec{\tau} \times \vec{e}_x)_y \vec{e}_y-(\vec{\tau} \times \vec{e}_x)_z \vec{e}_z]\biggr\}\nonumber\\
& = &  \frac{q \mathbb{E}_0 \, \d x }{2 k_x^3(\mathcal{E}+mc^2)}
     \biggr\{ [(\vec{\tau}\times\vec{k})\cdot\vec{e}_x] \, \vec{\eta} - k_x^2 (\vec{\tau} \times \vec{e}_x)\biggr\},
\end{eqnarray}
i.e.,
\begin{eqnarray}\label{eq:tt-1b}
    \Delta \vec{l}_{\rm t} & \approx &
 \frac{q \, \d x }{2 k_x^3(\mathcal{E}+mc^2)}
     \biggr\{ [(\vec{\tau}\times\vec{k})\cdot\vec{\mathbb{E}}] \, \vec{\eta} - k_x^2 (\vec{\tau} \times \vec{\mathbb{E}})\biggr\},
\end{eqnarray}
where we have used $\vec{\eta}=(k_y \vec{e}_y + k_z \vec{e}_z)=(0, k_y, k_z)=\vec{k}-k_x \,(\vec{\mathbb{E}}/\mathbb{E}_0)$, $\vec{\mathbb{E}}=\mathbb{E}_0\, \vec{e}_x$, and $k_x$ can be viewed as $k_x=\vec{k}\cdot (\vec{\mathbb{E}}/\mathbb{E}_0)$.

From Eq. (\ref{eq:tt-1b}) one may observe that the ETSG effect (or the spatial shift) is related to the interaction between spin and electric field. For example, the term $(\vec{\tau} \times \vec{k}) \cdot \vec{\mathbb{E}}$ represents the \emph{Rashba-like spin-orbit coupling} \cite{2015Manchon}, which can cause the well-known spin Hall effect in an electric field. We shall investigate this interesting topic in the future work.

\begin{remark} Here we come to estimate the magnitude of the spatial shift $\Delta \vec{l}_{\rm t}$. For simplicity, we consider the incident wave is in normal incidence with momentum $\vec{k}=(k,0,0)$ and the spin direction $\vec{\tau}=(0,1,0)$, because
\begin{eqnarray}
 (\vec{\tau}\times\vec{k})\cdot\vec{\mathbb{E}}=0, \;\;\; \vec{\tau} \times \vec{\mathbb{E}}=-\vec{e}_z, \;\;\; k_x=k,
\end{eqnarray}
then from Eq. (\ref{eq:tt-1b}) the shift becomes
\begin{eqnarray}
    \Delta \vec{l}_{\rm t}  \approx \frac{q  k_x^2{\mathbb{E}}_0\, \vec{e}_z\, \d x }{2 k_x^3(\mathcal{E}+mc^2)}= \frac{q  {\mathbb{E}}_0 \, \vec{e}_z \, \d x }{2 k(\mathcal{E}+mc^2)}.
\end{eqnarray}
Because
\begin{eqnarray}
k(x)=\dfrac{1}{c\hbar} \sqrt{{\mathcal{E}}^2 - m^2c^4},
\end{eqnarray}
\begin{eqnarray}
\mathcal{E}= \mathcal{E}_0 -\sum V = \mathcal{E}_0+ \int_{0}^{x} q \mathbb{E}_0 \d x=  \mathcal{E}_0+  q \mathbb{E}_0 x,
\end{eqnarray}
where $\mathcal{E}_0$ is the energy of the wave at $x=0$, then we have
\begin{eqnarray}
    \int_{0}^{x}  \Delta \vec{l}_{\rm t}  = \int_{\mathcal{E}_0}^{\mathcal{E}} \frac{c\hbar \vec{e}_z\,\d \mathcal{E} }{2 (\mathcal{E}+mc^2)\sqrt{{\mathcal{E}}^2 - m^2c^4}}.
\end{eqnarray}
If we designate $\dfrac{\mathcal{E}}{mc^2}=\xi$, we then have
\begin{eqnarray}
   \Delta \vec{\mathcal{L}} \equiv \int_{0}^{x}  \Delta \vec{l}_{\rm t} & = & \int_{\xi_0}^{\xi} \frac{c\hbar\, \vec{e}_z\d \xi }{2 mc^2(1+\xi)\sqrt{{\xi^2-1}}}
    = \vec{e}_z\,  \frac{\hbar }{2mc} \sqrt{\frac{\xi-1}{\xi+1}}\,\Biggr|_{\xi_0}^\xi \nonumber\\
    &=& \vec{e}_z\,  \frac{\hbar }{2mc} \left\{\sqrt{\frac{\mathcal{E}_0 +q\mathbb{E}_0 x -mc^2 }{\mathcal{E}_0 +q\mathbb{E}_0 x +mc^2}}-\sqrt{\frac{\mathcal{E}_0  -mc^2 }{\mathcal{E}_0 +mc^2}}\right\}.
\end{eqnarray}
When $x$ is small, the shift $|\Delta \vec{\mathcal{L}}|\approx 0$. When $x$ is sufficient large, and we select $\mathcal{E}_0 =3 mc^2$, then
\begin{eqnarray}
   |\Delta \vec{\mathcal{L}}|
    &=&  \frac{\hbar }{2mc} \left\{\sqrt{\frac{\mathcal{E}_0 +q\mathbb{E}_0 x -mc^2 }{\mathcal{E}_0 +q\mathbb{E}_0 x +mc^2}}-\sqrt{\frac{\mathcal{E}_0  -mc^2 }{\mathcal{E}_0 +mc^2}}\right\} \approx \frac{\hbar }{2mc}  \left(1- \frac{\sqrt{2}}{2}\right)\nonumber\\
    &=& \frac{1}{4\pi}  \left(1- \frac{\sqrt{2}}{2}\right) \frac{h}{mc} \approx 0.023 \times \frac{h}{mc},
\end{eqnarray}
i.e., the magnitude of the spatial shift is about $0.023$ Compton wavelength.
$\blacksquare$
\end{remark}
